\documentclass[aps,english,preprint,nofootinbib,preprintnumbers]{revtex4}
\usepackage{mathrsfs}
\usepackage{hyperref}
\usepackage{graphicx}
\usepackage{amsmath, amssymb}
\usepackage{babel}
\usepackage{color}
\usepackage{slashed}

\def\beqn{\begin{eqnarray}}
\def\eeqn{\end{eqnarray}}
\def\barr{\begin{array}}
\def\earr{\end{array}}
\def\btab{\begin{tabular}}
\def\etab{\end{tabular}}
\def\bite{\begin{itemize}}
\def\eite{\end{itemize}}
\def\bcen{\begin{center}}
\def\ecen{\end{center}}

\begin{document}

\title{Radiative corrections to semileptonic beta decays: Progress and challenges}

\author{Chien-Yeah Seng$^{1}$}

\affiliation{$^{1}$Helmholtz-Institut f\"{u}r Strahlen- und Kernphysik and Bethe Center for Theoretical Physics,\\
	Universit\"{a}t Bonn, 53115 Bonn, Germany}

\date{\today}

\begin{abstract}

We review some recent progress in the theory of electroweak radiative corrections in semileptonic decay processes. The resurrection of the so-called Sirlin's representation based on current algebra relations permits a clear separation between the perturbatively-calculable and incalculable pieces in the $\mathcal{O}(G_F\alpha)$ radiative corrections. The latter are expressed as compact hadronic matrix elements that allow systematic non-perturbative analysis such as dispersion relation and lattice QCD. This brings substantial improvements to the precision of the electroweak radiative corrections in semileptonic decays of pion, kaon, free neutron and $J^P=0^+$ nuclei that are important theory inputs in precision tests of the Standard Model. Unresolved issues and future prospects are discussed. 
\end{abstract}

\maketitle


\tableofcontents

\section{Introduction}

Beta decays are defined as decay processes $\phi_i\rightarrow\phi_f\ell\nu$ triggered by charged weak interactions, where a strong-interacting particle $\phi_i$ decays into another particle $\phi_f$, accompanied by the emission of a charged lepton $\ell$ and a neutrino $\nu$. They had played a historically important role in the shaping of the Standard Model (SM), which has been extremely successful in the description of all experimentally-observed phenomena involving strong, weak and electromagnetic interactions. Back in 1930, the observed continuous beta decay spectrum led to the neutrino postulation by Pauli~\cite{Brown:1978pb}, and later the establishment of the Fermi theory of beta decay~\cite{Fermi:1934hr}. In 1957, Wu's experiment of the beta decay of $^{60}$Co~\cite{Wu:1957my} provided the first experimental confirmation of the parity nonconservation in the weak interaction postulated by Lee and Yang~\cite{Lee:1956qn}, and the foundation of the vector-minus-axial (V-A) structure in the charged weak interaction~\cite{Feynman:1958ty,Sudarshan:1958vf}. In 1963, Cabibbo proposed a $2\times 2$ unitary matrix to mix different components of the charged weak current in order to explain the observed amplitude of the strangeness-changing weak decays~\cite{Cabibbo:1963yz}. The matrix was later generalized to $3\times 3$~\cite{Kobayashi:1973fv}, which is now known as the Cabibbo-Kobayashi-Maskawa (CKM) matrix, to account for the observed charged-conjugation times parity (CP)-violation in kaon decay~\cite{Christenson:1964fg}. This completed the construction of the charged weak interaction sector in SM.

SM is believed to be incomplete due to its failure to explain certain astrophysical phenomena, such as dark matter~\cite{Aghanim:2018eyx,Simon:2019nxf,Salucci:2018hqu,Allen:2011zs}, dark energy~\cite{Aghanim:2018eyx,Riess:1998cb,Perlmutter:1998np} and the matter-antimatter asymmetry~\cite{Sakharov:1967dj,Aghanim:2018eyx,Mossa:2020gjc}. However, since the discovery of the Higgs boson in year 2012~\cite{Chatrchyan:2012ufa,Aad:2012tfa}, all attempts to search for physics beyond the SM (BSM) in high-energy colliders have so far returned null results~\cite{ATLASPublic,CMSPublic}. Future discovery potential along this direction may rely either on major upgrades of the existing colliders, or the construction of next-generation colliders such as the Future Circular Collider (FCC)~\cite{Abada:2019lih,Abada:2019zxq,Benedikt:2018csr} and the Circular Electron Positron Collider (CEPC)~\cite{CEPCStudyGroup:2018rmc,CEPCStudyGroup:2018ghi}. On the other hand, low-energy precision tests, which aim to detect small deviations between precisely-measured observables and their corresponding SM predictions, become increasingly important in the quest for the BSM search, and beta decays of hadrons and nuclei are perfectly suited for this purpose. As a famous example, the unitarity of the top-row CKM matrix elements ($V_{ud}$, $V_{us}$, $V_{ub}$) is one of the most precisely tested predictions of the SM. With the $V_{ud}$ measured from superallowed $0^+\rightarrow 0^+$ nuclear decays and $V_{us}$ measured from kaon decays (while $V_{ub}$ is negligible), we currently observe~\cite{Zyla:2020zbs}:
\begin{equation}
\Delta_\mathrm{CKM}^u\equiv|V_{ud}|^2+|V_{us}|^2+|V_{ub}|^2-1=-0.0015(3)_{V_{ud}}(4)_{V_{us}},
\end{equation}
which shows a 3 standard deviation ($3\sigma$) tension to the SM prediction of $\Delta_\mathrm{CKM}^u=0$. Furthermore, as there also exists a disagreement between the $V_{us}$ measured from different channels of kaon decay, the actual tension could range from 3 to 5$\sigma$ depending on the choice of $V_{us}$. This makes the top-row CKM unitarity one of the most promising avenues for the discovery of BSM physics~\cite{Bryman:2019ssi,Bryman:2019bjg,Kirk:2020wdk,Grossman:2019bzp,Belfatto:2019swo,Cheung:2020vqm,Jho:2020jsa,Yue:2020wkj,Endo:2020tkb,Capdevila:2020rrl,Eberhardt:2020dat,Crivellin:2020lzu,Coutinho:2019aiy,Gonzalez-Alonso:2018omy,Falkowski:2019xoe,Cirgiliano:2019nyn,Falkowski:2020pma,Becirevic:2020rzi,Crivellin:2021njn,Tan:2019yqp,Crivellin:2021bkd,Crivellin:2020ebi,Crivellin:2020klg,Dekens:2021bro}, in addition to the muon $g-2$~\cite{Fermigm2,Aoyama:2020ynm,Miller:2007kk,Miller:2012opa,Jegerlehner:2009ry} and the B-decay anomalies~\cite{Aaij:2019wad,Aaij:2014ora,Aaij:2015yra,Aaij:2015oid}. Meanwhile, the various correlation coefficients in the beta decay spectrum shape also provide powerful constraints to the Wilson coefficients of possible BSM couplings in the effective field theory (EFT) representation~\cite{Gonzalez-Alonso:2018omy,Falkowski:2019xoe,Cirgiliano:2019nyn,Falkowski:2020pma,Becirevic:2020rzi,Crivellin:2021njn}.

Quantities such as $V_{ud}$ and $V_{us}$ are obtained through a combination of experimental measurements and theory inputs. As future beta decay experiments are generically aiming at a precision level of $10^{-4}$~\cite{Cirgiliano:2019nyn}, the corresponding theory inputs must also reach the same level of accuracy in order to make full use of the experimental results. This represents a real challenge because a large number of SM higher-order corrections enter at this level, and many of them are governed by the Quantum Chromodynamics (QCD) at the hadronic scale, where perturbation theory does not apply. Therefore, a careful combination between appropriate theory frameworks, experiments, nuclear structure calculations as well as lattice QCD is necessary to keep the theory uncertainties of all such corrections under control. In this review we focus on one particularly important correction, namely the electroweak radiative corrections (EWRCs) to semileptonic beta decays.  

Earliest attempts to understand the electromagnetic (EM) RCs to beta decays can be traced back to the 1930s, where the Fermi's function~\cite{Fermi:1934hr} was derived by solving the Dirac equation under a Coulomb potential to describe the Coulombic interaction between the final state nucleus and the electron. Later,
the establishment of quantum electrodynamics (QED) by Feynman, Schwinger and Tomonaga allowed a fully relativistic calculation of the EMRCs order-by-order. Based on this, Ref.\cite{Behrends:1955mb} performed the first comprehensive analysis of the RC in the decay of a fermion, with particular emphasis on the muon decay. In that time, the tree-level decay processes was described by a parity-conserving four-fermion interaction. Parity-violating interactions were included after Wu's experiment in 1957~\cite{Lee:1956qn}, and the corresponding updates of the RC followed~\cite{Kinoshita:1957zz}. After the V-A structure in charged weak interaction was established in 1958, the calculation of its RC also followed~\cite{Kinoshita:1958ru}. Another important breakthrough was achieved by Sirlin in Ref.\cite{Sirlin:1967zza}, where the so-called model-independent terms in the $\mathcal{O}(G_F\alpha)$ RC to the beta decay of an unpolarized neutron were derived, with $G_F$ the Fermi's constant and $\alpha$ the fine-structure constant. Meanwhile, the remaining model-dependent corrections were included as a renormalization to the vector and axial coupling constants.
Following this idea (and also Ref.\cite{Kallen:1967wfa}), Wilkinson and Macefield~\cite{Wilkinson:1970cdv} divided the general $\mathcal{O}(G_F\alpha)$ RC to beta decays of near-degenerate systems (i.e. the parent and daughter particles are nearly degenerate) into ``outer'' and ``inner'' corrections; the former are model-independent and sensitive to the electron energy, while the latter are model-dependent and could be taken as constants once terms of order $\alpha E_e/M$ are neglected, where $E_e$ is the electron energy and $M$ is the mass of the parent/daughter particle.~\footnote{Throughout this review, we use lower-case $m$ for masses of fermions, and upper-case $M$ for masses of bosons and particles with unspecified spin.}

There was no mean to calculate the inner corrections until the establishment of the ultraviolet (UV)-complete theory of Fermi's interaction, namely the electroweak theory based on the SU(2)$_\mathrm{L}\times$U(1)$_\mathrm{B}$ gauge symmetry~\cite{Glashow:1961tr,Weinberg:1967tq,Salam:1968rm}. Within this framework, Sirlin showed~\cite{Sirlin:1974ni} that the full EWRC to the vector coupling constant $g_V$ in beta decay processes is greatly simplified if the Fermi's constant $G_F$ is defined through the muon decay lifetime $\tau_\mu$:
\begin{equation}
\frac{1}{\tau_\mu}\equiv\frac{G_F^2m_\mu^5}{192\pi^3}F(x)(1+\delta_\mu)\label{eq:renormGF}
\end{equation}
where $m_\mu$ is the muon mass, $F(x)$ is a phase-space factor, and $\delta_\mu$ describes the (UV-finite) EMRC to the Fermi interaction.
This idea was brought to a more solid ground in Ref.\cite{Sirlin:1977sv} based on the current algebra formalism, which applies to a generic semileptonic decay process. A great advantage of this formalism is that one is able to express the inner corrections explicitly in terms of hadronic or nuclear matrix elements of electroweak currents. These matrix elements usually depend on physics at all scales. Their asymptotic properties at the scale $q\sim M_W$ ($M_W$ is the mass of the $W$-boson) are usually well-known, because QCD is asymptotically-free~\cite{Gross:1973id,Politzer:1973fx} and perturbatively-calculable at the UV-end. With this, the large electroweak logarithm that is common to all semileptonic decay processes was derived~\cite{Sirlin:1981ie}. However, the contributions from the physics at the intermediate scale, $q\sim 1$~GeV, are governed by non-perturbative QCD where no analytic solutions exist. This becomes the main obstacle of the theory prediction of the inner RCs at the $10^{-4}$ level.

The inner-outer separation of the $\mathcal{O}(G_F\alpha)$ RC does not work in non-degenerate beta decays, because the recoil correction on top of $\alpha/\pi$ is not small and must be treated as a whole. A typical example of this type is the kaon semileptonic decay $K\rightarrow \pi\ell\nu$ (known as $K_{\ell 3}$) which is an important avenue for the measurement of $|V_{us}|$. Due to the large recoil correction, the interest is not only on the RC to the full decay rate, but also its effect to the Dalitz plot spanned by the kinematic variables $y=2p_K\cdot p_\ell/M_K^2$ and $z=2p_K\cdot p_\pi/M_K^2$. 
Earliest studies of the RCs in $K_{\ell 3}$ can be traced back to the works by Ginsberg since the late 1960s~\cite{Ginsberg:1966zz,Ginsberg:1968pz,Ginsberg:1969jh,Ginsberg:1970vy}, where an effective UV cutoff $\Lambda\sim m_N$ ($m_N$ is the nucleon mass) was introduced to regularize the UV-divergences of the loop corrections. Follow-up works
either assumed specific models for the strong and electroweak interactions~\cite{Becherrawy:1970ah,Bytev:2002nx,Andre:2004tk}, or put more emphasis on the so-called ``model-independent'' piece in the
long-distance electromagnetic corrections which is analogous to the outer corrections in neutron and nuclear beta decays~\cite{Garcia:1981it,JuarezLeon:2010tj,Torres:2012ge,Neri:2015eba}. None of the works above were able to provide a rigorous quantification of the theory uncertainties due to the unknown hadron physics at $q\sim\Lambda$. 

Chiral Perturbation Theory (ChPT), as an effective field theory (EFT) of QCD, provided an excellent starting point to study the $K_{\ell 3}$ RC with a controllable error analysis. The most general chiral Lagrangian that involves dynamical photons~\cite{Urech:1994hd} and leptons~\cite{Knecht:1999ag} was written down, from which one-loop calculations were performed. All the dependences on the non-perturbative QCD at the chiral symmetry breaking scale $\Lambda_\chi\sim 4\pi F_\pi$ (where $F_\pi$ is the pion decay constant) are parameterized in terms of a few low energy constants (LECs) in the chiral Lagrangian that are not constrained by the chiral symmetry and must be estimated separately with phenomenological models~\cite{Ananthanarayan:2004qk,DescotesGenon:2005pw}. Within this theory framework, the RC to $\pi_{e3}$~\cite{Cirigliano:2002ng} and $K_{\ell 3}$~\cite{Cirigliano:2001mk,Cirigliano:2004pv,Cirigliano:2008wn} were both calculated to the order $\mathcal{O}(e^2p^2)$, with $e$ the positron charge and $p$ a typical small momentum scale in ChPT. Unknown higher-order chiral corrections and poorly-constrained LECs set a natural limit of $10^{-3}$ to the theory precision within such a framework.

From the descriptions above, it seems that the theory of EWRCs for near-degenerate and non-degenerate beta decays proceed along two completely different paths. It was not realized until very recently that they can be studied under a unified theory framework, namely the Sirlin's current algebra formalism. This allows us to express the LECs in the ChPT again in terms of well-defined hadronic matrix elements of electroweak currents. With that, they can be studied using the latest computational techniques. There are two extremely powerful tools for this purpose: The first is the dispersion relation analysis that relates the hadronic matrix elements to experimental measurables, in particular the various structure functions obtained in deep inelastic scattering (DIS) experiments. The second tool is the lattice QCD, which is currently the only available method to perform first-principles calculations of QCD observables in the non-perturbative regime. Several recent studies demonstrated that, bay appropriately combining the Sirlin's representation of EWRC and the aforementioned computational techniques, one is able to overcome the natural limitations of traditional frameworks, and
bring the theory precision of the RCs in both near-degenerate and non-degenerate systems to a $10^{-4}$ level.

We will report in this review several important progress in the study of the EWRCs in semileptonic decays of pion, kaon, free neutron and spinless, parity-even ($J^P=0^+$) nuclei that took place in the recent years. The structure of the paper is as follows. In Sec.\ref{sec:EWRCgeneric}
we introduce the universal framework for a generic $\mathcal{O}(G_F\alpha)$ EWRC. Within this framework, we make a careful separation of the exactly known terms, the perturbatively calculable terms and the non-perturbative terms; the main challenge to theorists is to calculate the latter to a satisfactory precision level. In Sec.\ref{sec:Sirlin} we introduce the Sirlin's current algebra representation that allows us to express the non-perturbative corrections in terms of well-defined hadronic matrix elements. These two sections set the stage for our later analysis. As a comparison, in Sec.\ref{sec:EFT} we briefly review the EFT representation of the RCs that is widely adopted in existing literature. After that, we proceed to discuss the application of the Sirlin's representation to different hadronic systems, namely pion (Sec.\ref{sec:pion}), generic isopin-half, spin-half ($I=J=1/2$) particles (Sec.\ref{sec:IJhalf}), free neutron (Sec.\ref{sec:freen}), $0^+$ nuclei (Sec.\ref{sec:superallowed}) and kaon (Sec.\ref{sec:Kl3}). In each individual case, we describe how the contributions from the physics at the hadron scale are constraint by lattice QCD, dispersion relation or other non-perturbative methodologies to achieve a precision level of $10^{-4}$. We also identify the major unresolved problems and outline the possible future directions towards their resolution, and close with a loose summary in Sec.\ref{sec:summary}. Some details of the calculations are provided in the Appendices.

\section{\label{sec:EWRCgeneric}EWRC in a generic semileptonic beta decay}

\subsection{Basic ingredients}

\begin{figure}
	\begin{centering}
		\includegraphics[scale=0.14]{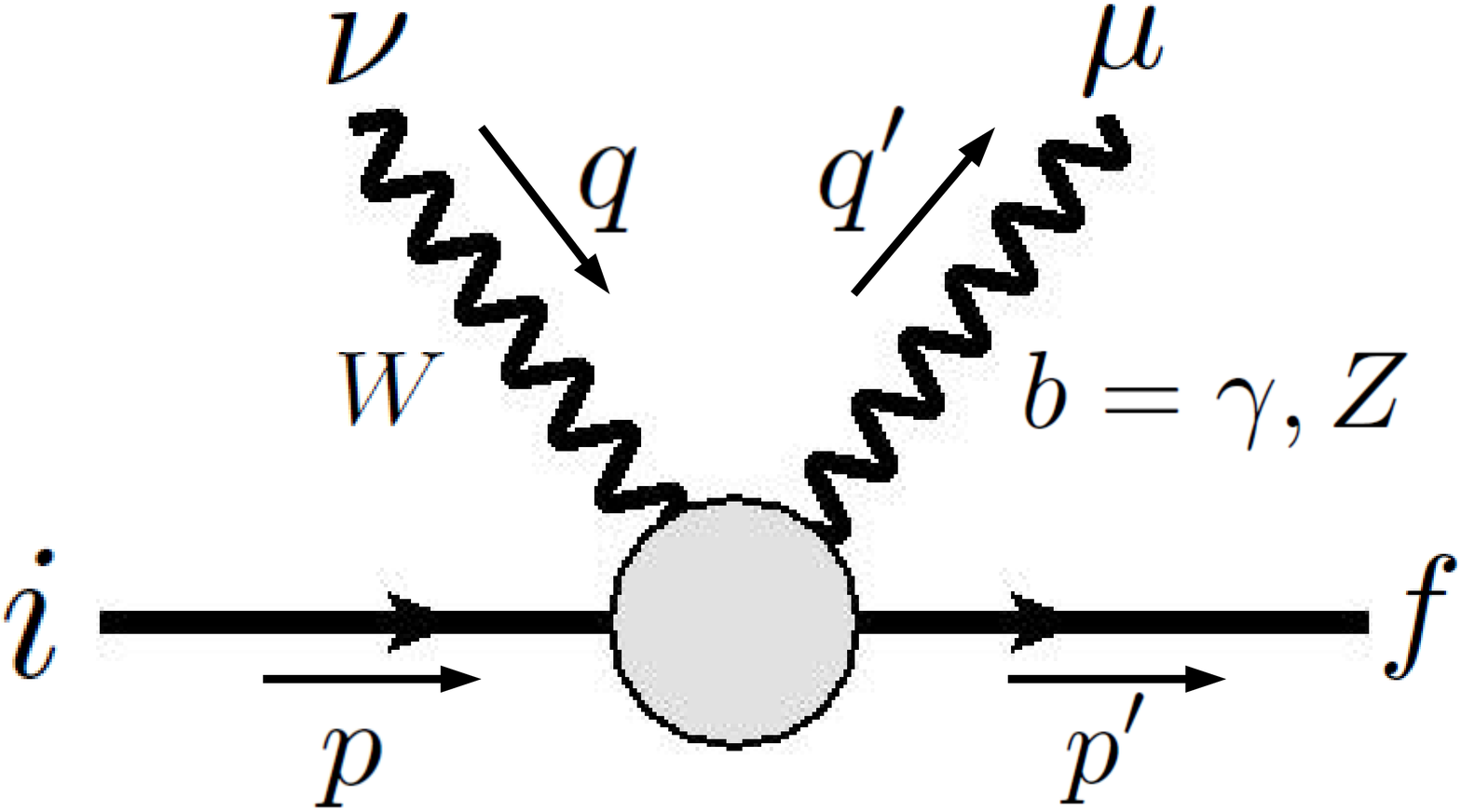}\hfill
		\par\end{centering}
	\caption{\label{fig:Tmunu}Diagrammatic representation of the generalized Compton tensor $T^{\mu\nu}_{(b)}(q';p',p)$.}
\end{figure}

We start by reviewing the part of the electroweak Lagrangian in the SM which is responsible for beta decays and their RCs.
The electroweak currents are defined through the interaction terms between the elementary fermions ($f$) and the electroweak gauge bosons ($b$):
\begin{equation}
\mathcal{L}_{fb}=-\frac{g}{2\sqrt{2}}\left(J_W^\mu W_\mu^++\mathrm{h.c.}\right)-eJ_\gamma^\mu A_\mu-\frac{g}{4c_W}J_Z^\mu Z_\mu~,
\end{equation}
with the usual notations $c_W=\cos\theta_W$, $s_W=\sin\theta_W$ where $\theta_W$ is the weak mixing angle, $g$ is the SU(2)$_\mathrm{L}$ coupling constant, and $W_\mu^\pm$, $Z_\mu$, $A_\mu$ are the $W^\pm$-boson, $Z$-boson and photon field respectively.
In the quark sector, the electroweak currents read:
\begin{eqnarray}
J_\gamma^\mu&=&e_i\bar{q}_i\gamma^\mu q_i\nonumber\\
J_W^\mu&=&\bar{u}_i V_{ij}\gamma^\mu(1-\gamma_5)d_j\nonumber\\
J_Z^\mu&=&\bar{u}_i\gamma^\mu(g_V^u+g_A^u\gamma_5)u_i+\bar{d}_i\gamma^\mu(g_V^d+g_A^d\gamma_5)d_i~,
\end{eqnarray}
where $e_u=2/3$, $e_d=-1/3$, $g_V^u=1-4e_us^2_W$, $g_V^d=-1-4e_ds^2_W$, $g_A^u=-1$, $g_A^d=1$. Here $J_\gamma^\mu$ is just an old notation for the electromagnetic current, and we will use it interchangeably with the more familiar notation $J_\mathrm{em}^\mu$ in the following.

We are interested in the generic beta decay $\phi_i(p)\rightarrow\phi_f(p')\ell(p_\ell)\nu(p_\nu)$. Since we will treat $\beta^+$ and $\beta^-$ decay (i.e. beta decay that emits a positively-charged or negatively-charged lepton) simultaneously, it is convenient to define the lepton spinor and the charged weak current as:
\begin{equation}
J^\mu=\left\{
\begin{array}{c}
J_W^{\mu\dagger}~,~\beta^+\\
J_W^\mu~,~\beta^-
\end{array}\right.\:\:,\:\:
L_\mu=\left\{
\begin{array}{c}
\bar{u}_\nu(p_\nu)\gamma_\mu(1-\gamma_5)v_\ell(p_\ell)~,~\beta^+\\
\bar{u}_\ell(p_\ell)\gamma_\mu(1-\gamma_5)v_\nu(p_\nu)~,~\beta^-
\end{array}\right.
\end{equation}
The electroweak currents satisfy the following equal-time commutation relations:
\begin{eqnarray}
\left[J^0(\vec{x},t),J_\mathrm{em}^\mu(\vec{y},t)\right]&=&\eta J^\mu(\vec{x},t)\delta^{(3)}(\vec{x}-\vec{y})\nonumber\\
\left[J^0(\vec{x},t),J_Z^\mu(\vec{y},t)\right]&=&4c_W^2\eta J^\mu(\vec{x},t)\delta^{(3)}(\vec{x}-\vec{y})~,\label{eq:CA}
\end{eqnarray}
where $\eta=+1(-1)$ for $\beta^+(\beta^-)$ decay.
They are known as the current algebra relations that are exact relations in QCD. 

Next, we define the so-called non-forward generalized Compton tensor $T^{\mu\nu}_{(b)}$ (where $b=\gamma,Z$)  and a related quantity $\Gamma^\mu_{(b)}$ involving a current derivative:
\begin{eqnarray}
T^{\mu\nu}_{(b)}(q';p',p)&=&\int d^4x e^{iq'\cdot x}\bigl\langle\phi_f(p')\bigr|T\left\{J_b^\mu(x)J^\nu(0)\right\}\bigl|\phi_i(p)\bigr\rangle\nonumber\\
\Gamma^{\mu}_{(b)}(q';p',p)&=&\int d^4x e^{iq'\cdot x}\bigl\langle\phi_f(p')\bigr|T\left\{J_b^\mu(x)\partial\cdot J(0)\right\}\bigl|\phi_i(p)\bigr\rangle~.\label{eq:tensors}
\end{eqnarray}
(There is also another vector involving  $\partial\cdot J_b$ that vanishes for $b=\gamma$, so we do not bother to define it separately.)
The physical meaning of $T^{\mu\nu}_{(b)}$ is simply given by Fig.\ref{fig:Tmunu}.
They satisfy the following Ward identities:
\begin{eqnarray}
q_\mu'T^{\mu\nu}_{(b)}(q';p',p)&=&-\eta A_b iF^\nu(p',p)+i\int d^4x e^{iq'\cdot x}\bigl\langle\phi_f(p')\bigr|T\left\{\partial\cdot J_b(x)J^\nu(0)\right\}\bigl|\phi_i(p)\bigr\rangle\nonumber\\
q_\nu T^{\mu\nu}_{(b)}(q';p',p)&=&-\eta A_b iF^\mu(p',p)-i\Gamma^\mu_{(b)}(q';p',p)~,\label{eq:Ward}
\end{eqnarray}
where $q\equiv p'-p+q'$, and $A_b=1(4c_W^2)$ for $b=\gamma (Z)$.
In the above we have defined the charged weak matrix element at the hadron side as:
\begin{equation}
F^\mu(p',p)=\bigl\langle\phi_f(p')\bigr|J^\mu(0)\bigl|\phi_i(p)\bigr\rangle~.
\end{equation}

The following leading-twist, free-field operator product expansion (OPE) relations at large $q^{\prime 2}$ are extremely useful:
\begin{eqnarray}
T_{(\gamma)}^{\mu\nu}(q';p',p)&\rightarrow&\frac{i}{q^{\prime 2}}\left\{\eta\left[g^{\mu\nu}q^{\prime\lambda}-g^{\nu\lambda}q^{\prime\mu}-g^{\mu\lambda}q^{\prime\nu}\right]-2i\bar{Q}\epsilon^{\mu\nu\alpha\lambda}q_\alpha'\right\}F_\lambda(p',p)\nonumber\\
T_{(Z)}^{\mu\nu}(q';p',p)&\rightarrow&\frac{4i}{q^{\prime 2}}\left\{c_W^2\eta\left[g^{\mu\nu}q^{\prime\lambda}-g^{\nu\lambda}q^{\prime\mu}-g^{\mu\lambda}q^{\prime\nu}\right]+2i\bar{Q}s_W^2\epsilon^{\mu\nu\alpha\lambda}q_\alpha'\right\}F_\lambda(p',p)\nonumber\\
\label{eq:OPE}
\end{eqnarray}
where $\bar{Q}=(e_u+e_d)/2$ is the average charge of the SU(2) doublet in the quark sector. It is important that we retain the explicit $e_q$-dependence in the formula above as it distinguishes between the RC in beta decays and muon decay. Our convention for the antisymmetric tensor is $\epsilon^{0123}=-1$. Notice that although we write Eq.\eqref{eq:OPE} in terms of $T^{\mu\nu}_{(b)}$ and $F_\lambda$, they are actually operator relations that are independent of the external states. We provide in Appendix~\ref{sec:OPEderive} a simple derivation of the relations above using the Wick's theorem of free fields.

\subsection{Weak corrections}

\begin{figure}
	\begin{centering}
		\includegraphics[scale=0.14]{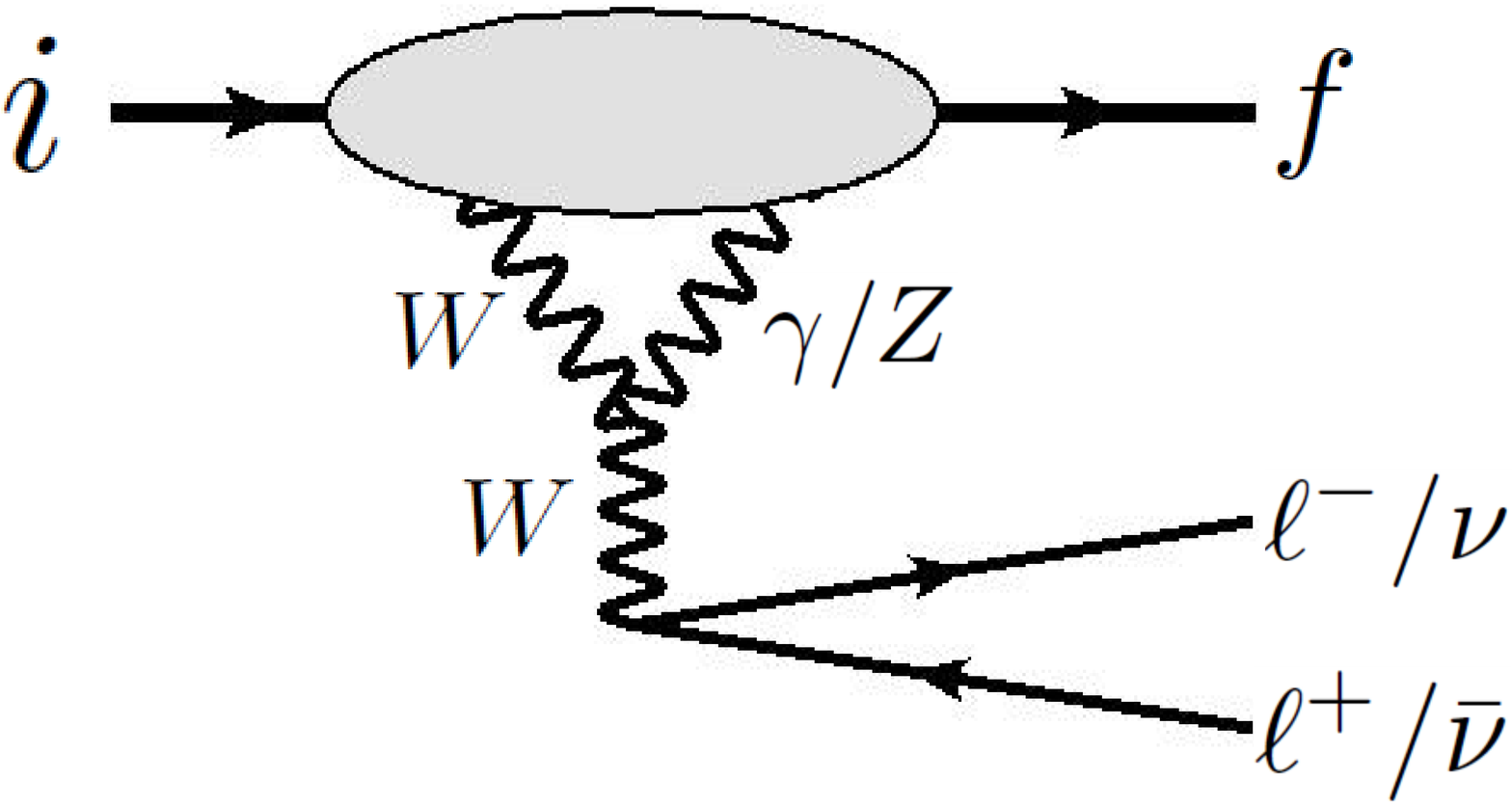}
		\includegraphics[scale=0.14]{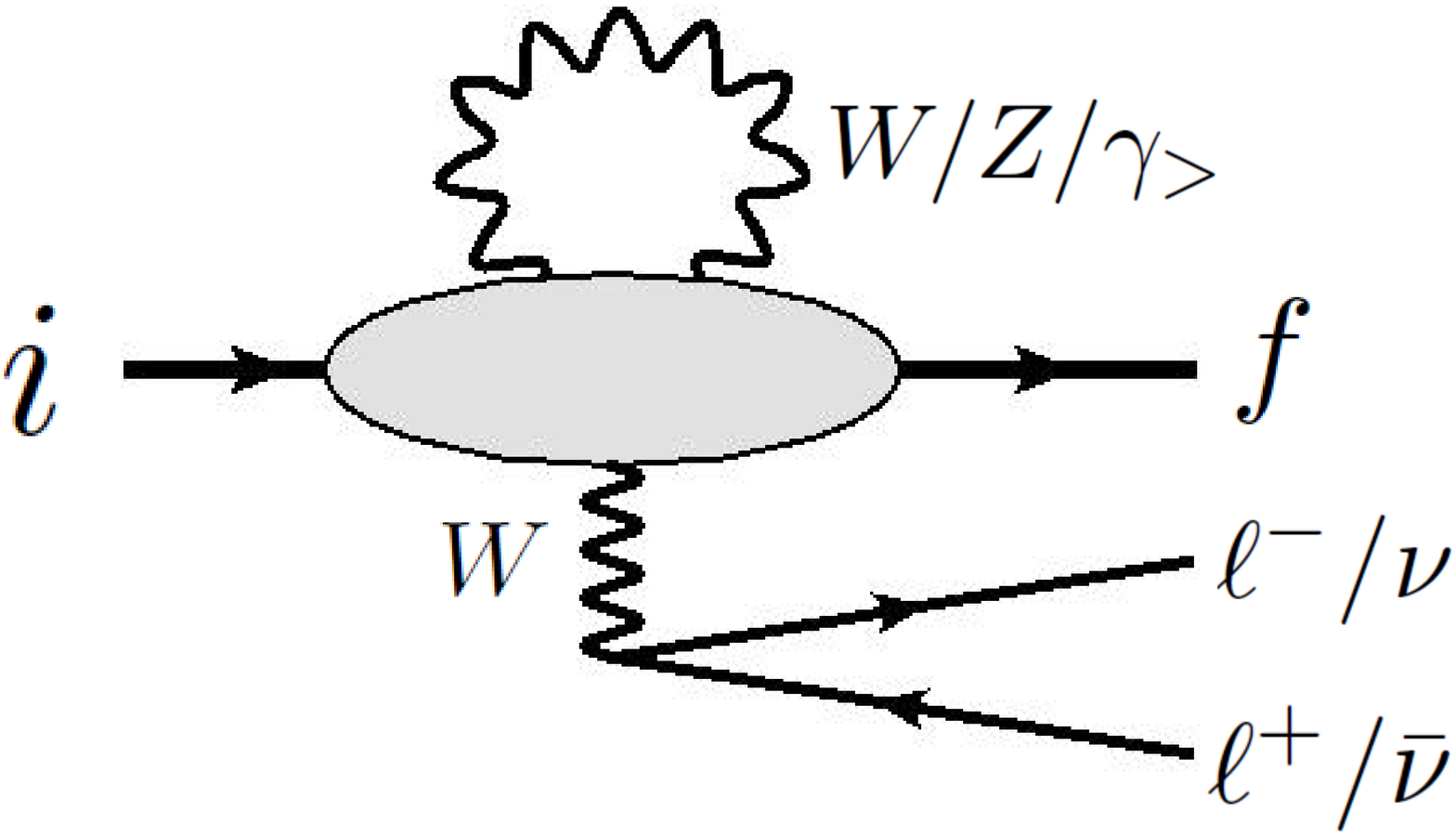}
		\includegraphics[scale=0.14]{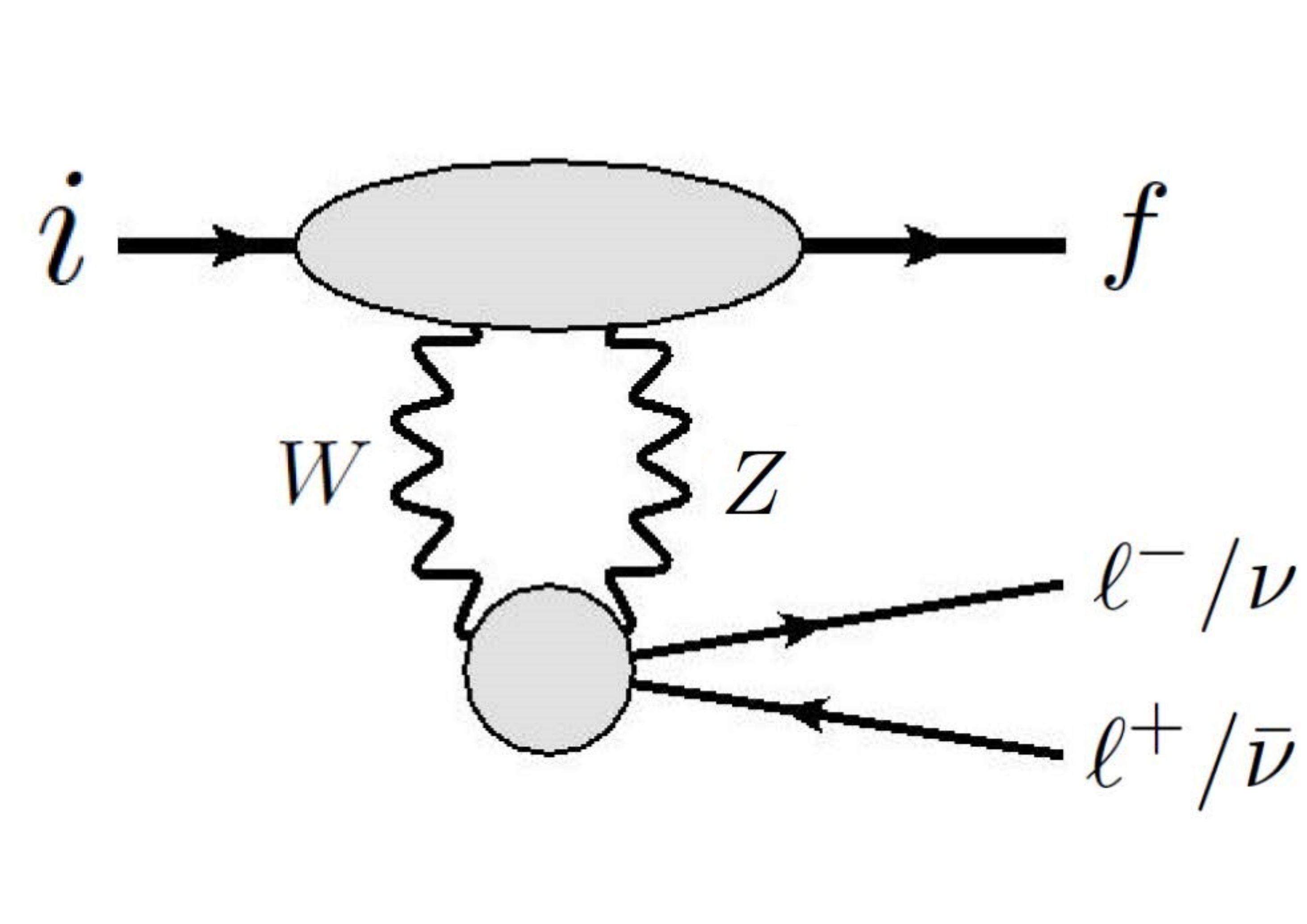}\hfill
		\par\end{centering}
	\caption{\label{fig:weak}The one-loop weak corrections to a generic semileptonic beta decay of hadron.}
\end{figure}

The tree-level beta decay amplitude for the generic beta decay process $\phi_i(p)\rightarrow\phi_f(p')\ell(p_\ell)\nu(p_\nu)$ can be written as:
\begin{equation}
\mathfrak{M}_0=-\frac{G_F^0}{\sqrt{2}}F^\lambda L_\lambda~,
\end{equation}
where $G_F^0=\sqrt{2}g^2/(8M_W^2)$ is the bare Fermi's constant.

We want to now discuss the $\mathcal{O}(G_F\alpha)$ EWRCs to the tree-level amplitude. It is most beneficial to separate them into two classes, namely the ``weak'' corrections and the EMRCs; the former depend only on physics at the scale $q'\sim M_W$, while the latter may probe both the UV- and infrared (IR)-physics. In this subsection we will concentrate on the weak corrections, which consist of the three diagrams in Fig.\ref{fig:weak}. Throughout this paper we adopt the Feynman gauge. 

\subsubsection{First diagram}

The first diagram in Fig.\ref{fig:weak} has at least two heavy propagators, so the only relevant region for the loop integral is $q'\sim M_W$, because otherwise the whole term will scale as $\mathcal{O}(G_F^2)$ which is negligible. Since $M_W\gg p,p'$, we can set $p,p'\rightarrow 0$ in the integrand. That leads to:
\begin{equation}
i\mathfrak{M}_{\mathrm{I}(b)}\approx -\eta\frac{iG_F^0}{\sqrt{2}}g_b^2L_\mu\int\frac{d^4q'}{(2\pi)^4}\frac{1}{q^{\prime 2}-M_W^2}\frac{1}{q^{\prime 2}-M_b^2}\left[q'_\lambda T^{\lambda\mu}_{(b)}+q'_\lambda T^{\mu\lambda}_{(b)}-2q^{\prime\mu}T^\lambda_{(b)\lambda}\right]~.\label{eq:iMb1}
\end{equation}
where $b=\gamma,Z$ is the neutral gauge boson attached to the hadron, and $g_\gamma=e$, $g_Z=g/2$.  

The first two terms in the square bracket can be simplified by the Ward identity~\eqref{eq:Ward}, again neglecting $\mathcal{O}(G_F^2)$ corrections:
\begin{equation}
q'_\lambda T^{\lambda\mu}_{(b)}\approx q'_\lambda T^{\mu\lambda}_{(b)}\approx-\eta A_b iF^\mu(p',p)~.\label{eq:WIresult}
\end{equation}  
Notice that the electroweak currents are either exactly or partially conserved, so we can set their total divergence to zero when $q'\rightarrow \infty$. The last term can be simplified using the free-field OPE~\eqref{eq:OPE}:
\begin{equation}
T^{\lambda}_{(b)\lambda}\rightarrow \eta\frac{2i}{q^{\prime 2}}A_b q'_\lambda F^\lambda(p',p)~.\label{eq:OPEresult1}
\end{equation}
After substituting Eqs.\eqref{eq:WIresult} and \eqref{eq:OPEresult1} into Eq.\eqref{eq:iMb1}, we find:
\begin{equation}
i\mathfrak{M}_{\mathrm{I}(b)}\approx -3g_b^2A_bi\int\frac{d^4q'}{(2\pi)^4}\frac{1}{q^{\prime 2}-M_W^2}\frac{1}{q^{\prime 2}-M_b^2}i\mathfrak{M}_0~,\label{eq:iMb1final}
\end{equation}
i.e. $i\mathfrak{M}_{\mathrm{I}(b)}\propto i\mathfrak{M}_0$ with the (UV-divergent) proportionality constant independent of $\eta$ and $\bar{Q}$. This means the same proportionality constant will appear in the correction to the muon decay by the same diagram. Therefore, its effect is simply reabsorbed into the definition of renormalized Fermi's constant $G_F$ defined in Eq.\eqref{eq:renormGF}. In other words, when we replace $G_F^0\rightarrow G_F$ at the tree-level amplitude, we should at the same time replace $\mathfrak{M}_{\mathrm{I}(b)}$ by $\mathfrak{M}_{\mathrm{I}(b),\mathrm{eff}}$, where:
\begin{equation}
\mathfrak{M}_{\mathrm{I}(b),\mathrm{eff}}\approx 0~.
\end{equation}

Finally, since the derivation of above makes use of the physics at $q'\sim M_W$ where the gauge bosons are essentially probing a free quark, we can equally reproduce Eq.\eqref{eq:iMb1final} by taking $\phi_i,\phi_f$ as free quarks.

\subsubsection{Second diagram} 

The second diagram in Fig.\ref{fig:weak} represents the loop correction to the hadron vertex. The gauge boson in the loop can be $W$, $Z$ or $\gamma$. In particular, we do the separation
\begin{equation}
\frac{1}{q^{\prime 2}}=\frac{1}{q^{\prime 2}-M_W^2}+\frac{M_W^2}{M_W^2-q^{\prime 2}}\frac{1}{q^{\prime 2}}\label{eq:separate}
\end{equation}
to the photon propagator. With this, we can schematically ``split'' the photon into two pieces, $\gamma=\gamma_>+\gamma_<$, with the propagator given by the first and the second term at the right hand side respectively (i.e. $\gamma_>$ represents a photon with mass $M_W$, while $\gamma_<$ is a massless photon with propagator attached to a Pauli-Villars regulator $M_W^2/(M_W^2-q^{\prime 2})$). For the ``weak'' corrections we include only the contributions of $W$, $Z$ and $\gamma_>$, namely the massive gauge bosons. 

Similar to the first one, this diagram also contains two heavy propagators and hence to the order $\mathcal{O}(G_F\alpha)$ it can only probe the physics at $q'\sim M_W$. To derive its amplitude in terms of $T^{\mu\nu}_{(b)}$ requires another piece of theory apparatus known as the on-mass-shell perturbation formula~\cite{Brown:1970dd} which we will introduce only when we discuss the EMRC later. We simply present the final result, and interested readers may refer to Ref.\cite{Sirlin:1977sv} for the derivation:
\begin{eqnarray}
i\mathfrak{M}_{\mathrm{II}}&\approx &-\eta\sqrt{2}iG_F^0e^2L_\mu\int\frac{d^4q'}{(2\pi)^4}\frac{q^{\prime\mu}}{(q^{\prime 2}-M_W^2)^2}T^\lambda_{(\gamma)\lambda}\nonumber\\
&&-\eta\frac{iG_F^0}{4\sqrt{2}}g^2L_\mu\int\frac{d^4q'}{(2\pi)^4}\left[\frac{1}{(q^{\prime 2}-M_W^2)^2}+\frac{1}{(q^{\prime 2}-M_Z^2)^2}\right]q^{\prime\mu}T^\lambda_{(Z)\lambda}~.
\end{eqnarray} 
Substituting the free-field OPE for $T^\lambda_{(b)\lambda}$, we see again that $i\mathfrak{M}_{\mathrm{II}}\propto i\mathfrak{M}_0$ with the proportionality constant independent of $\eta$ and $\bar{Q}$, so it is simply reabsorbed into the definition of $G_F$, i.e.,
\begin{equation}
\mathfrak{M}_{\mathrm{II},\mathrm{eff}}\approx 0~.
\end{equation} 
The same conclusion can also be achieved through a calculation with $\phi_i,\phi_f$ as free quarks.

\subsubsection{Third diagram}

The third piece of weak correction comes from the $WZ$-box diagram in Fig.\ref{fig:weak}, which is again sensitive only to $q'\sim M_W$. Using the following Dirac matrix identity:
\begin{equation}
\gamma_\mu\gamma_\nu\gamma_\alpha=g_{\mu\nu}\gamma_\alpha-g_{\mu\alpha}\gamma_\nu+g_{\nu\alpha}\gamma_\mu-i\epsilon_{\mu\nu\alpha\beta}\gamma^\beta\gamma_5~,\label{eq:3Gamma}
\end{equation}
we obtain, up to $\mathcal{O}(G_F\alpha)$,
\begin{eqnarray}
i\mathfrak{M}_{\mathrm{III}}&\approx &-\frac{ig^4}{128c_W^2}L_\beta\int\frac{d^4q'}{(2\pi)^4}\frac{1}{q^{\prime 2}(q^{\prime 2}-M_W^2)(q^{\prime 2}-M_Z^2)}\Bigl\{-4c_W^2\eta q'_\mu\left(T^{\mu\beta}_{(Z)}+T^{\beta\mu}_{(Z)}\right)\nonumber\\
&&+4c_W^2\eta q^{\prime\beta}T^\lambda_{(Z)\lambda}-4is_W^2\epsilon^{\mu\alpha\nu\beta}q'_\alpha T^{(Z)}_{\mu\nu}\Bigr\}~.
\end{eqnarray} 
The first two terms in the curly bracket can be simplified using the Ward identities modulo $\mathcal{O}(G_F^2)$ corrections. The third and fourth terms can be simplified using the free-field OPE, namely Eq.\eqref{eq:OPEresult1} and
\begin{equation}
\epsilon_{\mu\alpha\nu\beta}q^{\prime\alpha}T^{\mu\nu}_{(Z)}\rightarrow -16\bar{Q}s_W^2\frac{1}{q^{\prime 2}}\left[q^{\prime 2}g_{\beta\lambda}-q'_\beta q'_\lambda\right]F^\lambda~.\label{eq:OPEresult2}
\end{equation}

With all the above, we can now perform the $q'$ integral. Unlike in diagram 1 and 2, here the loop integral is UV-finite, and the outcome reads:
\begin{equation}
\mathfrak{M}_{\mathrm{III}}\approx-\frac{\alpha}{4\pi}\frac{c_W^2}{s_W^2}\ln c_W^2\left\{\frac{5}{2}\frac{c_W^2}{s_W^2}+3\frac{s_W^2}{c_W^2}\bar{Q}\right\}\mathfrak{M}_0~,
\end{equation}
where we have used the tree-level relation $c_W^2=M_W^2/M_Z^2$. We observe that there is a term that proportional to $\bar{Q}$, so only this term distinguishes between beta decays and muon decay. Therefore, upon the renormalization of $G_F$, only the $\bar{Q}$ term in $\mathfrak{M}_{\mathrm{III}}$ is retained, and we have to replace $\bar{Q}$ by $\bar{Q}+1/2$ because -1/2 is the average charge of the SU(2) doublet in the lepton sector. This gives us:
\begin{equation}
\mathfrak{M}_{\mathrm{III},\mathrm{eff}}\approx -\frac{\alpha}{2\pi}\ln\frac{M_W^2}{M_Z^2}\mathfrak{M}_0~,
\end{equation}
where we have substituted $1/6$ for $\bar{Q}$ from now on.

Of course apart from the three diagrams Fig.\ref{fig:weak}, there are also other weak corrections that do not involve the hadron sector. But those corrections are obviously the same in the muon decay and are simply reabsorbed into $G_F$.

\subsubsection{pQCD corrections}

In deriving the expressions above, we have made use of the Ward identities~\eqref{eq:Ward}
and the OPE formula~\eqref{eq:OPE}. The former are exact relations, but the latter is only the leading-twist free-field approximation that works at large $q^{\prime 2}$. In order to achieve a safe $10^{-4}$ precision, we need to add the $\mathcal{O}(\alpha_s)$ perturbative QCD (pQCD) on top of it, with $\alpha_s$ the strong coupling constant. Higher-order pQCD corrections as well as higher-twist corrections are not necessary because the weak corrections are only probing the $q'\sim M_W$ region.

The $\mathcal{O}(\alpha_s)$ correction to the OPE of the full $T^{\mu\nu}_{(b)}$ (i.e. Eq.\eqref{eq:OPE}) was studied in an Abelian theory by Adler and Tung~\cite{Adler:1969ei}, and later generalized to non-Abelian theory by Sirlin~\cite{Sirlin:1977sv}. 
It takes a rather simple form for the combinations $T^\lambda_{(b)\lambda}$ and $\epsilon_{\mu\alpha\nu\beta}T^{\mu\nu}_{(Z)}$: One simply multiplies the right hand side of Eqs.\eqref{eq:OPEresult1}, \eqref{eq:OPEresult2} by a factor $1-\alpha_s/\pi$. With this, one can show that the total $\mathcal{O}(\alpha_s)$ pQCD correction to the amplitudes in Fig.\ref{fig:weak} is given by~\cite{Seng:2019lxf,Seng:2020jtz}:
\begin{equation}
\mathfrak{M}_\mathrm{weak,pQCD}=-\frac{3\alpha}{8\pi}a_\mathrm{pQCD}\mathfrak{M}_0~,
\end{equation}
where
\begin{eqnarray}
a_\mathrm{pQCD}&=&\frac{1}{3}\int dQ^2\Biggl\{\frac{2M_W^2}{(M_W^2+Q^2)^2}-\frac{M_W^2(M_Z^2-M_W^2)Q^2}{(M_W^2+Q^2)^2(M_Z^2+Q^2)^2}\nonumber\\
&&+\frac{M_W^2}{(M_W^2+Q^2)(M_Z^2+Q^2)}\left[\frac{c_W^2}{s_W^2}+\frac{s_W^2}{c_W^2}\right]\Biggr\}\frac{\alpha_s(Q^2)}{\pi}\nonumber\\
&\approx &0.068~.
\end{eqnarray}

\subsection{General theory for beta decays}

\begin{figure}
	\begin{centering}
		\includegraphics[scale=0.14]{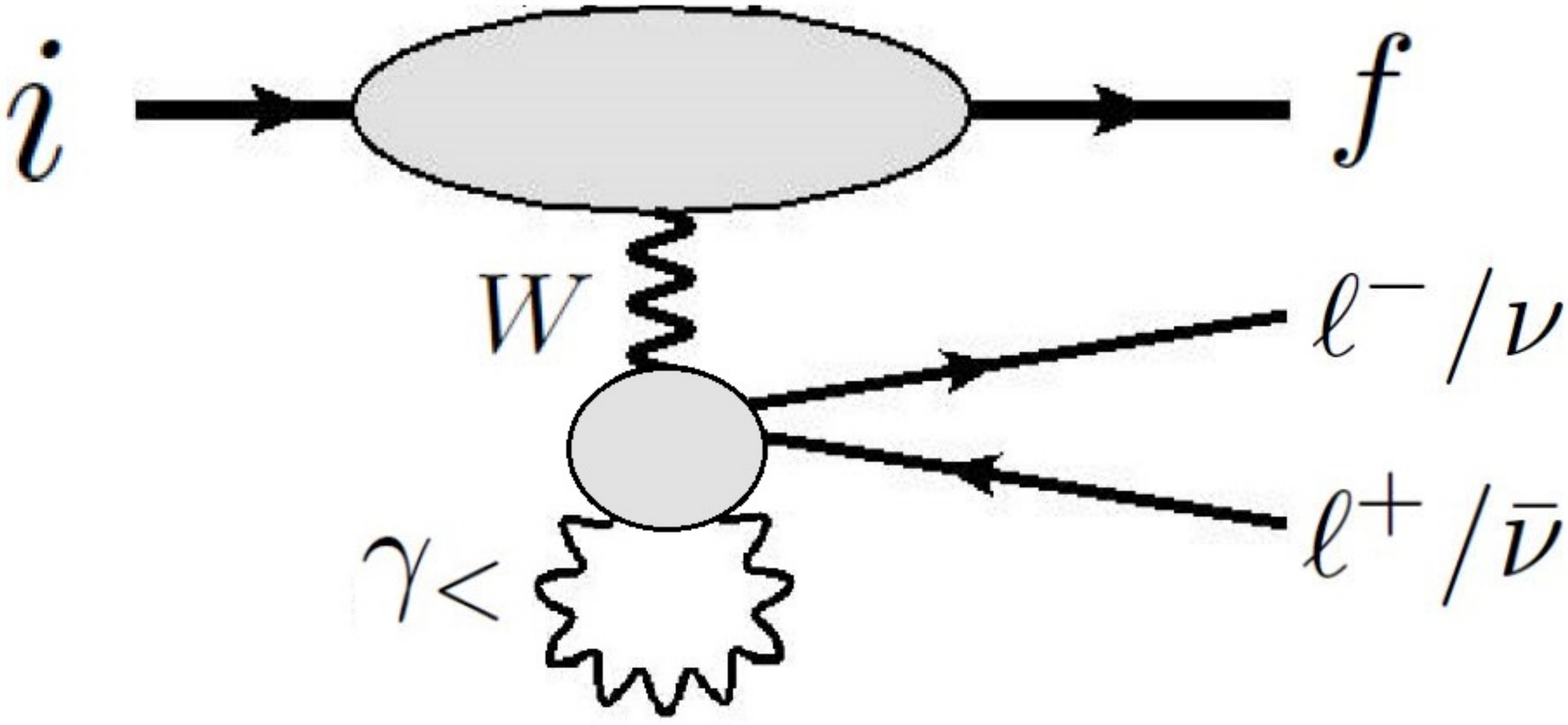}
		\includegraphics[scale=0.14]{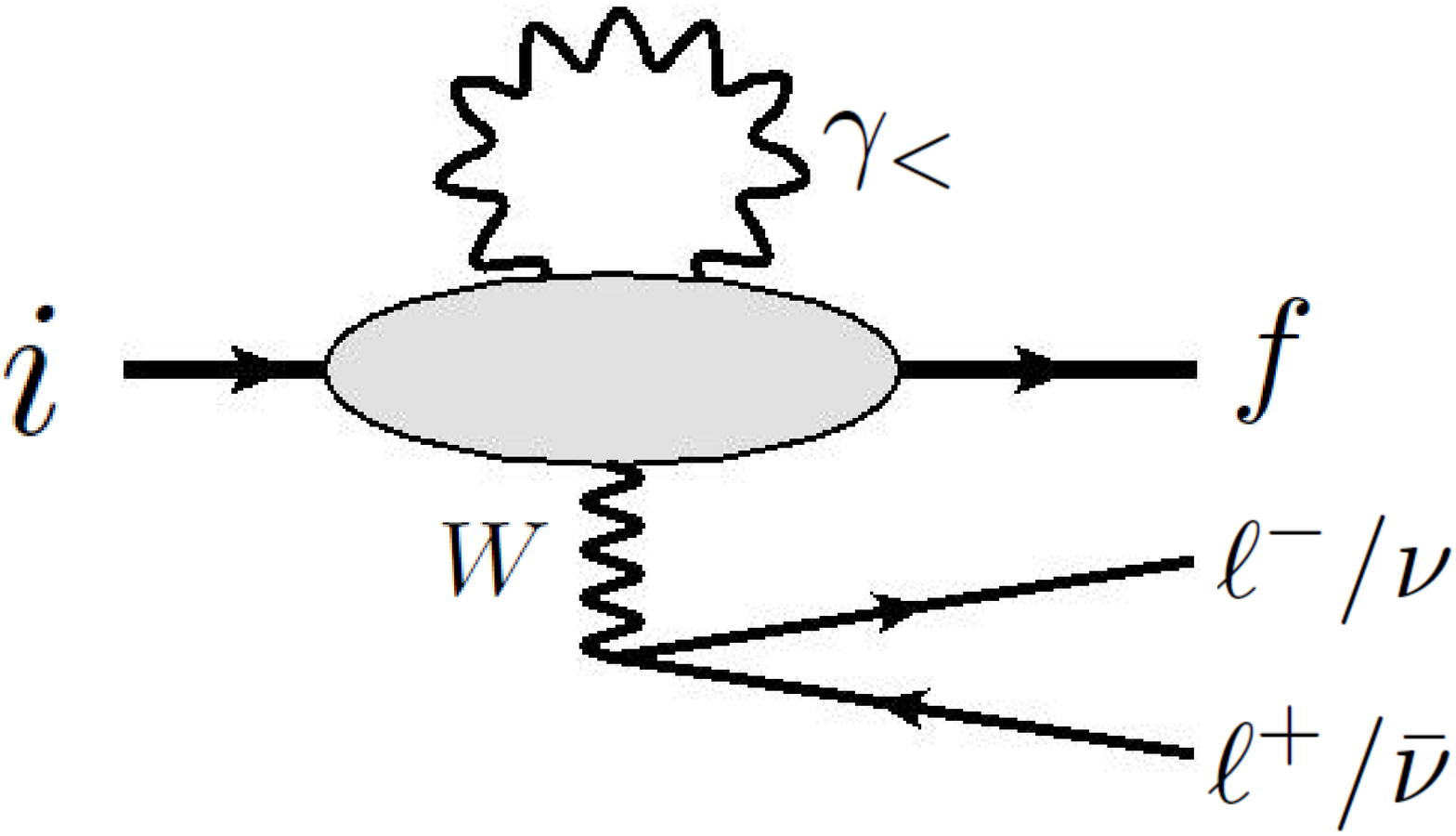}
		\includegraphics[scale=0.14]{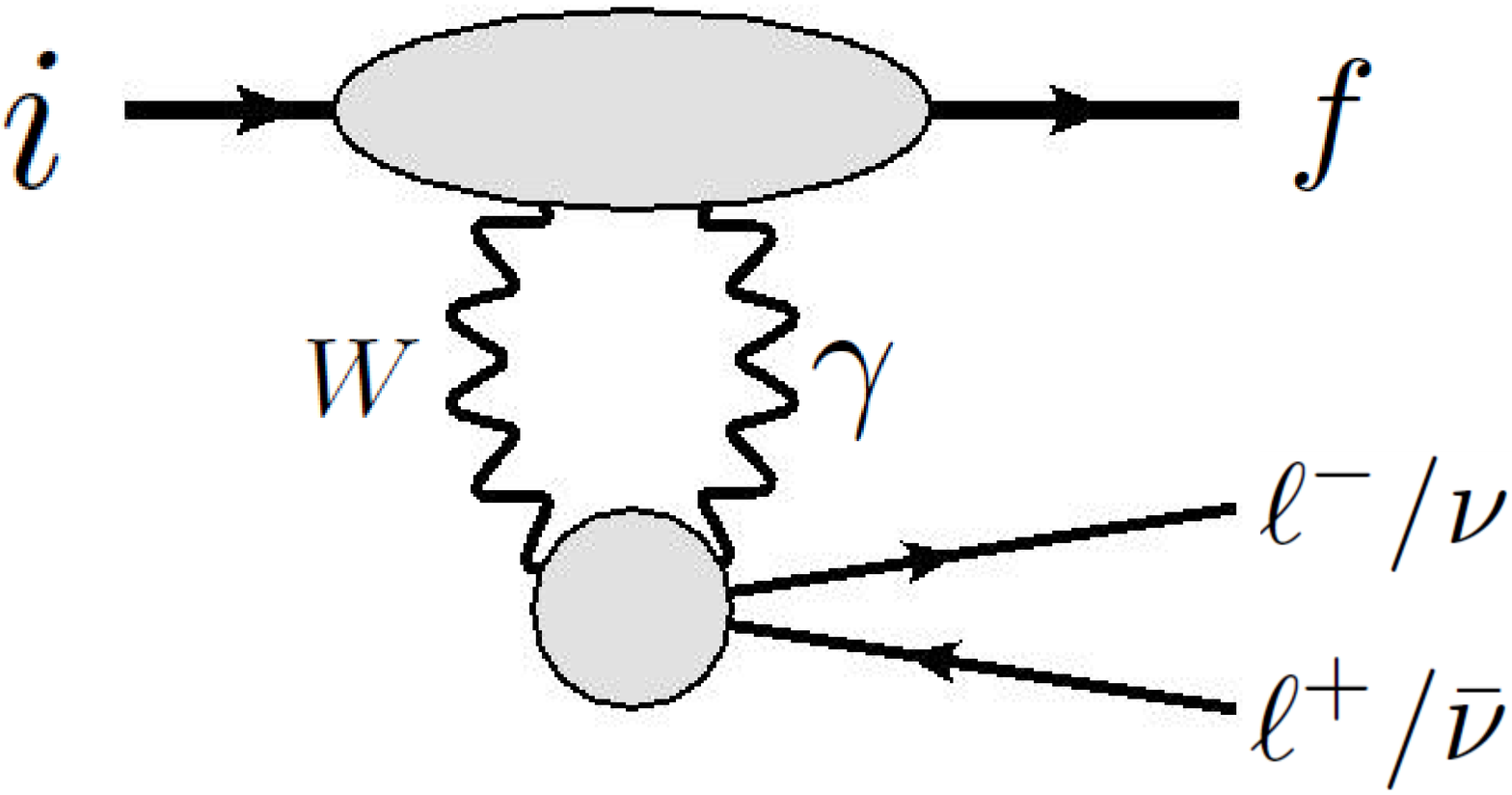}\hfill
		\par\end{centering}
	\caption{\label{fig:EM}One-loop EMRCs to a generic semileptonic beta decay. }
\end{figure}

The remaining $\mathcal{O}(G_F\alpha)$ EWRCs that are not yet analyzed above are summarized in Fig.\ref{fig:EM}. They involve only photonic loops and therefore shall be known as the EMRCs to beta decays.
Among the three diagrams, the first two involve the ``$<$'' photon of which propagator is multiplied by a Pauli-Villars regularization factor $M_W^2/(M_W^2-q^{\prime 2})$, see Eq.\eqref{eq:separate} (notice that although the loop in the first diagram is on the lepton side, its effect clearly cannot be reabsorbed into the definition of $G_F$ because it is IR-divergent). The third term involves the full photon, but at the same time there is also a $W$-propagator in the loop. Upon neglecting the dependence of external momenta in the $W$-propagator which effects are $\mathcal{O}(G_F^2)$, this diagram is equivalent to a $\gamma_<$-loop correction to the Fermi interaction. 

To summarize, the description of a generic semileptonic (or leptonic) beta decay process up to $\mathcal{O}(G_F\alpha)$ may be casted in terms of a renormalized Fermi interaction and its QED loop corrections involving $\gamma_<$, i.e. the full Lagrangian of interest is: 
\begin{equation}
\mathcal{L}=\mathcal{L}_\mathrm{QCD}+\mathcal{L}_{\mathrm{QED},\gamma_<}+\mathcal{L}_{4f}'~.\label{eq:Lfull}
\end{equation}
Here, $\mathcal{L}_\mathrm{QCD}$ is the full QCD Lagrangian while $\mathcal{L}_{\mathrm{QED},\gamma_<}$ is the usual QED Lagrangian except that the photon propagator is always multiplied by a regularization factor $M_W^2/(M_W^2-q^{\prime 2})$. The renormalized Fermi's interaction reads:
\begin{equation}
\mathcal{L}_{4f}'=-\left\{1-\frac{\alpha}{2\pi}\left[\ln\frac{M_W^2}{M_Z^2}+\frac{3}{4}a_\mathrm{pQCD}\right]\right\}\frac{G_F}{\sqrt{2}}J_W^\lambda\bar{\ell}\gamma_\lambda(1-\gamma_5)\nu_\ell+\mathrm{h.c.}~,\label{eq:L4f}
\end{equation}
where $G_F=1.1663787(6)$~GeV$^{-2}$ is the physical Fermi's constant obtained from muon decay which reabsorbs most of the weak RCs. The $\ln(M_W^2/M_Z^2)$ and $a_\mathrm{pQCD}$ terms are the residual corrections originate from the $WZ$-box diagram and the pQCD corrections to the weak RCs, respectively. Notice also that, although we reached Eq.\eqref{eq:Lfull} by using the Feynman gauge, but since $\mathcal{L}_{\mathrm{QED},\gamma_<}$ is gauge-invariant, we conclude that the result is in fact also gauge-invariant.

It is tempting to interpret Eq.\eqref{eq:Lfull} as a low-energy EFT of beta decay, which is not strictly correct. An EFT describes only the physics at low scale, i.e. $q\ll M_W$, but $\mathcal{L}$ still probes the physics all the way up to $q\sim M_W$. All the loop integrals, however, are UV-finite so no counterterms are needed and no scale-dependence is introduced, which are very different from ordinary EFTs. 

From now onwards we are only dealing with EM corrections, so we will suppress the label $(b)$ in $T^{\mu\nu}$ and $\Gamma^\mu$ for simplicity, knowing that it always refers to $b=\gamma$ unless otherwise mentioned.

\section{\label{sec:Sirlin}Sirlin's representation}

We may now start to discuss the EMRCs in Fig.\ref{fig:EM}:
\begin{itemize}
	\item The first diagram is simply the wavefunction renormalization of the charged lepton. Elementary calculation gives:
	\begin{equation}
	Z_\ell=1-\frac{\alpha}{4\pi}\left[\ln\frac{M_W^2}{m_\ell^2}+\frac{9}{2}-2\ln\frac{m_\ell^2}{M_\gamma^2}\right]~,
	\end{equation}
	where a small photon mass $M_\gamma$ was introduced to regularize the IR-divergence.
	\item The second diagram represents the EMRC to the hadronic charged weak matrix element: $F^\lambda\rightarrow F^\lambda+\delta F^\lambda$. We will discuss more about it later.
	\item The third diagram is the famous $\gamma W$-box correction. For future convenience, we split it into two pieces: $\delta \mathfrak{M}_{\gamma W}=\delta \mathfrak{M}_{\gamma W}^a+\delta \mathfrak{M}_{\gamma W}^b$ by applying the Dirac matrix identity~\eqref{eq:3Gamma} to the lepton structure:
	\begin{eqnarray}
	\delta \mathfrak{M}_{\gamma W}^a&=&\eta\frac{G_Fe^2}{\sqrt{2}}L_\lambda\int\frac{d^4q'}{(2\pi)^4}\frac{M_W^2}{M_W^2-q^{\prime 2}}\frac{2g^{\nu\lambda}p_\ell^\mu-g^{\mu\lambda}q^{\prime\nu}-g^{\nu\lambda}q^{\prime\mu}+g^{\mu\nu}q^{\prime\lambda}}{[(p_\ell-q')^2-m_\ell^2][q^{\prime 2}-M_\gamma^2]}T_{\mu\nu}\nonumber\\
	\delta \mathfrak{M}_{\gamma W}^b&=&-i\frac{G_Fe^2}{\sqrt{2}}L_\lambda\int\frac{d^4q'}{(2\pi)^4}\frac{M_W^2}{M_W^2-q^{\prime 2}}\frac{\epsilon^{\mu\nu\alpha\lambda}q_\alpha'}{[(p_\ell-q')^2-m_\ell^2]q^{\prime 2}}T_{\mu\nu}~.\label{eq:MgammaW}
	\end{eqnarray}
	In particular, using the Ward identities~\eqref{eq:Ward}, we are able to isolate a part of $\delta \mathfrak{M}_{\gamma W}^a$ which is proportional to $\mathfrak{M}_0$ and is exactly integrable:
	\begin{eqnarray}
	\delta \mathfrak{M}_{\gamma W}^a&=&\frac{\alpha}{2\pi}\left[\ln\frac{M_W^2}{m_\ell^2}+1\right]\mathfrak{M}_0+\eta\frac{G_Fe^2}{\sqrt{2}}L_\lambda\int\frac{d^4q'}{(2\pi)^4}\frac{M_W^2}{M_W^2-q^{\prime 2}}\nonumber\\
	&&\times\frac{q^{\prime\lambda}T^\mu_{\:\:\mu}+2p_{\ell\mu}T^{\mu\lambda}-(p-p')_\mu T^{\lambda\mu}+i\Gamma^\lambda}{[(p_\ell-q')^2-m_\ell^2][q^{\prime 2}-M_\gamma^2]}~.\label{eq:MgammaWa}
	\end{eqnarray}
	As usual, terms of $\mathcal{O}(G_F^2)$ are discarded.
\end{itemize}

Combining everything above, we obtain following amplitude for the $\phi_i\rightarrow\phi_f\ell\nu$ decay:
\begin{eqnarray}
\mathfrak{M}&=&-\frac{G_F}{\sqrt{2}}F^\lambda L_\lambda\left\{1-\frac{\alpha}{2\pi}\left[\ln\frac{M_W^2}{M_Z^2}+\frac{1}{4}\ln\frac{M_W^2}{m_\ell^2}-\frac{1}{2}\ln\frac{m_\ell^2}{M_\gamma^2}+\frac{9}{8}+\frac{3}{4}a_\mathrm{pQCD}\right]+\frac{1}{2}\delta_\mathrm{HO}^\mathrm{QED}\right\}\nonumber\\
&&-\frac{G_F}{\sqrt{2}}\delta F^\lambda L_\lambda+\delta \mathfrak{M}_{\gamma W}^a+\delta \mathfrak{M}_{\gamma W}^b~.\label{eq:deltaMvir}
\end{eqnarray}
Notice that we have also included an extra correction $\delta_\mathrm{HO}^\mathrm{QED}$ which describes the higher-order (HO) QED effects. It is numerically sizable and must included for a $10^{-4}$ precision. We will discuss its value in a few subsections later.

\subsection{On-mass-shell perturbation formula and Ward identity}

An important result in Ref.\cite{Sirlin:1977sv} is the representation of the hadronic vertex correction $\delta F^\lambda$ in terms of well-defined matrix elements, which we shall describe in this subsection. 

The derivation first made use of the so-called on-mas-shell (OMS) perturbation formula by Brown~\cite{Brown:1970dd}. It states the following: the first-order correction to the matrix element $F^\mu(p',p)= \bigl\langle\phi_f(p')\bigr|J^\mu(0)\bigl|\phi_i(p)\bigr\rangle$ due to the perturbation Lagrangian $\delta\mathcal{L}$ is given by:
\begin{equation}
\delta F^\mu(p',p)=\lim_{\delta\bar{p}\rightarrow\delta p}iT^\mu(\delta\bar{p};p',p)\equiv\lim_{\delta\bar{p}\rightarrow\delta p}\left\{i\bar{T}^\mu(\delta\bar{p};p',p)-iB^\mu(\delta\bar{p};p',p)\right\}~,\label{eq:OMS}
\end{equation}
where $\delta p=p-p'$ (we avoid using $q$ because it is used as the loop momentum), and
\begin{eqnarray}
\bar{T}^\mu(\delta\bar{p};p',p)&=&\int d^4xe^{i\delta\bar{p}\cdot x}\bigl\langle\phi_f(p')\bigr|T\left\{J^\mu(x)\delta\mathcal{L}(0)\right\}\bigl|\phi_i(p)\bigr\rangle\nonumber\\
B^\mu(\delta\bar{p};p',p)&=&-\frac{i\delta M_f^2}{(p-\delta\bar{p})^2-M_f^2}F^\mu(p-\delta\bar{p},p)-F^\mu(p',p'+\delta\bar{p})\frac{i\delta M_i^2}{(p'+\delta\bar{p})^2-M_i^2}~,\nonumber\\
\end{eqnarray}
where $\delta M_{i,f}^2$ are the first-order perturbation to the squared masses $M_{i,f}^2$ from $\delta\mathcal{L}$. Both $\bar{T}^\mu$ and $B^\mu$ are singular at $\delta\bar{p}=\delta p$, but their singularities cancel each other which makes $T^\mu$ well-defined in the $\delta\bar{p}\rightarrow\delta p$ limit. The equations above apply to spinless external particles $\phi_{i,f}$. For spin 1/2 systems assuming CP-invariance~\cite{Sirlin:1977sv}, the results are exactly the same except that now $B^\mu$ reads:
\begin{equation}
B^\mu(\delta\bar{p};p',p)=-\bar{u}_f(p')\left[\frac{i\delta M_f}{\slashed{p}-\delta\bar{\slashed{p}}-M_f}\mathfrak{T}^\mu(\delta\bar{p})+\mathfrak{T}^\mu(\delta\bar{p})\frac{i\delta M_i}{\slashed{p}'+\delta\bar{\slashed{p}}-M_i}\right]u_i(p)~,
\end{equation}
where the vertex matrix $\mathfrak{T}^\mu$ is defined through:
\begin{equation}
F^\mu(p',p)=\bar{u}_f(p')\mathfrak{T}^\mu(\delta p)u_i(p)~.
\end{equation}

In the EMRC, we are dealing with a non-local perturbation Lagrangian:
\begin{equation}
\delta\mathcal{L}(0)=\frac{e^2}{2}\int\frac{d^4q'}{(2\pi)^4}\int d^4xe^{iq'\cdot x}\frac{M_W^2}{M_W^2-q^{\prime 2}}\frac{1}{q^{\prime 2}-M_\gamma^2}T\left\{J_\mathrm{em}^\lambda(x)J^\mathrm{em}_\lambda(0)\right\}~,
\end{equation}
and subsequently,
\begin{eqnarray}
\bar{T}^\mu(\delta\bar{p};p',p)&=&\frac{e^2}{2}\int\frac{d^4q'}{(2\pi)^4}\frac{M_W^2}{M_W^2-q^{\prime 2}}\frac{1}{q^{\prime 2}-M_\gamma^2}\int d^4xe^{i\delta\bar{p}\cdot x}\int d^4y e^{iq'\cdot y}\nonumber\\
&&\times\bigl\langle \phi_f(p')\bigr|T\left\{J^\mu(x)J_\mathrm{em}^\nu(y)J_\nu^\mathrm{em}(0)\right\}\bigl|\phi_i(p)\bigr\rangle~.\label{eq:Tbarmu3pt}
\end{eqnarray}
To further simplify the expression, Sirlin developed a Ward identity treatment on top of the OMS formula. It starts from the following mathematical identity:
\begin{equation}
iT^\mu(\delta\bar{p};p',p)=-\delta\bar{p}_\nu\frac{\partial}{\partial\delta\bar{p}_\mu}iT^\nu(\delta\bar{p};p',p)+\frac{\partial}{\partial\delta\bar{p}_\mu}\left[\delta\bar{p}_\nu iT^\nu(\delta\bar{p};p',p)\right]~.\label{eq:OMSWard}
\end{equation}
Next, using the current algebra relation \eqref{eq:CA} one may demonstrate that,
\begin{eqnarray}
\partial_{x,\mu}T\{J^\mu (x)J_\mathrm{em}^\lambda(y)J_\lambda^\mathrm{em}(0)\}&=&T\{\partial\cdot J(x)J_\mathrm{em}^\lambda(y)J_\lambda^\mathrm{em}(0)+\delta^4(x-y)\eta J^\lambda(x)J_\lambda^\mathrm{em}(0)\nonumber\\
&&+\delta^4(x)\eta J_\mathrm{em}^\lambda(y)J_\lambda(0)\}~,
\end{eqnarray}
from which the following relation is derived:
\begin{eqnarray}
\delta\bar{p}_\nu i\bar{T}^\nu(\delta\bar{p};p',p)&=&iD(\delta\bar{p};p',p)-\frac{e^2}{2}\eta\int\frac{d^4q'}{(2\pi)^4}\int d^4xe^{iq'\cdot x}\frac{M_W^2}{M_W^2-q^{\prime 2}}\frac{1}{q^{\prime 2}-M_\gamma^2}\nonumber\\
&&\times \bigl\langle\phi_f(p')\bigr|T\{e^{i\delta\bar{p}\cdot x}J^\lambda(x)J_\lambda^\mathrm{em}(0)+J_\mathrm{em}^\lambda(x)J_\lambda(0)\}\bigl|\phi_i(p)\bigr\rangle~,\label{eq:OMS2}
\end{eqnarray}
where 
\begin{eqnarray}
D(\delta\bar{p};p',p)&\equiv& \frac{ie^2}{2}\int\frac{d^4q'}{(2\pi)^4}\frac{M_W^2}{M_W^2-q^{\prime 2}}\frac{1}{q^{\prime 2}-M_\gamma^2}\int d^4x e^{i\delta\bar{p}\cdot x}\int d^4ye^{iq'\cdot y}\nonumber\\
&&\times\bigl\langle\phi_f(p')\bigr|T\{\partial\cdot J(x)J_\mathrm{em}^\lambda(y)J_\lambda^\mathrm{em}(0)\}\bigl|\phi_i(p)\bigr\rangle~.\label{eq:D3pt}
\end{eqnarray}
Finally, taking derivatives with respect to $\delta\bar{p}_\mu$ at both sides of Eq.\eqref{eq:OMS2} in the $\delta\bar{p}\rightarrow\delta p$ limit gives:
\begin{equation}
\lim_{\delta\bar{p}\rightarrow\delta p}\frac{\partial}{\partial\delta\bar{p}_\mu}\left(\delta\bar{p}_\nu i\bar{T}^\nu-iD\right)=-\frac{e^2}{2}\eta\int\frac{d^4q'}{(2\pi)^4}\frac{\partial}{\partial q'_\mu}\left(\frac{M_W^2}{M_W^2-q^{\prime 2}}\frac{1}{q^{\prime 2}-M_\gamma^2}\right)T^\lambda_{\:\:\lambda}(q';p',p)~.\label{eq:OMS3}
\end{equation}

Now, combining Eqs.\eqref{eq:OMS}, \eqref{eq:OMSWard} and \eqref{eq:OMS3},
one could write the hadron vertex correction $\delta F^\lambda$ as a sum of two terms:
\begin{equation}
\delta F^\lambda=\delta F_2^\lambda+\delta F_3^\lambda~,
\end{equation}
(and correspondingly, $\delta \mathfrak{M}_{2,3}\equiv -(G_F/\sqrt{2})\delta F^\lambda_{2,3}L_\lambda$), where:
\begin{eqnarray}
\delta F_2^\lambda(p',p)&\equiv&-\frac{e^2}{2}\eta\int\frac{d^4q'}{(2\pi)^4}\frac{\partial}{\partial q'_\lambda}\left(\frac{M_W^2}{M_W^2-q^{\prime 2}}\frac{1}{q^{\prime 2}-M_\gamma^2}\right)T^\mu_{\:\:\mu}(q';p',p)\nonumber\\
\delta F_3^\lambda(p',p)&\equiv&-\lim_{\delta\bar{p}\rightarrow\delta p}i\delta\bar{p}_\nu\frac{\partial}{\partial \delta\bar{p}_\lambda}\left[\bar{T}^\nu(\delta\bar{p};p',p)-B^\nu(\delta\bar{p};p',p)\right]\nonumber\\
&&+\lim_{\delta\bar{p}\rightarrow\delta p}i\frac{\partial}{\partial\delta\bar{p}_\lambda}\left[D(\delta\bar{p};p',p)-\delta\bar{p}\cdot B(\delta\bar{p};p',p)\right]~.\label{eq:F2F3}
\end{eqnarray}
We find that $\delta F_2^\lambda$ depends only on a two-current matrix element, so it should be named as a ``two-point function''. Meanwhile, $\delta F_3^\lambda$ has a further dependence on three-current matrix elements and should be named as a ``three-point function''. We also observe that, the two terms in $\delta F_3^\lambda$ (represented by the two limits) possess the following feature respectively: (1) the first term must vanish in the forward limit (i.e. $\delta p\rightarrow 0$), and (2) the second term must vanish in the symmetry limit (i.e. $\partial\cdot J\rightarrow 0$). We will see that these properties lead to a great simplification in near-degenerate beta decay processes.

\subsection{Partial cancellation between the hadronic vertex correction and the $\gamma W$-box diagram}

Performing the $q'$-derivative in $\delta F_2^\lambda$ returns two terms:
\begin{equation}
\delta F_2^\lambda(p',p)=-e^2\eta\int\frac{d^4q'}{(2\pi)^4}\left[\frac{M_W^2}{(M_W^2-q^{\prime 2})^2}\frac{q^{\prime\lambda}}{q^{\prime 2}-M_\gamma^2}-\frac{M_W^2}{M_W^2-q^{\prime 2}}\frac{q^{\prime\lambda}}{(q^{\prime 2}-M_\gamma^2)^2}\right]T^\mu_{\:\:\mu}~.
\end{equation}
We observe that the first term at the right hand side possesses one extra heavy propagator, and therefore can only depends on physics at $q'\sim M_W$ for our required level of precision. Therefore, we can simply apply the OPE formula in Eq.\eqref{eq:OPEresult1}, and include the $\mathcal{O}(\alpha_s)$ correction again through a simple factor of $1-\alpha_s/\pi$. On the other hand, the second term probes simultaneously the IR- and UV-physics. However, at $q'\sim M_W$ it cancels with a similar term in $\delta \mathfrak{M}_{\gamma W}^a$ that is also proportional to $T^\mu_{\:\:\mu}$. In fact, upon combining $\delta \mathfrak{M}_2$ and $\delta \mathfrak{M}_{\gamma W}^a$ we obtain:
\begin{equation}
\delta \mathfrak{M}_2+\delta \mathfrak{M}_{\gamma W}^a=\frac{\alpha}{2\pi}\left[\ln\frac{M_W^2}{m_\ell^2}+\frac{3}{4}+\frac{1}{2}\tilde{a}_g^\mathrm{res}\right]\mathfrak{M}_0+\left(\delta \mathfrak{M}_2+\delta \mathfrak{M}_{\gamma W}^a\right)_\mathrm{int}~,
\end{equation}
where
\begin{equation}
\tilde{a}_g^\mathrm{res}=\frac{1}{2}\int dQ^2\frac{M_W^2}{(M_W^2+Q^2)^2}\frac{\alpha_s(Q^2)}{\pi}\approx 0.019
\end{equation}
is the $\mathcal{O}(\alpha_s)$ pQCD correction to the first term in $\delta F_2^\lambda$, and
\begin{eqnarray}
\left(\delta \mathfrak{M}_2+\delta \mathfrak{M}_{\gamma W}^a\right)_\mathrm{int}&=&\eta\frac{G_Fe^2}{\sqrt{2}}L_\lambda\int\frac{d^4q'}{(2\pi)^4}\frac{M_W^2}{M_W^2-q^{\prime 2}}\frac{1}{(p_\ell-q')^2-m_\ell^2}\frac{1}{q^{\prime 2}-M_\gamma^2}\nonumber\\
&&\times\left\{\frac{2p_\ell\cdot q'q^{\prime\lambda}}{q^{\prime 2}-M_\gamma^2}T^\mu_{\:\:\mu}+2p_{\ell\mu}T^{\mu\lambda}-(p-p')_\mu T^{\lambda\mu}+i\Gamma^\lambda\right\}\label{eq:resintegral}
\end{eqnarray}
is the ``residual integral'' term that cannot be simply reduced to something proportional to $\mathfrak{M}_0$. 

Collecting everything we derived so far in this section, we obtain the following representation of the $\phi_i\rightarrow\phi_f\ell\nu$ decay amplitude:
\begin{eqnarray}
\mathfrak{M}&=&-\frac{G_F}{\sqrt{2}}F^\lambda L_\lambda\left\{1+\frac{\alpha}{2\pi}\left[\ln\frac{M_Z^2}{m_\ell^2}-\frac{1}{4}\ln\frac{M_W^2}{m_\ell^2}+\frac{1}{2}\ln\frac{m_\ell^2}{M_\gamma^2}-\frac{3}{8}+\frac{1}{2}\tilde{a}_g\right]+\frac{1}{2}\delta_\mathrm{HO}^\mathrm{QED}\right\}\nonumber\\
&&+\left(\delta \mathfrak{M}_2+\delta \mathfrak{M}_{\gamma W}^a\right)_\mathrm{int}-\frac{G_F}{\sqrt{2}}\delta F_3^\lambda L_\lambda+\delta \mathfrak{M}_{\gamma W}^b~,\label{eq:Sirlinrep}
\end{eqnarray}
where $\tilde{a}_g\equiv-(3/2)a_\mathrm{pQCD}+\tilde{a}_g^\mathrm{res}\approx -0.083$. At $\mathcal{O}(G_F\alpha)$ there are only three quantities remained to be evaluated: $\left(\delta \mathfrak{M}_2+\delta \mathfrak{M}_{\gamma W}^a\right)_\mathrm{int}$ in Eq.\eqref{eq:resintegral}, $\delta F_3^\lambda$ in Eq.\eqref{eq:F2F3}, and $\delta \mathfrak{M}_{\gamma W}^b$ in Eq.\eqref{eq:MgammaW}, which are all well-defined hadronic matrix elements of electroweak currents. We shall name Eq.\eqref{eq:Sirlinrep} as the ``Sirlin's representation'' of the beta decay amplitude, which serves as an important starting point for all the following discussions.

\subsection{Bremsstrahlung}

\begin{figure}
	\begin{centering}
		\includegraphics[scale=0.14]{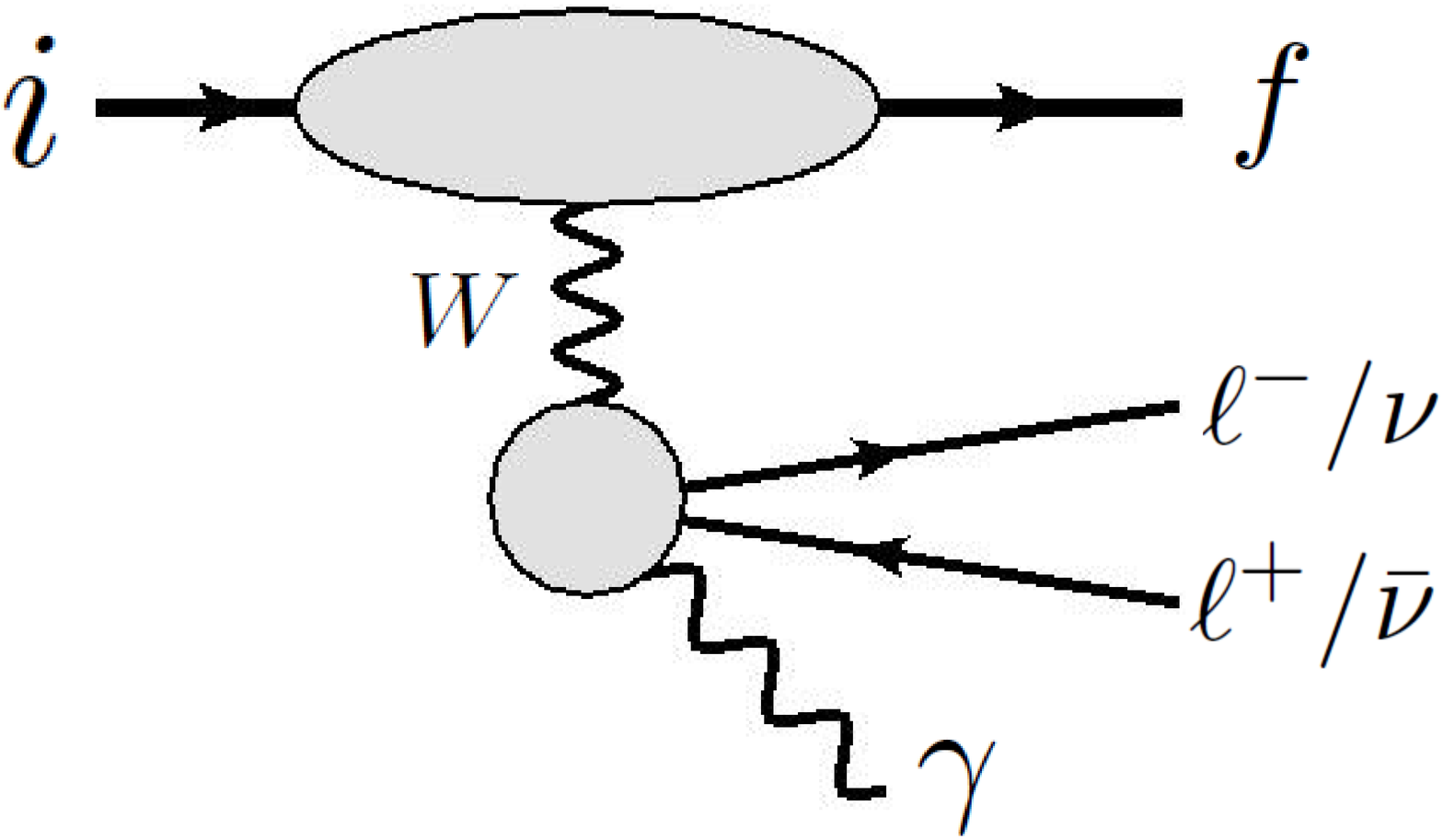}
		\includegraphics[scale=0.14]{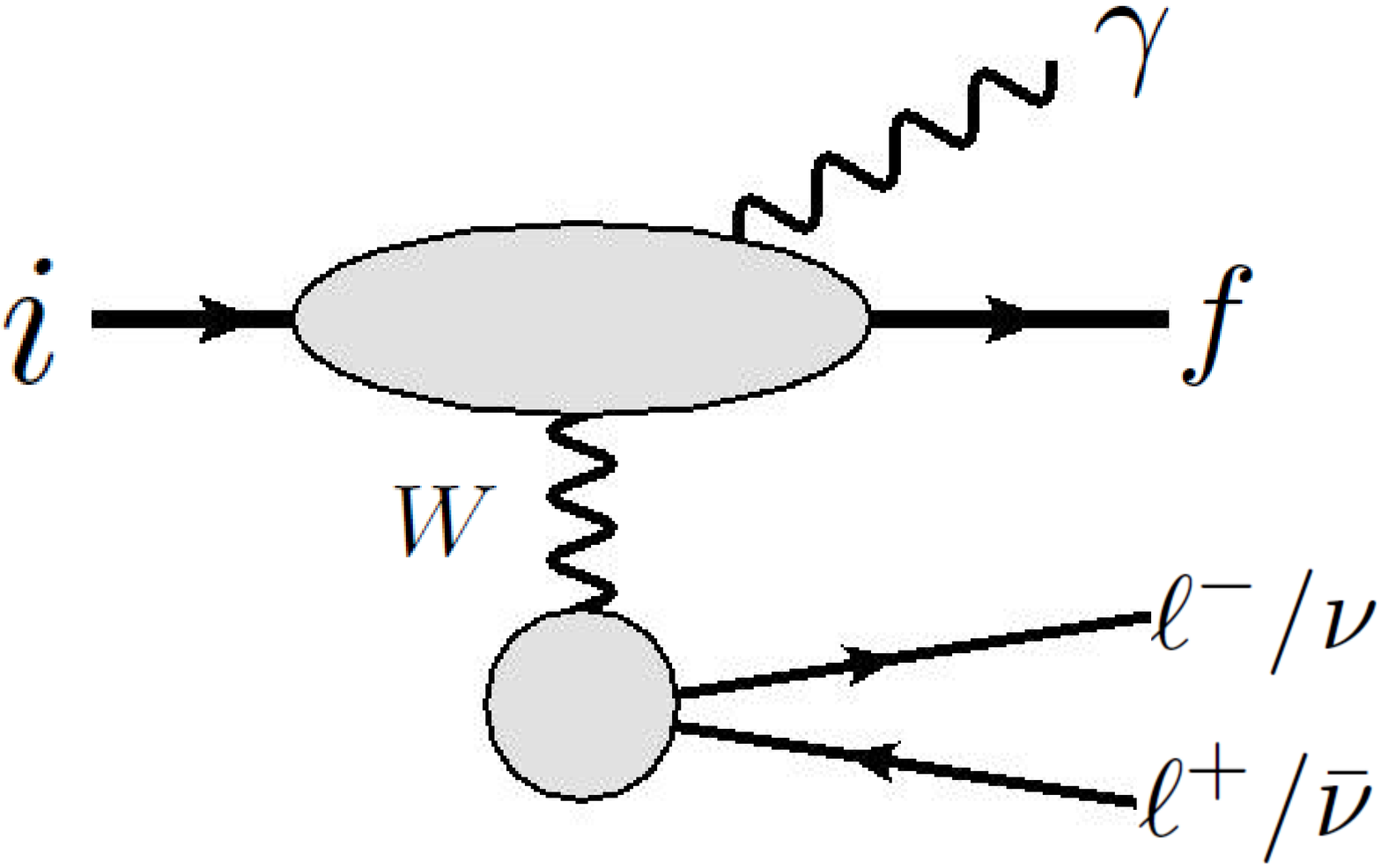}\hfill
		\par\end{centering}
	\caption{\label{fig:brem}Bremsstrahlung diagrams. We do not consider a photon emitted by the $W$ because that gives an extra $1/M_W^2$ suppression.}
\end{figure}

The absolute square of Eq.\eqref{eq:Sirlinrep} has an IR-divergence which must be canceled by the squared amplitude
of the bremsstrahlung process $\phi_i(p)\rightarrow \phi_f(p')\ell(p_\ell)\nu(p_\nu)\gamma(k)$ described by Fig.\ref{fig:brem}according to the KLN theorem~\cite{Kinoshita:1962ur,Lee:1964is}. The amplitude reads:
\begin{eqnarray}
\mathfrak{M}_\mathrm{brems}&=&-\frac{G_Fe}{\sqrt{2}}\frac{1}{2p_\ell\cdot k}\left\{\eta\left[2p_\ell\cdot \varepsilon^*g^{\mu\nu}+k^\mu\varepsilon^{*\nu}-k^\nu\varepsilon^{*\mu}\right]+i\epsilon^{\mu\nu\alpha\beta}k_\alpha\varepsilon_\beta^*\right\}F_\mu L_\nu\nonumber\\
&&+\frac{iG_Fe}{\sqrt{2}}\varepsilon^{*\mu}L^\nu T_{\mu\nu}(k;p',p)~.\label{eq:Mbrems}
\end{eqnarray}
We see that the generalized Compton tensor $T_{\mu\nu}$ appears again, except that now it depends on an on-shell photon momentum $k$.

\subsection{\label{sec:EWlog}Large electroweak logarithms and the higher-order QED effects}

Among the unevaluated one-loop virtual corrections, $(\delta \mathfrak{M}_2+\delta \mathfrak{M}_{\gamma W}^a)_\mathrm{int}$ and $\delta F_3^\lambda$ clearly do not depend on physics at the UV-scale. In fact, only $\delta M_{\gamma W}^b$ can probe the $q'\sim M_W$ scale. 
With the free-field OPE we can obtain the large electroweak logarithm in the latter:
\begin{equation}
\delta \mathfrak{M}_{\gamma W}^b=\frac{\alpha}{8\pi}\ln\frac{M_W^2}{\Lambda^2}\mathfrak{M}_0+...,\label{eq:MbUV}
\end{equation}
where $\Lambda$ is an arbitrary IR-cutoff scale. 
Adding this with other electroweak logs in Eq.\eqref{eq:Sirlinrep} gives the total electroweak logarithm in the one-loop amplitude:
\begin{equation}
\delta \mathfrak{M}_\mathrm{vir}=\frac{\alpha}{2\pi}\ln\frac{M_Z^2}{\Lambda^2}\mathfrak{M}_0+...
\end{equation}
which is a well-known result~\cite{Sirlin:1981ie}. This large logarithm is process-independent, and is usually scaled out as a common multiplicative factor as we will see later.

The expression above also allows us to discuss a numerically significant component of the RC, namely the higher-order QED effects to the decay rate:
\begin{equation}
\Gamma\rightarrow\left(1+\delta_\mathrm{HO}^\mathrm{QED}\right)\Gamma~,
\end{equation}
which we have first introduced in Eq.\eqref{eq:deltaMvir}.
So far we are working strictly at the order $\mathcal{O}(G_F\alpha)$ corrections to the Fermi interaction, but in order to reach the precision level of $10^{-4}$ (which is particularly important for the extraction of $V_{ud}$), leading effects of higher order in $\alpha$ are non-negligible. The largest among them comes from the resummation of electroweak logarithms: $\alpha^n\ln^n(M_Z/\Lambda)$, which can be straightforwardly implemented by considering the running of the fine structure constant. To do so, we first realize that the electroweak logarithm in the decay rate can be effectively reproduced by the following integral:
\begin{equation}
\frac{\alpha}{\pi}\ln\frac{M_Z^2}{\Lambda^2}=\int_{\Lambda^2}^{\infty}\frac{dQ^2}{Q^2}\frac{M_Z^2}{M_Z^2+Q^2}\frac{\alpha}{\pi}~.
\end{equation}
From here we promote the $Q^2$ dependence of the fine structure constant, $\alpha\rightarrow\alpha(Q^2)$, where $\alpha(0)=1/137.035999084(21)$. The numerical contribution due to the running is then:
\begin{equation}
\int_{\Lambda^2}^{\infty}\frac{dQ^2}{Q^2}\frac{M_Z^2}{M_Z^2+Q^2}\left[\frac{\alpha(Q^2)}{\pi}-\frac{\alpha(0)}{\pi}\right]\approx 0.0010~,
\end{equation}
where we have taken $\Lambda=M_\rho$ in the numerical evaluation, which is a common choice in standard literature. The analysis above is obviously incomplete: It only includes the resummation of large logs in the UV-region (i.e. $Q^2>\Lambda^2$), but not in the IR-region. It also does not cover other possibly important $\mathcal{O}(\alpha^2)$ effects. These unconsidered effects are generally process-dependent. Following Ref.\cite{Erler:2002mv}, we assign an uncertainty of $3\times 10^{-4}$ for these missing effects, and quote a final value of: 
\begin{equation}
\delta_\mathrm{HO}^\mathrm{QED}=0.0010(3)~.
\end{equation} 
Notice also that
Czarnecki, Marciano and Sirlin has performed a detailed study of the higher-order QED effects for the case of neutron in Ref.\cite{Czarnecki:2004cw}, where they considered the resummation of QED logs in both the UV and IR-region, as well as $\mathcal{O}(\alpha^2\ln(M_Z/m_p))$ and other important $\mathcal{O}(\alpha^2)$ effects. They obtained $\delta_\mathrm{HO}^\mathrm{QED}(\mathrm{neutron})=0.0013$ which is consistent with our estimation. This serves as a justification of our error assignment in $\delta_\mathrm{HO}^\mathrm{QED}$ based on Ref.\cite{Erler:2002mv}.   

It is customary to collect the large electroweak logarithm, the $\mathcal{O}(\alpha_s)$ pQCD correction on top of it (which comes predominantly from $\delta \mathfrak{M}_{\gamma W}^b$ as it is the only piece which extends down to $Q^2\sim \Lambda^2$) and $\delta_\mathrm{HO}^\mathrm{QED}$ to define a process-independent, short-distance electroweak factor $S_\mathrm{EW}$. One may write it schematically as~\cite{Cirigliano:2011ny}:
\begin{equation}
S_\mathrm{EW}\risingdotseq 1+\frac{2\alpha}{\pi}\left(1-\frac{\alpha_s}{4\pi}\right)\ln\frac{M_Z}{M_\rho}+\delta_\mathrm{HO}^\mathrm{QED}~,
\end{equation}
where the $M_\rho=770$~MeV appears as an IR scale. The expression above is schematic because one cannot directly infer the numerical value of $S_\mathrm{EW}$ from it; for instance, $\alpha_s$ is not really a constant but a function of $Q^2$ that participates actively in the loop integral. For nearly all practical purposes in existing literature, the value $S_\mathrm{EW}=1.0232(3)_\mathrm{HO}$ was always taken, where the major uncertainty comes from $\delta_\mathrm{HO}^\mathrm{QED}$. Therefore, it is more convenient to just take the numerical value above as a working definition of $S_\mathrm{EW}$. 

Finally, there is another potentially large higher-order QED effect that occurs in beta decays of heavy nuclei with charge numbers $Z_i,Z_f\gg1$. In most parts of the EMRCs (real and virtual), the contribution from the parent and daughter nucleus partially cancel each other so that the result depends only on $|Z_i-Z_f|=1$. There is only one exception that comes from $\delta M_{\gamma W}^a$, which gives the following correction to the squared amplitude:
\begin{equation}
\delta |\mathfrak{M}|^2\approx -\frac{\pi\alpha Z_f}{\beta}\eta|\mathfrak{M}_0|^2~,\label{eq:Fermioneloop}
\end{equation}
where $\beta=\sqrt{1-m_e^2 M^2/(p\cdot p_e)^2}$ is the electron's speed in the nuclei's rest frame (here we assume heavy nuclei so $p'\approx p$). Physically, it represents the final-state Coulomb interaction between the electron and the static daughter nucleus. Since $Z_f\gg 1$, the coefficient $\pi\alpha Z_f/\beta$ may not be small so the fixed-order QED correction is not a good approximation. Fortunately, the full Coulomb interaction is exactly calculable by solving the Dirac equation with a Coulomb potential. To implement this effect, we first start from the usual one-loop field theory calculation, and remove Eq.\eqref{eq:Fermioneloop} from the result. As a compensation, we multiply the full differential decay rate (including real and virtual RCs) by the following Fermi's function~\cite{Fermi:1934hr,Wilkinson:1970cdv}:
\begin{equation}
F(Z_f,\beta)=\frac{2(1+\gamma)}{\Gamma(2\gamma+1)^2}\left(\frac{2m_e\beta R}{\sqrt{1-\beta^2}}\right)^{2(\gamma-1)}e^{-\eta\pi\alpha Z_f/\beta}|\Gamma(\gamma+i\alpha Z_f/\beta)|^2,\label{eq:Fermifun}
\end{equation}
where $\gamma=\sqrt{1-(\alpha Z_f)^2}$, and $R$ is the nuclear radius. Expansion in powers of $\alpha$ gives:
\begin{equation}
F(Z_f,\beta)=1-\frac{\pi\alpha Z_f}{\beta}\eta+\mathcal{O}(Z_f^2\alpha^2)~,
\end{equation}
which reproduces the correction in Eq.\eqref{eq:Fermioneloop}.

\section{\label{sec:EFT}Effective field theory representation}

Despite being completely general, the Sirlin's representation was not historically the most widely adopted starting point for EWRCs. Possible reasons are that the expressions of some components, e.g. $\delta F_3^\lambda$, are rather complicated and do not offer an immediately transparent physical picture; in fact, until a few years ago, the only occasions where it was used were nearly-degenerate decay processes where $\delta F_3^\lambda$ is negligible. Also, Eq.\eqref{eq:Sirlinrep} is not a Lagrangian from which one obtains everything simply using Feynman rules and Feynman diagrams, as most theorists are more used to. Furthermore, it is tailored exclusively to deal with the RCs, and is not convenient (although not prohibited as well) for the study of other SM corrections, such as the isospin-breaking corrections and the recoil corrections, on the same footing.   

The EFT representation provides a satisfactory solution to the aforementioned issues. Under this formalism one  writes down the most general Lagrangian that consists of hadronic degrees of freedom (DOFs) and is compatible with all the symmetry requirements of the underlying theory (i.e. electroweak + QCD). One then calculates decay amplitudes through Feynman diagrams, using the Feynman rules derived from the EFT Lagrangian.
The predictive power is guaranteed upon agreeing on a specific power counting scheme which is assigned to all small parameters (e.g. $\alpha$, recoil corrections and isospin-breaking corrections). This allows us to  treat all of them simultaneously under a single, unified and model-independent framework. 

The EFT which is most relevant to the study of RCs in strongly-interacting systems is ChPT. It is a (or ``the''?) well-developed low-energy EFT of QCD stemmed from the idea of the spontaneously-broken chiral symmetry in QCD. 
In this review we only briefly introduce the relevant notations and the basic framework. Interested readers may refer to nicely-written textbooks, e.g. Ref.\cite{Scherer:2012xha}, or articles~\cite{Bernard:1995dp,Bernard:2007zu} (baryon sector), for more details.

\subsection{Spontaneously-broken chiral symmetry}

We introduce the basic concepts of ChPT by first examining the symmetry properties of a 3-flavor (i.e. $n_f=3$) QCD. Its Lagrangian reads:
\begin{equation}
\mathcal{L}_\mathrm{QCD}=\bar{\psi}_Li\slashed{D}\psi_L+\bar{\psi}_Ri\slashed{D}\psi_R-\bar{\psi}_LM_q^\dagger\psi_R-\bar{\psi}_RM_q\psi_L-\frac{1}{4}G_{\mu\nu}^aG^{a\mu\nu}~,\label{eq:LQCD}
\end{equation}
where $\psi=(u\:d\:s)^\mathrm{T}$ is the quark field and $M_q=\mathrm{diag}(m_u,m_d,m_s)$ is the quark mass matrix. Apart from the local SU(3)$_c$ symmetry, the Lagrangian above possesses another global symmetry known as the ``chiral symmetry'' in the limit of massless quarks, which means the following: If we take $M_q\rightarrow 0$, then the Lagrangian is invariant under separate SU(3) transformations of the left- and right-handed quark fields:
\begin{equation}
\psi_L\rightarrow L\psi_L~,\quad \psi_R\rightarrow R\psi_R~,
\end{equation} 
where $L\in \mathrm{SU}(3)_\mathrm{L}$ and $R\in \mathrm{SU}(3)_\mathrm{R}$ are SU(3) matrices in the flavor space. When quark masses are non-zero, this symmetry is explicitly broken but a residual symmetry retains in the $m_u=m_d=m_s$ limit, namely:
\begin{equation}
\psi\rightarrow V\psi~,
\end{equation}
where $V\in \mathrm{SU}(3)_\mathrm{V}$ is again an SU(3) matrix in the flavor space. The latter is the ``vector'' SU(3) symmetry that leads to the famous eightfold way picture first proposed by Gell-Mann and Ne'eman~\cite{Gell-Mann:1961omu,Neeman:1961jhl}. If one concentrates only on two flavors, then it reduces to the well-known isospin symmetry.  

Since light quark masses are very small, one naturally expects the flavor symmetries above to hold, at least approximately, in strongly-interacting systems. Experimental measurements of the hadronic mass spectra show that this is indeed the case for SU(2)$_\mathrm{V}$ and SU(3)$_\mathrm{V}$. However, to one's surprise, it was observed that the full chiral symmetry is not respected at all by hadrons. A clear evidence is the non-observation of the so-called ``parity doubling'' effect, which requires each hadron to have a corresponding degenerate hadron with opposite parity if chiral symmetry holds. In reality one does not observe such a doubling, even approximately. For instance, the lightest $J^p=(1/2)^+$ baryon, i.e. proton, has a mass of 938~MeV. On the other hand, the lightest $J^p=(1/2)^-$ baryon has a mass around 1535~MeV which can never be interpreted as a parity counterpart of the proton. 

The observed phenomena can be explained if QCD vacuum is not annihilated by the SU(3) axial charges, i.e.
\begin{equation}
\hat{Q}_\mathrm{A}^a|0\rangle\neq 0~,
\end{equation}
where $\hat{Q}_\mathrm{A}^a=\int d^3x \bar{\psi}\gamma^0\gamma_5T^a\psi$ ($a=1,...,8$). This implies the full SU(3) chiral symmetry group undergoes a spontaneous symmetry breaking to a smaller vector SU(3) symmetry group: $\mathrm{SU(3)_L}\times\mathrm{SU(3)_R}\rightarrow\mathrm{SU(3)_V}$. According to the Nambu-Goldstone theorem~\cite{Nambu:1960tm,Goldstone:1961eq}, the spontaneously-broken chiral symmetry leads to the emergence of massless bosons, which later gain small masses due to the quark mass matrix that further introduces an explicit breaking of the symmetry. These particles are therefore known as the pseudo-Nambu-Goldstone bosons (pNGBs) and can be easily identified as the pseudoscalar meson octet ($\pi, K,\eta$) that are known to to be much lighter than all other hadrons. 

\subsection{pNGBs and the chiral power counting}

ChPT (without external sources) consists of the most general effective Lagrangian that satisfy the following criteria:
\begin{itemize}
	\item It is invariant under $\mathrm{SU(3)_L}\times\mathrm{SU(3)_R}$ in the limit of massless quarks.
	\item The chiral symmetry is spontaneously broken to SU(3)$_\mathrm{V}$, and the pseudoscalar octets appear as the pNGBs.
	\item The chiral symmetry is explicitly broken by the insertion of the quark mass matrix $M_q$. 
\end{itemize} 
It is thus a theory with the pNGBs as dynamical DOFs. The latter often appears through a non-linear realization. The most convenient parameterization suitable for a 3-flavor ChPT is the following exponential representation:
\begin{equation}
U=\exp\left\{i\frac{\lambda_i\phi_i}{F_0}\right\}~,
\end{equation}
where $\{\phi_i\}$ are the pNGBs, $\{\lambda_i\}$ are the Gell-Mann matrices and $F_0$ is the pion decay constant in the chiral limit. It transforms as $U\rightarrow RU L^\dagger$ under a chiral rotation. Such a representation guarantees that the  interactions of the pNGBs must contain either derivatives or insertions of the quark mass matrix $M_q$.

The smallness of the pNGB masses provides a natural small scale in the theory. One may therefore define a power counting rule such that terms in ChPT are arranged according to an increasing power of $p/\Lambda_\chi$, where $p$ is either the pNGB masses or their momenta, while $\Lambda_\chi\sim 4\pi F_\pi\sim 1$~GeV is the chiral symmetry breaking scale. Notice that quark masses scale as $\mathcal{O}(p^{1/2})$ because $M_\phi^2\propto m_q$. In this way, with any given precision goal only a finite number of terms in the Lagrangian and a finite number of loops in the Feynman diagrams are needed. 

We illustrate the ideas above by writing down the leading order (LO) chiral Lagrangian in the mesonic sector without external sources. It scales as $\mathcal{O}(p^2)$ and consists of two terms:
\begin{equation}
\mathcal{L}^{p^2}=\frac{F_0^2}{4}\left\langle \partial_\mu U(\partial^\mu U)^\dagger\right\rangle+\frac{F_0^2}{4}\left\langle\chi U^\dagger+U\chi^\dagger\right\rangle~,\label{eq:Lp2QCD}
\end{equation} 
where $\chi=2B_0M_q$ with $B_0$ a constant number, and $\langle...\rangle$ is the trace over the flavor space. The first term at the right hand side preserves the chiral symmetry, whereas the second term breaks the symmetry in the same way as QCD does, namely: If the quark mass matrix would transform as $M_q\rightarrow RM_qL^\dagger$ under the chiral rotation, then the symmetry would be restored. This way of implementing symmetry-breaking effects is known as the spurion method, and $M_q$ acts as a ``spurion field''. 

With Eq.\eqref{eq:Lp2QCD} one could then perform tree-level field theory calculations by expanding $U$ in powers of pNGB fields. For instance, expansion to $\mathcal{O}(\phi^2)$ yields the kinetic and mass term of the pNGBs, and higher expansions give the interaction vertices. 

\subsection{External sources}

To study beta decays of strongly-interacting systems and the RCs, the pure-QCD Lagrangian \eqref{eq:LQCD} is obviously not enough, and we need to consider its coupling to external sources and dynamical photons. The coupling terms read:
\begin{equation}
\mathcal{L}_\mathrm{ext}=\bar{\psi}_L\gamma^\mu(l_\mu+q_LA_\mu)\psi_L+\bar{\psi}_R\gamma^\mu(r_\mu+q_RA_\mu)\psi_R~,\label{eq:Lext}
\end{equation}
where $A_\mu$ is the dynamical photon field (notice that it is the full field instead of the ``$<$'' component), and $\{l_\mu,r_\mu,q_L,q_R\}$ are spurion fields which  will later be identified to quantities in the SM electroweak sector. 

The first question is how to introduce the spurion fields in Eq.\eqref{eq:Lext} into the EFT. The principle is the same as $M_q$: They should break the chiral symmetry in the EFT in exactly the same way they do in the underlying theory. Since $\{l_\mu,r_\mu\}$ are currents, it is most convenient to discuss their symmetry-breaking pattern by considering a \textit{local} SU(3)$_\mathrm{L}\times$SU(3)$_\mathrm{R}$ chiral symmetry. We observe that the full Lagrangian $\mathcal{L}_\mathrm{QCD}+\mathcal{L}_\mathrm{ext}$ would be invariant under a local chiral transformation, if the spurion fields would transform as:
\begin{equation}
l_\mu\rightarrow Ll_\mu L^\dagger-i(\partial_\mu L)L^\dagger~,\quad r_\mu\rightarrow Rr_\mu R^\dagger-i(\partial_\mu R)R^\dagger~,\quad q_L\rightarrow Lq_LL^\dagger~,\quad q_R\rightarrow Rq_LL^\dagger
\end{equation}  
accordingly.

We now construct the building blocks for the most general EFT that satisfies the \textit{local} chiral symmetry (assuming the transformation rules of the spurions). First, the ordinary derivative on $U$ should be promoted to a chiral covariant derivative:
\begin{equation}
D_\mu U\equiv \partial_\mu U-i(r_\mu+q_RA_\mu)U+iU(l_\mu+q_LA_\mu)~,
\end{equation}
such that it transforms as $D_\mu U\rightarrow R(D_\mu U)L^\dagger$ under the local chiral rotation. 
Next, for the description of interactions with leptons, it is also convenient to define $u\equiv\sqrt{U}$, which transforms as:
\begin{equation}
u\rightarrow KuL^\dagger=RuK^\dagger~,
\end{equation}
where $K$ is a field-dependent unitary matrix. With this we define the so-called ``chiral vielbein'':
\begin{equation}
u_\mu\equiv iu^\dagger(D_\mu U)u^\dagger=i[u^\dagger(\partial_\mu-ir_\mu-iq_RA_\mu)u-u(\partial_\mu-il_\mu-iq_LA_\mu)u^\dagger]
\end{equation}
which is an axial vector and transforms as $u_\mu\rightarrow Ku_\mu K^\dagger$. We also define 
the covariant derivate on the spurion fields:
\begin{equation}
\nabla_\mu q_R=\partial_\mu q_R-i[r_\mu,q_R]~,\quad \nabla_\mu q_L=\partial_\mu q_L-i[l_\mu,q_L]~,
\end{equation}
such that they transform as $\nabla_\mu q_R\rightarrow R(\nabla_\mu q_R)R^\dagger$, $\nabla_\mu q_L\rightarrow L(\nabla_\mu q_L)L^\dagger$. Finally, we define
the field-strength tensors corresponding to the external currents:
\begin{equation}
f_{R\mu\nu}=\partial_\mu r_\nu-\partial_\nu r_\mu-i[r_\mu,r_\nu]~,\quad f_{L\mu\nu}=\partial_\mu l_\nu-\partial_\nu l_\mu-i[l_\mu,l_\nu]~.
\end{equation}
They transform as $f_{R\mu\nu}\rightarrow Rf_{R\mu\nu}R^\dagger$, $f_{L\mu\nu}\rightarrow Lf_{L\mu\nu}L^\dagger$.

In order to connect Eq.\eqref{eq:Lext} to the EM and Fermi interaction that we are interested in, we need to identify the external currents and spurions as:
\begin{equation}
q_R=q_L=-eQ^\mathrm{em}~,\quad l_\mu=\sum_{\ell}\left(\bar{\ell}_L\gamma_\mu\nu_{\ell L}Q_\mathrm{L}^\mathrm{w}+\mathrm{h.c.}\right)~,\quad r_\mu=0~,
\end{equation}
where~\footnote{Notice that in $Q_\mathrm{L}^\mathrm{w}$ we use the physical Fermi's constant $G_F$, which is just a matter of choice; one could choose $G_F^0$ instead, and the only consequence is that the numerical value of some of the $\mathcal{O}(e^2p^2)$ counterterms that we will introduce later will shift accordingly to absorb this finite difference.}
\begin{equation}
Q^{\mathrm{em}}=\left(\begin{array}{ccc}
2/3 & 0 & 0\\
0 & -1/3 & 0\\
0 & 0 & -1/3
\end{array}\right)~,\quad Q_{\mathrm{L}}^{\mathrm{w}}=-2\sqrt{2}G_{F}\left(\begin{array}{ccc}
0 & V_{ud} & V_{us}\\
0 & 0 & 0\\
0 & 0 & 0
\end{array}\right)~.
\end{equation}
The EM interaction introduces a new expansion parameter $e$, which needs to be considered simultaneously with $p$ in the chiral expansion. Following usual convention, we take $e\sim p$, i.e. they count as the same order in the chiral power counting. 

\subsection{\label{sec:EFTmeson}Mesonic ChPT with external sources} 

We are now ready write down the full Lagrangian with leptons, photon and pNGBs as dynamical DOFs:
\begin{equation}
\mathcal{L}=\mathcal{L}_\mathrm{lepton}+\mathcal{L}_\gamma+\mathcal{L}_\mathrm{ChPT}~.
\end{equation}
The pure lepton and photon Lagrangian are simply:
\begin{eqnarray}
\mathcal{L}_\mathrm{lepton}&=&\sum_{\ell}\left[\bar{\ell}(i\slashed{\partial}+e\slashed{A}-m_\ell)\ell+\bar{\nu}_{\ell L}i\slashed{\partial}\nu_{\ell L}\right]\nonumber\\
\mathcal{L}_\gamma&=&-\frac{1}{4}F_{\mu\nu}F^{\mu\nu}-\frac{1}{2\xi}(\partial\cdot A)^2+\frac{1}{2}M_\gamma^2A_\mu A^\mu~,
\end{eqnarray}
where  $\xi$ is the EM gauge parameter which is always chosen as 1 (i.e. Feynman gauge) in existing calculations. Also, an infinitesimal photon mass $M_\gamma$ is introduced to regularize the IR-divergences. In the meantime, the ChPT Lagrangian is arranged by the increasing power of the chiral order:
\begin{equation}
\mathcal{L}_\mathrm{ChPT}=\mathcal{L}^{(2)}+\mathcal{L}^{(4)}+...~.
\end{equation} 
The LO chiral Lagrangian consists of two types: $\mathcal{L}^{(2)}=\mathcal{L}^{p^2}+\mathcal{L}^{e^2}$, where
\begin{equation}
\mathcal{L}^{p^2}=\frac{F_0^2}{4}\left\langle D_\mu U(D^\mu U)^\dagger+\chi U^\dagger+U\chi^\dagger\right\rangle~,\quad \mathcal{L}^{e^2}=ZF_0^4\left\langle q_LU^\dagger q_RU\right\rangle~.\label{eq:Lp2}
\end{equation}
The $\mathcal{O}(p^2)$ term is just a simple generalization of Eq.\eqref{eq:Lp2QCD} to include the external sources, while the $\mathcal{O}(e^2)$ term
represents the short-distance EM effect and $Z\approx 0.8$ is obtained from the $\pi^\pm-\pi^0$ mass splitting. 

Applying $\mathcal{L}^{(2)}$ to one loop produces UV-divergences that are regulated using dimensional regularization. They are 
then reabsorbed by the LECs in the next-to-leading-order (NLO) chiral Lagrangian: $\mathcal{L}^{(4)}=\mathcal{L}^{p^4}+\mathcal{L}^{e^2p^2}+\mathcal{L}^{e^4}$. The last term is purely EM and does not contribute to beta decays at tree-level so we may discard it. The first term is the well-known Gasser-Leutwyler Lagrangian:~\cite{Gasser:1983yg,Gasser:1984gg}:
\begin{eqnarray}
\mathcal{L}^{p^4}&=&L_1\left\langle D_\mu U(D^\mu U)^\dagger\right\rangle^2+L_2\left\langle D_\mu U(D_\nu U)^\dagger\right\rangle \left\langle D^\mu U(D^\nu U)^\dagger\right\rangle\nonumber\\
&&+L_3\left\langle D_\mu U(D^\mu U)^\dagger D_\nu U(D^\nu U)^\dagger\right\rangle+L_4\left\langle D_\mu U(D^\mu U)^\dagger\right\rangle\left\langle\chi U^\dagger+U\chi^\dagger\right\rangle\nonumber\\
&&+L_5\left\langle D_\mu U(D^\mu U)^\dagger(\chi U^\dagger+U\chi^\dagger)\right\rangle+L_6\left\langle \chi U^\dagger+U\chi^\dagger\right\rangle^2\nonumber\\
&&+L_7\left\langle \chi U^\dagger-U\chi^\dagger\right\rangle^2+L_8\left\langle U\chi^\dagger U\chi^\dagger+\chi U^\dagger\chi U^\dagger\right\rangle\nonumber\\
&&-iL_9\left\langle f_{\mu\nu}^RD^\mu U(D^\nu U)^\dagger+f_{\mu\nu}^L(D^\mu U)^\dagger D^\nu U\right\rangle+L_{10}\left\langle Uf_{\mu\nu}^L U^\dagger f_R^{\mu\nu}\right\rangle~.
\end{eqnarray}
Meanwhile, the $\mathcal{O}(e^2p^2)$ terms that involve only dynamical photons were first written down by Urech~\cite{Urech:1994hd}:
\begin{eqnarray}
\mathcal{L}^{e^2p^2}_{\{K\}}&=&F_0^2\left\{\frac{1}{2}K_1\left\langle D^\mu U(D_\mu U)^\dagger\right\rangle\left\langle q_Rq_R+q_Lq_L\right\rangle+K_2\left\langle D^\mu U(D_\mu U)^\dagger\right\rangle\left\langle q_R Uq_L U^\dagger\right\rangle\right.\nonumber\\
&&+K_3\left(\left\langle(D^\mu U)^\dagger q_R U\right\rangle\left\langle(D_\mu U)^\dagger q_R U\right\rangle+\left\langle D^\mu Uq_L U^\dagger\right\rangle\left\langle D_\mu Uq_L U^\dagger\right\rangle\right)\nonumber\\
&&K_4\left\langle(D^\mu U)^\dagger q_R U\right\rangle\left\langle D_\mu Uq_L U^\dagger\right\rangle+K_5\left\langle q_Lq_L(D^\mu U)^\dagger D_\mu U+q_Rq_RD^\mu U(D_\mu U)^\dagger\right\rangle\nonumber\\
&&+K_6\left\langle(D^\mu U)^\dagger D_\mu Uq_L U^\dagger q_R U+D^\mu U(D_\mu U)^\dagger q_R Uq_LU^\dagger\right\rangle\nonumber\\
&&+\frac{1}{2}K_7\left\langle\chi^\dagger U+U^\dagger\chi\right\rangle\left\langle q_Rq_R+q_Lq_L\right\rangle+K_8\left\langle\chi^\dagger U+U^\dagger \chi\right\rangle\left\langle q_RUq_LU^\dagger\right\rangle\nonumber\\
&&+K_9\left\langle(\chi^\dagger U+U^\dagger \chi)q_Lq_L+(\chi U^\dagger+U\chi^\dagger)q_Rq_R\right\rangle\nonumber\\
&&+K_{10}\left\langle(\chi^\dagger U+U^\dagger\chi)q_L U^\dagger q_R U+(\chi U^\dagger+U\chi^\dagger)q_R Uq_LU^\dagger\right\rangle\nonumber\\
&&+K_{11}\left\langle(\chi^\dagger U-U^\dagger \chi)q_LU^\dagger q_R U+(\chi U^\dagger-U\chi^\dagger)q_R Uq_LU^\dagger\right\rangle\nonumber\\
&&+K_{12}\left\langle(D^\mu U)^\dagger[\nabla_\mu q_R,q_R]U+D^\mu U[\nabla_\mu q_L,q_L]U^\dagger\right\rangle\nonumber\\
&&\bigl.+K_{13}\left\langle\nabla^\mu q_R U\nabla_\mu q_L U^\dagger\right\rangle+K_{14}\left\langle\nabla^\mu q_R\nabla_\mu q_R+\nabla^\mu q_L\nabla_\mu q_L\right\rangle\biggr\}~,\label{eq:LKi}
\end{eqnarray}
whereas the other terms that further include dynamical leptons were introduced by Knecht et al.~\cite{Knecht:1999ag}:
\begin{eqnarray}
\mathcal{L}^{e^2p^2 }_{\{X\}}&=&e^2F_0^2\sum_\ell\left\{X_1\bar{\ell}\gamma_\mu \nu_{\ell L}\left\langle u^\mu\left\{\mathcal{Q}_\mathrm{R}^\mathrm{em},\mathcal{Q}_\mathrm{L}^\mathrm{w}\right\}\right\rangle+X_2\bar{\ell}\gamma_\mu \nu_{\ell L}\left\langle u^\mu\left[\mathcal{Q}_\mathrm{R}^\mathrm{em},\mathcal{Q}_\mathrm{L}^\mathrm{w}\right]\right\rangle\right.\nonumber\\
&&\left.+X_3m_\ell\bar{\ell}\nu_{\ell L}\left\langle\mathcal{Q}_\mathrm{L}^\mathrm{w}\mathcal{Q}^\mathrm{em}_\mathrm{R}\right\rangle+h.c.\right\}+e^2\sum_\ell X_6\bar{\ell}(i\slashed{\partial}+e\slashed{A})\ell~,
\end{eqnarray}
where $\mathcal{Q}_\mathrm{R}^\mathrm{em}=u^\dagger Q^\mathrm{em}u$ and $\mathcal{Q}_\mathrm{L}^\mathrm{w}=
uQ_\mathrm{L}^\mathrm{w}u^\dagger$.

The bare LECs $\{L_i,K_i,X_i\}$ are UV-divergent and scale-independent. They are related to the scale-dependent, UV-finite renormalized LECs as follows:
\begin{equation}
L_i^r(\mu)=L_i-\Gamma_i\lambda~,\quad K_i^r(\mu)=K_i-\Sigma_i\lambda~,\quad X_i^r(\mu)=X_i-\Xi_i\lambda~,
\end{equation}
where $\mu$ is the renormalization scale, and
\begin{equation}
\lambda=\frac{\mu^{d-4}}{16\pi^2}\left(\frac{1}{d-4}-\frac{1}{2}[\ln 4\pi-\gamma_E+1]\right)~,
\end{equation}
with $d$ the spacetime-dimension in the dimensional regularization approach.
The constant coefficients $\{\Gamma_i,\Sigma_i,\Xi_i\}$ can be evaluated with the heat kernel method and were given in Refs.\cite{Gasser:1984gg}, \cite{Urech:1994hd} and \cite{Knecht:1999ag} respectively.

One could in principle proceed further to construct the NNLO chiral Lagrangian $\mathcal{L}^{(6)}$ in the same way, but the appearance of too many new LECs render this step less useful. The $\mathcal{O}(p^6)$ Lagrangian was studied in Refs.\cite{Fearing:1994ga,Bijnens:1999sh} and was found to contain 90+4 independent terms for $n_f=3$. The $\mathcal{O}(e^2p^4)$ Lagrangian has not yet been investigated. 

\subsection{\label{sec:HBChPT}Nucleon sector}

The nucleon doublet $\psi_N=(p\:n)^\mathrm{T}$ can be introduced to a two-flavor ChPT as a matter field that transforms as $\psi_N\rightarrow K\psi_N$ under the chiral rotation. However, by doing so the chiral Lagrangian now contains a heavy DOF. Therefore, when calculating loops, one not only obtains terms that are suppressed by $p/\Lambda_\chi\ll 1$, but also encounters terms like $m_N/\Lambda_\chi\sim 1$ that are not suppressed at all. The appearance of such terms implies a breakdown of the chiral power counting.

There are several prescriptions to restore the chiral power counting in the presence of nucleons, e.g. the 
heavy baryon chiral perturbation theory (HBChPT)~\cite{Jenkins:1990jv,Bernard:1992qa}, the infrared regularization~\cite{Becher:1999he} and the extended on-mass-shell scheme~\cite{Fuchs:2003qc,Gegelia:1999gf}; here we will only introduce the first method, namely the heavy baryon approach. To understand it, we first write the nucleon momentum as:
\begin{equation}
p^\mu=m_Nv^\mu+k^\mu~,
\end{equation}
which defines the velocity vector $v^\mu$ that satisfies $v^2=1$. The residual momentum $k$ is small so $k/m_N\ll 1$ is taken as another small expansion parameter. With this we can define the following projection operators:
\begin{equation}
P_{v\pm}=\frac{1\pm\slashed{v}}{2}~,
\end{equation}
which project the full nucleon field $\psi_N$ into the ``light'' and ``heavy'' component respectively:
\begin{equation}
N_v(x)\equiv e^{im_N v\cdot x}P_{v+}\psi_N(x)~,\quad H_v(x)\equiv e^{im_Nv\cdot x}P_{v-}\psi_N(x)~.
\end{equation}
One observes immediately from the free Lagrangian that $N_v$ resembles a massless particle, whereas $H_v$ has an effective mass $2m_N$. The latter can then be integrated out from the theory, which leaves the light component $N_v$ as the only dynamical DOF. In the $\{N_v,\bar{N}_v\}$ subspace, the nucleon propagator reduces to:
\begin{equation}
\frac{i(\slashed{p}+m_N)}{p^2-m_N^2+i\epsilon}\rightarrow\frac{i}{v\cdot k+i\epsilon}~.
\end{equation} 
Since the heavy mass scale $m_N$ is now absent in the propagator, the issue of the power counting violation does not appear anymore. Another advantage is that the independent Dirac structures reduce to just 1 and $S_v^\mu$, where
\begin{equation}
S_v^\mu\equiv \frac{i}{2}\sigma^{\mu\nu}\gamma_5 v_\nu
\end{equation}
is the spin matrix satisfying $S_v\cdot v=0$. 

EFT treatment of the free neutron beta decay based on HBChPT was introduced by Ando et al.~\cite{Ando:2004rk} and Bernard el al.~\cite{Bernard:2004cm}. 
The effective Lagrangian up to $\mathcal{O}(1/m_N)$ reads:
\begin{equation}
\mathcal{L}_\beta=\mathcal{L}_{e\nu\gamma}+\mathcal{L}_{NN\gamma}+\mathcal{L}_{e\nu NN}~,
\end{equation}
with (here we use $\psi_e$ instead of $e$ to represent the electron field, in order to avoid confusions with the electric charge $e$)
\begin{eqnarray}
\mathcal{L}_{e\nu\gamma}&=&-\frac{1}{4}F^{\mu\nu}F_{\mu\nu}-\frac{1}{2\xi}(\partial\cdot A)^2+\left(1+\frac{\alpha}{4\pi}e_1\right)\bar{\psi}_ei\slashed{D}\psi_e-m_e\bar{\psi}_e\psi_e+\bar{\nu}_{eL} i\slashed{\partial}\nu_{eL}\nonumber\\
\mathcal{L}_{NN\gamma}&=&\bar{N}_v\left[1+\frac{\alpha}{8\pi}e_2(1+\tau_3)\right]iv\cdot DN_v\nonumber\\
\mathcal{L}_{e\nu NN}&=&-\frac{G_F^0V_{ud}}{\sqrt{2}}\bar{\psi}_{e}\gamma_\mu(1-\gamma_5)\nu_{e}\left\{\bar{N}_v\tau^+\left[\left(1+\frac{\alpha}{4\pi}e_V\right)v^\mu+2\mathring{g}_A\left(1+\frac{\alpha}{4\pi}e_A\right)S_v^\mu\right]N_v\right.\nonumber\\
&&+\frac{1}{2m_N}\bar{N}_v\tau^+\left[i(v^\mu v^\nu-g^{\mu\nu})(\overleftarrow{\partial}-\overrightarrow{\partial})_\nu-2i\mathring{\mu}_V[S_v^\mu,S_v\cdot(\overleftarrow{\partial}+\overrightarrow{\partial})]\right.\nonumber\\
&&\left.\left.+2i\mathring{g}_Av^\mu S_v\cdot(\overleftarrow{\partial}-\overrightarrow{\partial})\right]N_v\right\}~.
\end{eqnarray}
The pure-QCD LECs in the Lagrangian are as follows: $\mathring{g}_A$ is the nucleon axial coupling constant (we choose the sign convention $\mathring{g}_A<0$ throughout this review) and $\mathring{\mu}_V=\mathring{\mu}_p-\mathring{\mu}_n$ is the isovector magnetic moment.
Meanwhile, $\{e_1,e_2,e_V,e_A\}$ are unknown LECs that characterize the short-distance EMRC effects, and Ref.\cite{Ando:2004rk} showed that only two combinations out of the four LECs are physically relevant:
\begin{equation}
e_{V,A}^R(\mu)=e_{V,A}-\frac{1}{2}(e_1+e_2)+\frac{3}{2}\left[\frac{2}{4-d}-\gamma_E+\ln 4\pi+1\right]+3\ln\left(\frac{\mu}{m_N}\right)~.
\end{equation}
The effective Lagrangian above allows for a simultaneous treatment of the $\mathcal{O}(\alpha)$ EMRC and the $\mathcal{O}(1/m_N)$ recoil corrections to the free neutron beta decay.  

\section{\label{sec:pion}pion semileptonic beta decay}

After reviewing all the major theory frameworks, we shall start to work on solid examples.  
The pion semileptonic decay $\pi^+\rightarrow \pi^0 e^+\nu_e$ (denoted as $\pi_{e3}$) is arguably the simplest beta decay process that serves as an ideal prototype for high-precision study of the EWRCs. It possesses three major advantages:
\begin{itemize}
	\item It is spinless, so at tree level only the vector component of the charged weak current contributes.
	\item It is near-degenerate, i.e. $M_{\pi^+}-M_{\pi^0}\ll M_\pi$, which simplifies the discussion a lot upon neglecting recoil corrections on top of the RCs.
	\item It does not suffer from nuclear structure uncertainties.
\end{itemize}
Therefore, we shall discuss its RC in some detail.

\subsection{Tree-level analysis}

To facilitate our discussion, we define:
\begin{equation}
\Delta\equiv M_{\pi^+}-M_{\pi^0}~,\quad \varepsilon\equiv \left(\frac{m_e}{\Delta}\right)^2~.
\end{equation}
Both $\Delta/M_{\pi^+}\approx 0.0329$ and $\varepsilon\approx 0.0124$ can be treated as small expansion parameters.

At tree-level, the matrix element of the charged weak current for spinless systems can be parameterized as:
\begin{eqnarray}
\langle\phi_f(p')|J^\mu(0)|\phi_i(p)\rangle&=&F_+^{if}(t)(p+p')^\mu+F_-^{if}(t)(p-p')^\mu\nonumber\\
&=&V_{ab}^{(*)}\left[f_+^{if}(t)(p+p')^\mu+f_-^{if}(t)(p-p')^\mu\right]~,\label{eq:FFspinless}
\end{eqnarray}
where $t=(p-p')^2$, and $V_{ab}^{(*)}$ is the relevant CKM matrix element that enters $F_\pm^{if}$ (the complex conjugate sign applies to $\beta^+$ decay). Applying the above to $\pi_{e3}$, in the isospin limit we have $f_+^{\pi^+\pi^0}(0)=-\sqrt{2}$ and $f_-^{\pi^+\pi^0}(0)=0$. Therefore, the entire $f_-^{\pi^+\pi^0}$ originates from the strong isospin-breaking effect which, according to the Behrends-Sirlin-Ademollo-Gatto theorem~\cite{Behrends:1960nf,Ademollo:1964sr}, scales as $\mathcal{O}(m_d-m_u)^2$ and is negligible to our precision goal of $10^{-4}$. Furthermore, a simple $\rho$-dominance picture suggests that the $t$-dependence in $f_+^{\pi^+\pi^0}(t)$ scales as $t/M_\rho^2<10^{-4}$ and is negligible; this conclusion was supported by later studies~\cite{Cirigliano:2002ng}. Therefore, one may approximate:
\begin{equation}
\langle\pi^0(p')|J^\mu(0)|\pi^+(p)\rangle\approx -\sqrt{2}V_{ud}^*(p+p')^\mu~.
\end{equation}

With the above, one could perform analytically the phase-space integral and obtain the following decay rate at tree level,
\begin{equation}
\left(\Gamma_{\pi_{e3}}\right)_\mathrm{tree}=\frac{G_F^2|V_{ud}|^2M_{\pi^+}^5}{64\pi^3}I_\pi~,
\end{equation}
where the phase space factor (which includes all the recoil effects except those in $f_+^{\pi^+\pi^0}(t)$ that we neglected) reads:
\begin{equation}
I_\pi=\frac{32}{15}\left(\frac{\Delta}{M_{\pi^+}}\right)^5\left(1-\frac{\Delta}{2M_{\pi^+}}\right)^3\mathbf{f}(\varepsilon,\Delta)~,
\end{equation}
with 
\begin{eqnarray}
\mathbf{f}(\varepsilon,\Delta)&=&\mathbf{f}^{(0)}(\varepsilon)+\left(\frac{\Delta}{M_{\pi^+}}\right)^2\mathbf{f}^{(2)}(\varepsilon)+\mathcal{O}(\Delta^3)\nonumber\\
\mathbf{f}^{(0)}(\varepsilon)&=&\sqrt{1-\varepsilon}\left(1-\frac{9}{2}\varepsilon-4\varepsilon^2\right)+\frac{15\varepsilon^2}{2}\ln\left(\frac{1+\sqrt{1-\varepsilon}}{\sqrt{\varepsilon}}\right)\nonumber\\
\mathbf{f}^{(2)}(\varepsilon)&=&\frac{1}{112(\sqrt{1-\varepsilon}+1)^2}\biggl[210(1+\varepsilon)\varepsilon^2(1+\sqrt{1-\varepsilon})^2\ln\left(\frac{1+\sqrt{1-\varepsilon}}{\sqrt{\varepsilon}}\right)\nonumber\\
&&+2(\varepsilon\sqrt{1-\varepsilon}+2\varepsilon-2\sqrt{1-\varepsilon}-2)(36\varepsilon^3+179\varepsilon^2-11\varepsilon+6)\biggr]~.
\end{eqnarray}
Since $\varepsilon$ is small, to our precision goal we simply take $\mathbf{f}^{(2)}(\varepsilon)\rightarrow \mathbf{f}^{(2)}(0)=-3/28$. This gives:
\begin{equation}
\mathbf{f}(\varepsilon,\Delta)\approx\sqrt{1-\varepsilon}\left(1-\frac{9}{2}\varepsilon-4\varepsilon^2\right)+\frac{15\varepsilon^2}{2}\ln\left(\frac{1+\sqrt{1-\varepsilon}}{\sqrt{\varepsilon}}\right)-\frac{3}{7}\frac{\Delta^2}{(M_{\pi^+}+M_{\pi^0})^2}
\end{equation}
as displayed in Refs.\cite{Sirlin:1977sv,Pocanic:2003pf,Czarnecki:2019iwz}. Numerically, $I_\pi=7.376(1)\times 10^{-8}$. 

\subsection{\label{sec:pionEWRC}EWRCs}

\begin{figure}
	\begin{centering}
		\includegraphics[scale=0.14]{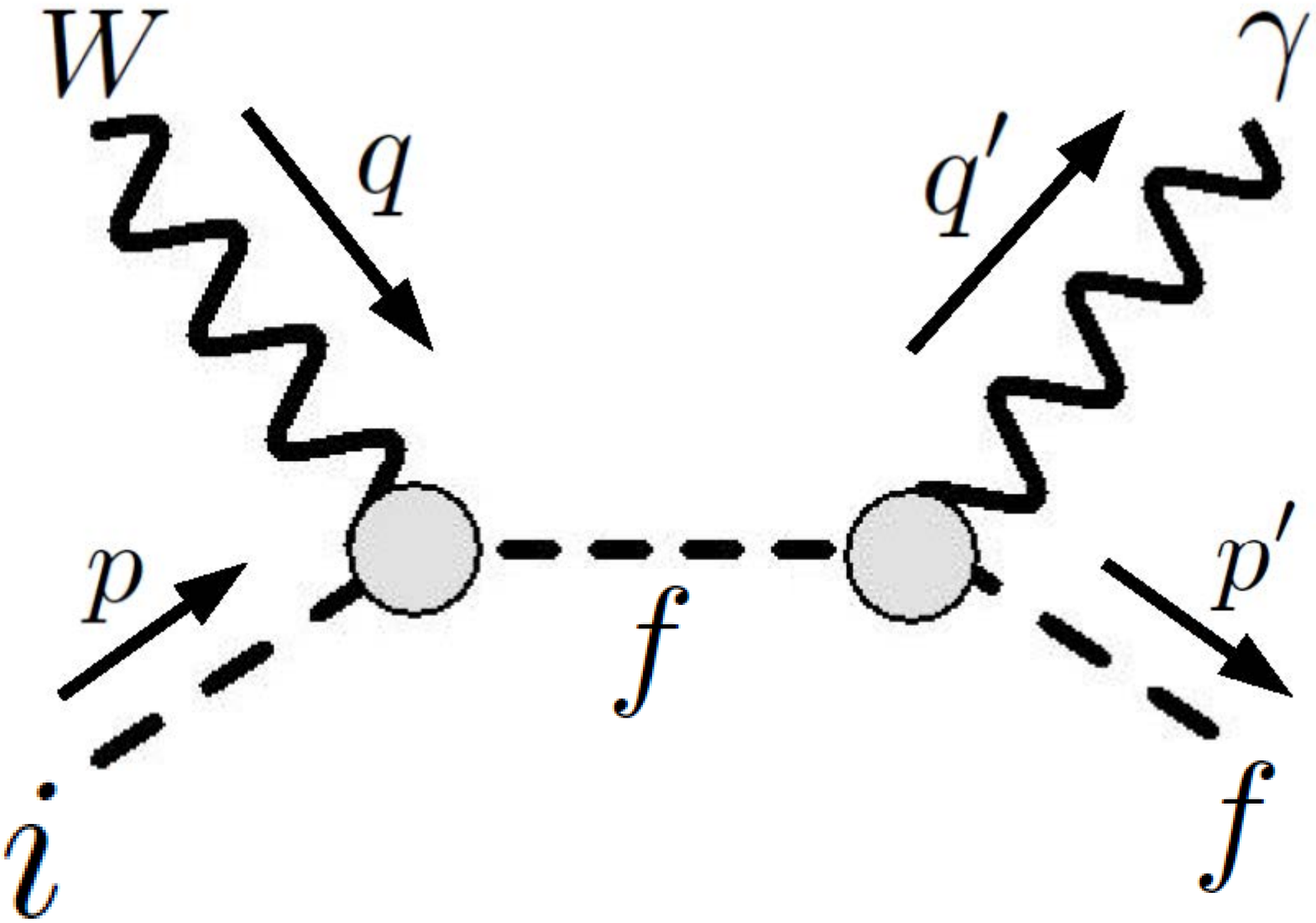}
		\includegraphics[scale=0.14]{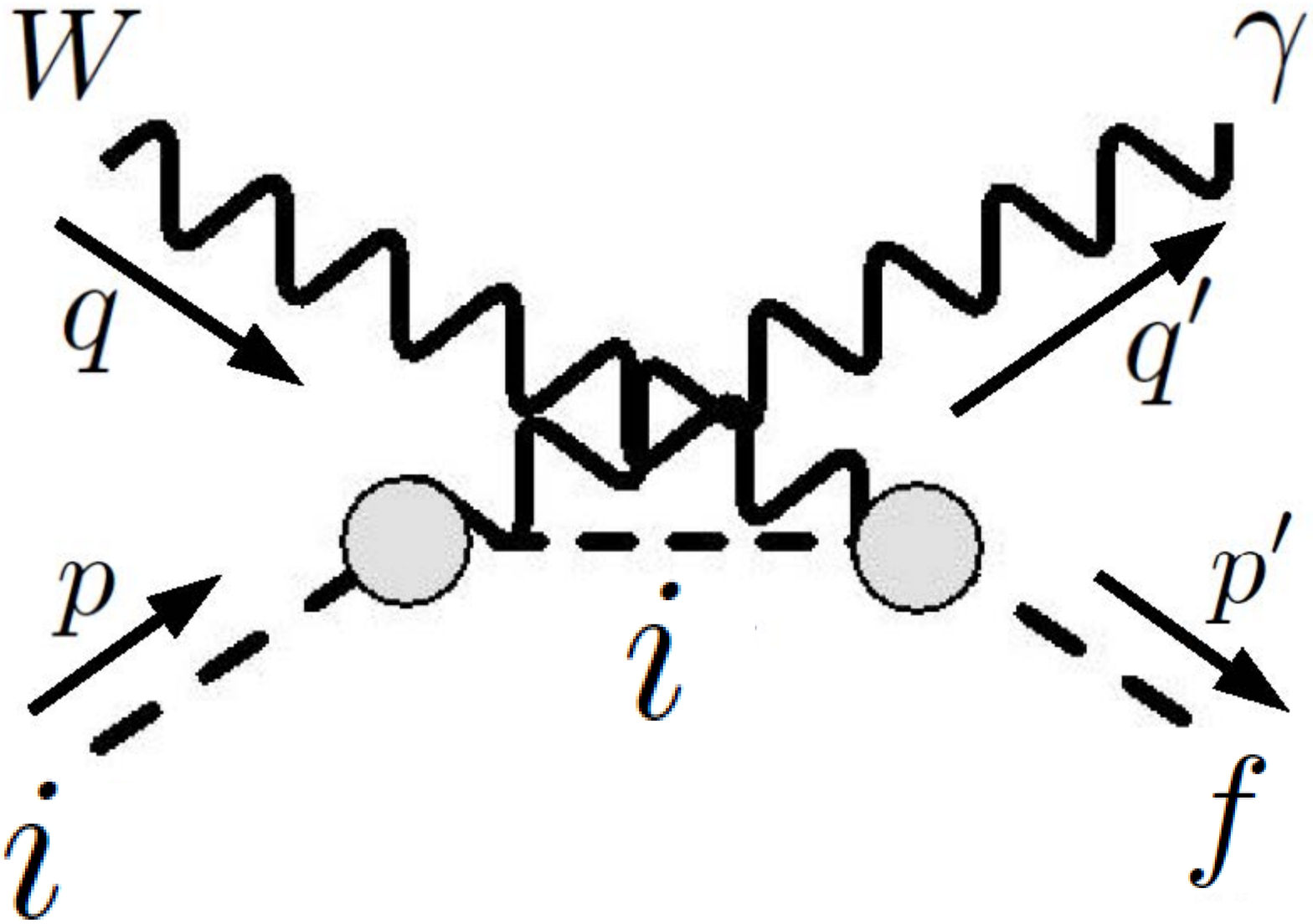}\hfill
		\par\end{centering}
	\caption{\label{fig:convection}Pole diagrams that give rise to the convection term in $T^{\mu\nu}$.}
\end{figure}

The tree-level decay rate above is corrected by EWRCs:
\begin{equation}
\left(\Gamma_{\pi_{e3}}\right)_\mathrm{tree}\rightarrow \Gamma_{\pi_{e3}}=\frac{G_F^2|V_{ud}|^2M_{\pi^+}^5}{64\pi^3}(1+\delta_\pi)I_\pi~,
\end{equation}
which effect is contained in the quantity $\delta_\pi$. 

The best starting point for the discussion of $\delta_\pi$ is the Sirlin's representation \eqref{eq:Sirlinrep}. First of all, we can split the charged weak current into the vector and axial components:
\begin{equation}
J^\mu=J_V^\mu+J_A^\mu~.\label{eq:Jsplit}
\end{equation}
Since pions are spinless, the matrix elements in $\delta F^\lambda_3$ can only involve $J_V^\mu$ and not $J_A^\mu$. As $\partial\cdot J_V=0$ in the isospin limit, we conclude that $\delta F_3^\lambda$ vanishes if we neglect the recoil corrections and strong isospin breaking effects on top of the RC, according to our discussions after Eq.\eqref{eq:F2F3}.

Next, one needs to evaluate the contributions from $\left(\delta \mathfrak{M}_2+\delta\mathfrak{M}_{\gamma W}^a\right)_\mathrm{int}$ and $\mathfrak{M}_\mathrm{brems}$. Due to the near-degeneracy between $\pi^+$ and $\pi^0$, it is easy to see that these terms are only probe the physics at the IR-end: $q'\sim E_e$. Therefore, the only relevant contribution to $T^{\mu\nu}$ comes from the pole diagrams in Fig.\ref{fig:convection}. In fact, further neglecting the $q'$-dependence in the electromagnetic and charged weak vertices give us the so-called ``convection term'' contribution to $T^{\mu\nu}$~\cite{Meister:1963zz}:
\begin{equation}
T^{\mu\nu}_\mathrm{conv}(q';p',p)=\frac{iZ_f(2p'+q')^\mu F^\nu(p',p)}{(p'+q')^2-M_f^2}+\frac{iZ_i(2p-q')^\mu F^\nu(p',p)}{(p-q')^2-M_i^2}\label{eq:Tmunuconv}
\end{equation} 
which is the simplest structure that satisfies the exact EM Ward identity (the first line in Eq.\eqref{eq:Ward}), and thus gives the full IR-divergent structures in both the loop and phase-space integrals. As a conclusion, to evaluate $\left(\delta \mathfrak{M}_2+\delta\mathfrak{M}_{\gamma W}^a\right)_\mathrm{int}$ and $\mathfrak{M}_\mathrm{brems}$ for near-degenerate decays it is sufficient to replace $T^{\mu\nu}$ by $T^{\mu\nu}_\mathrm{conv}$. By doing so, the charged weak matrix element $F^\nu$ is independent of the photon momentum, so the integrals can be performed analytically. 

Finally, we need also to evaluate $\delta \mathfrak{M}_{\gamma W}^b$. This term is UV- and IR-finite and contributes generically to a $\sim 10^{-3}$ correction to the tree-level decay rate. For near-degenerate beta decays, one may further simplify the integral by taking the ``forward limit'', i.e. setting $M_f=M_i=M$, $p'=p$ (with $p^2=M^2$) and $m_e,E_e\rightarrow 0$. The induced error of such an approximation scales generically as $E_e/\Delta E$ where $\Delta E$ is the energy splitting between the ground state and the first excited state that contribute to $T^{\mu\nu}$. When $\phi_{i,f}$ are hadrons (e.g. pion, kaon or nucleon), $\Delta E$ is at least $M_\pi$ so the $E_e/\Delta E$ corrections are negligible; however, more care needs to be taken for nuclear beta decays as nuclear excitations may have much smaller energy gaps. In any case, in the forward limit, $\delta \mathfrak{M}_{\gamma W}^b$ for spinless system reads:
\begin{equation}
\delta \mathfrak{M}_{\gamma W}^b=\Box_{\gamma W}(\phi_i,\phi_f,M) \mathfrak{M}_0~,
\end{equation}
where
\begin{equation}
\Box_{\gamma W}(\phi_i,\phi_f,M)\equiv \frac{ie^2}{2M^2}\int\frac{d^4q}{(2\pi)^4}\frac{M_W^2}{M_W^2-q^{2}}\frac{1}{q^{2}}\epsilon^{\mu\nu\alpha\beta}q_\alpha p_{\beta}\frac{T_{\mu\nu}^{if}(q;p,p)}{F_+^{if}(0)}~.\label{eq:forwardbox}
\end{equation}
The quantity $\Box_{\gamma W}$ is the only component in the $\mathcal{O}(G_F\alpha)$ RCs to near-degenerate beta decays that contains large hadronic uncertainties. In $\pi_{e3}$, we simply substitute $\phi_i=\pi^+$, $\phi_f=\pi^0$ and $M=M_\pi$.  

Summarizing everything above, we can write $\delta_\pi$ as:
\begin{equation}
\delta_\pi=\frac{\alpha}{2\pi}\left[\bar{g}(E_m)+3\ln\frac{M_Z}{m_p}+\ln\frac{M_Z}{M_W}+\tilde{a}_g\right]+\delta_\mathrm{HO}^\mathrm{QED}+2\Box_{\gamma W}(\pi^+,\pi^0,M_\pi)~,\label{eq:deltapi}
\end{equation}
where~\cite{Wilkinson:1970cdv}
\begin{eqnarray}
\bar{g}(E_m)&=&m_e^{-5}f^{-1}\int_{m_e}^{E_m}\sqrt{E_e^2-m_e^2}E_e(E_m-E_e)^2F(Z_f,\beta)g(E_e,E_m)dE_e\nonumber\\
f&=&m_e^{-5}\int_{m_e}^{E_m}\sqrt{E_e^2-m_e^2}E_e(E_m-E_e)^2F(Z_f,\beta)dE_e~,
\end{eqnarray}
(notice that the Fermi's function $F(Z_f,m_e)$ is unity in $\pi_{e3}$ because $Z_f=0$) with $E_m=(M_i^2-M_f^2+m_e^2)/(2M_i)$ the electron's end-point energy, and~\cite{Sirlin:1967zza}
\begin{eqnarray}
g(E_e,E_m)&=&3\ln\frac{m_p}{m_e}-\frac{3}{4}+4\left(\frac{1}{\beta}\tanh^{-1}\beta-1\right)\left(\ln\frac{2(E_\mathrm{m}-E_e)}{m_e}+\frac{E_m-E_e}{3E_e}-\frac{3}{2}\right)\nonumber\\
&&-\frac{4}{\beta}\mathrm{Li}_2\left(\frac{2\beta}{1+\beta}\right)+\frac{1}{\beta}\tanh^{-1}\beta\left(2+2\beta^2+\frac{(E_m-E_e)^2}{6E_e^2}-4\tanh^{-1}\beta\right)\nonumber\\\label{eq:Sirlinfunction}
\end{eqnarray}
is known customarily as the Sirlin's function. The appearance of the proton mass $m_p$ in $g(E_e,E_m)$ is nothing but a convention, as it cancels with the $3\ln(M_Z/m_p)$ term in $\delta_\pi$. 

From Eq.\eqref{eq:deltapi} one finds that the only unknown quantity is $\Box_{\gamma W}(\pi^+,\pi^0,M_\pi)$, which has taken more than 40 years for the physics community to finally achieve a high-precision determination, which we will describe in the following.

\subsection{Early numerical estimation}

The first numerical estimation of $\delta_\pi$ was done in 1978 by Ref.\cite{Sirlin:1977sv}. Given the precision goal at that time, both  $\delta_\mathrm{HO}^\mathrm{QED}$ and $\tilde{a}_g$ were discarded. Meanwhile, not much was known about $\Box_{\gamma W}$ except its electroweak logarithm structure as we described in Eq.\eqref{eq:MbUV}.
With this, $\Box_{\gamma W}$ was parameterized as:
\begin{equation}
\Box_{\gamma W}(\pi^+,\pi^0,M_\pi)=\frac{\alpha}{2\pi}\left[\frac{1}{2}\ln\frac{M_W}{M_A}+\frac{1}{2}a_{\bar{g}}+C\right]~,\label{eq:pionboxSirlin}
\end{equation}
where $M_A$ is an IR-cutoff scale below which the free-field OPE fails to work, $a_{\bar{g}}$ is the pQCD correction to the electroweak log, and $C$ represents the contribution from the non-perturbative QCD at $Q^2<M_A^2$. With this parameterization, one can write:
\begin{equation}
\delta_\pi\approx\frac{\alpha}{2\pi}\left[\bar{g}(E_m)+3\ln\frac{M_Z}{m_p}+\ln\frac{M_Z}{M_A}+2C+a_{\bar{g}}\right]~,
\end{equation}
which is a type of parameterization that frequently appeared in early literature. 
Ref.\cite{Sirlin:1977sv} did a rough numerical estimation with a (outdated) $Z$-boson mass of $78.2$~GeV together with $M_A=1.3$~GeV and $C\approx a_{\bar{g}}\approx 0$, and obtained $\delta_\pi \approx 3.1\times 10^{-2}$ without any estimation of the theory uncertainty. 

\subsection{ChPT treatment}

\begin{figure}
	\begin{centering}
		\includegraphics[scale=0.12]{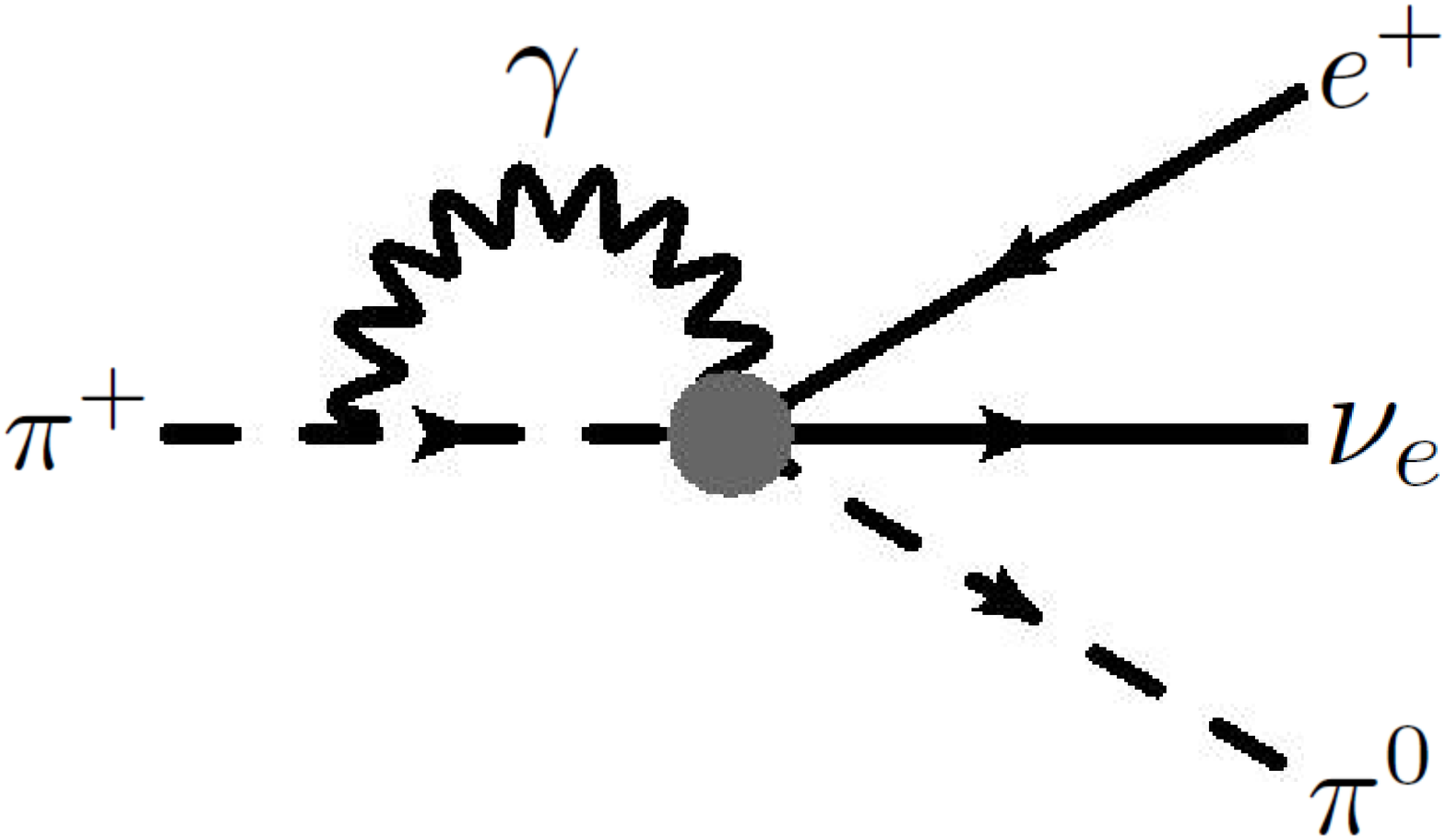}
		\includegraphics[scale=0.12]{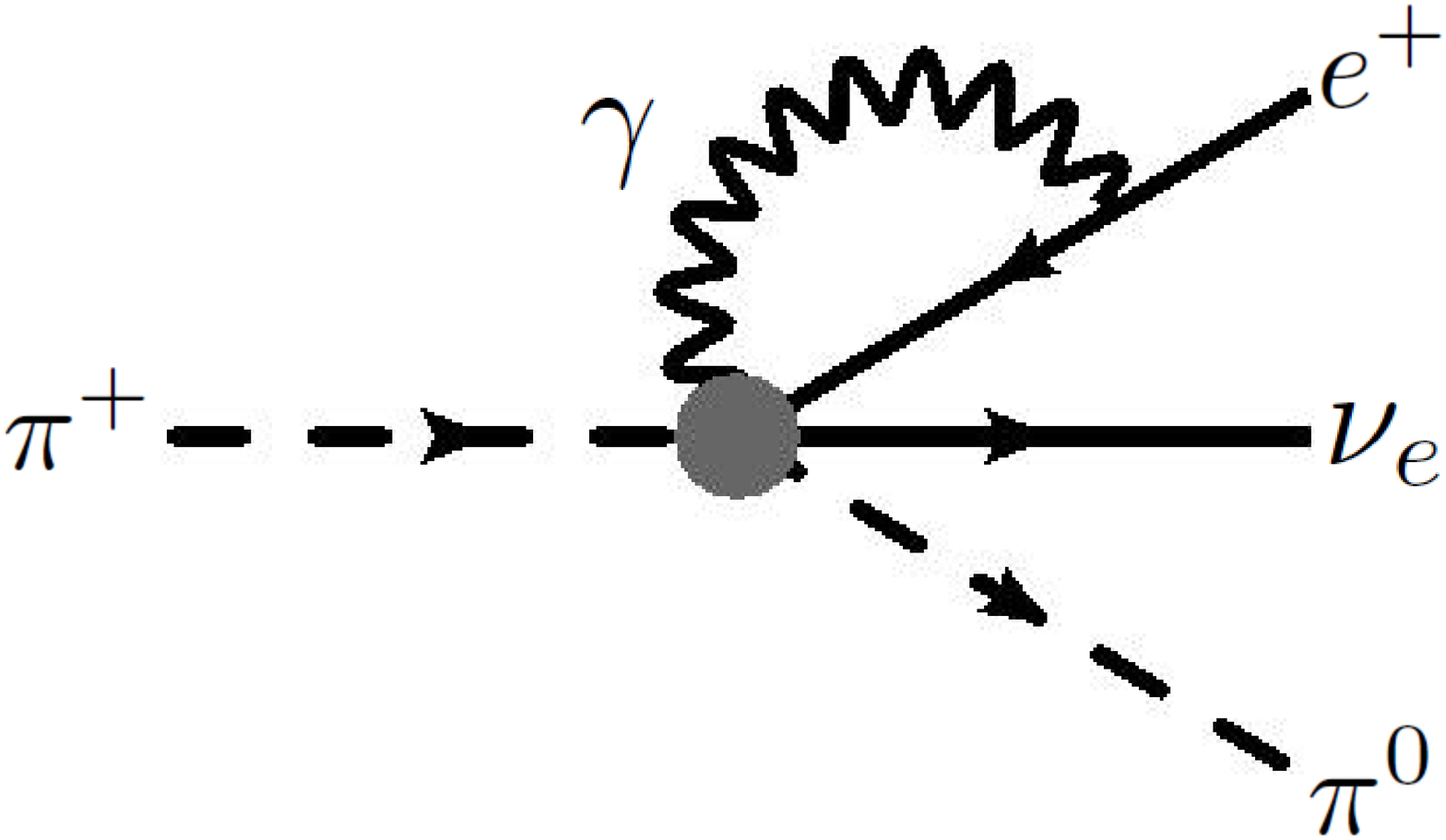}
		\includegraphics[scale=0.12]{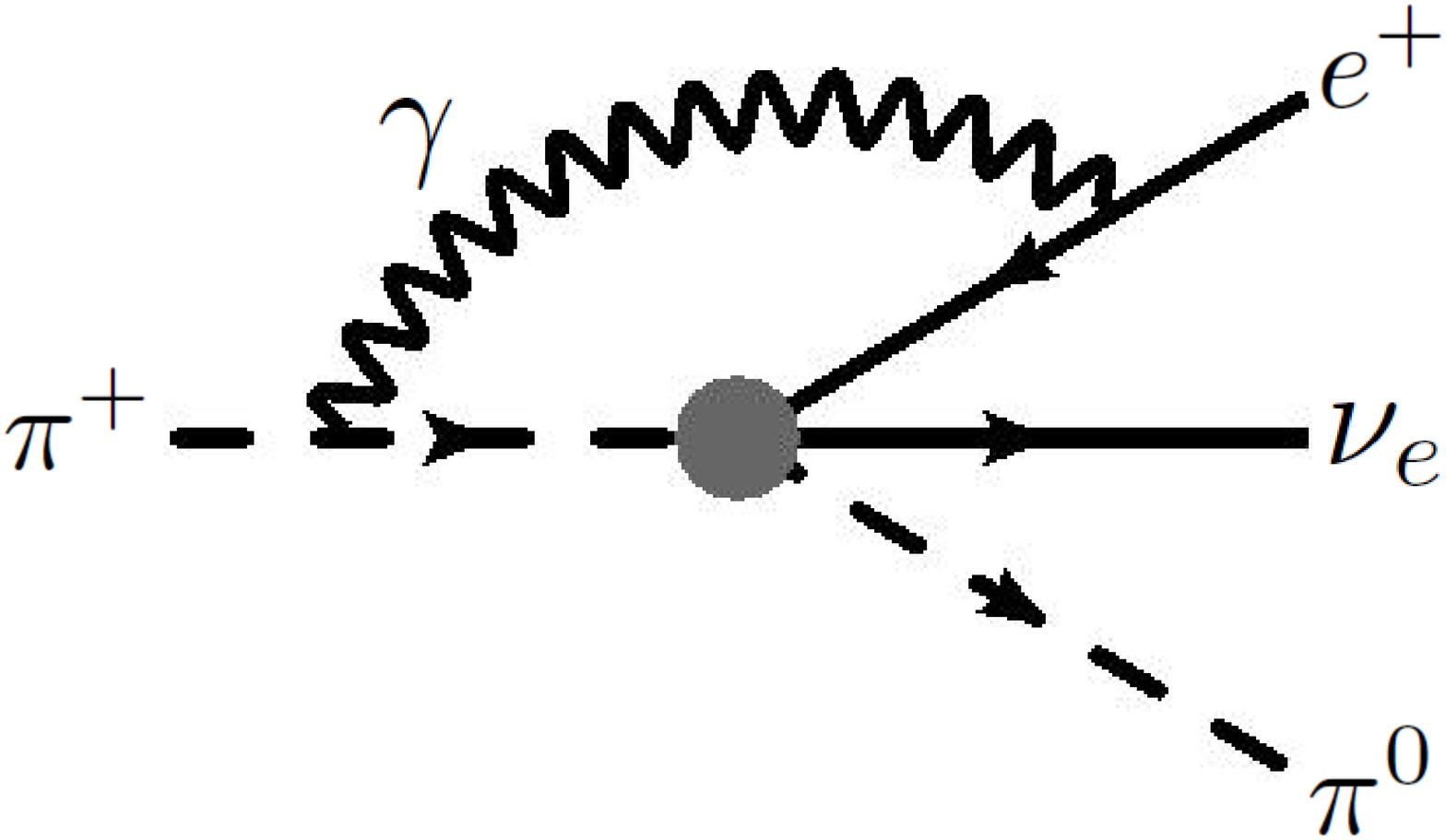}\hfill
		\par\end{centering}
	\caption{\label{fig:pion}1PI diagrams in the $\pi_{e3}$ EMRC within ChPT. }
\end{figure}

Independent studies of $\delta_\pi$ using ChPT was performed at the beginning of the 21st century~\cite{Cirigliano:2002ng,Cirigliano:2003yr}. Instead of the Sirlin's representation, the calculation was based on the chiral Lagrangian in Sec.\ref{sec:EFT}, where the dynamical DOFs are the pNGBs, leptons and the (full) photon. One-loop photonic contribution to the $\pi^+$ and $e^+$ wavefunction renormalization as well as the one-particle irreducible diagrams (1PIs) in Fig.\ref{fig:pion} and the bremsstrahlung process were calculated, with the inclusion of the LECs $\{K_i,X_i\}$ that cancel the UV-divergences. The procedures above led to a theory prediction of $\delta_\pi$ to $\mathcal{O}(e^2p^2)$.  

The ChPT parameterization of the $\pi_{e3}$ EWRCs is as follows:
\begin{equation}
\delta_\pi=\left(S_\mathrm{EW}-1\right)+2\delta_\mathrm{EM}^\pi+\Delta I_\pi~.
\end{equation}
There are three terms at the right hand side. The first term describes the universal short-distance electroweak corrections, with $S_\mathrm{EW}$ explained in Sec.\ref{sec:EWlog}. In the language of ChPT, the large electroweak logarithm and its associated pQCD corrections is contained in the following combination of LECs (we take the $\rho$-meson mass as the renormalization scale, i.e. $\mu=M_\rho$, following the standard choice):
\begin{eqnarray}
X_6^\mathrm{phys}(M_\rho)&\equiv& X_6^r(M_\rho)-4K_{12}^r(M_\rho)\nonumber\\
&\risingdotseq&\tilde{X}_6^\mathrm{phys}(M_\rho)-\frac{1}{2\pi^2}\left(1-\frac{\alpha_s}{4\pi}\right)\ln\frac{M_Z}{M_\rho}~,\label{eq:X6phys}
\end{eqnarray}
where $\tilde{X}_6^\mathrm{phy}$ is expected to be of natural size. Meanwhile, $\delta_\mathrm{HO}^\mathrm{QED}$ is usually not defined as a part of the LECs, and should be added separately to the decay rate. The remaining two terms encode the effects of the long-distance EMRC: $\delta_\mathrm{EM}^\pi$ describes the correction to $f_+^{\pi^+\pi^0}(0)$, while $\Delta I_\pi$ describes the correction to the phase space factor $I_\pi$. 

As in all fixed-order ChPT calculations, the theory uncertainties in the $\delta_\pi$ determined above come from (1) higher-order chiral corrections and (2) LECs. Since here we are working with a two-flavor theory, the higher-order corrections are expected to scale as $M_\pi^2/\Lambda_\chi^2\sim 10^{-2}$ which is small, so the theory uncertainty comes mainly from the LECs that goes into $\delta_\mathrm{EM}^\pi$:
\begin{equation}
\delta_\mathrm{EM}^\pi=e^2\left[-\frac{2}{3}X_1-\frac{1}{2}\tilde{X}_6^\mathrm{phys}(M_\rho)\right]+...~.
\end{equation}
(Notice that $X_1$ is scale-independent, so $X_1^r=X_1$) Combining simple model calculations of $K_{12}^r$~\cite{Moussallam:1997xx} and 
rough estimations of the upper bounds of $X_1,X_6^r$ using na\"{\i}ve dimensional analysis, Ref.\cite{Cirigliano:2003yr} quoted the following result:
\begin{equation}
\delta_\mathrm{EM}^\pi=(0.46\pm 0.05)\%~,\:\:\Delta I_\pi=0.1\%~,
\end{equation}
which leads to $\delta_\pi=0.0334(3)_\mathrm{HO}(10)_\mathrm{LEC}$~\cite{Knecht:2004xr}.

From the discussions in Sec.\ref{sec:pionEWRC}, it is clear that the uncertainties of the aforementioned LECs must originate from the $\gamma W$-box diagram. Therefore, a high-precision determination of $\Box_{\gamma W}(\pi^+,\pi^0,M_\pi)$ will similarly pin down $(2/3)X_1+(1/2)\tilde{X}_6^\mathrm{phys}$. This was only made possible very recently through a combination of pQCD and lattice calculation we will describe below. 

\subsection{\label{sec:latticeQCD}First-principles calculation}

An important breakthrough was achieved in Ref.\cite{Feng:2020zdc}, where $\delta_\pi$ was calculated to a $10^{-4}$ precision with a first-principles approach. Here we briefly describe the procedure. The first step is to write $\Box_{\gamma W}$ as an integral with respect to $Q^2=-q^2$:
\begin{equation}
\Box_{\gamma W}(\pi^+,\pi^0,M_\pi)=\frac{3\alpha}{2\pi}\int_0^\infty\frac{dQ^2}{Q^2}\frac{M_W^2}{M_W^2+Q^2}\mathbb{M}_\pi(Q^2)~,
\end{equation}
so to precisely determine the box diagram one needs to know the function $\mathbb{M}_\pi(Q^2)$ at all values of $Q^2$. The key to proceed is to identify a separation scale $Q_\mathrm{cut}^2$, above which the partonic description of $T^{\mu\nu}$ (with pQCD corrections) works reasonably well. As we will see later, both the direct lattice calculation and the experimental analysis of the Gross-Llewellyn-Smith (GLS) sum rule~\cite{Gross:1969jf} suggest that $Q_\mathrm{cut}^2=2$~GeV$^2$ is a valid choice. One therefore splits $\Box_{\gamma W}$ into two pieces:
\begin{equation}
\Box_{\gamma W}(\pi^+,\pi^0,M_\pi)=\Box_{\gamma W}^>+\Box_{\gamma W}^<(\pi^+,\pi^0,M_\pi)
\end{equation}
which represent the contribution from $Q^2>Q_\mathrm{cut}^2$ and $Q^2<Q_\mathrm{cut}^2$ respectively, and the former is process-independent. We shall discuss these two terms separately.

\subsubsection{Large-$Q^2$ contribution}

From the discussions in the previous sections, we know that 
\begin{equation}
\mathbb{M}_\pi(Q^2)\rightarrow \frac{1}{12}\left(1-\frac{\alpha_s(Q^2)}{\pi}+\mathcal{O}(\alpha_s^2)\right)
\end{equation}
at large $Q^2$. However, knowing that the size of $\alpha_s$ increases with decreasing $Q^2$, the $\mathcal{O}(\alpha_s)$ correction itself is not sufficient to achieve the required precision goal in extending $Q^2$ down to $Q^2_\mathrm{cut}$. Higher-order pQCD corrections are needed.

Ref.\cite{Marciano:2005ec} made an important observation that, in the chiral limit, the pQCD correction to $\Box_{\gamma W}^>$  is the same to all orders as the pQCD correction to the polarized Bjorken sum rule~\cite{Bjorken:1966jh,Bjorken:1969mm}. This allows us to directly apply the existing theory analysis of the latter to our case. A proof of the statement above is provided in Appendix.\ref{sec:pQCDmatch}.
With such, we can write:
\begin{equation}
\Box_{\gamma W}^>=\frac{3\alpha}{2\pi}\int_{Q_\mathrm{cut}^2}^\infty\frac{dQ^2}{Q^2}\frac{M_W^2}{M_W^2+Q^2}\frac{1}{12}C_\mathrm{Bj}(Q^2)~.
\end{equation}
The pQCD correction factor $C_\mathrm{Bj}(Q^2)$ is ordered in increasing powers of $\alpha_s$:
\begin{equation}
C_\mathrm{Bj}(Q^2)=1-\sum_{n=1}^\infty c_n\left(\frac{\alpha_S}{\pi}\right)^n~,\label{eq:BjpQCD}
\end{equation}
and the expansion coefficients are currently determined up to $n=4$~\cite{Baikov:2010iw,Baikov:2010je}:
\begin{eqnarray}
c_1&=&1\nonumber\\
c_2&=&4.583-0.3333n_f\nonumber\\
c_3&=&41.44-7.607n_f+0.1775n_f^2\nonumber\\
c_4&=&479.4-123.4n_f+7.697n_f^2-0.1037n_f^3~,
\end{eqnarray} 
with $n_f$ the number of active quark flavors. With that, one may obtain $C_\mathrm{Bj}$ as a function of $Q^2$ by substituting the running strong coupling constant $\alpha_S(Q^2)$ in the $\overline{\mathrm{MS}}$ scheme; well-coded programs for the latter, e.g. the RunDec package~\cite{Chetyrkin:2000yt}, is available for public. Integrating over $Q^2$ yields $\Box_{\gamma W}^>=2.16\times 10^{-3}$ at $Q_\mathrm{cut}^2=2$~GeV$^2$. The theory uncertainty due the higher-order pQCD corrections was estimated from the difference between the $\mathcal
{O}(\alpha_s^3)$ and $\mathcal{O}(\alpha_s^4)$ results, and was found to be less than $1\times10^{-5}$.

\subsubsection{Small-$Q^2$ contribution}

\begin{figure}
	\begin{centering}
		\includegraphics[scale=0.27]{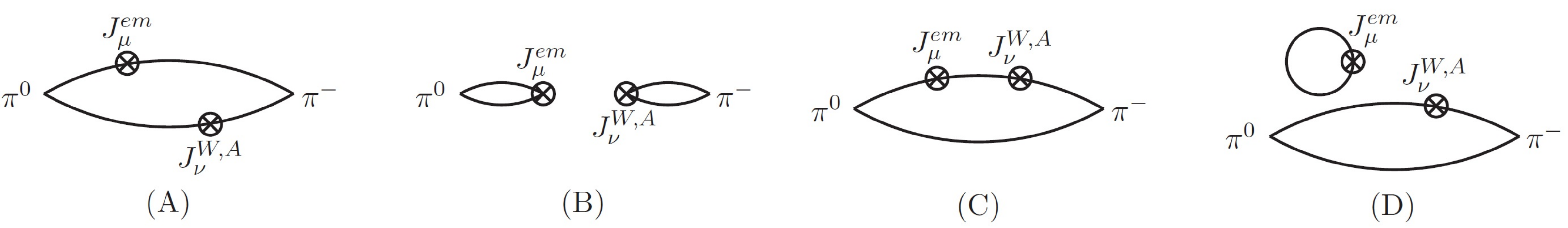}\hfill
		\par\end{centering}
	\caption{\label{fig:contraction}Contraction diagrams for $T^{\mu\nu}$ calculated on lattice. Diagram (D) vanishes in the flavor SU(3) limit ans was discarded in the calculation. Figure taken from Ref.\cite{Feng:2020zdc}. }
\end{figure}

\begin{figure}
	\begin{centering}
		\includegraphics[scale=0.3]{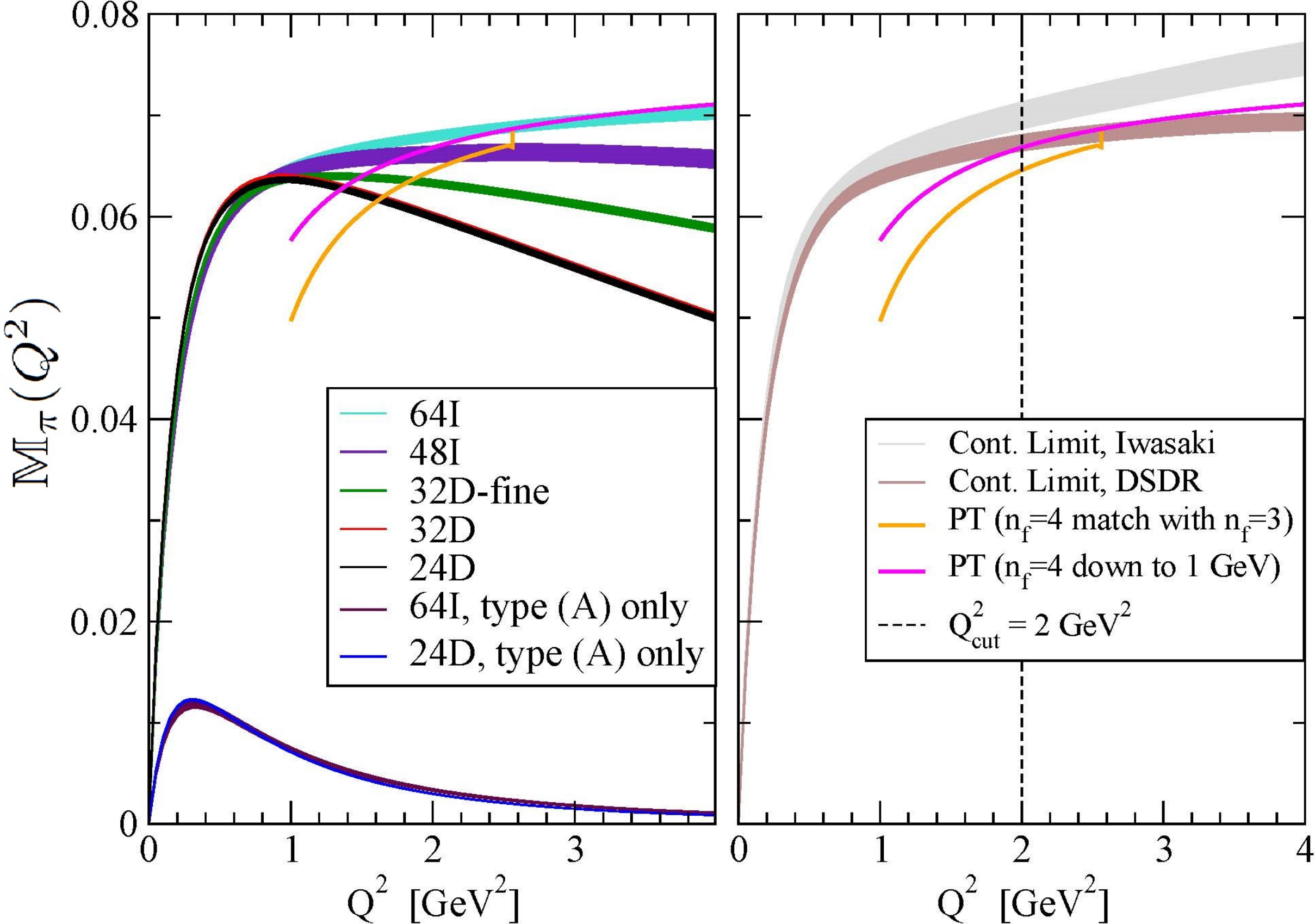}\hfill
		\par\end{centering}
	\caption{\label{fig:latresult}Main outcomes for the lattice calculation of $\mathbb{M}_\pi(Q^2)$ in Ref.\cite{Feng:2020zdc} before (left) and after (right) continuum extrapolation.}
\end{figure}

At $Q^2<Q_\mathrm{cut}^2$, reliable theory inputs for $\mathbb{M}_\pi(Q^2)$ comes from the direct lattice calculation of $T^{\mu\nu}(q;p,p)$ that involves four-point correlation functions depicted in Fig.\ref{fig:contraction}. Ref.\cite{Feng:2020zdc} (see also. Ref.\cite{Seng:2021qdx} for a summary) performed the first calculation of such kind with five lattice QCD gauge ensembles (DSDR and Iwasaki gauge actions) at the physical pion mass using 2+1 flavor domain wall fermions. The main outcomes are presented in Fig.\ref{fig:latresult}. One observes that, after continuum extrapolation, the lattice result of $\mathbb{M}_\pi(Q^2)$ continues smoothly to the pQCD theory prediction around $Q^2=2~$GeV$^2$, which justifies our choice of $Q_\mathrm{cut}^2$. The outcome at low-$Q^2$ reads:
\begin{equation}
\Box_{\gamma W}^<(\pi^+,\pi^0,M_\pi)=\left\{ \begin{array}{ll}
0.671(11)\times10^{-3}~,&\mathrm{Iwasaki}\\
0.647(7)\times10^{-3}~,&\mathrm{DSDR}
\end{array}\right.
\end{equation}
where the main uncertainty comes from the lattice discretization effect. Combining the ``$>$'' and ``$<$'' contributions, Ref.\cite{Feng:2020zdc} quoted the following final result:
\begin{equation}
\Box_{\gamma W}(\pi^+,\pi^0,M_\pi)=2.830(11)_\mathrm{stat}(26)_\mathrm{syst}\times 10^{-3}~,
\end{equation}
with the total uncertainty controlled at the level of $\sim$1\%. This translated into 
\begin{equation}
\delta_\pi=0.0332(1)_{\gamma W}(3)_\mathrm{HO}~,
\end{equation}
which is currently the best determination of $\delta_\pi$. It is consistent with the ChPT determination, but 3 times better in precision.

It was later pointed out in Ref.\cite{Seng:2020jtz} that, equating the expressions of $\delta_\pi$ between the ChPT representation at $\mathcal{O}(e^2p^2)$ and the Sirlin's representation provides a matching relation between the LECs and $\Box_{\gamma W}$. In the case of pion, the matching reads:
\begin{eqnarray}
\frac{4}{3}X_1+\bar{X}_6^\mathrm{phys}(M_\rho)&=&-\frac{1}{2\pi\alpha}\left(\Box_{\gamma W}(\pi^+,\pi^0,M_\pi)-\frac{\alpha}{8\pi}\ln\frac{M_W^2}{M_\rho^2}\right)+\frac{1}{8\pi^2}\left(\frac{5}{4}-\tilde{a}_g\right)\nonumber\\
&=&0.0140(6)_{\gamma W}(8)_\mathrm{ChPT}\label{eq:pimatch}
\end{eqnarray}
where we have defined
\begin{equation}
\bar{X}_6^\mathrm{phys}(M_\rho)\equiv X_6^\mathrm{phys}(M_\rho)+\frac{1}{2\pi^2}\ln\frac{M_Z}{M_\rho}
\end{equation}
that removes only the large electroweak logarithm but retain the pQCD corrections to all orders.  Eq.\eqref{eq:pimatch} serves as the first rigorous determination of the $\mathcal{O}(e^2p^2)$ LECs with controlled theory uncertainties, which come from the $\gamma W$-box diagram (lattice) and the neglected higher-order ChPT corrections.

We end with a short comment on the future prospects. With the new lattice result above, $\pi_{e3}$ has formally turned into the theoretically cleanest avenue for the measurement of the CKM matrix element $|V_{ud}|$. The main limitation, however, stems from the smallness of the $\pi_{e3}$ branching ratio $\sim 10^{-8}$ which also implies a large experimental uncertainty in its measurement. The current best measurement of BR($\pi_{e3}$) was obtained from the PIBETA experiment~\cite{Pocanic:2003pf}, and the deduced $|V_{ud}|$ is ten times less precise than that from superallowed nuclear decays. A next-generation experiment for rare pion decays known as PIENUX at TRIUMF was recently proposed~\cite{ArevaloSnowmass}, which aims to improve, among other things, the BR($\pi_{e3}$) precision by an order of magnitude. This will eventually make it the best avenue to extract $|V_{ud}|$.

\section{\label{sec:IJhalf}Beta decay of $I=J=1/2$ particles}

The second example we study is the beta decay of generic $I=J=1/2$ particles, including the free neutron and nuclear mirrors. It is of great interest for the  measurement of the matrix element $|V_{ud}|$ as well as the search of BSM physics. All relevant decays of such kind involve nearly-degenerate parent and daughter nuclei, so we may define a small scale  $\Delta\sim p-p'\sim p_e\ll m_{i,f}$ that quantifies the size of the recoil corrections.

We first define the electromagnetic and charged weak form factors assuming the absence of second-class currents~\cite{Weinberg:1958ut}:
\begin{eqnarray}
\langle \phi(p',s')|J^\mathrm{em}_\mu(0)|\phi(p,s)\rangle &=&\bar{u}_{s'}(p')\left[f_1^\phi \gamma_\mu-\frac{if_2^\phi}{2m}\sigma_{\mu\nu}(p-p')^\nu\right]u_s(p)\nonumber\\
\langle \phi_f(p',s')|J_\mu(0)|\phi_i(p,s)\rangle &=&V_{ud}^{(*)}\bar{u}_{s'}(p')\left[G_V \gamma_\mu-\frac{iG_M}{2m}\sigma_{\mu\nu}(p-p')^\nu\right.\nonumber\\
&&\left.+G_A\gamma_\mu\gamma_5+\frac{G_P}{2m}\gamma_5(p-p')_\mu\right]u_s(p)~,\label{eq:FFs}
\end{eqnarray}
where $m=(m_i+m_f)/2$. All form factors are functions of $-(p-p')^2$. Among all the charged weak form factors, the vector ($G_V$) and axial ($G_A$) form factors are the leading ones, and we are particularly interested in their values at $t=0$: $\mathring{g}_V\equiv G_V(0)$, $\mathring{g}_A\equiv G_A(0)$. 
Meanwhile, the weak magnetism ($G_M$) and the pseudoscalar ($G_P$) form factors are suppressed by $\Delta/m$ at tree-level.

\subsection{Outer and inner corrections}

We consider the beta decay of a polarized spin-1/2 particle to unpolarized final states. At tree-level, the differential decay rate possesses the following structures~\cite{Jackson:1957zz}:
\begin{equation}
\frac{d\Gamma}{d\Omega_ed\Omega_\nu dE_e}\propto 1+a_0\frac{\vec{p}_e\cdot\vec{p}_\nu}{E_eE_\nu}+\hat{s}\cdot\left[A_0\frac{\vec{p}_e}{E_e}+B_0\frac{\vec{p}_\nu}{E_\nu}\right]
\end{equation}
where $\hat{s}$ is the unit polarization vector of the parent nucleus, and 
\begin{equation}
a_0=\frac{1-\lambda_0^2}{1+3\lambda_0^2}~,\:A_0=\frac{-2(\lambda_0-\lambda_0^2\eta)}{1+3\lambda_0^2}~,\:B_0=\frac{-2(\lambda_0+\lambda_0^2\eta)}{1+3\lambda_0^2}~,
\end{equation}
with $\lambda_0=\mathring{g}_A/\mathring{g}_V$ the bare axial-to-vector coupling ratio.
The expression above is however modified by recoil corrections and EWRCs. The former is well-studied~\cite{Holstein:1974zf,Wilkinson:1982hu,Ando:2004rk,Gudkov:2008pf,Ivanov:2012qe,Ivanov:2020ybx} and we shall focus exclusively on the latter. 

The initial attempts to study the $\mathcal{O}(G_F\alpha)$ corrections relied on the calculation of elementary Feynman diagrams of one-loop QED corrections and bremsstrahlung~\cite{Sirlin:1967zza,Garcia:1981it}. The loop integrals were exactly calculable upon replacing $T^{\mu\nu}$ by its convection term~\eqref{eq:Tmunuconv}, while a Pauli-Villars regulator was introduced to regularize the UV-divergence. However, it turns out that the UV-divergences from the vertex corrections and the box diagram cancel each other, rendering the total result UV-finite. The corrected squared amplitude reads: 
\begin{eqnarray}
|\mathfrak{M}|^2_{\phi_i\rightarrow \phi_fe\nu(\gamma)}&=&16G_F^2|V_{ud}|^2m_im_fE_e(E_m-E_e)\mathring{g}_V^2(1+3\lambda_0^2)F(Z_f,\beta)\left(1+\frac{\alpha}{2\pi}\delta^{(1)}\right)\nonumber\\
&&\times\left\{1+\left(1+\frac{\alpha}{2\pi}\delta^{(2)}\right)a_0\frac{\vec{p}_e\cdot\vec{p}_\nu}{E_eE_\nu}+\hat{s}\cdot\left[\left(1+\frac{\alpha}{2\pi}\delta^{(2)}\right)A_0\frac{\vec{p}_e}{E_e}+B_0\frac{\vec{p}_\nu}{E_\nu}\right]\right\}\nonumber\\
&&+\mathcal{O}(\Delta^3)~,\label{eq:outer}
\end{eqnarray}
where
\begin{eqnarray}
\delta^{(1)}&=&g(E_e,E_m)\nonumber\\
\delta^{(2)}&=&2\left(\frac{1-\beta^2}{\beta}\right)\tanh^{-1}\beta+\frac{4(E_m-E_e)(1-\beta^2)}{3\beta^2E_e}\left(\frac{1}{\beta}\tanh^{-1}\beta-1\right)\nonumber\\
&&+\frac{(E_m-E_e)^2}{6\beta^2E_e^2}\left(\frac{1-\beta^2}{\beta}\tanh^{-1}\beta-1\right)~,
\end{eqnarray}
with $g(E_e,E_m)$ defined in Eq.\eqref{eq:Sirlinfunction}, and $F(Z_f,\beta)$ the Fermi's function given in Eq.\eqref{eq:Fermifun}.

The functions $\delta^{(1)}$, $\delta^{(2)}$ and $F(Z_f,\beta)$ \textit{do not} fully describe the $\mathcal{O}(G_F\alpha)$ EWRCs, but only capture the model-independent contributions (known as the ``outer corrections'') that are sensitive to physics at $q\sim E_e$ and are non-trivial functions of $E_e$. The remaining RCs are sensitive to the details of hadron structures, and are constants upon neglecting recoil corrections. In fact, they simply renormalize the vector and axial coupling constants~\cite{Sirlin:1967zza}:
\begin{equation}
g_V= \mathring{g}_V\left(1+\frac{\alpha}{2\pi}c\right)~,\quad g_A= \mathring{g}_A\left(1+\frac{\alpha}{2\pi}d\right)~.
\end{equation}
Therefore, the squared amplitude including the full $\mathcal{O}(G_F\alpha)$ RC is given by Eq.\eqref{eq:outer}, upon replacing $\mathring{g}_V\rightarrow g_V$ and $\mathring{g}_A\rightarrow g_A$. This includes replacing $\lambda$ by its renormalized version: $\lambda_0\rightarrow\lambda= g_A/g_V$.  

The constants $c$ and $d$ represent the so-called ``inner corrections'' and were only quantifiable after the establishment of the full SM. Given the near-degenerate nature of the decay, the best starting point is again the Sirlin's representation we introduced in Sec.\ref{sec:Sirlin}. Among all the unevaluated one-loop corrections in Eq.\eqref{eq:Sirlinrep}, $(\delta\mathfrak{M}_2+\delta\mathfrak{M}_{\gamma W}^a)_\mathrm{int}$ and $\delta F_3^\lambda$ probe only the physics at the IR-region and could be calculated model-independently; in particular, the $J^\mu_V$ contribution to $\delta F_3^\lambda$ vanishes in the degenerate limit and is practically negligible. One could verify that the Sirlin's representation reproduces exactly the same outer corrections as in Eq.\eqref{eq:outer}, and at the same time also provides a rigorous definition of the inner corrections in terms of hadronic matrix elements of electroweak currents. It reads:
\begin{eqnarray}
g_V&=&\mathring{g}_V\left\{1+\frac{\alpha}{4\pi}\left[3\ln\frac{M_Z}{m_p}+\ln\frac{M_Z}{M_W}+\tilde{a}_g\right]+\frac{1}{2}\delta_\mathrm{HO}^\mathrm{QED}+\Box_{\gamma W}^V\right\}\nonumber\\
g_A&=&\mathring{g}_A\left\{1+\frac{\alpha}{4\pi}\left[3\ln\frac{M_Z}{m_p}+\ln\frac{M_Z}{M_W}+\tilde{a}_g\right]+\frac{1}{2}\delta_\mathrm{HO}^\mathrm{QED}+\Box_{\gamma W}^A\right\}~,\label{eq:innerRC}
\end{eqnarray}
and as a consequence, the renormalized axial-to-vector coupling ratio reads:
\begin{equation}
\lambda=\frac{\mathring{g}_A}{\mathring{g}_V}\left(1+\Box_{\gamma W}^A-\Box_{\gamma W}^V\right)~.\label{eq:lambdaRC}
\end{equation}
The only unknown quantities in the equation above are $\Box_{\gamma W}^V$ and $\Box_{\gamma W}^A$, which originates from $\delta \mathfrak{M}_{\gamma W}^b$:
\begin{equation}
\delta \mathfrak{M}_{\gamma W}^b\equiv-\frac{G_F}{\sqrt{2}}L_\lambda\bar{u}_{s'}(p)\gamma^\lambda\left[\mathring{g}_V\Box_{\gamma W}^V+\gamma_5\mathring{g}_A\Box_{\gamma W}^A\right]u_s(p)~.
\end{equation} 
Knowing that $\delta\mathfrak{M}_{\gamma W}^b$ is IR-finite, we have set set $m_i=m_f=m$, $m_e\rightarrow 0$ and $p'\rightarrow p$ at both sides. To process further, we may set $s'=s$ at both sides, and make use of the following spinor identities:
\begin{equation}
\bar{u}_s(p)\gamma^\mu u_s(p)=2p^\mu~,\quad \bar{u}_s(p)\gamma^\mu\gamma_5 u_s(p)=2s^\mu~,
\end{equation} 
which defines the spin vector $s^\mu$ that satisfies $s^2=m^2$ and $p\cdot s=0$. This leads to:
\begin{eqnarray}
\Box_{\gamma W}^V&=&\frac{ie^2}{2m^2\mathring{g}_V}\int\frac{d^4q}{(2\pi)^4}\frac{M_W^2}{M_W^2-q^2}\frac{\epsilon^{\mu\nu\alpha\lambda}q_\alpha p_\lambda}{(q^2)^2}T_{\mu\nu}^{ss}(q;p,p)\nonumber\\
\Box_{\gamma W}^A&=&-\frac{ie^2}{2m^2\mathring{g}_A}\int\frac{d^4q}{(2\pi)^4}\frac{M_W^2}{M_W^2-q^2}\frac{\epsilon^{\mu\nu\alpha\lambda}q_\alpha s_\lambda}{(q^2)^2}T_{\mu\nu}^{ss}(q;p,p)~,\label{eq:BoxVBoxA}
\end{eqnarray}
where we have displayed explicitly the momenta and spins in the generalized Compton tensor to remind the readers about the limits taken. Therefore, the study of the $\mathcal{O}(G_F\alpha)$ inner RCs boil down to the calculation of the two well-defined hadronic matrix elements at the right hand side of the equation above.

\subsection{\label{sec:DRrep}Dispersive representation}

The precise calculation of the full integrals in Eq.\eqref{eq:BoxVBoxA} is highly non-trivial as they probe the hadron physics contained in $T_{\mu\nu}^{ss}$ at all values of $q$. In the absence of analytic solutions of QCD in the non-perturbative regime, there are only two ways to proceed with controlled theory uncertainties:
\begin{itemize}
	\item First-principles calculation with lattice QCD, in analogy to the calculation of $\Box_{\gamma W}(\pi^+,\pi^0,M_\pi)$ described in Sec.\ref{sec:latticeQCD};
	\item Data-driven analysis that relates the hadronic matrix elements to experimental observables. 
\end{itemize} 
In this subsection we introduce the dispersive representation of $\Box_{\gamma W}^V$ and $\Box_{\gamma W}^A$, which is the starting point for a fully data-driven analysis. 

Our first important observation is that, the component of $T_{\mu\nu}^{ss}$ that contributes to the integral in Eq.\eqref{eq:BoxVBoxA} must contain an antisymmetric tensor. Such a component can be parameterized as follows:
\begin{equation}
V_{ud}^{(*)-1}T_{\mu\nu}^{ss}(q;p,p)=-i\epsilon_{\mu\nu\alpha\beta}\frac{q^\alpha p^\beta}{2p\cdot q}T_3+i\epsilon_{\mu\nu\alpha\beta}\frac{q^\alpha}{p\cdot q}\left[s^\beta S_1+\left(s^\beta-\frac{s\cdot q}{p\cdot q}p^\beta\right)S_2\right]+...~,\label{eq:invariant}
\end{equation}
which defines the invariant amplitudes $T_3$, $S_1$ and $S_2$ that are functions of $Q^2=-q^2$ and $\nu=p\cdot q/m$. Plugging Eq.\eqref{eq:invariant} into Eq.\eqref{eq:BoxVBoxA} yields:
\begin{eqnarray}
\Box_{\gamma W}^V&=& \frac{e^2}{2m\mathring{g}_V}\int\frac{d^4q}{(2\pi)^4}\frac{M_W^2}{M_W^2+Q^2}\frac{1}{(Q^2)^2}\frac{\nu^2+Q^2}{\nu}T_3(\nu,Q^2)\nonumber\\
\Box_{\gamma W}^A&=& \frac{e^2}{m\mathring{g}_A}\int\frac{d^4q}{(2\pi)^4}\frac{M_W^2}{M_W^2+Q^2}\frac{1}{(Q^2)^2}\left\{\frac{\nu^2-2Q^2}{3\nu}S_1(\nu,Q^2)-\frac{Q^2}{\nu}S_2(\nu,Q^2)\right\}~,\nonumber\\
\label{eq:gVgAinv}
\end{eqnarray}
from which we made the second important observation, namely only components of the invariant amplitudes that are odd with respect to $\nu\rightarrow -\nu$ can contribute to the integral. Using isospin symmetry, one can show that they can only be contributed by the isoscalar component of the electromagnetic current (see Appendix \ref{sec:crossing} for a derivation), so we may add a superscript (0) to the invariant amplitudes.

\begin{figure}
	\begin{centering}
		\includegraphics[scale=0.5]{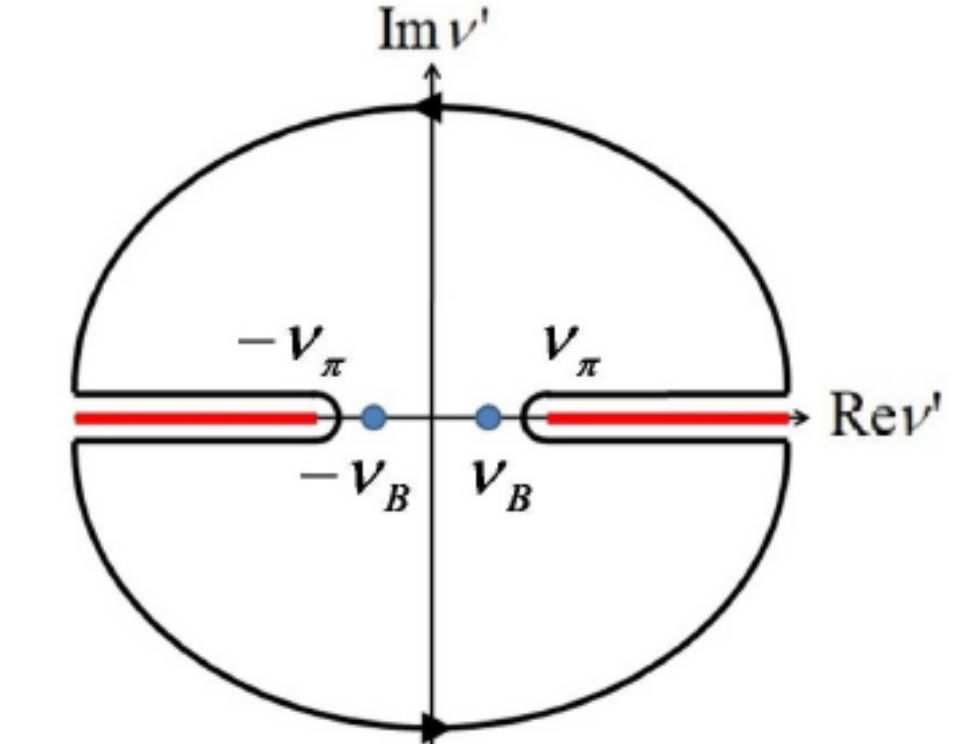}\hfill
		\par\end{centering}
	\caption{\label{fig:contour}Positions of the poles and cuts of the single-nucleon invariant amplitudes.}
\end{figure}

We want to derive a dispersion relation for these amplitudes, and to do so we need to know the positions of their singularities. First of all, there are obviously the elastic (Born) poles located at $\nu=\pm \nu_B=\pm Q^2/(2m)$. After that, there are poles due to excited states and cuts due to multi-particle intermediate states. For example, for the invariant amplitudes of a single nucleon, the cuts start at the pion production threshold: $\nu=\pm \nu_\pi=(2m_N M_\pi+M_\pi^2+Q^2)/(2m_N)$ (see Fig.\ref{fig:contour}). 

In the physical region (i.e. $\nu>\nu_B$), the discontinuity of $T_{\mu\nu}^{ss}$ with respect to the variable $\nu$ is given by:
\begin{equation}
\mathrm{Disc}T_{\mu\nu}^{ss}(\nu)\equiv T_{\mu\nu}^{ss}(\nu+i\epsilon)-T_{\mu\nu}^{ss}(\nu-i\epsilon)=4\pi W_{\mu\nu}
\end{equation}
where 
\begin{eqnarray}
V_{ud}^{(*)-1}W_{\mu\nu}&=&\frac{V_{ud}^{(*)-1}}{4\pi}\int d^4x e^{iq\cdot x}\langle \phi_f(p,s)|[J_\mu^\mathrm{em}(x),J_\nu(0)]|\phi_i(p,s)\rangle\nonumber\\
&=&\frac{V_{ud}^{(*)-1}}{4\pi}\sum_X (2\pi)^4\delta^{(4)}(p+q-p_X)\langle \phi_f(p,s)|J_\mu^\mathrm{em}(0)|X\rangle \langle X|J_\nu(0)|\phi_i(p,s)\rangle\nonumber\\
&=&-i\epsilon_{\mu\nu\alpha\beta}\frac{q^\alpha p^\beta}{2p\cdot q}F_3+i\epsilon_{\mu\nu\alpha\beta}\frac{q^\alpha}{p\cdot q}\left[s^\beta g_1+\left(s^\beta-\frac{s\cdot q}{p\cdot q}p^\beta\right)g_2\right]+...\label{eq:Wmunu}
\end{eqnarray}
with $F_3$, $g_1$, $g_2$ the structure functions. Again, we only need their components contributed by the isosinglet electromagnetic current, which will be labeled by a superscript (0). We therefore obtain the following unsubtracted dispersion relation (DR):
\begin{eqnarray}
T_3^{(0)}(\nu,Q^2)&=&-4i\nu\int_{\nu_B}^{\infty}d\nu'\frac{F_3^{(0)}(\nu',Q^2)}{\nu^{\prime 2}-\nu^2}\nonumber\\
S_{1}^{(0)}(\nu,Q^2)&=&-4i\nu\int_{\nu_B}^{\infty}d\nu'\frac{g_{1}^{(0)}(\nu',Q^2)}{\nu^{\prime 2}-\nu^2}\nonumber\\
S_{2}^{(0)}(\nu,Q^2)&=&-4i\nu^3\int_{\nu_B}^{\infty}d\nu'\frac{g_{2}^{(0)}(\nu',Q^2)}{\nu^{\prime 2}(\nu^{\prime 2}-\nu^2)}~.
\end{eqnarray}
Notice that we have simplified the DR of $S_2$ using the Burkhardt-Cottingham sum rule~\cite{Burkhardt:1970ti}:
\begin{equation}
\int_{\nu_B}^{\infty}\frac{d\nu'}{\nu^{\prime 2}}g_2(\nu',Q^2)=0~,
\end{equation}
which is a superconvergence relation and is expected to hold at all $Q^2$.
Plugging the DRs into Eq.\eqref{eq:gVgAinv} returns:
\begin{eqnarray}
\Box_{\gamma W}^V&=&\frac{\alpha}{\pi\mathring{g}_V}\int_0^\infty \frac{dQ^2}{Q^2}\frac{M_W^2}{M_W^2+Q^2}\int_0^1 dx\frac{1+2r}{\left(1+r\right)^2}F_3^{(0)}(x,Q^2)\nonumber\\
\Box_{\gamma W}^A&=&-\frac{2\alpha}{\pi\mathring{g}_A}\int_0^\infty \frac{dQ^2}{Q^2}\frac{M_W^2}{M_W^2+Q^2}\int_0^1\frac{dx}{(1+r)^2}\Bigg\{\frac{5+4r}{3}g_1^{(0)}(x,Q^2)-\frac{4m^2x^2}{Q^2}g_2^{(0)}(x,Q^2)\Bigg\}~,\nonumber\\\label{eq:DRformula}
\end{eqnarray}
where we have replaced the $\nu$-dependence of the structure functions by $x=Q^2/(2m\nu)$ which is the Bjorken variable, and have defined $r=\sqrt{1+4m^2x^2/Q^2}$, which contains the effect of the target mass $m$.  

Eq.\eqref{eq:DRformula} is the desired dispersive representation of the inner corrections. It expresses the integrands in terms of structure functions that depend on on-shell hadronic matrix elements, which makes it possible to determine the integral using experimental inputs. Below we discuss several generic features that apply to all external states, namely the asymptotic and Born contribution, as well as exact isospin relations that relate the integrands to charge-diagonal matrix elements.

\subsection{Asymptotic contribution}

We define the contribution from $Q^2>\Lambda^2$, where $\Lambda$ is a scale above which the free parton picture works, as the ``asymptotic contribution''. In this region, the structure functions satisfy the following sum rules assuming the free parton model:
\begin{equation}
\int_0^1dx F_3^{(0)}(x,Q^2)=\frac{\mathring{g}_V}{6}~,\:\:\int_0^1dx g_1^{(0)}(x,Q^2)=-\frac{\mathring{g}_A}{12}~.\quad Q^2\rightarrow\infty
\end{equation}
In particular, the second relation is the exactly the polarized Bjorken sum rule~\cite{Bjorken:1966jh,Bjorken:1969mm}.
Therefore, the asymptotic contribution to $\Box_{\gamma W}^{V,A}$ is:
\begin{equation}
\Box_{\gamma W}^{V,A}=\frac{\alpha}{8\pi}\ln\frac{M_W^2}{\Lambda^2}+...
\end{equation}
which is exactly the same result as Eq.\eqref{eq:MbUV} derived from the leading-twist OPE to $T_{\mu\nu}$.

The expression above assumes free parton, and therefore should be modified upon the included of pQCD corrections. As demonstrated in Appendix \ref{sec:pQCDmatch}, this correction is identical to the pQCD correction to the polarized Bjorken sum rule for both $F_3^{(0)}$ and $g_1^{(0)}$. Therefore, the pQCD-corrected asymptotic contribution to the inner RCs is given by:
\begin{equation}
\left(\Box_{\gamma W}^{V,A}\right)_\mathrm{asym}\approx\frac{\alpha}{8\pi}\int_{\Lambda^2}^{\infty}\frac{dQ^2}{Q^2}\frac{M_W^2}{M_W^2+Q^2}C_\mathrm{Bj}(Q^2)~,
\end{equation}
with $C_\mathrm{Bj}(Q^2)$ defined in Eq.\eqref{eq:BjpQCD}.

\subsection{Born contribution}

The Born contribution exhibits itself as a delta function in the structure functions at $x=1$. Plugging $X=\phi_f$ into Eq.\eqref{eq:Wmunu}, one obtains its contribution to the structure functions in terms of the form factors in Eq.\eqref{eq:FFs}:
\begin{eqnarray}
F_{3,B}^{(0)}(x,Q^2)&=&-\frac{1}{2}\left[f_1^S(Q^2)+f_2^S(Q^2)\right]G_A(Q^2)\delta(x-1)\nonumber\\
g_{1,B}^{(0)}(x,Q^2)&=&\frac{1}{8}\left[f_1^S(Q^2)\left(2G_V(Q^2)+G_M(Q^2)\right)+f_2^S(Q^2)G_V(Q^2)\right]\delta(x-1)\nonumber\\
g_{2,B}^{(0)}(x,Q^2)&=&-\frac{Q^2}{32m^2}\left[f_1^S(Q^2)G_M(Q^2)+f_2^S(Q^2)\left(G_V(Q^2)+2G_M(Q^2)\right)\right]\delta(x-1)~,\nonumber\\\label{eq:Born}
\end{eqnarray}
where $f_{1,2}^S=f_{1,2}^{\phi_i}+f_{1,2}^{\phi_f}$ are the isoscalar electromagnetic form factors. With the above we obtain:
\begin{eqnarray}
\left(\Box_{\gamma W}^V\right)_B&=&-\frac{\alpha}{2\pi\mathring{g}_V}\int_0^\infty \frac{dQ^2}{Q^2}\frac{1+2r_B}{\left(1+r_B\right)^2}\left[f_1^S(Q^2)+f_2^S(Q^2)\right]G_A(Q^2)\nonumber\\
\left(\Box_{\gamma W}^A\right)_B&=&-\frac{\alpha}{4\pi\mathring{g}_A}\int_0^\infty \frac{dQ^2}{Q^2}\frac{1}{(1+r_B)^2}\Bigg\{\frac{5+4r_B}{3}\left[f_1^S(Q^2)\left(2G_V(Q^2)+G_M(Q^2)\right)\right.\nonumber\\
&&\left.+f_2^S(Q^2)G_V(Q^2)\right]+f_1^S(Q^2)G_M(Q^2)+f_2^S(Q^2)\left(G_V(Q^2)+2G_M(Q^2)\right)\Bigg\}~,\nonumber\\
\label{eq:BoxBorn}
\end{eqnarray}
where $r_B\equiv r|_{x=1}$. Notice that we have set $M_W^2/(M_W^2+Q^2)\rightarrow 1$ as the form factors only survive at $Q^2\ll M_W^2$. 

\subsection{\label{sec:isospin}Exact isospin relations}

The structure functions $F_3^{(0)}$ and $S_{1,2}^{(0)}$ are not directly measurable, because (1) they come from the hadronic tensor $W_{\mu\nu}$ that involves different initial and final states, and (2) the isosinglet electromagnetic current $J_\mathrm{em}^{(0)}$ does not exist in nature. Fortunately, we can relate them to physically measurable structure functions through isospin symmetry.  
For $F_3^{(0)}$, we have the following isospin relation:
\begin{equation}
2F_3^{(0)}=F_{3,\phi_f}^{\gamma Z}-F_{3,\phi_i}^{\gamma Z}~,
\end{equation}
where 
\begin{equation}
\frac{1}{4\pi}\int d^4x e^{iq\cdot x}\langle \phi(p,s)|[J_\mathrm{em}^\mu(x),J_Z^\nu(0)]+[J_Z^\mu(x),J_\mathrm{em}^\nu(0)]|\phi(p,s)\rangle=-i\epsilon^{\mu\nu\alpha\beta}\frac{q_\alpha p_\beta}{2p\cdot q}F_{3,\phi}^{\gamma Z}+...
\end{equation}
The structure function $F_{3,\phi}^{\gamma Z}$ is in principle measurable through the parity-odd observables in the inclusive $e\phi$ scattering experiments. 

Meanwhile, for $g_{j}^{(0)}$ ($j=1,2$), we have the following isospin relation:
\begin{equation}
2g_{j}^{(0)}=g_{j,\phi_f}^\gamma-g_{j,\phi_i}^\gamma~,
\end{equation}
where
\begin{equation}
\frac{1}{4\pi}\int d^4x e^{iq\cdot x}\langle \phi(p,s)|[J_\mathrm{em}^\mu(x),J_\mathrm{em}^\nu(0)]|\phi(p,s)\rangle=i\epsilon^{\mu\nu\alpha\beta}\frac{q_\alpha}{p\cdot q}\left[s_\beta g_{1,\phi}^\gamma+\left(s_\beta-\frac{s\cdot q}{p\cdot q}p_\beta\right)g_{2,\phi}^\gamma\right]+...
\end{equation}
The polarized structure functions $g_{1,\phi}^{\gamma}$ and $g_{2,\phi}^{\gamma}$ are well-measured quantities in ordinary DIS experiments with nucleon and light nuclei. 

To summarize, we presented in this section the dispersive representation of the inner RC to the beta decay of a generic $I=J=1/2$ particle, and derived analytic formulas and relations that hold in general. The remaining unanalyzed pieces are process-dependent and have to be studied case-by-case. In the next section we will use a particularly important example, namely the free neutron, to demonstrate how these process-dependent contributions can be pinned down with appropriate experimental inputs.  

\section{\label{sec:freen}Free neutron}

The beta decay of free neutron currently provides the second best measurement of $|V_{ud}|$ after superallowed beta decays. The master formula reads~\cite{Zyla:2020zbs}:
\begin{equation}
|V_{ud}|^2=\frac{5024.7~\mathrm{s}}{\tau_n(1+3\lambda^2)(1+\Delta_R^V)}
\end{equation}
where $\tau_n$ is the neutron lifetime and $\lambda$ is the renormalized axial-to-vector coupling ratio. The numerator at the right hand side includes the effects from recoil corrections and the outer RCs, while $\Delta_R^V$ is the so-called ``nucleus-dependent RC'' to the Fermi matrix element which takes the following form in the Sirlin's representation:
\begin{equation}
\Delta_R^V=\frac{\alpha}{2\pi}\left[3\ln\frac{M_Z}{m_p}+\ln\frac{M_Z}{M_W}+\tilde{a}_g\right]+\delta_\mathrm{HO}^\mathrm{QED}+2\Box_{\gamma W}^V~.
\end{equation}
It is nothing but the inner correction to $g_V^2$ as we discussed in Eq.\eqref{eq:innerRC}. 
The current limiting factor to the extraction of $|V_{ud}|$ from the neutron beta decay comes from experiment rather than theory. The experimental determination of the neutron lifetime $\tau_n$ suffers from the well-known beam-bottle discrepancy~\cite{Bowman:2014nsk}, while the measurements of $\lambda$ before~\cite{Bopp:1986rt,Erozolimsky:1997wi,Liaud:1997vu,Mostovoi:2001ye} and after 2002~\cite{Schumann:2007hz,Mund:2012fq,Darius:2017arh,Brown:2017mhw,Markisch:2018ndu} also show a large systematic disagreement (see Ref.\cite{Czarnecki:2018okw} for more discussions). Several experiments are under construction to measure $\tau_n$ with a precision better than 0.4~s and resolve the beam-bottle discrepancy~\cite{Pattie:2019brb,Wietfeldt:2014gia}, as well as to reach an accuracy level of $10^{-4}$ in the $\lambda$ measurement~\cite{Fry:2018kvq,Dubbers:2007st,Wang:2019pts}. 

Despite being limited by the experimental precision, the EWRC in free neutron has long been a main focus in precision physics. In particular, the quantity $\Delta_R^V$ is of great interest because it not only appears in the neutron, but also in nuclear beta decays as the single-nucleon contribution to the RCs. For instance, it has long been the major source of uncertainty in the most precise extraction of $|V_{ud}|$ from superallowed nuclear beta decays. In this section, we will briefly review the past attempts to pin down the EWRCs in free neutron, and discuss the recent progress based on the dispersive representation.

\subsection{Earlier attempts}

Similar to Eq.\eqref{eq:pionboxSirlin} for pion, a famous parameterization of $\Box_{\gamma W}^V$ for neutron in the early days reads~\cite{Sirlin:1977sv}:
\begin{equation}
\Box_{\gamma W}^V=\frac{\alpha}{2\pi}\left[\frac{1}{2}\ln\frac{M_W}{M_A}+C_\mathrm{Born}+\frac{1}{2}A_g\right]~,
\end{equation} 
where the first term at the right hand side represents the large electroweak logarithm with $M_A$ an effective IR cutoff scale, $C_\mathrm{Born}$ is the Born contribution, and $A_g$ is the pQCD correction. 
While $C_\mathrm{Born}$ and $A_g$ are easily calculable (see Sec.\ref{sec:IJhalf}), there was no unique method to determine the scale $M_A$. Based on a simple vector-meson-dominance (VMD) picture, Ref.\cite{Marciano:1985pd} set $M_A\approx 1.2$~GeV, i.e. the mass of the $A_1$ resonance, and estimated the error by varying $M_A$ up or down by a factor 2. Such a simple estimation returned $\Delta_R^V=0.0240(8)$~\cite{Hardy:2004id}.

Ref.\cite{Marciano:2005ec} improved upon the determination above by adopting an interpolation function approach. First, they wrote:
\begin{equation}
\Box_{\gamma W}^V=\frac{\alpha}{8\pi}\int_0^\infty dQ^2\frac{M_W^2}{M_W^2+Q^2}F(Q^2)~,\label{eq:MSintegral}
\end{equation} 
and assumed the following dominant physics that contribute to $F(Q^2)$ at different regions of $Q^2$: 
\begin{itemize}
	\item $0<Q^2<Q_1^2$ (long distances): Pure Born contribution, which is completely fixed by nucleon form factors, dominates.
	\item $Q_1^2<Q^2<(1.5~\mathrm{GeV})^2$ (intermediate distances): A VMD-inspired interpolating function is constructed:
	\begin{equation}
	F(Q^2)=\frac{c_\rho}{Q^2+M_\rho^2}+\frac{c_A}{Q^2+M_A^2}+\frac{c_{\rho'}}{Q^2+M_{\rho'}^2}~,\label{eq:MSVMD}
	\end{equation}
	with $M_\rho=0.776$~GeV, $M_A=1.230$~GeV, $M_{\rho'}$=1.465~GeV.
	\item $Q^2>(1.5~\mathrm{GeV})^2$ (short distances): $F(Q^2)$ is given by the leading-twist OPE + pQCD correction.
\end{itemize}
There are four free parameters in this parameterization, namely the coefficients $\{c_\rho,c_A,c_{\rho'}\}$ in the interpolating function, and the matching scale $Q_1^2$ between the long and intermediate distances. Ref.\cite{Marciano:2005ec} fixed these four parameters with the four criteria as follows (we will see later than some of them are invalidated by more recent studies):
\begin{enumerate}
	\item The result of the integral \eqref{eq:MSintegral} at $Q^2>(1.5~\mathrm{GeV})^2$ is required to be the same using the VMD parameterization and the asymptotic expression.
	\item In the large-$Q^2$ limit, the coefficient of the $1/Q^4$ term in Eq.\eqref{eq:MSVMD} is required to vanish by chiral symmetry.
	\item Eq.\eqref{eq:MSVMD} is required to vanish at $Q^2=0$ by ChPT.
	\item Finally, the connection scale $Q_1^2$ is chosen through the matching of $F_\mathrm{Born}$ and $F_\mathrm{Interpolation}$ at $Q^2=Q_1^2$. 
\end{enumerate}
With the above, Ref.\cite{Marciano:2005ec} obtained $c_\rho=-1.490$, $c_A=6.855$, $c_{\rho'}=-4.414$ and $Q_1^2=(0.823~\mathrm{GeV})^2$, which give $\Delta_R^V=0.02361(38)$ in total. This value was taken as the state-of-the-art determination of $\Delta_R^V$ until 2018.

In the meantime, neutron beta decay had also been studied within the HBChPT framework we described in Sec.\ref{sec:HBChPT}. The outcome to one-loop agreed naturally with the Sirlin's representation in terms of the outer corrections, and a comparison of the remaining terms between the two formalisms yields the following matching relation between the renormalized LECs in the EFT and the inner correction~\cite{Ando:2004rk},
\begin{equation}
e_V^R\approx-\frac{5}{4}+\frac{2\pi}{\alpha}\Delta_R^V~.
\end{equation}
Of course, since chiral symmetry does not impose any constraint on $e_V^R$, the EFT framework itself does not improve our understanding of the inner RCs. 

\subsection{$\Box_{\gamma W}^V$: DR analysis}

\begin{figure}
	\begin{centering}
		\includegraphics[scale=0.3]{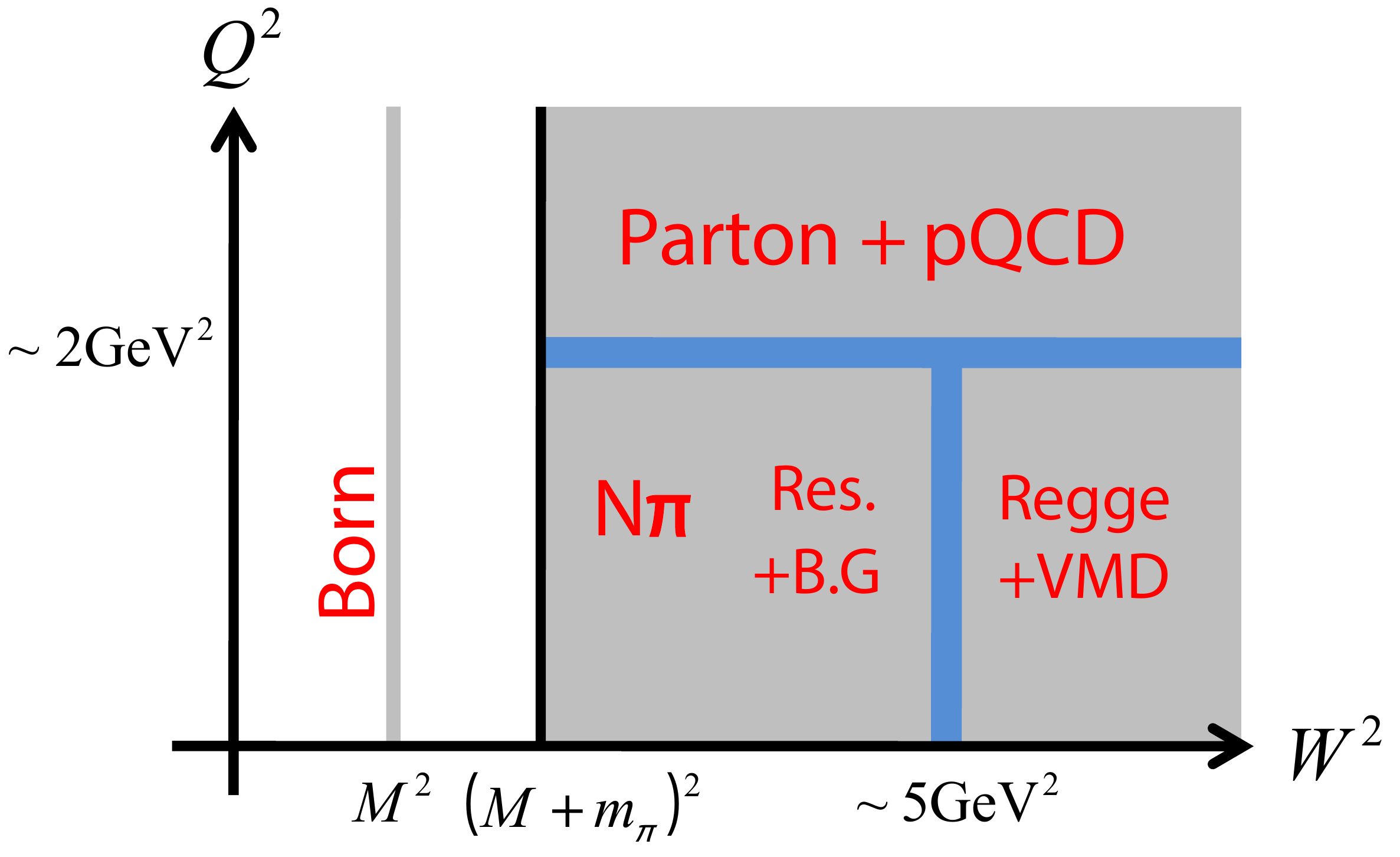}\hfill
		\par\end{centering}
	\caption{\label{fig:phase}The $W^2-Q^2$ phase diagram for free neutron, taken from Ref.\cite{Seng:2018yzq}. }
\end{figure}

A next important breakthrough occurred in 2018 when the dispersive representation of $\Box_{\gamma W}^V$ was first introduced in Ref.\cite{Seng:2018yzq}. Within this formalism, the function $F(Q^2)$ defined in Ref.\cite{Marciano:2005ec} is expressed in terms of an integral with respect to the P-odd structure function $F_3^{(0)}(x,Q^2)$ (notice that $\mathring{g}_V=1$ for free neutron):
\begin{equation}
F(Q^2)=\frac{8}{Q^2}\int_0^1dx\frac{1+2r}{(1+r)^2}F_3^{(0)}(x,Q^2)=\frac{6}{Q^2}M_3^{(0)}(1,Q^2)~,
\end{equation}
where we have defined the so-called ``first Nachtmann moment'' of $F_3^{(0)}$~\cite{Nachtmann:1973mr,Nachtmann:1974aj}:
\begin{equation}
M_3^{(0)}(1,Q^2)\equiv\frac{4}{3}\int_0^1dx\frac{1+2r}{(1+r)^2}F_3^{(0)}(x,Q^2)~,
\end{equation}
which reduces to the simple Mellin moment at large $Q^2$.

The main challenge now is to identify all the dominant on-shell intermediate states $|X\rangle$ in Eq.\eqref{eq:Wmunu} that contribute to $F_3^{(0)}$ at different regions of $\{x,Q^2\}$.
They are most conveniently represented by the phase diagram in Fig.\ref{fig:phase} (where $W^2=(p+q)^2$): First, at $W^2=m_N^2$ there is an isolated Born contribution. The continuum contribution starts at the pion production threshold $W^2=(m_N+M_\pi)^2$, which should be further separated to regions of high and low $Q^2$. At $Q^2>2$~GeV$^2$ (which we will justify later), the parton + pQCD description is valid so the contribution from this region can be computed to high accuracy. The situation is more complicated at low $Q^2$: When $W^2$ is small, what we mainly observe are baryon resonances (res) on top of a smooth background; approaching large $W^2$, the contributions from multi-hadron intermediate states start to dominate, which can be described economically by a Regge exchange picture ($\mathbb{R}$). 

None of the discussions above would really be necessary if we had precise experimental data of $F_{3,N}^{\gamma Z}$ at all values of $\{x,Q^2\}$ because they could then be related to $F_3^{(0)}$ using the isospin relations in Sec.\ref{sec:isospin}. Unfortunately, no such data exists to our knowledge (there are some measurement of $F_3^{\gamma Z}$ at very large $Q^2$~\cite{Abramowicz:2015mha,Argento:1983dj}), although they are in principle measurable in parity-violating electron-nucleus scattering experiments, e.g. in Jefferson Lab. Therefore, we are forced to look at other possible experimental inputs. We
consider the inclusive $\nu p(\bar{\nu}p)$ scattering, of which differential cross section is given by:
\begin{equation}
\frac{d^2\sigma^{\nu p(\bar{\nu}p)}}{dxdy}=\frac{G_F^2m_N E}{\pi(1+Q^2/M_W^2)^2}\left[xy^2 F_1^{\nu p(\bar{\nu}p)}+\left(1-y-\frac{m_N xy}{2E}\right)F_2^{\nu p(\bar{\nu}p)}\pm x\left(y-\frac{y^2}{2}F_3^{\nu p(\bar{\nu}p)}\right)\right]
\end{equation}
where $y=\nu/E$ with $E$ the initial neutrino energy. The parity-odd, spin-independent structure functions probed in these processes are~\cite{Onengut:2005kv}:
\begin{eqnarray}
\frac{|V_{ud}|^{-2}}{4\pi}\sum_X(2\pi)^4\delta^{(4)}(p+q-p_X)\langle p|(J_W^\mu)^\dagger |X\rangle \langle X|J_W^\nu |p\rangle &=&-\frac{i\epsilon^{\mu\nu\alpha\beta}q_\alpha p_\beta}{2p\cdot q}F_3^{\nu p}+...,\nonumber\\
\frac{|V_{ud}|^{-2}}{4\pi}\sum_X(2\pi)^4\delta^{(4)}(p+q-p_X)\langle p|J_W^\mu |X\rangle \langle X|(J_W^\nu)^\dagger |p\rangle &=&-\frac{i\epsilon^{\mu\nu\alpha\beta}q_\alpha p_\beta}{2p\cdot q}F_3^{\bar{\nu} p}+...,\nonumber\\
\end{eqnarray} 
and we can define their average, $F_3^{\nu p+\bar{\nu}p}=(F_3^{\nu p}+F_3^{\bar{\nu}p})/2$, which is obtained from the difference between the neutrino and antineutrino cross sections. 

$F_3^{\nu p+\bar{\nu}p}$ resides in a different isospin channel and is not directly related to $F_3^{(0)}$ through a simple isospin rotation. However, we could parameterize both of them based on the physical picture in Fig.\ref{fig:phase} as follows:
\begin{equation}\label{eq:ourparam}
F_3=F_{3,B}+
\begin{cases}
F_{3,\text{pQCD}} & Q^2 \gtrsim 2\text{ GeV}^2 \\[2mm] 
F_{3,\pi N}+F_{3,\text{res}}+F_{3,\mathbb{R}} & Q^2 \lesssim 2\text{ GeV}^2~.
\end{cases}
\end{equation}
We may separate all the terms at the RHS into two classes: 
\begin{itemize}
	\item The ``non-asymptotic'' pieces (Born, low-energy continuum, resonances): They are clearly different for different $F_3$, and need to be calculated case-by-case.
	\item The ``asymptotic'' pieces ($F_{3,\text{pQCD}}$ at large $Q^2$ and $F_{3,\mathbb{R}}$ at large $W^2$): They are largely universal for different $F_3$ (up to multiplicative factors), so we can either calculate them explicitly ($F_{3,\text{pQCD}}$), or infer one from the other. 
\end{itemize}

Let us investigate the non-asymptotic pieces in more detail. First, the Born contribution to $F_3^{(0)}$ is given in Eq.\eqref{eq:Born}, while 
\begin{equation}
F_{3,B}^{\nu p+\bar{\nu}p}(x,Q^2)=-\left[f_1^V(Q^2)+f_2^V(Q^2)\right]G_A(Q^2)\delta(x-1)~,
\end{equation}
where $f_i^V=f_i^p-f_i^n$ are the isovector EM form factors. Both $F_{3,B}^{(0)}$ and $F_{3,B}^{\nu p+\bar{\nu}p}$ can be fixed to high precision using the experimental measurements of the nucleon electromagnetic~\cite{Ye:2017gyb,Lorenz:2012tm,Lorenz:2014yda,Lin:2021umk,Lin:2021umz} and axial form factors~\cite{Bernard:2001rs,Bhattacharya:2011ah}. Next, the $N\pi$ contribution can be calculated using ChPT, and is found to be small~\cite{Seng:2018qru}. In doing so, we find that the simplest inelastic (i.e. $N\pi$) contribution to $F(Q^2)$ does not vanish at $Q^2=0$, which invalidates one of the four criteria imposed in Ref.\cite{Marciano:2005ec} as we advertised in the previous subsection.
Finally, $F_{3,\text{res}}^{\nu p+\bar{\nu}p}$ is dominated by the $\Delta$-resonance, which can be calculated using the parameterizations in Ref.\cite{Lalakulich:2005cs}. Meanwhile, since $F_3^{(0)}$ only probes $I=1/2$ intermediate states, the $\Delta$-resonance cannot contribute and thus $F_{3,\text{res}}^{(0)}$ is very small.

Next we turn to the asymptotic pieces. First, at large $Q^2$, $F_3^{(0)}$ obeys the pQCD-corrected polarized Bjorken sum rule as we have already discussed:
\begin{equation}
M_3^{(0)}(1,Q^2)=3\left[1-\sum_{n=1}^\infty c_n\left(\frac{\alpha_S}{\pi}\right)^n\right]~.
\end{equation}
Meanwhile, $F_3^{\nu p+\bar{\nu}p}$ follows the pQCD-corrected GLS sum rule~\cite{Gross:1969jf}:  
\begin{equation}
M_3^{\nu p+\bar{\nu}p}(1,Q^2)=3\left[1-\sum_{n=1}^\infty\tilde{c}_n\left(\frac{\alpha_S}{\pi}\right)^n\right]~,
\end{equation}
where the coefficients $\tilde{c}_n$ are again known to $n=4$~\cite{Baikov:2010iw,Baikov:2010je,Baikov:2012zn}:
\begin{eqnarray}
\tilde{c}_1&=&1\nonumber\\
\tilde{c}_2&=&4.583-0.333n_f\nonumber\\
\tilde{c}_3&=&41.44-8.020n_f+0.1775n_f^2\nonumber\\
\tilde{c}_4&=&479.4-129.2n_f+7.930n_f^2-0.1037n_f^3~.
\end{eqnarray} 
They are very similar to the pQCD corrections to the polarized Bjorken sum rule, except for very small differences starting from $\mathcal{O}((\alpha_S/\pi)^3)$ due to singlet (disconnected) diagrams. Neglecting these small differences give us the following approximate matching at large $Q^2$:
\begin{equation}
M_3^{(0)}(1,Q^2)\approx \frac{1}{18}M_3^{\nu p+\bar{\nu}p}(1,Q^2)~.\quad\text{Large $Q^2$}\label{eq:largeQ2ratio}
\end{equation}

\begin{figure}
	\begin{centering}
		\includegraphics[scale=0.2]{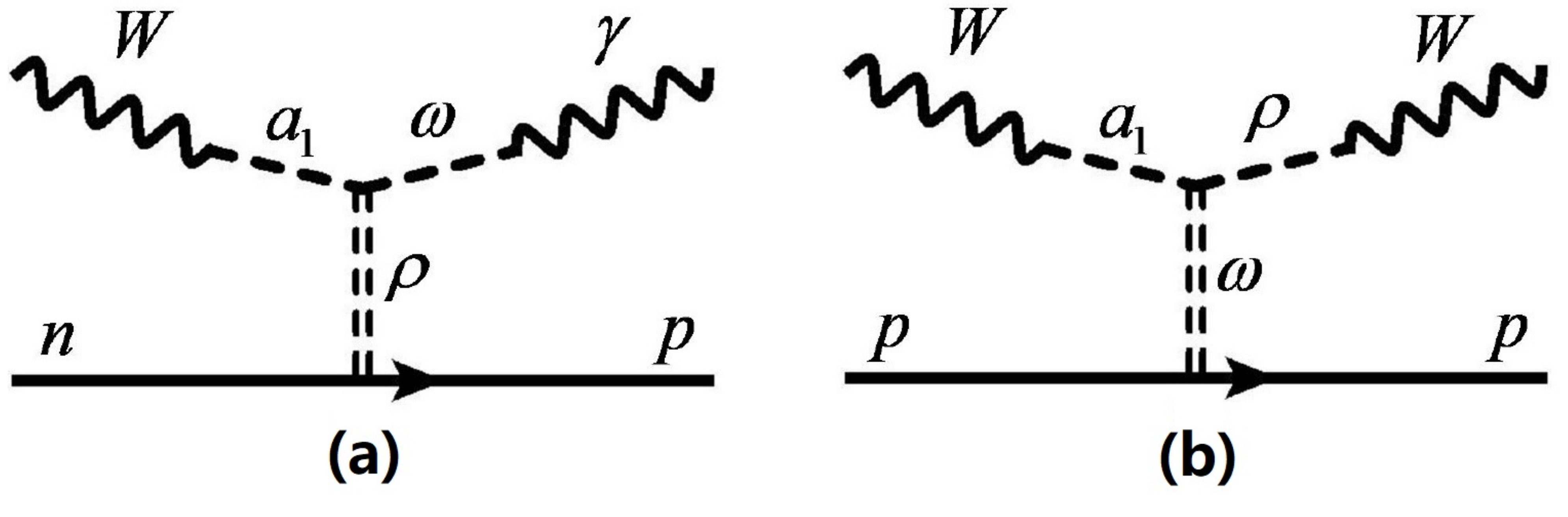}\hfill
		\par\end{centering}
	\caption{\label{fig:Regge}The Regge exchange picture for (a) $F_3^{(0)}$ and (b) $F_3^{\nu p+\bar{\nu}p}$, taken from Ref.\cite{Seng:2018yzq}.}
\end{figure}

We next turn to the small-$Q^2$, large-$W^2$ region. The leading Regge-exchange contributions to $F_3^{(0)}$ and $F_3^{\nu p+\bar{\nu}p}$ are depicted in Fig.\ref{fig:Regge}, where the gauge bosons first fluctuate into vector/axial mesons, which in turn couple to the nucleon through the exchange of a $t$-channel Regge trajectory. We may adopt the following simple parameterization of the Regge contribution~\cite{Seng:2018yzq,Seng:2018qru}:
\begin{eqnarray}
F_{3,\mathbb{R}}^{(0)}&=&f_\mathrm{th}C_{\gamma W}\frac{M_\omega^2}{M_\omega^2+Q^2}\frac{M_{a_1}^2}{M_{a_1}^2+Q^2}\left(\frac{\nu}{\nu_0}\right)^{\alpha_0^\rho}\nonumber\\
F_{3,\mathbb{R}}^{\nu p+\bar{\nu}p}&=&f_\mathrm{th}C_{W W}\frac{M_\rho^2}{M_\rho^2+Q^2}\frac{M_{a_1}^2}{M_{a_1}^2+Q^2}\left(\frac{\nu}{\nu_0}\right)^{\alpha_0^\omega}~.
\end{eqnarray}
Let us explain the notations above: $\alpha_0^\phi$ is the Regge trajectory intercept, with $\alpha_0^\rho\approx \alpha_0^\omega\approx 0.477$~\cite{Kashevarov:2017vyl}, $\nu_0=1$~GeV is a threshold scale, $f_\mathrm{th}$ is a threshold function that approaches 1(0) when $W^2$ approaches $\infty(0)$, and $C_{\gamma W}$, $C_{WW}$ are functions of $Q^2$ that account for the residual $Q^2$-dependence not captured by the VMD picture.

We assume that the ratio between $F_{3,\mathbb{R}}^{(0)}$ and $F_{3,\mathbb{R}}^{\nu p+\bar{\nu}p}$ at $Q^2=0$ and $Q^2=2$~GeV$^2$ follow the prediction of VMD~\cite{Lichard:1997ya,deSwart:1963pdg} and the parton picture (i.e. Eq.\eqref{eq:largeQ2ratio}) respectively. Given that the $\rho$ and $\omega$ trajectories are almost degenerate~\cite{Kashevarov:2017vyl}, it turns out that the two conditions predict the same ratio:
\begin{equation}
\frac{M_3^{(0)}(1,0)}{M_3^{\nu p+\bar{\nu}p}(1,0)}\approx \frac{M_3^{(0)}(1,2~\mathrm{GeV}^2)}{M_3^{\nu p+\bar{\nu}p}(1,2~\mathrm{GeV}^2)}\approx\frac{1}{18}~,
\end{equation}
which suggests the following simple relation between the coefficient functions:
\begin{equation}
C_{\gamma W}(Q^2)\approx \frac{1}{18}C_{WW}(Q^2)~.\label{eq:Cratio}
\end{equation}

\begin{figure}
	\begin{centering}
		\includegraphics[scale=0.3]{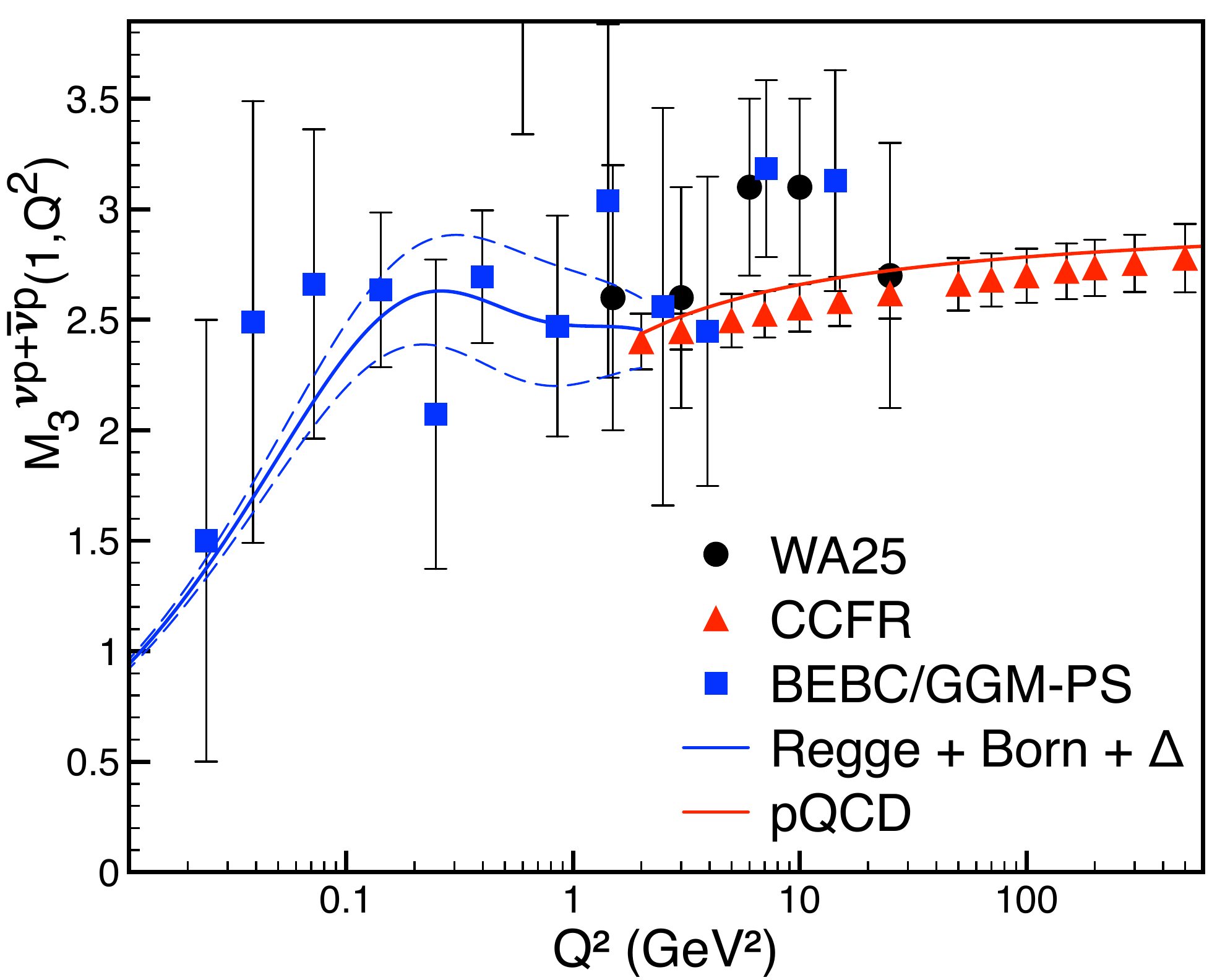}
		\includegraphics[scale=0.3]{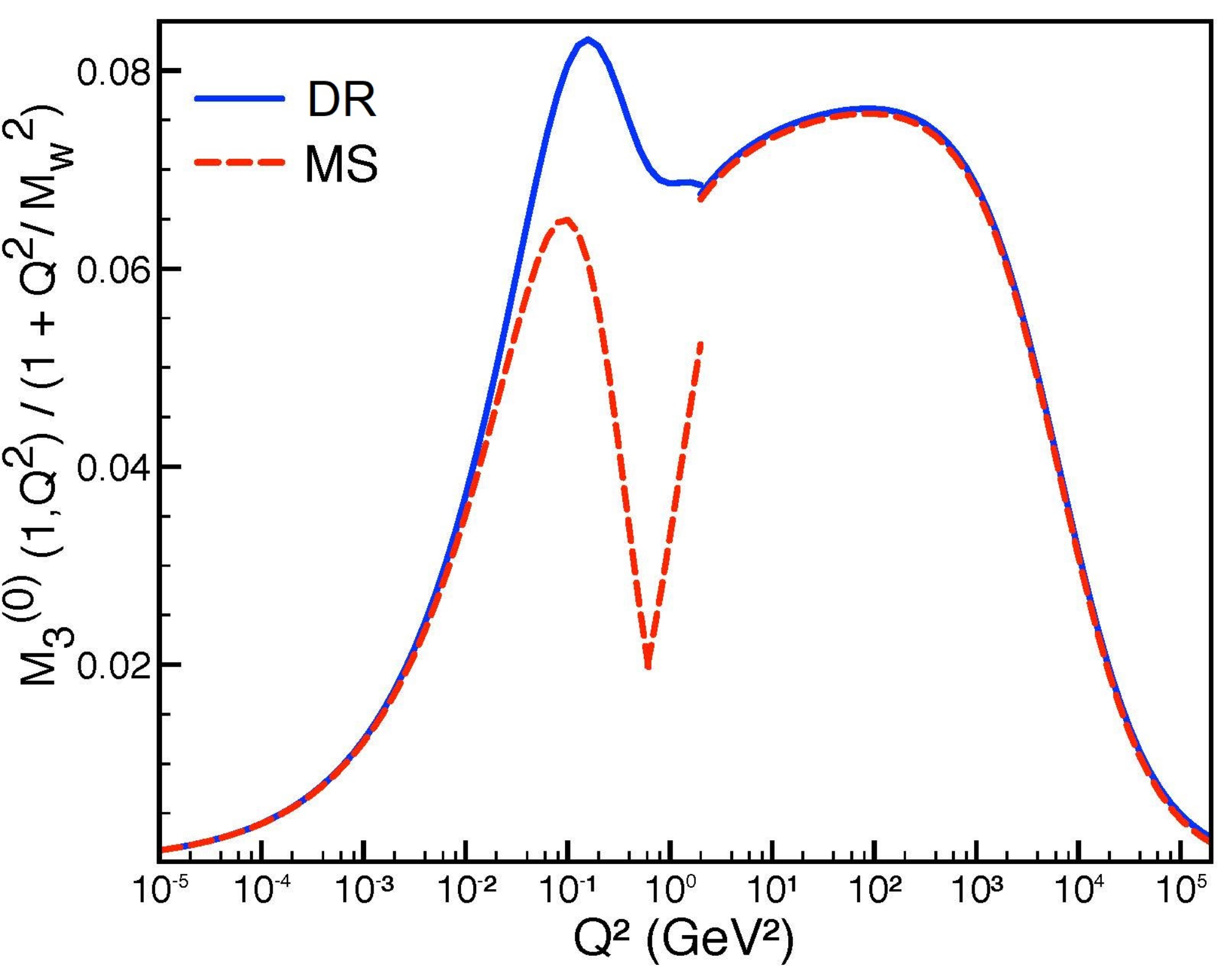}\hfill
		\par\end{centering}
	\caption{\label{fig:GLSSR}Left panel: Fitting $M_3^{\nu p+\bar{\nu}p}(1,Q^2)$ to experimental data. Right panel: Comparison between the theory predictions of $M_3^{(0)}(1,Q^2)$ using the DR method~\cite{Seng:2018yzq} (blue) and the method in Ref.\cite{Marciano:2005ec} (red). Figures are taken from Ref.\cite{Seng:2018yzq}.}
\end{figure}

Eq.\eqref{eq:Cratio} provides a systematic procedure to infer the values of $M_3^{(0)}(1,Q^2)$ at small-$Q^2$. First, the full $M_3^{\nu p+\bar{\nu}p}(1,Q^2)$ was measured by the CCFR~\cite{Kataev:1994rj,Kim:1998kia}, BEBC and Gargamelle~\cite{Bolognese:1982zd}, and WA25~\cite{Allasia:1985hw} collaborations over 0.15~GeV$^2<Q^2<$600~GeV$^2$, which are summarized in the left panel of Fig.\ref{fig:GLSSR}. In particular, the high-precision measurement from CCFR agrees with the parton + pQCD prediction at $Q^2>2~$GeV$^2$, which justifies our choice of $Q_\mathrm{cut}^2=2~$GeV$^2$ as the separation scale between the perturbative and non-perturbative regions. One may then subtract the Born, $N\pi$ and resonance contribution from the experimental result to obtain $M_{3,\mathbb{R}}^{\nu p+\bar{\nu}p}(1,Q^2)$ at low-$Q^2$, which fixes the function $C_{WW}(Q^2)$. Then, through Eq.\eqref{eq:Cratio} we obtain $M_{3,\mathbb{R}}^{(0)}(1,Q^2)$. Combining this with theory calculations of other contributions give us the full $M_3^{(0)}(1,Q^2)$. Readers should be aware that Eq.\eqref{eq:Cratio} is strictly speaking just a model and is subject to a residual theory uncertainty; however, given the rather poor quality of the data of $M_3^{\nu p+\bar{\nu}p}(1,Q^2)$ at low $Q^2$, we may assume at this stage that the final uncertainty in $M_3^{(0)}(1,Q^2)$ is predominantly from experiments. 

Refs.\cite{Seng:2018yzq,Seng:2018qru} quoted a new result of $\Delta_R^V=0.02467(22)$ based on the analysis above. While reducing the existing theory uncertainty in Ref.\cite{Marciano:2005ec} by a half, the new result also showed a significant shift in the central value. The origin of such difference is most clearly explained in the right panel of Fig.\ref{fig:GLSSR}. One observes an unphysical dip in the red curve around $Q^2= (0.823~\mathrm{GeV})^2$ due to the incorrect small-$Q^2$ behavior of the interpolating function in Ref.\cite{Marciano:2005ec}, and the incorrect assumption that only the Born contribution exists at $Q^2\leq (0.823~\mathrm{GeV})^2$. Also, there is a discontinuity in the red curve at the UV-matching point due to the requirement that the integral of the interpolating function, instead of the interpolating function itself, matches with the partonic description at large-$Q^2$. 
The disagreement between the DR and interpolating function approach is thus apparent because $\Box_{\gamma W}^V$ is proportional to the area under the curve.

The increase of the size of $\Delta_R^V$ led to a reduction of $|V_{ud}|$ and eventually the appearance of a deficit in the top-row CKM matrix unitarity as it is well-known by now. Due to this reason, the DR result must be carefully scrutinized. As a response, Ref.\cite{Czarnecki:2019mwq} improved the original interpolating function method by abandoning the incorrect $Q^2=0$ constraint, and using a holographic QCD model to describe the physics at intermediate $Q^2$. This treatment returned a somewhat smaller value $\Delta_R^V=0.02426(32)$. However, it was pointed out shortly~\cite{Hayen:2020cxh,Hayen:2021iga} that this treatment missed the target mass correction at low-$Q^2$ which, after adding back, gives $\Delta_R^V=0.02473(27)$ which is now consistent with Refs.\cite{Seng:2018yzq,Seng:2018qru}. Several later, DR-based calculations also reported consistent results~\cite{Seng:2020wjq,Shiells:2020fqp}.   

\subsection{$\Box_{\gamma W}^A$: DR analysis}

$\Box_{\gamma W}^A$ is responsible for the  renormalization of the axial coupling constant $g_A$. As we explained before, in beta decay experiments one does not have direct access to the bare coupling constant $\mathring{g}_A$, but rather the renormalized axial-to-vector ratio $\lambda=g_A/g_V$. And in fact, since the latter is directly measured through various vector-vector correlations in the differential decay rate (e.g. $\vec{p}_e\cdot\vec{p}_\nu$, $\hat{s}\cdot\vec{p}_e$ and $\hat{s}\cdot\vec{p}_\nu$; see Eq.\eqref{eq:outer}), the quantity $\Box_{\gamma W}^A$ plays no role in the determination of $|V_{ud}|$ from the free-neutron beta decay and hence had received much less attention than $\Box_{\gamma W}^V$ from the theory side. 

The situation is changed following the increased precision of the first-principles calculations of the bare coupling constant $\mathring{g}_A$ with lattice QCD~\cite{Khan:2006de,Lin:2008uz,Capitani:2012gj,Horsley:2013ayv,Bali:2014nma,Abdel-Rehim:2015owa,Alexandrou:2017hac,Capitani:2017qpc,Edwards:2005ym,Yamazaki:2008py,Yamazaki:2009zq,Bratt:2010jn,Green:2012ud,Yamanaka:2018uud,Liang:2018pis,Ishikawa:2018rew,Ottnad:2018fri,Bhattacharya:2016zcn,Berkowitz:2017gql,Chang:2018uxx,Gupta:2018qil}. In particular, the authors in Ref.\cite{Chang:2018uxx} reported a percent-level determination of $\mathring{g}_A=-1.271(10)(7)$ and aimed for sub-percent precision in the near future~\cite{Walker-Loud:2019cif}. Once the lattice precision reaches $10^{-3}$ or above, it would be possible to search for signals of BSM physics, in particular the right-handed currents~\cite{Gonzalez-Alonso:2016etj,Alioli:2017ces,Gonzalez-Alonso:2018omy,Falkowski:2020pma}, by comparing the experimental measurement of $\lambda$ and the SM prediction in the right hand side of Eq.\eqref{eq:lambdaRC}. Due to this reason, one needs to understand the difference between $\Box_{\gamma W}^A$ and $\Box_{\gamma W}^V$. The first comprehensive analysis of this inner RC was done in Refs.\cite{Hayen:2020cxh,Hayen:2021iga} based on the holographic QCD model inspired by Ref.\cite{Czarnecki:2019mwq}. The model-dependent nature of this method, however, makes it challenging to have a rigorous estimation of the associated theoretical uncertainty.  

As we detailed in Sec.\ref{sec:DRrep}, the DR analysis of $\Box_{\gamma W}^A$ may proceed in exactly the same was as $\Box_{\gamma W}^V$, and is in fact much more promising than the latter. This is because its dispersive integral depends on the polarized structure functions $g_{1,N}^\gamma$ and $g_{2,N}^\gamma$ upon an isospin rotation (see Sec.\ref{sec:isospin}). These structure functions are directly measurable in ordinary DIS processes, which means the low-$Q^2$ part in the dispersive integral of $\Box_{\gamma W}^A$ can in principle be fixed completely by experiment, without invoking any extra modelings. The $g_{1,N}^\gamma$ structure function, in particular, is a well-measured quantity over the past 30 years~\cite{Anthony:1993uf,Abe:1994cp,Abe:1997cx,Adams:1994zd,Alexakhin:2006oza,Alekseev:2010hc,Aghasyan:2017vck,Ackerstaff:1997ws,Deur:2004ti,Wesselmann:2006mw,Deur:2008ej,Guler:2015hsw,Fersch:2017qrq,Zheng:2021yrn}.  The data on $g_{2,N}^\gamma$ are more scarce~\cite{Anthony:1999py,Anthony:2002hy,Amarian:2003jy,Wesselmann:2006mw,Kramer:2005qe,Fersch:2017qrq}, but the size of its contribution to the inner correction is also generic two orders of magnitude smaller.  

\begin{figure}
	\begin{centering}
		\includegraphics[scale=0.35]{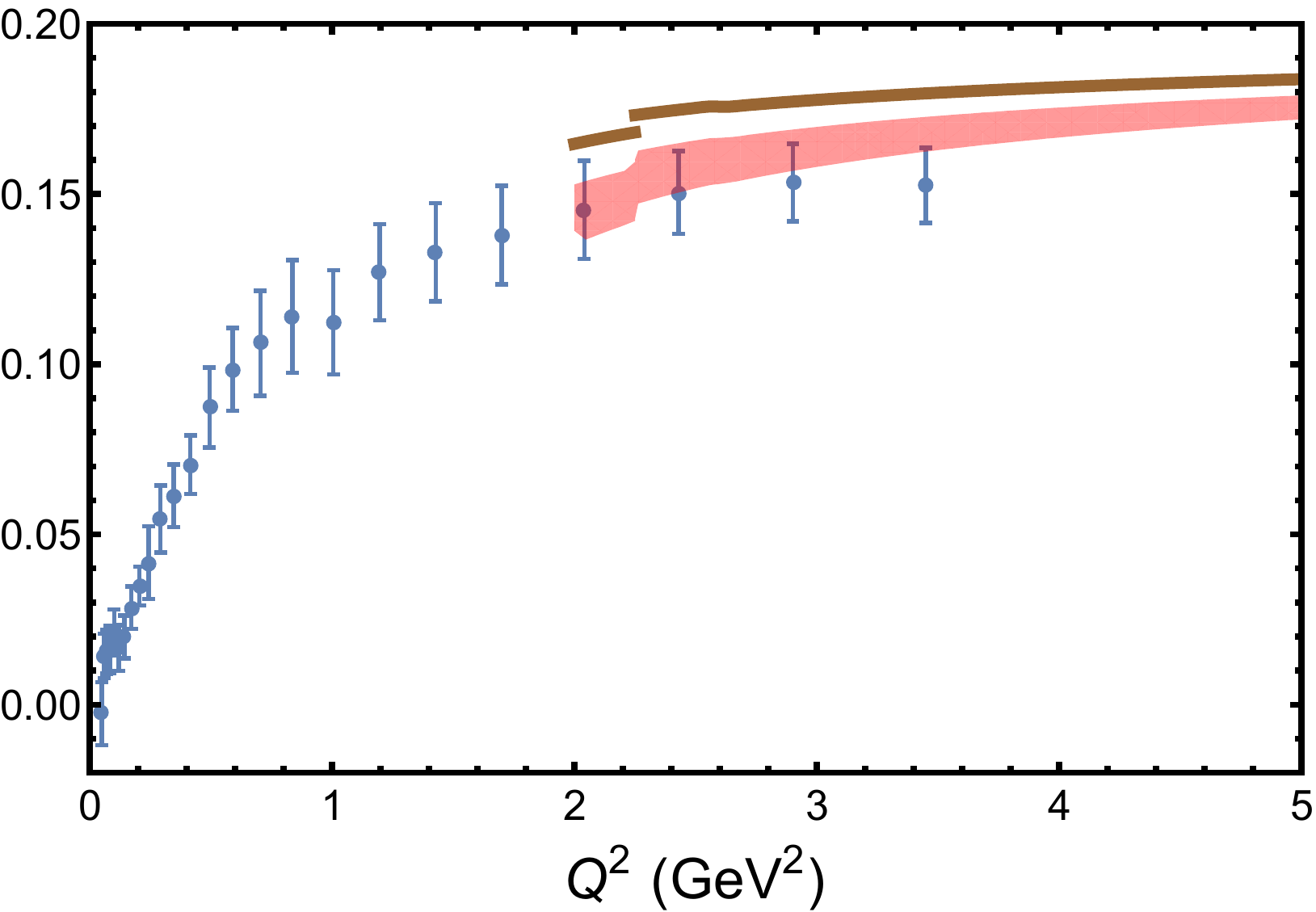}
		\hfill
		\par\end{centering}
	\caption{\label{fig:data} Data points of $\bar{\Gamma}_1^{p-n}$ reconstructed from the EG1b experiment versus the theory prediction with (red band) and without (brown curve) the HT correction. Figure taken from Ref.\cite{Gorchtein:2021qdf}.}
\end{figure}

The first complete DR analysis of $\Box_{\gamma W}^A$ was performed in  Ref.\cite{Gorchtein:2021qdf}, and we briefly outline the main strategies here. First, it is separated into three pieces:
\begin{equation}
\Box_{\gamma W}^A=\frac{\alpha}{2\pi}\left[d_B+d_1+d_2\right]
\end{equation}  
which represent the Born contribution, the inelastic contribution from $g_1$ and the inelastic contribution from $g_2$, respectively. They are evaluated as follows:
\begin{itemize}
	\item The Born contribution is just given by Eq.\eqref{eq:BoxBorn}. Isospin symmetry requires $G_V(Q^2)=f_1^V(Q^2)$ and $G_M(Q^2)=f_2^V(Q^2)$, so the integral is completely fixed by the nucleon electromagnetic form factors.
	\item For the inelastic contribution from $g_1^\gamma$, what we need is the following quantity:
	\begin{equation}
	\bar{\Gamma}_1^{p-n}(Q^2)\equiv \int_0^{x_\pi}dx\frac{4(5+4r)}{9(1+r)^2}\left\{g_{1,p}^\gamma(x,Q^2)-g_{1,n}^\gamma(x,Q^2)\right\}
	\end{equation} 
	($x_\pi=Q^2/[(m_N+M_\pi)^2-m_N^2+Q^2]$ is the pion production threshold) 
	as a function of $Q^2$. At $Q^2>2$~GeV$^2$, one resorts again to the pQCD-corrected polarized Bjorken sum rule, and include additionally a small higher-twist (HT) correction which is required to match the theory prediction with the experimental data~\cite{Deur:2014vea,Kotlorz:2017wpu,Ayala:2018ulm}. At $Q^2<2$~GeV$^2$, data are taken from the EG1b experiment at JLab~\cite{Fersch:2017qrq,Guler:2015hsw} that provided the first three moments of $g_{1,N}^{\gamma}$ from 0.05~GeV$^2$ to 3.5~GeV$^2$, which allow for a precise reconstruction of $\bar{\Gamma}_1^{p-n}(Q^2)$ at any value of $Q^2$ within the range. Fig.\ref{fig:data} shows the combination of the experimental and theory prediction of $\bar{\Gamma}_1^{p-n}(Q^2)$. One then performs the $Q^2$-integral to obtain the full $g_1^\gamma$ contribution. 
	\item Finally, the $g_2^{\gamma}$ contribution is split into two pieces, namely the twist-two and twist-three (and higher) contributions. The former is related to $g_1^\gamma$ through the Wandzura-Wilczek relation~\cite{Wandzura:1977qf}, while the latter is related to the so-called nucleon color polarizability~\cite{Shuryak:1981pi,Jaffe:1989xx} calculated within the baryon chiral effective theory~\cite{Alarcon:2020icz}.
\end{itemize}
Combining everything above, Ref.\cite{Gorchtein:2021qdf} reported $\Box_{\gamma W}^A=3.96(6)\times 10^{-3}$. The theory uncertainty is much smaller than that of $\Box_{\gamma W}^V$, due to the much more precise experimental inputs at low and moderate $Q^2$. 

\subsection{Future prospects with lattice QCD}

\begin{figure}
	\begin{centering}
		\includegraphics[scale=0.2]{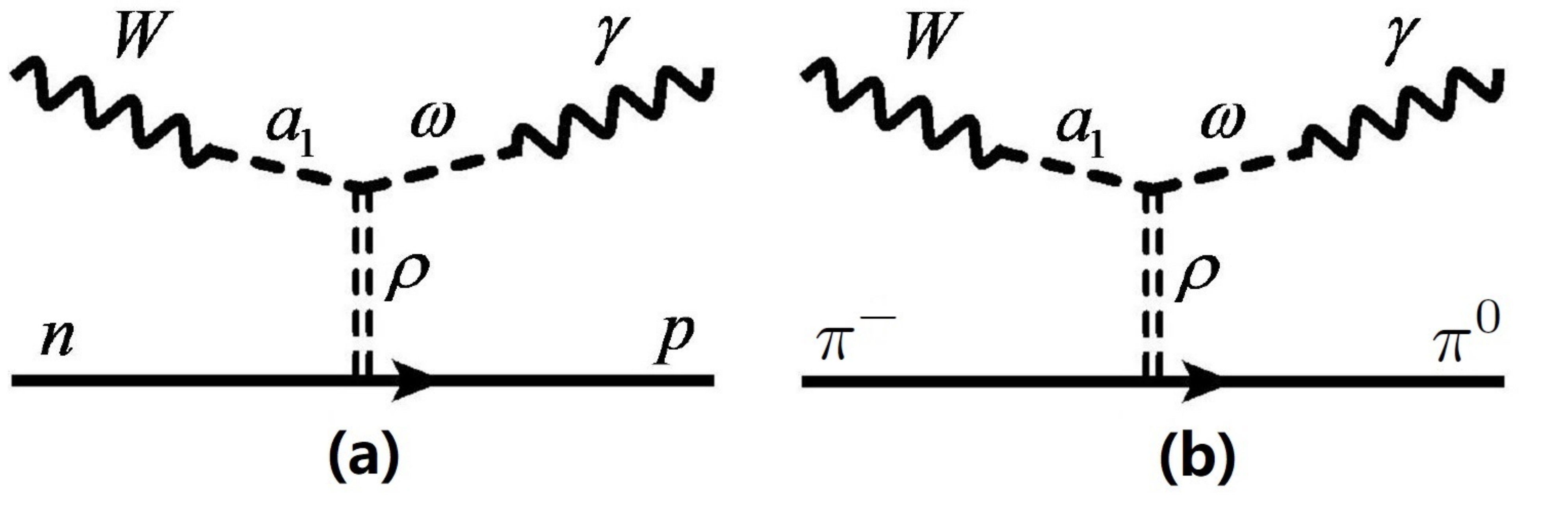}\hfill
		\par\end{centering}
	\caption{\label{fig:indirect} The comparison between the leading Regge contribution to the pion and nucleon $\gamma W$-box diagram. Figure taken from Ref.\cite{Seng:2020wjq}. }
\end{figure}

The major limiting factor in the DR analysis is the precision of the experimental inputs. This is more of an issue for $\Box_{\gamma W}^V$ because the existing $\nu(\bar{\nu})p$ scattering data are obsolete and contain large uncertainties in the non-perturbative region. The Deep Underground Neutrino Experiment (DUNE) conducted by Fermilab may start to provide new data with higher precision in the next decade~\cite{Acciarri:2016crz,Alvarez-Ruso:2017oui}; but even with that, one still needs to deal with the residual model-dependence between the matching of $M_3^{(0)}(1,Q^2)$ and $M_3^{\nu p+\bar{\nu}p}(1,Q^2)$. 

Lattice QCD is so far the only promising approach to solve the two issues above simultaneously. The lattice calculation of $\Box_{\gamma W}^<(\pi^+,\pi^0,M_\pi)$ we described in Sec.\ref{sec:latticeQCD} serves as a first prototype and has shown great success. In fact, apart from pinning down the inner RC in $\pi_{e3}$, this calculation also provides an indirect lattice input to the dispersion integral of $\Box_{\gamma W}^V$ for neutron and serves as a useful cross-check of the experimental data-based DR analysis~\cite{Seng:2020wjq,Seng:2021qdx}. The idea is most easily visualized in Fig.\ref{fig:indirect}: There is a one-to-one correspondence between the leading Regge contribution to the pion and nucleon $\gamma W$-box diagram, which allows us to obtain $M_{3,\mathbb{R}}^{(0)}(1,Q^2)$ through a simple rescaling of the lattice result of $\mathbb{M}_\pi(Q^2)$ (after subtracting the small resonance contribution to the latter). With the strategy above, Ref.\cite{Seng:2020wjq} reported $\Delta_R^V=0.02477(24)$ which is consistent with the original DR treatment. 

So far the direct lattice calculation of $M_3^{(0)}(1,Q^2)$ is not yet available but is under intense investigation. A way to proceed is to calculate the four-point correlation functions in analogy to Ref.\cite{Feng:2020zdc}. This is, however, more challenging than that in the pion sector because the quark contraction diagrams are more complicated, the lattice observables are much noisier, and the handling of excited-state contaminations are more challenging. Preliminary results are expected within the next couple of years~\cite{BhattacharyaSnowmass}. There is also an alternative proposal to calculate $M_3^{(0)}(1,Q^2)$ using the Feynman-Hellmann theorem (FHT)~\cite{Feynman:1939zza,Hellmann} on lattice~\cite{Seng:2019plg}, which states that the parity-odd structure function $F_3(x,Q^2)$ can be obtained from the second-order variation of the nucleon ground state energy upon the introduction of a periodic source term into the Hamiltonian:
\begin{equation}
H_\mathrm{scr}(t)=2\lambda_1\int d^3x\cos(\vec{q}\cdot\vec{x})J_\mathrm{em}^2(\vec{x},t)-2\lambda_2\int d^3x\sin(\vec{q}\cdot x)J_A^3(\vec{x},t)~.
\end{equation} 
The proposal was inspired by an earlier study of the parity-even structure functions~\cite{Chambers:2017dov}. Some other applications of the FHT on lattice include the calculation of hadronic charges and form factors~\cite{Chambers:2017tuf,Chang:2018uxx}, Compton scattering
amplitude~\cite{Agadjanov:2016cjc,Agadjanov:2018yxh}
and hadron resonances~\cite{RuizdeElvira:2017aet}.

Finally, it is worth pointing out that a future lattice calculation of $\Box_{\gamma W}^A$ is equally important, despite that it does not directly affect the $|V_{ud}|$ extraction. Given that $\Box_{\gamma W}^A$ is very precisely determined through the DR analysis, an extra lattice calculation of the same quantity serves as an excellent avenue to cross-check the validity of the lattice methodologies which are also applied to $\Box_{\gamma W}^V$. In particular, through this procedure one could check if there are any large systematic effects that are missed in the first-principles calculations. This is a highly non-trivial issue, as we have recently seen cases with significant disagreement between phenomenological and lattice determination of certain quantities, some of which have profound impacts on the interpretation of the results of precision experiments. A well-known example is the hadronic vacuum polarization contribution to $g_\mu-2$~\cite{Aoyama:2020ynm,Borsanyi:2020mff}. 

\section{\label{sec:superallowed}Superallowed beta decays}

Our next example is the beta decay of $J^P=0^+$ nuclei which is referred to as ``superallowed beta decay''. It is of particular interest as it provides by far the most precise measurement of $|V_{ud}|$. This is because: 
\begin{itemize}
	\item At tree level only the vector charged weak current is involved, whose matrix element is exactly known assuming isospin symmetry. 
	\item Experimental data of 23 superallowed transitions had been accumulated over five decades~\cite{Towner:1973yrc,Hardy:1975eq,Hardy:1990sz,Hardy:2004id,Hardy:2008gy,Towner:2010zz,Hardy:2014qxa,Hardy:2020qwl}, with 15 among them whose $ft$-value precision is 0.23\% or better; the large sample size leads to a huge gain in statistics~\cite{Hardy:2020qwl}. In fact, this makes it the only avenue where the experimental uncertainty in $|V_{ud}|$ is smaller than the theory uncertainty.
\end{itemize}
However, the price to pay is that one needs to carefully account for the nucleus-dependent corrections that enter each transition. A detailed description of all the nuclear theory calculations is beyond the scope of this review, and here we will just briefly outline the current situation.

The matrix element $|V_{ud}|$ is determined from superallowed decays through the following master formula:
\begin{equation}
|V_{ud}|^2=\frac{2984.432(3)\:\mathrm{s}}{\mathcal{F}t(1+\Delta_R^V)}~,\label{eq:Vudsuper}
\end{equation} 
where $\mathcal{F}t$ is a nucleus-independent quantity which comes from the nucleus-dependent $ft$-value after applying various nucleus-dependent corrections:
\begin{equation}
\mathcal{F}t=ft(1+\delta_\mathrm{R}')(1+\delta_\mathrm{NS}-\delta_\mathrm{C})~.
\end{equation}
In the equation above, $\delta_\mathrm{R}'$ represents the nucleus-dependent outer corrections, $\delta_\mathrm{NS}$ entails the nuclear modifications of the single-nucleon RC, and $\delta_\mathrm{C}$ is the isospin breaking corrections.
These three corrections together turn different nucleus-dependent $ft$-values into a nucleus-independent $\mathcal{F}t$-value. 
The only correction among them that is known to satisfactory precision is the nucleus-dependent outer correction $\delta_\mathrm{R}'$ which is a function of the charge $Z_f$ of the daughter nucleus and the electron end-point energy $E_m$, and is calculated to the order $Z_f^2\alpha^3$~\cite{Sirlin:1967zza,Sirlin:1987sy,Sirlin:1986cc,Towner:2007np}. On the other hand, existing calculations of the remaining two corrections are each  plagued by their respective ambiguities and will be discussed as follows.

\subsection{The nuclear structure correction}

To rigorously define the nuclear structure correction term $\delta_\mathrm{NS}$, we start from the nuclear $\gamma W$-box diagram in the inner correction to the Fermi matrix element, which is the only piece of the EWRCs that depends on details of the non-perturbative QCD. It can be written in a dispersive representation, in full analogy to Eq.\eqref{eq:DRformula}:~\footnote{There is a mistake in Ref.\cite{Seng:2018qru}: $\Box_{\gamma W}^\mathrm{nucl}$ should be  defined as the full nuclear $\gamma W$-box, instead of ``per nucleon''.}
\begin{equation}
\Box_{\gamma W}^\mathrm{nucl}=\frac{\alpha}{\pi\mathring{g}_V^\mathrm{nucl}}\int_0^\infty \frac{dQ^2}{Q^2}\frac{M_W^2}{M_W^2+Q^2}\int_0^1 dx\frac{1+2r}{\left(1+r\right)^2}F_{3,\mathrm{nucl}}^{(0)}(x,Q^2)~,\label{eq:nuclearboxDR}
\end{equation}
with $\mathring{g}_V^\mathrm{nucl}=G_V^\mathrm{nucl}(0)$. One then performs an artificial splitting of the nuclear axial $\gamma W$-box diagram into two pieces:
\begin{equation}
\Box_{\gamma W}^{\mathrm{nucl}}=\Box_{\gamma W}^V+\left[ \Box_{\gamma W}^{\mathrm{nucl}}- \Box_{\gamma W}^V\right]~,\label{eq:splitting}
\end{equation}
where $\Box_{\gamma W}^V$ is the single-nucleon $\gamma W$-box we studied in Sec.\ref{sec:freen}. With the splitting above, $\Box_{\gamma W}^V$ then enters $\Delta_R^V$ in Eq.\eqref{eq:Vudsuper}, whereas
\begin{equation}
\delta_\mathrm{NS}\equiv 2\left[\Box_{\gamma W}^{\mathrm{nucl}}-\Box_{\gamma W}^V\right]
\end{equation}
describes the nuclear modifications to the single-nucleon $\gamma W$-box diagram. The benefit of such a separation is apparent: The contributions to  $\Box_{\gamma W}^\mathrm{nucl}$ and $\Box_{\gamma W}^V$ at high energies are the same because in this region the gauge bosons mainly probe a single, free nucleon; for instance, the large-$Q^2$ contribution to both $\Box_{\gamma W}^\mathrm{nucl}$ and $\Box_{\gamma W}^V$ take the same form of $(\alpha/(8\pi))\ln(M_W^2/\Lambda^2)$. On the other hand, at low energies the two start to deviate because (a) the single-nucleon absorption spectrum is distorted in the nuclear medium, and (b) the two gauge bosons may probe two different nucleons in a nucleus. 

\subsubsection{\label{sec:earlyNS}Earlier treatments}

Early treatments of $\delta_\mathrm{NS}$ mainly consist of calculating effects (a) and (b) mentioned above with nuclear models, which add up to give:
\begin{equation}
\delta_\mathrm{NS}=\frac{\alpha}{\pi}\left[C_a+C_b\right]~.
\end{equation}

Refs.\cite{Towner:1994mw,Towner:2002rg,Towner:2007np} estimated the size of $C_a$ by considering the ``quenching effect'' of the free-nucleon nucleon axial and EM coupling constants for spin-flip processes in the nuclear medium~\cite{Brown:1983zzc,Brown:1987obh,Towner:1987zz}. That means, one starts from the Born contribution to the free-nucleon box diagram,
\begin{equation}
\Box_{\gamma W}^V=\frac{\alpha}{2\pi}C_B^\mathrm{free}+...~,
\end{equation}  
(where $C_B^\mathrm{free}$ is given in Eq.\eqref{eq:BoxBorn}, with the free-nucleon form factors) and apply a ``quenching factor'' $q$ which is different for different nucleus, to obtain:
\begin{equation}
\Box_{\gamma W}^\mathrm{nucl}=\frac{\alpha}{2\pi}qC_B^\mathrm{free}+...~,
\end{equation}
and so,
\begin{equation}
C_a\approx (q-1)C_B^\mathrm{free}~.
\end{equation}
Meanwhile, $C_b$ (denoted in earlier literature as ``$C_\mathrm{NS}$'') were calculated with nuclear shell models in Refs.\cite{Jaus:1989dh,Towner:1992xm,Barker:1991tw}, which was further improved in Ref.\cite{Towner:2007np} by incorporating the same quenching effect, i.e. $C_\mathrm{NS}\rightarrow C_\mathrm{NS}^\mathrm{quenched}$. The final numerical results of
\begin{equation}
\delta_\mathrm{NS}=\frac{\alpha}{\pi}\left[(q-1)C_B^\mathrm{free}+C_\mathrm{NS}^\mathrm{quenched}\right]
\end{equation}
for different superallowed transitions were summarized in Table VI of Ref.\cite{Towner:2007np} and appeared in several subsequent reviews by Hardy and Towner~\cite{Hardy:2008gy,Hardy:2014qxa}. In particular, with the theory inputs above, Ref.\cite{Hardy:2014qxa} quoted an average value of 
\begin{equation}
\overline{\mathcal{F}t}=3072.27\pm 0.72~\mathrm{s}
\end{equation}
based on the lifetimes of fourteen best-measured superallowed transitions from $^{10}$C to $^{74}$Rb.

\subsubsection{Recent developments}

\begin{figure}
	\begin{centering}
		\includegraphics[scale=0.25]{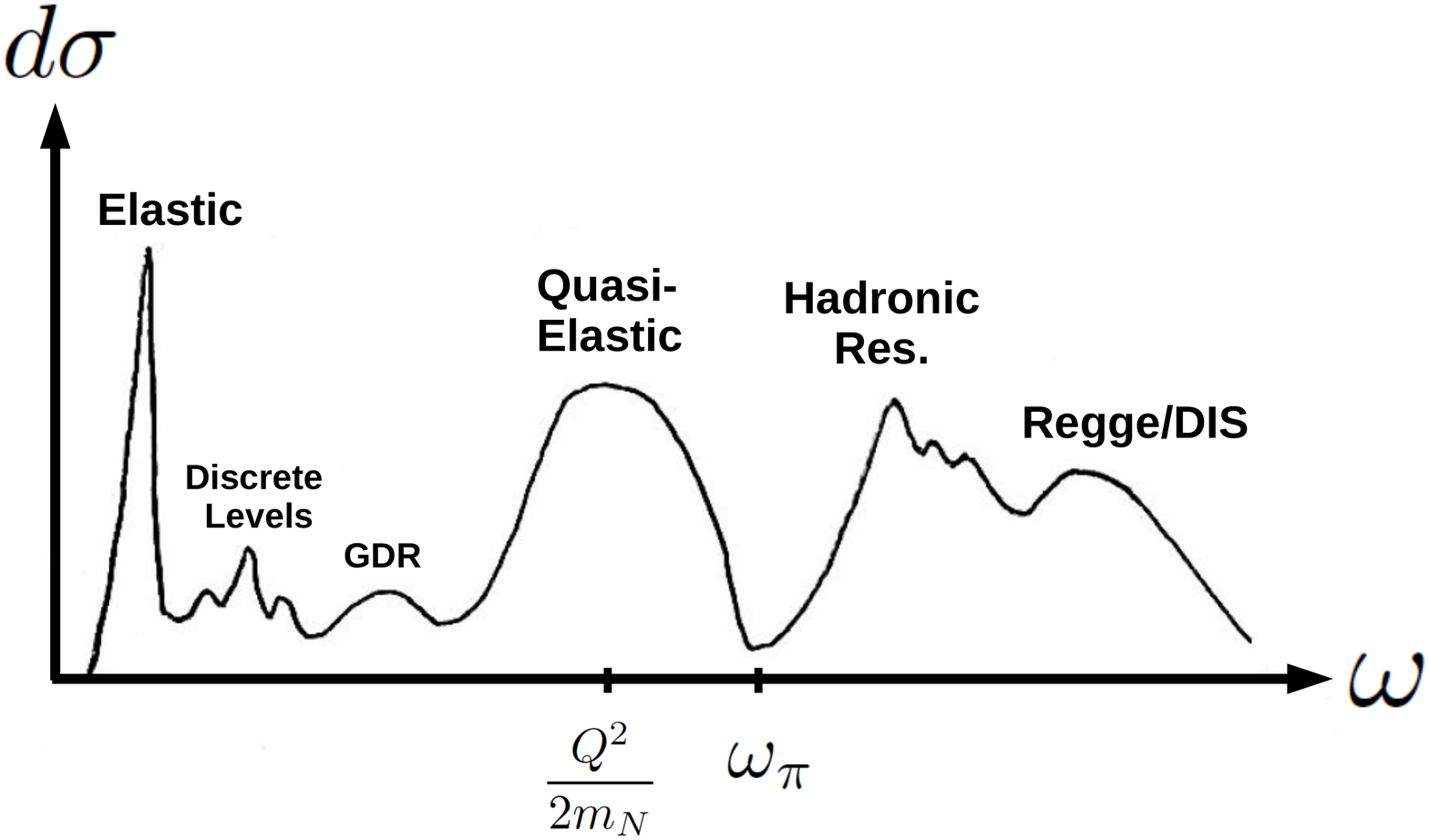}
		\hfill
		\par\end{centering}
	\caption{\label{fig:absoprtion} Idealized structure of the nuclear absorption spectrum at low energy.}
\end{figure}

Recent theory developments of the EWRC in beta decays, in particular the establishment of the DR formalism, cast doubts on the treatment above. This can be most clearly seen through the artificial splitting of the full nuclear $\gamma W$-box in Eq.\eqref{eq:splitting}. There is no approximation made in this splitting, however in practice the two terms at the right hand side are calculated with completely different fashions. The free nucleon term is evaluated under a fully relativistic framework of DR, making using inputs from intermediate- and high-energy data (or lattice inputs), whereas the $\delta_\mathrm{NS}$ term is calculated within non-relativistic shell models. The artificial distinction in the treatment of these two terms may lead to possible large systematic uncertainties.

Therefore, a more reasonable starting point is to treat $\Box_{\gamma W}^\mathrm{nucl}$ in the same footing with $\Box_{\gamma W}^V$, namely based on the fully relativistic dispersive representation in Eq.\eqref{eq:nuclearboxDR}. The dispersive integral depends on the  
the parity-odd nuclear structure function $F_{3,\mathrm{nucl}}^{(0)}(x,Q^2)$ which is defined through:
\begin{equation}
\frac{V_{ud}^{*-1}}{4\pi}\sum_X(2\pi)^4\delta^{(4)}(p+q-p_X)\langle \phi_f(p)|J_\mathrm{em}^{(0)\mu}|X\rangle \langle X|J_A^\nu|\phi_i(p)\rangle=-\frac{i\epsilon^{\mu\nu\alpha\beta}q_\alpha p_\beta}{2p\cdot q}F_{3,\mathrm{nucl}}^{(0)}(x,Q^2)~,\label{eq:F3nucl}
\end{equation}
where $\phi_{i,f}$ are the initial and final nuclear states in a given superallowed transition, and $\{X\}$ exhausts all on-shell intermediate nuclear states. Therefore, the study of $\delta_\mathrm{NS}$ boils down to the calculation of single-current nuclear matrix elements involving $\phi_{i,f}$ and the leading contributors within $\{X\}$.

A rigorous, model-independent study of the many-body effects in Eq.\eqref{eq:F3nucl} may involve sophisticated nuclear theory calculations, e.g. with chiral EFT~\cite{Weinberg:1990rz,Weinberg:1991um,Weinberg:1992yk,vanKolck:1999mw,Epelbaum:2005pn,Machleidt:2011zz,Bernard:2007sp,Bernard:2011zr,Girlanda:2011fh,Krebs:2012yv,Krebs:2013kha,Epelbaum:2014sza,Entem:2015xwa,Reinert:2017usi,Entem:2017gor,Hammer:2019poc} and its lattice implementations~\cite{Lee:2004si,Borasoy:2006qn,Lee:2008fa,Lahde:2019npb}. This is not yet attempted and is an important task that needs to be carried out in the future. However, here we are still able to get a flavor of some most important nuclear corrections by simply analyzing the idealized structure of the virtual photoabsorption spectrum on a nucleus as depicted in Fig.\ref{fig:absoprtion}. At low energies, a narrow peak is contributed by the elastic intermediate state and several smaller ones are from discrete energy levels and the giant dipole resonance (GDR). After that, a broad absorption peak around $\omega=Q^2/(2m_N)$ is observed; this is the so-called quasi-elastic (QE) peak that was supposed to be a delta function for free nucleon, but broadened by nuclear interactions. Above the pion production threshold $\omega_\pi=(Q^2+2m_N M_\pi+M_\pi^2)/(2m_N)$, the absorption spectrum starts to resemble that of a single nucleon, and so the contribution to $\delta_\mathrm{NS}$ vanishes.   

\begin{figure}
	\begin{centering}
		\includegraphics[scale=0.5]{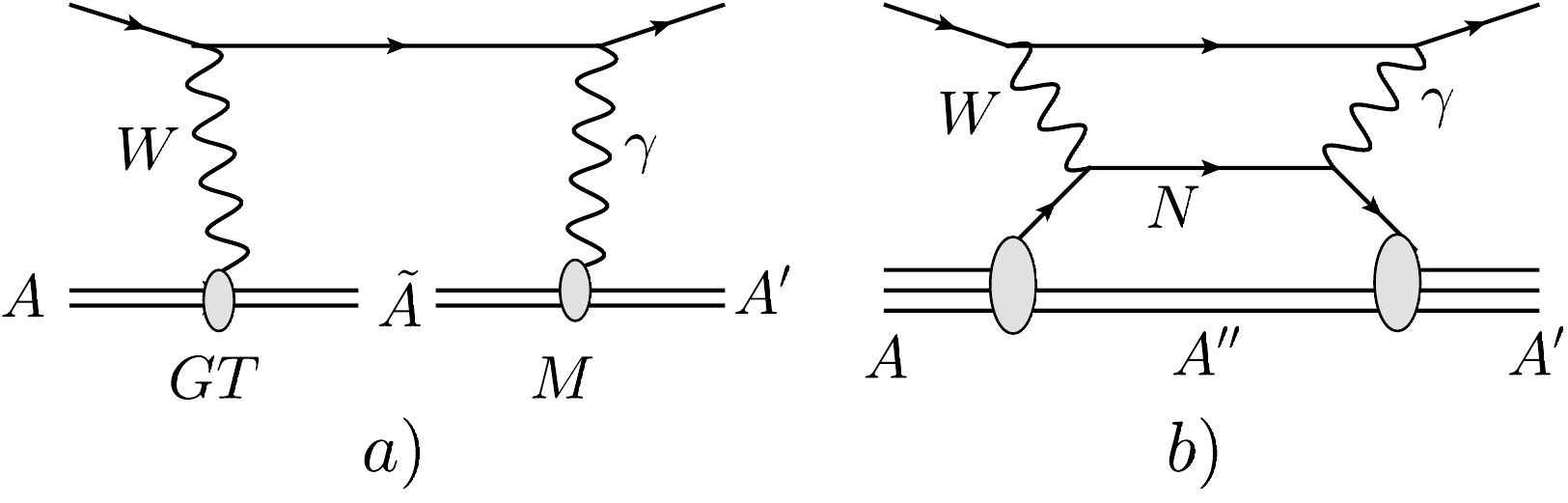}
		\hfill
		\par\end{centering}
	\caption{\label{fig:HTvsQE} Comparison between the contribution to $\Box_{\gamma W}^\mathrm{nucl}$ (a) effectively accounted for in the ``quenching mechanism'' and (b) from a QE nucleon in the nucleus. Figure taken from Ref.\cite{Seng:2018qru}.}
\end{figure}

With this picture, it is almost immediately apparent that the treatment in Sec.\ref{sec:earlyNS} did not account for the most important nuclear structure corrections. Since the quenching factors $q$ are obtained from spin-flip processes in Gamow-Teller (GT) transitions, the gauge bosons are effectively probing deeply off-shell nucleons in the discrete energy levels at the lower end of the nuclear absorption spectrum. On the other hand, the much more important contribution from the QE nucleon is missing in this treatment (see Fig.\ref{fig:HTvsQE}). Ref.\cite{Seng:2018qru} estimated, for the first time, the contribution from the latter using a simple Fermi gas model. 

Apart from the above, there is yet another nuclear correction which was never pointed out until recently~\cite{Gorchtein:2018fxl}. In evaluating the axial $\gamma W$-box diagrams for free neutron and nucleus, one always take $m_e,E_e\rightarrow 0$ and $p'\rightarrow p$. In free neutron, this approximation is well-justified because the energy splitting between the ground state and the on-shell intermediate state $X$ is at least $M_\pi$, so the energy-dependent effect is expected to scale as $(\alpha/\pi)E_e/M_\pi\sim 10^{-5}$ which is negligible at the present stage. However, in nucleus, there exists excited states with typical energy scale $\Lambda_\mathrm{nucl}$ of the order MeV, so $(\alpha/\pi) E_e/\Lambda_\mathrm{nucl}$ is no longer suppressed. In this case, a supposedly energy-independent nuclear ``inner'' correction $\delta_\mathrm{NS}$ generates an energy-dependent effect which is more commonly classified as an ``outer'' correction represented by $\delta_\mathrm{R}'$. 

The two novel effects above are now known as the new nuclear corrections (NNCs) to superallowed beta decays. Although there is currently no rigorous theory analysis of their sizes, simple estimations using a Fermi gas model suggest that they are individually sizable, but carry opposite signs and partially cancel each other. As a net effect, this leads to an increase of the theory uncertainty in $\mathcal{F}t$~\cite{Hardy:2020qwl}:
\begin{eqnarray}
\overline{\mathcal{F}t}&:& 3072.27\pm 0.72\:\mathrm{s}\rightarrow 3072.24\pm 1.85\:\mathrm{s}~.
\end{eqnarray} 
This large theory uncertainty is now the major limiting factor in the precise determination of $|V_{ud}|$. 

\subsection{The isospin-breaking correction}

Finally, we briefly discuss the isospin-breaking correction $\delta_\mathrm{C}$ although it is not a part of the EWRC. This correction
plays a central role in obtaining a universal $\mathcal{F}t$-value from different superallowed transitions, which is required by the conserved vector current (CVC) hypothesis. However, it turns out that among all existing nuclear theory calculations of $\delta_\mathrm{C}$, including shell-model with Woods-Saxon (WS) potential~\cite{Towner:2002rg,Towner:2007np,Hardy:2008gy}, Hartree-Fock wavefunctions~\cite{Ormand:1989hm,Ormand:1995df}, density functional theory~\cite{Satula:2011br}, random-phase approximation~\cite{Liang:2009pf}, isovector monopole resonance sum rule~\cite{Auerbach:2008ut} and the Damgaard model~\cite{Damgaard:1969yyx}, only the WS calculation is able to achieve such an alignment and is taken as the standard input for $\delta_\mathrm{C}$. Whether this indicates the success of this particular simplified nuclear picture, or rather the existence of some unexpected new physics in Nature (e.g. second-class currents~\cite{Weinberg:1958ut}), is not yet clear. 

It was recently pointed out~\cite{Koshchii:2020qkr} that it is possible to relate $\delta_\mathrm{C}$ to experimental observables generated by the same isospin-breaking effect, such as neutron skins and differences in nuclear charge radii within a given isomultiplet.
Charge radii of stable daughter nuclei are known to few parts in $10^4$~\cite{Angeli:2013epw}, whereas those for unstable parent nuclei can be measured at rare-isotope facilities, e.g. the Isotope Mass Separator On-Line Device facility (ISOLDE) and the Facility for Rare Isotopes Beams (FRIB). Meanwhile, the size of the parity-violating asymmetry in electron-nucleus (e.g. $^{12}$C) scattering at backward angles is largely affected by the neutron skin of the nucleus~\cite{Koshchii:2020qkr}, so in principle future parity-violating electron scattering experiments at Jefferson Lab~\cite{Souder:2016xcn} and at the Mainz Energy-Recovering Superconducting Accelerator (MESA)~\cite{Becker:2018ggl} will allow one to extract neutron skins of selected stable superallowed daughter nuclei to few per mille and therefore better constrain $\delta_\mathrm{C}$ in a fully model-independent manner, provided that the theoretical relations between these quantities are well-established.

\section{\label{sec:Kl3}Kaon semileptonic decays}

So far we only discussed semileptonic decay processes that involve a $u\leftrightarrow d$ transition and are attached to $V_{ud}$. In this section we switch to the simplest semileptonic decay process that involves a $u\leftrightarrow s$ transition, namely the $K\rightarrow \pi\ell \nu$ decay. It is commonly denoted as $K_{\ell 3}$, which represents collectively six types of decay processes: $K_{e3}^L$, $K_{e3}^S$, $K_{e3}^+$, $K_{\mu 3}^L$, $K_{\mu 3}^S$ and $K_{\mu 3}^+$. They are of great interest to both theorists and experimentalists as they provide one of the most precise measurements of $|V_{us}|$. This is done via the following master formula:
\begin{equation}
\Gamma_{K_{\ell 3}}=\frac{G_F^2|V_{us}|^2M_K^5C_K^2}{192\pi^3}S_\mathrm{EW}|f_+^{K^0\pi^-}(0)|^2I_{K\ell}^{(0)}\left(1+\delta_\mathrm{EM}^{K\ell}+\delta_\mathrm{SU(2)}^{K\pi}\right)~.\label{eq:Kl3master}
\end{equation}

The focus of this review is on the EWRCs to the $K_{\ell 3}$ decay rate, which are contained in the short-distance electroweak factor $S_\mathrm{EW}$ (discussed in  Sec.\ref{sec:EWlog}) and long-distance EM correction $\delta_\mathrm{EM}^{K\ell}$. Before doing so, we shall briefly discuss 
all the other inputs that are needed to extract $|V_{us}|$ from Eq.\eqref{eq:Kl3master}:
\begin{itemize}
	\item $\Gamma_{K_{\l3}}$ is the $K_{\ell 3}$ decay width. The kaon lifetimes, branching ratios and differential decay widths are measured by BNL E865~\cite{Sher:2003fb}, KTeV~\cite{Alexopoulos:2004sw,Alexopoulos:2004sx,Alexopoulos:2004sy,Abouzaid:2006ir,KTeV:2010sng}, KLOE~\cite{Ambrosino:2005vx,Ambrosino:2006up,Ambrosino:2005ec,Ambrosino:2006si,Sciascia:2005zz,Ambrosino:2006gn,Ambrosino:2007ac,KLOE:2007jte,KLOE:2010yit}. KLOE-2~\cite{KLOE-2:2019rev}, NA48~\cite{Lai:2004bt,Lai:2006cf,Batley:2006cj,Lai:2004kb} and ISTRA+~\cite{Romanovsky:2007qb,Yushchenko:2004zs} over the past two decades. 
	\item $f_+^{K^0\pi^-}(0)$ is the $K^0\pi^-$ charged weak form factor in the (unphysical) $t\rightarrow 0$ limit, with the general definition of the charged weak form factors in spinless systems given in Eq.\eqref{eq:FFspinless}. High-precision lattice calculations of this quantity were performed over the past decade by the FNAL/MILC~\cite{Gamiz:2013xxa,Bazavov:2013maa,Bazavov:2018kjg} and the ETM~\cite{Carrasco:2016kpy} collaborations and showed perfect mutual consistencies and a steady improvement in precision. The most recent Flavor Lattice Averaging Group (FLAG) online review (updated from the 2019 version~\cite{FlavourLatticeAveragingGroup:2019iem}) quoted the following average:
	\begin{equation}
	|f_+^{K^0\pi^-}(0)|=0.9698(17)~.\quad n_f=2+1+1\quad\text{Refs.\cite{Carrasco:2016kpy,Bazavov:2018kjg}}
	\end{equation}
	However, a recent calculation from the PACS collaboration based on a single lattice spacing returned a somewhat smaller value~\cite{Kakazu:2019ltq}.  
	\item $C_K$ is a simple isospin factor defined through the value of $f_+^{K\pi}(0)$ in the SU(3) limit:
	\begin{equation}
	|C_{K}|=|f_{+}^{K\pi}(0)|_{m_{u}=m_{d}=m_{s}}=\left\{ \begin{array}{cc}
	1~, & K^{0}\\
	1/\sqrt{2}~,  & K^{+}
	\end{array}\right.~.
	\end{equation}
	\item $I_{K\ell}^{(0)}$ is the tree-level phase space factor defined as:~\footnote{Notice that a number of important review papers, e.g. Refs.\cite{Antonelli:2010yf,Cirigliano:2011ny}, contain a typo in their formula for $I_{K\ell}^{(0)}$.}
	\begin{equation}
	I_{K\ell}^{(0)}=\int_{m_\ell^2}^{(M_K-M_\pi)^2}\frac{dt}{M_K^8}\bar{\lambda}^{3/2}\left(1+\frac{m_\ell^2}{2t}\right)\left(1-\frac{m_\ell^2}{t}\right)^2\left[\bar{f}_+^2(t)+\frac{3m_\ell^2\Delta_{K\pi}^2}{\left(2t+m_\ell^2\right)\bar{\lambda}}\bar{f}_0^2(t)\right]~,
	\end{equation}
	with $\bar{\lambda}=\left[t-(M_K+M_\pi)^2\right]\left[t-(M_K-M_\pi)^2\right]$ and $\Delta_{K\pi}=M_K^2-M_\pi^2$. Here we have defined the ``scalar'' form factor
	\begin{equation}
	f_0^{K\pi}(t)\equiv f_+^{K\pi}(t)+\frac{t}{M_K^2-M_\pi^2}f_-^{K\pi}(t) 
	\end{equation}
	which coincides with the ``vector'' form factor $f_+^{K\pi}$ at $t=0$. The normalized form factors $\bar{f}_{+,0}$ are defined by factoring out their values at $t=0$:
	$\bar{f}_{+,0}(t)\equiv f_{+,0}^{K\pi}(t)/f_{+}^{K\pi}(0)$. These form factors are obtained by fitting the $K_{\ell 3}$ differential decay rate to a specific parameterization. These  parameterizations generally fall into two classes~\cite{Abouzaid:2009ry}: The Class II parameterization that is based on a systematic mathematical expansion (e.g. the Taylor expansion~\cite{Antonelli:2010yf} and the $z$-parameterization~\cite{Hill:2006bq}) and the Class I parameterization that imposes additional physical constraints to reduce the number of independent parameters (e.g. the pole~\cite{Lichard:1997ya} and dispersive parameterization~\cite{Bernard:2006gy,Bernard:2009zm,Abouzaid:2009ry}). A detailed summary of the results from different parameterizations are given in Ref.\cite{Antonelli:2010yf}, and a recent update is provided by the NA48/2 collaboration~\cite{NA482:2018rgv}.
	\item Finally, $\delta_\mathrm{SU(2)}^{K\pi}$ is defined as the isospin breaking correction to the $f_+^{K\pi}(0)$ form factor at $t=0$:~\footnote{The appearance of the factor $C_{K^0}/C_K$ is simply due to our choice of normalization of the form factors in Eq.\eqref{eq:FFspinless}.}
	\begin{equation}
	\delta_\mathrm{SU(2)}^{K\pi}\equiv\left(\frac{C_{K^0}}{C_K}\frac{f_+^{K\pi}(0)}{f_+^{K^0\pi^-}(0)}\right)^2-1~,
	\end{equation}
	which only resides in $K_{\ell 3}^+$ by construction. Upon neglecting small EM contributions, it is completely fixed by the QCD quark mass parameters $m_s/\hat{m}$ and $\mathcal{Q}\equiv (m_s^2-\hat{m}^2)/(m_d^2-m_u^2)$.
	The value frequently quoted in reviews in the early 2010s~\cite{Antonelli:2010yf,Cirigliano:2011ny} is $\delta_\mathrm{SU(2)}^{K^+\pi^0}=0.058(8)$; with the latest lattice inputs one obtains $\delta_\mathrm{SU(2)}^{K^+\pi^0}=0.0457(20)$~\cite{Seng:2021nar}. At the same time, one also observes a slight discrepancy between the determinations from lattice QCD and  phenomenology~\cite{Colangelo:2018jxw}.
\end{itemize}

\subsection{Kinematics}

The main difference between $K_{\ell 3}$ and $\pi_{e3}$ or superallowed decays is that the initial and final hadron are not nearly-degenerate. This means many terms that are suppressed in the $\pi_{e3}$ and superallowed decays may contribute significantly in $K_{\ell 3}$, at both tree level and loops. Thus, to facilitate the discussion of the EWRC, we first briefly review the kinematics in $K_{\ell 3}$. 

We are interested in the decay process $K(p)\rightarrow \pi(p')+\ell^+(p_\ell)+\nu_\ell(p_\nu)+n\gamma$, where $\ell=e,\mu$ and $n\geq 0$. The inclusion of processes with emissions of real photons is necessary to ensure the cancellation of IR divergences. When all the massless particles (neutrino and photons) as well as the lepton spins are not undetected, the differential decay rate depends only on three variables:
\begin{equation}
x\equiv \frac{P^2}{M_K^2}~,\quad y\equiv \frac{2p\cdot p_\ell}{M_K^2}~,\quad z\equiv \frac{2p\cdot p'}{M_K^2}~,
\end{equation}
with $P\equiv p-p'-p_\ell$. Notice that for $n=0$ one must have $x=0$ due to momentum conservation. Another possible set of independent variables are the usual Mandelstam variables: $s\equiv (p'+p_\ell)^2$, $t\equiv (p-p')^2$, $u\equiv (p-p_\ell)^2$. We also define $r_\pi\equiv M_\pi^2/M_K^2$ and $r_\ell\equiv m_\ell^2/M_K^2$ for future convenience, and denote the decay amplitude with $n$ real photons in the final state as $\mathfrak{M}^{(n)}$. 

To $\mathcal{O}(G_F\alpha)$ precision, one needs only to include the $n=0$ and $n=1$ processes; the former is calculated to one-loop and the latter at tree level. The total decay rate is then given by:
\begin{eqnarray}
\Gamma_{K_{\ell 3}}&=&\frac{M_K}{256\pi^3}\int_{\mathcal{D}_3}dydz\left\{|\mathfrak{M}^{(0)}|^2+\frac{M_K^2}{32\pi^3}\int_0^{\alpha_+}dx\int\frac{d^3k}{E_k}\frac{d^3p_\nu}{E_\nu}\delta^{(4)}(P-k-p_\nu)|\mathfrak{M}^{(1)}|^2\right\}\nonumber\\
&&+\frac{M_K^3}{8192\pi^6}\int_{\mathcal{D}_{4-3}}dydz\int_{\alpha_-}^{\alpha_+}dx\int\frac{d^3k}{E_k}\frac{d^3p_\nu}{E_\nu}\delta^{(4)}(P-k-p_\nu)|\mathfrak{M}^{(1)}|^2~,\label{eq:Kl3int}
\end{eqnarray}
with $k$ the momentum of the real photon in $\mathfrak{M}^{(1)}$, and
\begin{equation}
\alpha_{\pm}\equiv 1-y-z+r_\pi+r_\ell+\frac{yz}{2}\pm\frac{1}{2}\sqrt{y^2-4r_\ell}\sqrt{z^2-4r_\pi}~.
\end{equation}
We have also defined two different regions in the $y-z$ plane, namely:
\begin{eqnarray}
\mathcal{D}_3&:&a(y)-b(y)<z<a(y)+b(y)~,\quad 2\sqrt{r_\ell}<y<1+r_\ell-r_\pi\nonumber\\
\mathcal{D}_{4-3}&:&2\sqrt{r_\pi}<z<a(y)-b(y)~,\quad 2\sqrt{r_\ell}<y<1-\sqrt{r_\pi}+\frac{r_\ell}{1-\sqrt{r_\pi}}~,
\end{eqnarray}
where
\begin{equation}
a(y)=\frac{(2-y)(1+r_\pi+r_\ell-y)}{2(1+r_\ell-y)}~,\quad b(y)=\frac{\sqrt{y^2-4r_\ell}(1+r_\ell-r_\pi-y)}{2(1+r_\ell-y)}~.
\end{equation}
We see that $|\mathfrak{M}^{(0)}|^2$ only resides within $\mathcal{D}_3$, while $|\mathfrak{M}^{(1)}|^2$ resides within a larger region $\mathcal{D}_4=\mathcal{D}_3+\mathcal{D}_{4-3}$. Both squared amplitudes contain IR-divergences in $\mathcal{D}_3$, which cancel each other in the first line of Eq.\eqref{eq:Kl3int}; meanwhile $|\mathfrak{M}^{(1)}|^2$ is IR-finite in $\mathcal{D}_{4-3}$. 

A lot can be learned from the squared amplitude of the $n=0$ process at tree level:
\begin{eqnarray}
|\mathfrak{M}_0^{(0)}|^2&=&2G_F^2|V_{us}|^2M_K^4\left\{A_1^{(0)}(y,z)|f_+^{K\pi}(t)|^2+A_2^{(0)}(y,z)f_+^{K\pi}(t)f_-^{K\pi}(t)\right.\nonumber\\
&&\left.+A_3^{(0)}(y,z)|f_-^{K\pi}(t)|^2\right\}
\end{eqnarray}
where
\begin{eqnarray}
A_1^{(0)}(y,z)&=&4(1-y)(y+z-1)+r_\ell (4y+3z-3)-4r_\pi+r_\ell(r_\pi-r_\ell)\nonumber\\
A_2^{(0)}(y,z)&=&2r_\ell(3-2y-z+r_\ell-r_\pi)\nonumber\\
A_3^{(0)}(y,z)&=&r_\ell(1-z+r_\pi-r_\ell)~.
\end{eqnarray}
We find that $A_2^{(0)}$ and $A_3^{(0)}$ are explicitly suppressed by a factor of $r_\ell$. Recall that $r_e\approx 1\times 10^{-6}$ and $r_\mu\approx 0.05$, we see immediately that the contribution from $f_-^{K\pi}$ to $|\mathfrak{M}_0^{(0)}|^2$ is negligible in the case of $K_{e3}$ at tree level. The same is for the EWRC: One can show that the one-loop virtual corrections can always be recast into a shift of the form factors:
\begin{equation}
f_\pm^{K\pi}(t)\rightarrow f_\pm^{K\pi}(t)+\delta f_\pm^{K\pi}(y,z)~,
\end{equation}
so in $K_{e3}$ the only relevant quantity is $\delta f_+^{K\pi}(y,z)$. This is an important observation when we discuss later the recent progress of the EWRC based on Sirlin's representation.

\subsection{\label{sec:Kl3ChPT}ChPT treatment of the EWRC}

ChPT provided the first analysis of the EWRC in $K_{\ell 3}$ with a rigorous quantification of theory uncertainties~\cite{Cirigliano:2001mk,Cirigliano:2004pv,Cirigliano:2008wn}. Here we will briefly discuss the main results without diving into any detail.  

\begin{figure}
	\begin{centering}
		\includegraphics[scale=0.10]{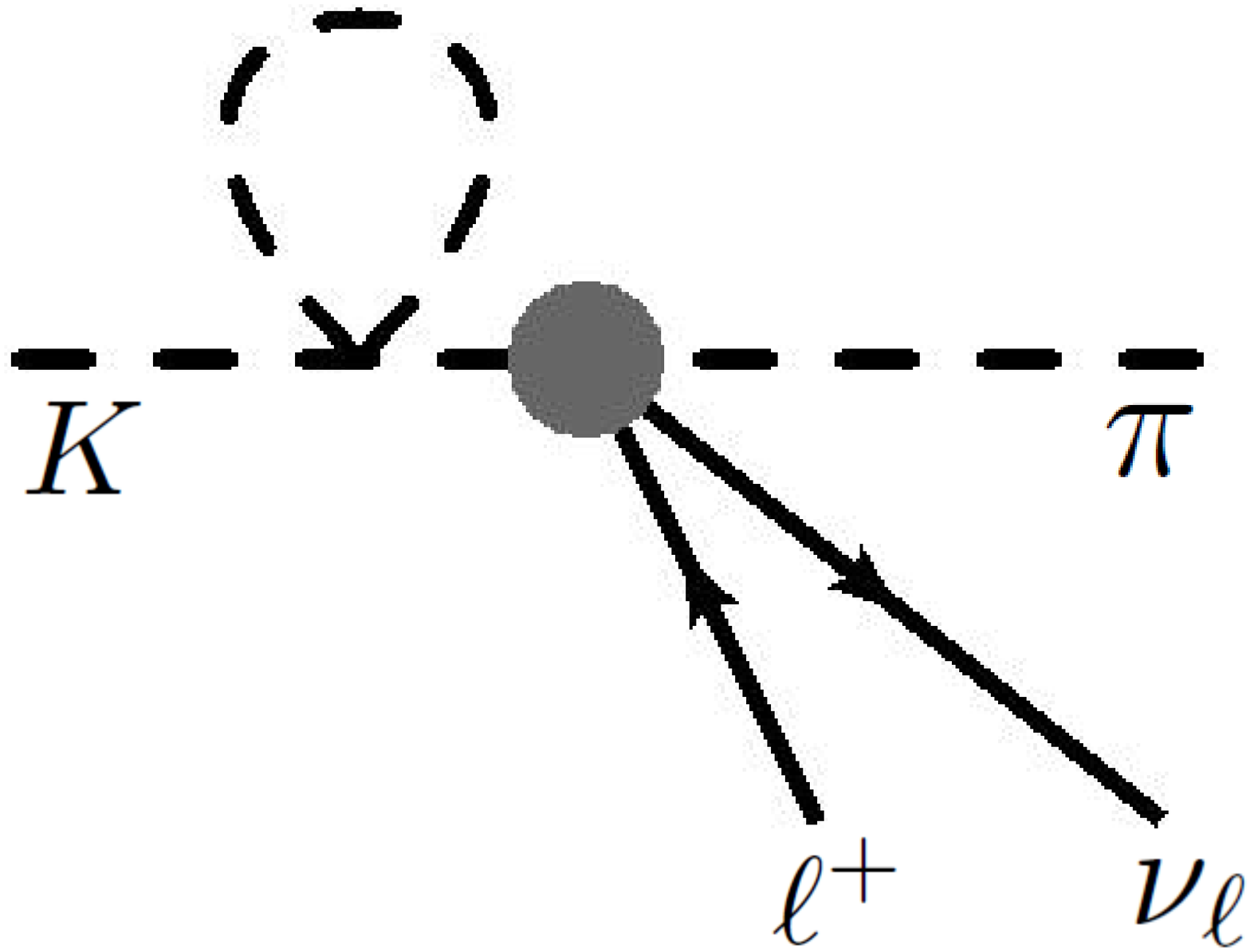}
		\includegraphics[scale=0.10]{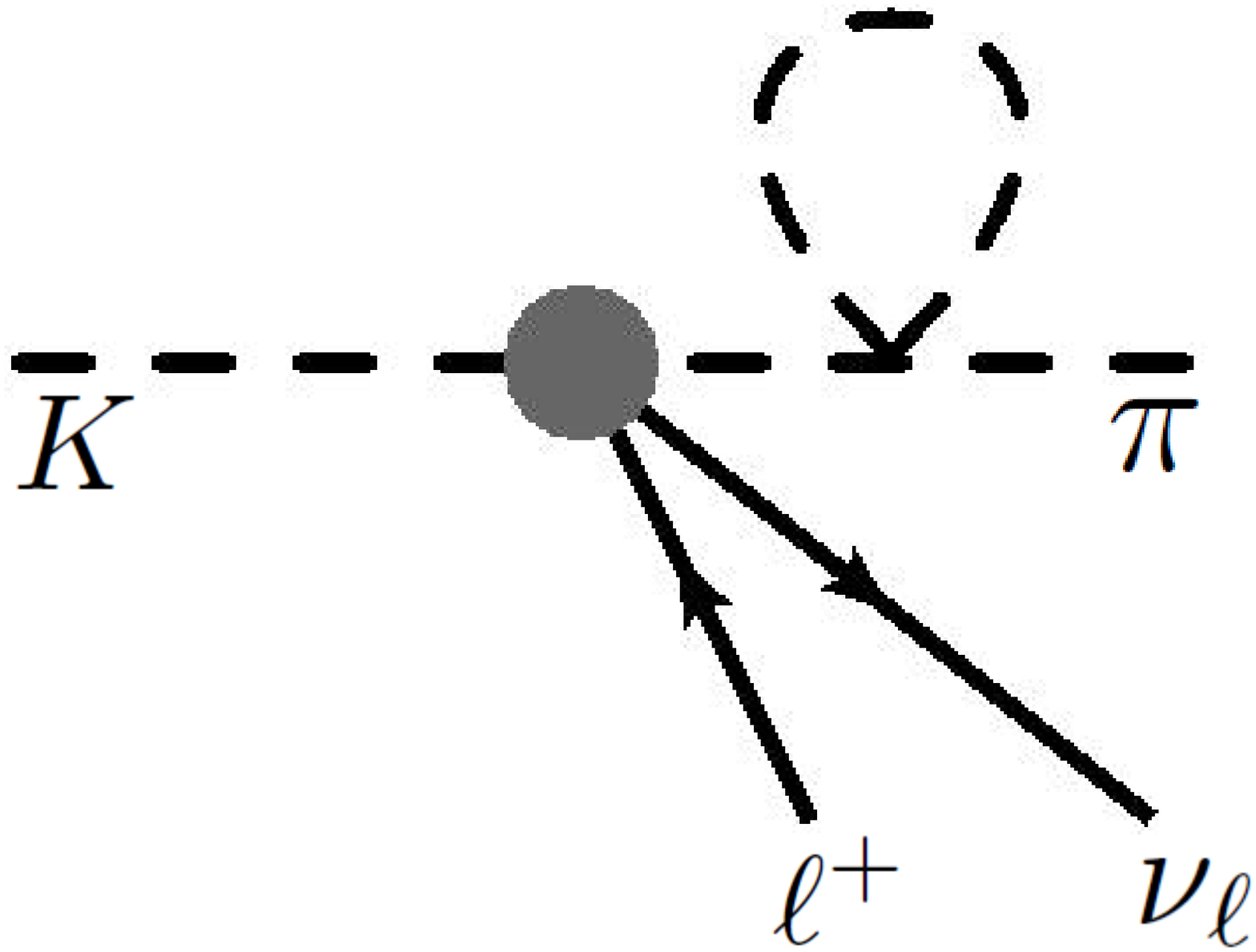}
		\includegraphics[scale=0.10]{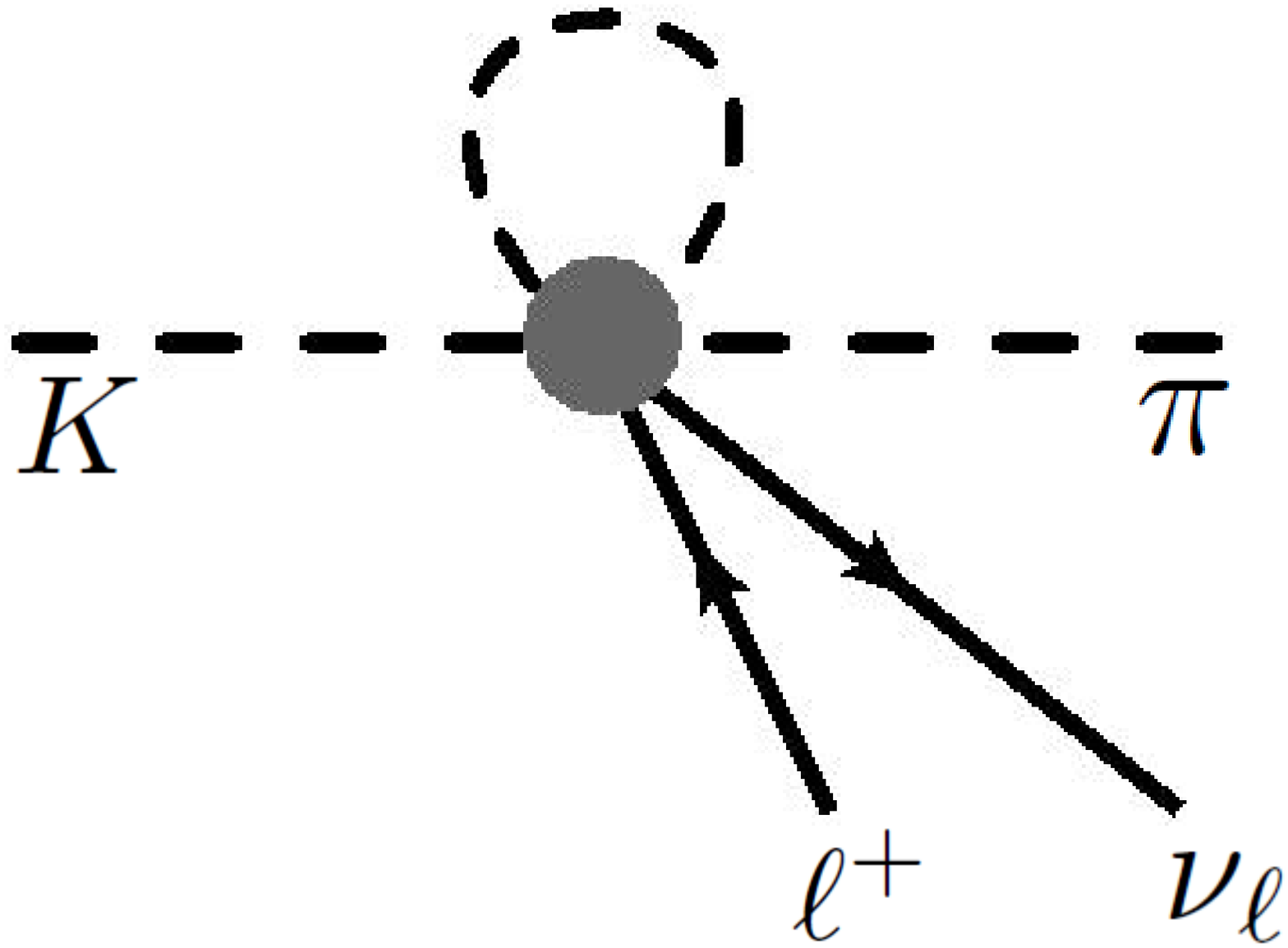}
		\includegraphics[scale=0.10]{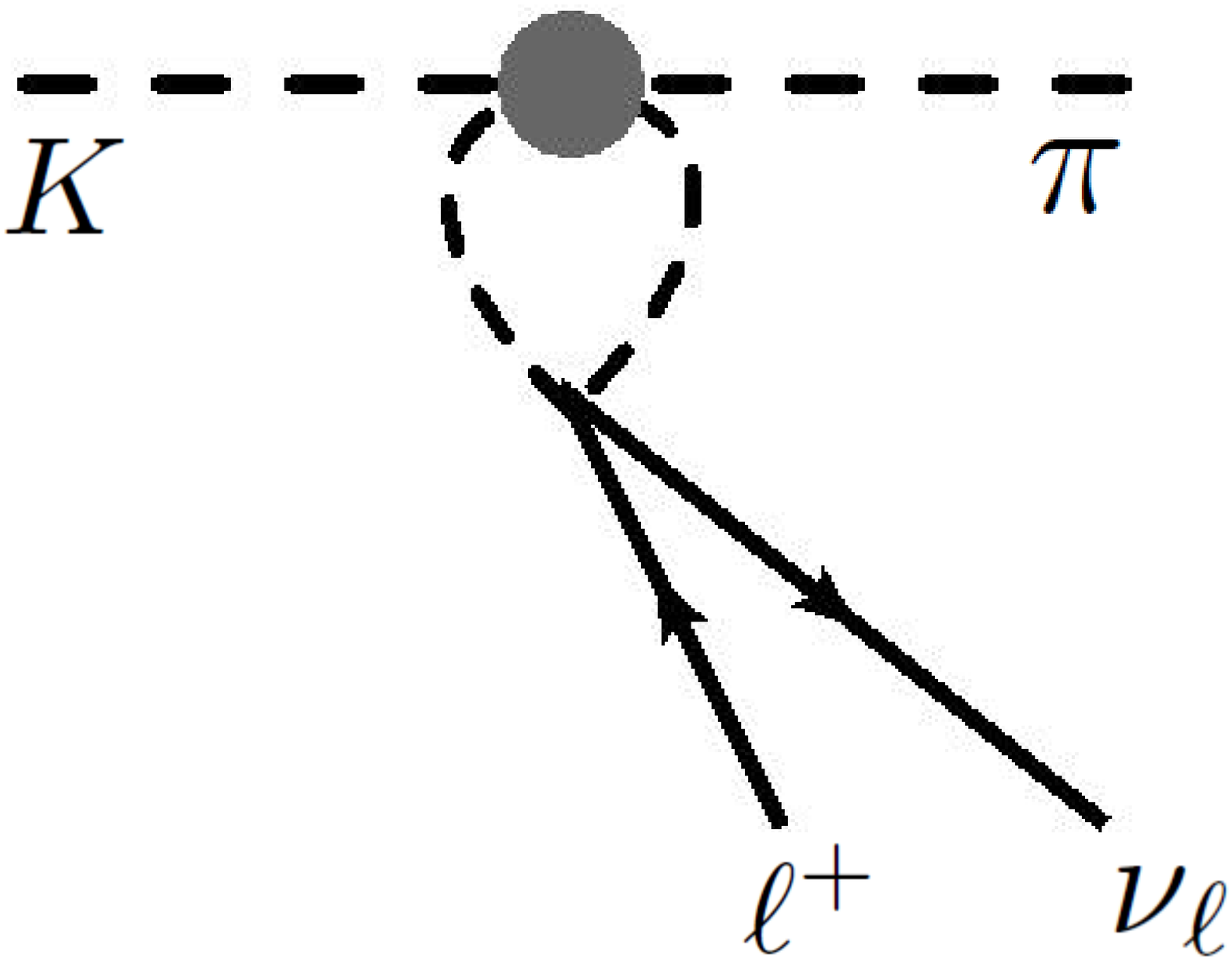}\hfill
		\par\end{centering}
	\caption{\label{fig:mesonloop}Meson loop corrections to $K_{\ell 3}$ in ChPT. }
\end{figure}

The starting point of the ChPT calculation is the effective chiral Lagrangian we described in Sec.\ref{sec:EFTmeson}. The active DOFs are the pNGBs, leptons and photons, while the contributions from the heavy gauge bosons are contained implicitly in the LECs. Applying the framework to $\mathcal{O}(e^2p^2)$, one encounters two kinds of loop diagrams:
\begin{itemize}
	\item The mesonic loop diagrams depicted in Fig.\ref{fig:mesonloop}. They correct the form factors $f_\pm^{K\pi}(t)$ that enter the phase space factor $I_{K\ell}^{(0)}$. 
	\item The photonic loop diagrams similar to those in $\pi_{e3}$ (see Fig.\ref{fig:pion}). These diagrams are IR-divergent and must be combined with the bremsstrahlung contributions in Fig.\ref{fig:brem}.
\end{itemize}
The loops above contain UV-divergences that are regulated using dimensional regularization and reabsorbed into the $\mathcal{O}(p^4)$ LECs $\{L_i^r\}$ as well as the $\mathcal{O}(e^2p^2)$ LECs $\{K_i^r,X_i^r\}$. There is a technical detail here which is not fully intuitive and is worth pointing out, namely: The meson loops not only describe pure QCD corrections, but also include short-distance QED corrections from the $\mathcal{O}(e^2)$ chiral Lagrangian (i.e. the term proportional to $Z$ in Eq.\eqref{eq:Lp2}), which effects enter through the meson mass splitting and the four-meson vertex. 
Similarly, one would na\"{\i}vely think that all $\{K_i^r\}$ are EM corrections and should all enter $\delta_\mathrm{EM}^{K\ell}$, but in practice, a subset of them which describes the EM-induced $\pi^0-\eta$ mixing is categorized $\delta_\mathrm{SU(2)}^{K^+\pi^0}$ instead. Interested readers may refer to Refs.\cite{Cirigliano:2001mk,Cirigliano:2011ny} for more details.

The EWRCs are contained in two quantities: $S_\mathrm{EW}$ and $\delta_\mathrm{EM}^{K\ell}$. The former, $S_\mathrm{EW}=1.0232(3)$ encodes the short-distance EWRC that is process-independent, whereas $\delta_\mathrm{EM}^{K\ell}$ describes the process-dependent, long-range EMRC.
The full analytic expression of the latter to $\mathcal{O}(e^2p^2)$ can be found in Ref.\cite{Cirigliano:2001mk} and will not be repeated here. There is, however, a slight difference in its numerical implementation in different literature: Refs.\cite{Cirigliano:2001mk,Cirigliano:2004pv} adopted a partial resummation scheme where the structure-independent corrections from soft photons are explicitly factored out. As a consequence, the resulting amplitude includes an incomplete resummation of $\mathcal{O}(e^2p^{2n})$ EM corrections with $n\geq 2$. On the other hand, Ref.\cite{Cirigliano:2008wn} presented the EM corrections at fixed order, $\mathcal{O}(e^2p^2)$. The two methods agree within error bars, but the fixed-order result is more frequently quoted in recent reviews. 

The two sources of errors in the calculation above are (1) the neglected terms of the order $\mathcal{O}(e^2p^4)$, and (2) the poorly-constrained LECs. To estimate the theory uncertainties and the correlations from the first source, Ref.\cite{Cirigliano:2008wn} decomposed $\delta_\mathrm{EM}^{K\ell}$ in the four independent (in terms of EMRCs) channels as:
\begin{eqnarray}
\delta_\mathrm{EM}^{K^0e}&=&\delta_1+\delta_2+\delta_3+\delta_4\nonumber\\
\delta_\mathrm{EM}^{K^0\mu}&=&\delta_1+\delta_2-\delta_3-\delta_4\nonumber\\
\delta_\mathrm{EM}^{K^+e}&=&\delta_1-\delta_2+\delta_3-\delta_4\nonumber\\
\delta_\mathrm{EM}^{K^+\mu}&=&\delta_1-\delta_2-\delta_3+\delta_4~,
\end{eqnarray}
where $\delta_1$ is common to all channels, $\delta_2$ is lepton-flavor independent but anti-correlated in kaon isospin, $\delta_3$ is isospin-independent but anti-correlated in lepton flavor, and $\delta_4$ are anti-correlated to both the isospin and the lepton flavor. Theory uncertainties from higher-order chiral corrections to different $\delta_i$ are assumed to be uncorrelated, and are obtained by multiplying the respective central values with $M_K^2/(4\pi F_\pi)^2\sim 0.2$.

In the meantime, the $\mathcal{O}(e^2p^2)$ LECs that enter $\delta_\mathrm{EM}^{K\ell}$ are $X_{1-3}^r$, $\tilde{X}_6^\mathrm{phys}$ and $K_{3-6}^r$ (see their definitions in Sec.\ref{sec:EFTmeson} and Eq.\eqref{eq:X6phys}), where only $X_1$ and $\tilde{X}_6^\mathrm{phys}$ contribute to $\delta f_+^{K\pi}$. These LECs are not constrained by chiral symmetry, and earlier estimations of their sizes were based on large-$N_c$ expansion~\cite{Bijnens:1996kk} or resonance models~\cite{Moussallam:1997xx,Ananthanarayan:2004qk,DescotesGenon:2005pw}. Recent reviews of mesonic LECs (e.g. Ref.\cite{Bijnens:2014lea}) often quote the results from Refs.\cite{Ananthanarayan:2004qk,DescotesGenon:2005pw}:
\begin{equation}
K_3^r=2.7,~K_4^r=1.4,~K_5^r=11.6,~K_6^r=2.8,~X_1=-3.7,~X_2^r=3.6,~X_3^r=5.0,~\tilde{X}_6^\mathrm{phys}=10.4,
\end{equation}
in units of $10^{-3}$ at $\mu=0.77~$GeV. None of the above is accompanied by a rigorously-estimated error, and in practice these values are used by assuming a 100\% uncertainty. Also, different LECs are assumed to be uncorrelated. 

\begin{table}
	\begin{centering}
		\begin{tabular}{|c|c|}
			\hline 
			& $\delta_{\mathrm{EM}}^{K\ell}(\%)$\tabularnewline
			\hline 
			\hline 
			$K_{e3}^{0}$ & $0.99\pm0.19_{e^{2}p^{4}}\pm0.11_{\mathrm{LEC}}$\tabularnewline
			\hline 
			$K_{e3}^{\pm}$ & $0.10\pm0.19_{e^{2}p^{4}}\pm0.16_{\mathrm{LEC}}$\tabularnewline
			\hline 
			$K_{\mu3}^{0}$ & $1.40\pm0.19_{e^{2}p^{4}}\pm0.11_{\mathrm{LEC}}$\tabularnewline
			\hline 
			$K_{\mu3}^{\pm}$ & $0.016\pm0.19_{e^{2}p^{4}}\pm0.16_{\mathrm{LEC}}$\tabularnewline
			\hline 
		\end{tabular}
		\par\end{centering}
	\caption{\label{tab:deltaChPT}Results of $\delta_{\mathrm{EM}}^{K\ell}$ based on fixed-order ChPT calculations in Ref.\cite{Cirigliano:2008wn}.}	
\end{table}

With the procedures above, Ref.\cite{Cirigliano:2008wn} presented the most updated ChPT determination of $\delta_\mathrm{EM}^{K\ell}$, as summarized in Table~\ref{tab:deltaChPT}. We observe that both the LECs and the higher-order chiral corrections contribute to uncertainties of order $10^{-3}$, which sets a natural limitation to the theory precision within the traditional EFT framework.
There are recently some proposals to compute the full $K_{\ell 3}$ RC, including both the virtual and real corrections, directly from lattice QCD~\cite{Giusti:2018guw,Sachrajda:2019uhh,Cirigliano:2019jig} using the similar approach that was proved successful in the kaon leptonic decay $K\rightarrow \mu\nu$ ($K_{\mu 2}$)~\cite{Carrasco:2015xwa,Lubicz:2016xro,Giusti:2017dwk,DiCarlo:2019thl}. However, the generalization from $K_{\mu 2}$ to $K_{\ell 3}$ that involves one more hadron at the final state greatly complicates the problem. A complete calculation with an overall per mille level precision is expected to take ten years or more~\cite{BoyleSnowmass}.   

\subsection{$K_{e3}$ EWRC in Sirlin's representation}

With the recent progress of the EWRC theory in pion, neutron and nuclei, it becomes apparent that the Sirlin's representation we described in Sec.\ref{sec:Sirlin} is also applicable to $K_{\ell 3}$ which enables one to overcome the natural limitations of the traditional ChPT framework. The underlying principle is rather straightforward: The ChPT treatment contains a large $\mathcal{O}(e^2p^4)$ uncertainty because the \textit{entire} RC, including both the numerically large and small pieces, is subject to a chiral expansion. On the other hand, in the Sirlin's representation the most important pieces in the one-loop EMRC, namely $\left(\delta\mathfrak{M}_2+\delta\mathfrak{M}_{\gamma W}^a\right)_\mathrm{int}$ and $\delta\mathfrak{M}_{\gamma W}^b$, are expressed in terms well-defined hadronic matrix elements involving a two-current product. This allows us to evaluate them non-perturbatively without resorting to a chiral expansion, and thus remove a large portion of the chiral uncertainty. Furthermore, the closed form of $\delta\mathfrak{M}_{\gamma W}^b$ permits a straightforward first-principles study using lattice QCD. This removes substantially the theory uncertainty from the unknown hadron physics at $q\sim \Lambda_\chi$, which translates into the LEC uncertainties in the ChPT description. 

The idea above was first pioneered in Refs.\cite{Seng:2019lxf,Seng:2020jtz}, and subsequently applied to the study of the EWRC in $K_{e3}$~\cite{Seng:2021boy,Seng:2021wcf}. In what follows, we briefly outline the procedures and discuss the main results. 

\subsubsection{$\left(\delta\mathfrak{M}_2+\delta \mathfrak{M}_{\gamma W}^a\right)_\mathrm{int}$ and $\delta\mathfrak{M}_{\gamma W}^{b,V}$}

First, a generic one-loop correction to the $K_{e3}$ decay amplitude reads:
\begin{equation}
\delta \mathfrak{M}_\mathrm{vir}=-\frac{G_F}{\sqrt{2}}L_\lambda I^\lambda~.
\end{equation}
The loop integrals are contained in the quantity $I^\lambda$. Upon contracting with the lepton piece and using the on-shell conditions, it simply translates into corrections to the form factors:
\begin{equation}
I^\lambda \rightarrow \delta f_+^{K\pi}(y,z)(p+p')^\lambda+\delta f_-^{K\pi}(y,z)(p-p')^\lambda~.
\end{equation} 
As we advertised before, in $K_{e3}$ only the correction $\delta f_+^{K\pi}$ is relevant. 

The important hadronic quantities in $I^\lambda$ are $T^{\mu\nu}$ and $\Gamma^\mu$ defined in Eq.\eqref{eq:tensors}, with $b=\gamma$, $\phi_i=K$ and $\phi_f=\pi$. Recall that in Eq.\eqref{eq:Jsplit} we split the charged weak current into the vector and axial piece. Accordingly, we may split:
\begin{equation}
T^{\mu\nu}=(T^{\mu\nu})_V+(T^{\mu\nu})_A~,\quad \delta\mathfrak{M}_{\gamma W}^b=\delta\mathfrak{M}_{\gamma W}^{b,V}+\delta\mathfrak{M}_{\gamma W}^{b,A}~.
\end{equation}
From parity, $(T^{\mu\nu})_A$ must contain an antisymmetric $\epsilon$-tensor, while $(T^{\mu\nu})_V$ does not have such a tensor. It is easy to see that, in near-degenerate decays (e.g. $\pi_{e3}$ and superallowed beta decays), $\delta\mathfrak{M}_{\gamma W}^{b,V}$ is negligible because when $p'\approx p$ and $p_e\approx 0$, there is no structure that survives after contracting with the $\epsilon$-tensor in the integrand. This is obviously not the case in $K_{e3}$ so this term must be retained. 

\begin{figure}
	\begin{centering}
		\includegraphics[scale=0.3]{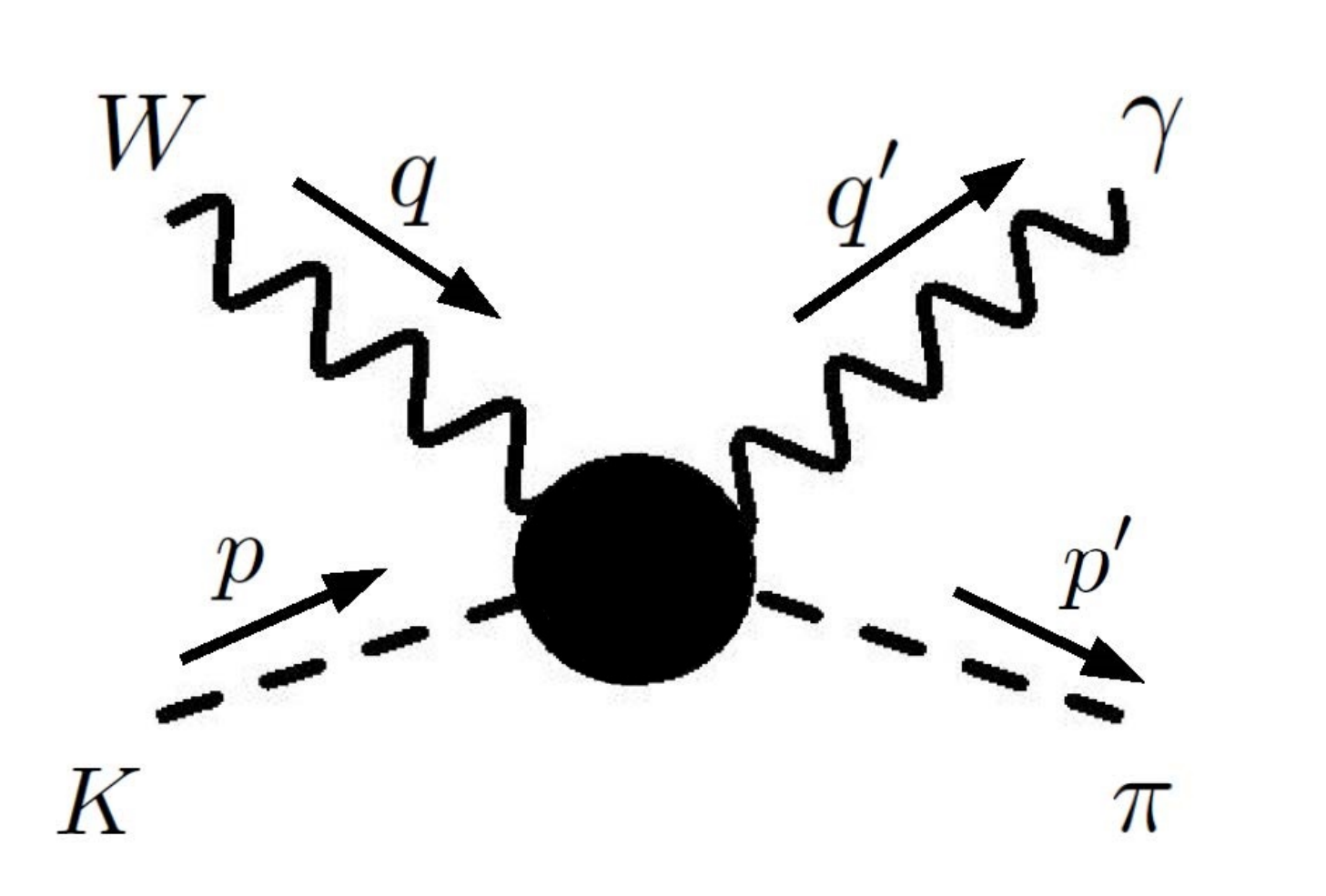}\hfill
		\par\end{centering}
	\caption{\label{fig:seagull}The seagull diagram in $T^{\mu\nu}$ and $\Gamma^\mu$. The black dot includes all possible diagrammatic structures except for a single $K$ or $\pi$ propagator.}
\end{figure}

We start by discussing the loop integrals that only probe the physics at $q'\sim p-p'\sim p_e$, namely $\left(\delta\mathfrak{M}_2+\delta \mathfrak{M}_{\gamma W}^a\right)_\mathrm{int}+\delta\mathfrak{M}_{\gamma W}^{b,V}$. Their combined effect reads:
\begin{eqnarray}
I_\mathfrak{A}^\lambda&=&-e^2\int\frac{d^4q'}{(2\pi)^4}\frac{1}{[(p_e-q')^2-m_e^2][q^{\prime 2}-M_\gamma^2]}\biggl\{\frac{2p_e\cdot q'q^{\prime\lambda}}{q^{\prime 2}-M_\gamma^2}T^\mu_{\:\:\mu}+2p_{e\mu}T^{\mu\lambda}\nonumber\\
&&-(p-p')_\mu T^{\lambda\mu}+i\Gamma^\lambda-i\epsilon^{\mu\nu\alpha\lambda}q'_\alpha(T_{\mu\nu})_V\biggr\}~.
\end{eqnarray}
We may now ask: What are the hadron physics inputs needed to evaluate the integral above to a satisfactory precision? In the most general terms, $T^{\mu\nu}$ and $\Gamma^\mu$ are fully described by two types of Feynman diagrams, namely (1) the pole diagrams Fig.\ref{fig:convection} and (2) the seagull diagram in Fig.\ref{fig:seagull}. The former are completely fixed by the $K$ and $\pi$ electromagnetic and charged weak form factors (remember $q=p'-p+q'$):~\footnote{We neglect the contribution from the electromagnetic form factor of $K^0$ which is numerically insignificant.}
\begin{eqnarray}
\left(T_{K^0\pi^-}^{\mu\nu}\right)_\mathrm{pole}&=&iV_{us}^* F_\mathrm{em}^{\pi^-}(q^{\prime 2})\frac{(2p'+q')^\mu}{(p'+q')^2-M_\pi^2}\left[f_+^{K^0\pi^-}(q^2)(2p+q)^\nu-f_-^{K^0\pi^-}(q^2)q^\nu\right]\nonumber\\
\left(T_{K^+\pi^0}^{\mu\nu}\right)_\mathrm{pole}&=&iV_{us}^* F_\mathrm{em}^{K^+}(q^{\prime 2})\frac{(2p-q')^\mu}{(p-q')^2-M_K^2}\left[f_+^{K^+\pi^0}(q^2)(2p'-q)^\nu-f_-^{K^+\pi^0}(q^2)q^\nu\right]\nonumber\\
\left(\Gamma_{K^0\pi^-}^\mu\right)_\mathrm{pole}&=&V_{us}^*\frac{M_K^2-M_\pi^2}{(p'+q')^2-M_\pi^2}(2p'+q')^\mu F_\mathrm{em}^{\pi^-}(q^{\prime 2})f_0^{K^0\pi^-}(q^2)\nonumber\\
\left(\Gamma_{K^+\pi^0}^\mu\right)_\mathrm{pole}&=&V_{us}^*\frac{M_K^2-M_\pi^2}{(p-q')^2-M_K^2}(2p-q')^\mu F_\mathrm{em}^{K^+}(q^{\prime 2})f_0^{K^+\pi^0}(q^2)~,
\end{eqnarray}
where 
\begin{equation}
\langle \phi(p')|J_\mathrm{em}^\mu |\phi(p)\rangle =F_\mathrm{em}^\phi(t)(p+p')^\mu
\end{equation}
defines the electromagnetic form factor of spinless particles, with $F_\mathrm{em}^{K^+}(0)=1$ and $F_\mathrm{em}^{\pi^-}(0)=-1$. With the above, the pole contribution to $I_\mathfrak{A}^\lambda$ can be straightforwardly calculated using the experimental parameterizations of the form factors \cite{Amendolia:1986wj,Amendolia:1986ui,Ananthanarayan:2017efc,Colangelo:2018mtw,Lazzeroni:2018glh,Zyla:2020zbs}. In particular, a monopole parameterization of the form factors allows one to perform the loop integral analytically using the standard Passarino-Veltman reduction~\cite{Passarino:1978jh}. Notice that $(\delta f_+^{K\pi})_{\mathfrak{A},\mathrm{pole}}$ is UV-finite even without the form factors, so the inclusion of the full form factors only give rise to power-suppressed (instead of logarithmic) corrections. 

The seagull diagram is required to ensure the gauge invariance, and it represents physically the collective effect from all non-elastic intermediate states. The precise evaluation of its effect seems to be highly non-trivial, but fortunately, since $I_\mathfrak{A}^\lambda$ only probes the small-$q'$ region, one needs to only consider some of the lowest intermediate states. For instance, one may estimate the contribution from the $1^{--}$ and $1^{++}$ resonances using the resonance chiral theory~\cite{Ecker:1988te,Ecker:1989yg,Cirigliano:2006hb}, and it effect is found to be less than $10^{-4}$.

To conclude, through the procedures above one can evaluate $(\delta f_+^{K\pi})_{\mathfrak{A}}$ to a high precision. The main theory uncertainties come from the form factors and a conservative estimation of the size of the seagull term contribution, which are both of the order $10^{-4}$.  

\subsubsection{$\delta\mathfrak{M}_{\gamma W}^{b,A}$}

The second non-trivial integral is:
\begin{equation}
I_\mathfrak{B}^\lambda = ie^2\int\frac{d^4q'}{(2\pi)^4}\frac{M_W^2}{M_W^2-q^{\prime 2}}\frac{\epsilon^{\mu\nu\alpha\lambda}q'_\alpha(T_{\mu\nu})_A}{[(p_e-q')^2-m_e^2]q^{\prime 2}}~,
\end{equation}
which comes from $\delta \mathfrak{M}_{\gamma W}^{b,A}$. It is IR-finite, but probes the physics from $q'=0$ all the way up to $q'\sim M_W$. In fact, this is nothing but the generalization of the forward axial $\gamma W$-box in Eq.\eqref{eq:forwardbox} to include non-forward effects due to the large $K-\pi$ mass splitting. Therefore, an efficient way to proceed is to first calculate the forward $K\pi$ $\gamma W$-box on lattice; this single step captures all the dominant physics in $I_\mathfrak{B}^\lambda$. After that, the uncertainty due to the non-forward effect may be estimated through standard power counting arguments.  
Following the same procedure in Ref.\cite{Feng:2020zdc}, Ref.\cite{Ma:2021azh} performed the first calculation of the forward $K\pi$ $\gamma W$-box diagram (recall the notations in Sec.\ref{sec:latticeQCD}):
\begin{equation}
\Box_{\gamma W}^<(K^0,\pi^-,M_\pi)=0.28(4)_\mathrm{lat}\times 10^{-3}~,
\end{equation}
from which $\Box_{\gamma W}^<(K^+,\pi^0,M_\pi)=1.06(7)_\mathrm{lat}\times 10^{-3}$ can also be obtained through a simple chiral matching~\cite{Seng:2020jtz}. 

To connect the result above with $I_\mathfrak{B}^\lambda$, we first split the latter as well as $(f_+^{K\pi})_\mathfrak{B}$ into the ``$>$'' and ``$<$'' components, which correspond to the integral at $Q^2>Q_\mathrm{cut}^2$ and $Q^2<Q_\mathrm{cut}^2$ respectively. The matching in the ``$>$'' region is rather trivial:
\begin{equation}
(\delta f_+^{K\pi})_\mathfrak{B}^>=\Box_{\gamma W}^>f_+^{K\pi}(t)~.
\end{equation}
Meanwhile, in the ``$<$'' region, we parameterize the integral as:
\begin{eqnarray}
(I_\mathfrak{B}^\lambda)^<&=&V_{us}^*[g_+(M_K^2,M_\pi^2,m_e^2,s,u)(p+p')^\lambda+g_-(M_K^2,M_\pi^2,m_e^2,s,u)(p-p')^\lambda\nonumber\\
&&+g_e(M_K^2,M_\pi^2,m_e^2,s,u)p_e^\lambda]~,
\end{eqnarray}
and we are only interested in $(\delta f_+^{K\pi})_\mathfrak{B}^<=g_+(M_K^2,M_\pi^2,m_e^2,s,u)$. We observe that, $g_+$ is simply related to the lattice calculation upon taking the forward limit:
\begin{equation}
g_+(M_\pi^2,M_\pi^2,0,M_\pi^2,M_\pi^2)=\Box_{\gamma W}^<(K,\pi,M_\pi)f_+^{K\pi}(0)~.
\end{equation}
Incorporating the non-forward corrections, we may write:
\begin{equation}
g_+(M_K^2,M_\pi^2,m_e^2,s,u)=g_+(M_\pi^2,M_\pi^2,0,M_\pi^2,M_\pi^2)+\mathcal{O}\left(\frac{E^2}{\Lambda_\chi^2}\right)
\end{equation}
with $E$ an energy scale that characterizes the non-forward kinematics; we may take $E=M_K^2$ as an upper limit. That leads to:
\begin{equation}
(\delta f_+^{K\pi})_\mathfrak{B}^<=\left[\Box_{\gamma W}^<(K,\pi,M_\pi)+\mathcal{O}\left(\frac{M_K^2}{\Lambda_\chi^2}\right)\right]f_+^{K\pi}(t)~,
\end{equation}
where the uncertainty from non-forward corrections is estimated by multiplying the central value with $M_K^2/\Lambda_\chi^2$. Utilizing the procedures above, one is able to fix $(\delta f_+^{K\pi})_\mathfrak{B}$ to high precision, with the theory uncertainty from both the lattice inputs and the non-forward effects controlled at the $10^{-4}$ level.

Finally, in analogy to the discussion at the end of Sec.\ref{sec:latticeQCD}, the lattice calculation of the $K\pi$-box diagram also allows us to determine a specific combination of the $\mathcal{O}(e^2p^2)$ LECs to high precision~\cite{Ma:2021azh}:
\begin{eqnarray}
-\frac{8}{3}X_1+\bar{X}_6^\mathrm{phys}(M_\rho)&=&-\frac{1}{2\pi\alpha}\left(\Box_{\gamma W}(K^0,\pi^-,M_\pi)-\frac{\alpha}{8\pi}\ln\frac{M_W^2}{M_\rho^2}\right)+\frac{1}{8\pi^2}\left(\frac{5}{4}-\tilde{a}_g\right)\nonumber\\
&=&22.6(1.0)_\mathrm{lat}\times 10^{-3}~.\label{eq:Kpimatch}
\end{eqnarray}

\subsubsection{$\delta F_3^\lambda(p',p)$}

The next required input is $\delta F_3^\lambda(p',p)$ defined in Eq.\eqref{eq:F2F3}. To evaluate this quantity we need to know the hadronic matrix elements $\bar{T}^\mu(\delta\bar{p};p',p)$ and $D(\delta\bar{p};p',p)$ which, according to their quark-level definitions in Eqs.\eqref{eq:Tbarmu3pt} and \eqref{eq:D3pt}, involve a product of three currents which makes them very challenging for first-principles studies. Fortunately, its contribution to $\delta f_+^{K\pi}$ is also smaller (apart from the IR-divergent pieces), so one may resort to an EFT calculation without suffering from a huge loss in precision. 

\begin{figure}
	\begin{centering}
		\includegraphics[scale=0.5]{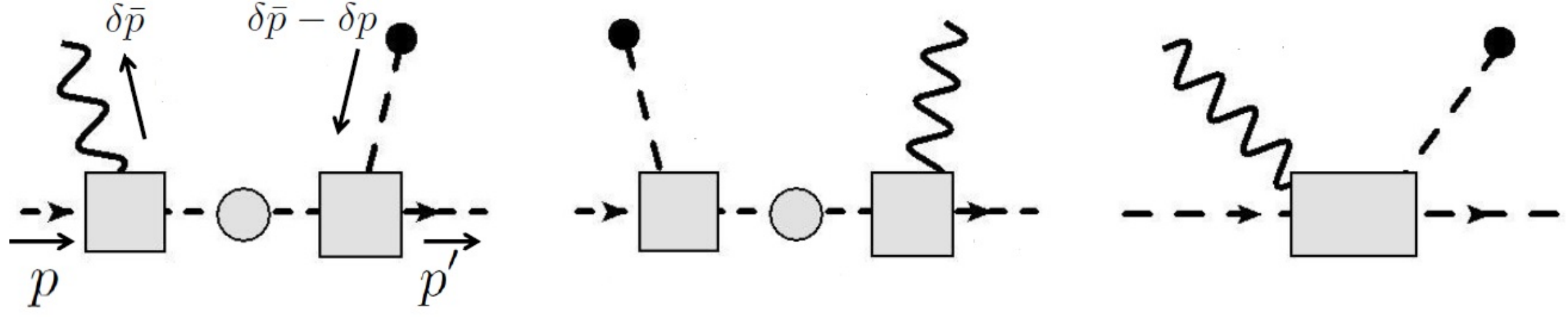}\hfill
		\par\end{centering}
	\caption{\label{fig:EFT}Diagrammatic representation of $i\bar{T}^\mu(\delta\bar{p};p',p)$ and $D(\delta\bar{p};p',p)$. An external wiggle line represents an insertion of $J^\mu$ or $\partial\cdot J$, an external dashed line ended with a black dot represents an insertion of $i\delta\mathcal{L}$, and internal dashed line with gray blob represents a full single-particle propagator. Figures taken from Ref.\cite{Seng:2019lxf}.}
\end{figure}

Ref.\cite{Seng:2019lxf} outlined the full procedure to calculate $i\bar{T}^\mu(\delta\bar{p};p',p)$ and $D(\delta\bar{p};p',p)$ in ChPT. The starting point is the following matrix elements:
\begin{eqnarray}
i\bar{T}^\mu(\delta\bar{p};p',p)&=&\int d^4xe^{i\delta\bar{p}\cdot x}\bigl\langle\pi(p')\bigr|T\left\{J^\mu(x)i\delta\mathcal{L}(0)\right\}\bigl|K(p)\bigr\rangle\nonumber\\
D(\delta\bar{p};p',p)&=&\int d^4xe^{i\delta\bar{p}\cdot x}\bigl\langle\pi(p')\bigr|T\left\{\partial\cdot J(x)i\delta\mathcal{L}(0)\right\}\bigl|K(p)\bigr\rangle~,
\end{eqnarray}
where the QED correction is represented schematically by $i\delta\mathcal{L}(0)$ instead of the explicit product of two EM currents. The diagrammatic representation of these matrix elements are given in Fig.\ref{fig:EFT}. 
To calculate them, we need the ChPT representation of the charged weak current and its divergence, which can be derived from the mesonic chiral Lagrangian in Sec.\ref{sec:EFTmeson}. The result reads:
\begin{eqnarray}
J^\mu &=&-\frac{iF_0^2}{8\sqrt{2}G_F}\left\langle Q_{\mathrm{L}}^{\mathrm{w}\dagger}\left([D^\mu U^\dagger,U]+[D^\mu U,U^\dagger]\right)\right\rangle+...\nonumber\\
\partial\cdot J &=&-\frac{iF_0^2}{8\sqrt{2}G_F}\left\langle Q_{\mathrm{L}}^{\mathrm{w}\dagger}\left([\chi^\dagger,U]+[\chi,U^\dagger]\right)\right\rangle+...~,
\end{eqnarray}
where the ellipses denote terms of higher chiral order. Here we have rewritten $\partial\cdot J$ using the equations of motion (EOMs), so that it satisfies the current conservation in the SU(3) limit at the operator level. This is a necessary step to avoid inconsistencies in the implementation of the Feynman rules (see discussions in Ref.\cite{Seng:2019lxf}). Meanwhile, the effect of $i\delta\mathcal{L}(0)$ is obtained by applying the Feynman rules derived from the chiral Lagrangian with dynamical photons. Notice that it does not necessarily involve a photon loop, but can as well be a local interaction that originates from short-distance QED effects (i.e. $\mathcal{L}^{e^2}$ and $\mathcal{L}^{e^2p^2}_{\{K\}}$).

As usual, $\delta F_3^{\lambda}(p',p)$ is expressed in terms of corrections to the two form factors:
\begin{equation}
\delta F_3^{\lambda}(p',p)=V_{us}^*\left[\delta f_{+,3}^{K\pi}(t)(p+p')^\lambda+\delta f_{-,3}^{K\pi}(t)(p-p')^\lambda\right]~,
\end{equation} 
and in $K_{e3}$ one only needs $\delta f_{+,3}^{K\pi}$. It is useful to split it into the IR-divergent and IR-finite terms:
\begin{equation}
\delta f_{+,3}^{K\pi}=\left(\delta f_{+,3}^{K\pi}\right)^\mathrm{IR}+\left(\delta f_{+,3}^{K\pi}\right)^\mathrm{fin}~.
\end{equation}
Both terms were calculated to $\mathcal{O}(e^2p^2)$ in Ref.\cite{Seng:2019lxf}. The results read:
\begin{eqnarray}
\left(\delta f_{+,3}^{K^0\pi^-}\right)^\mathrm{IR}_{e^2p^2}&=&\frac{\alpha}{4\pi}\frac{p'\cdot(p-p')}{2M_\pi^2}\left[\ln\frac{M_\pi^2}{M_\gamma^2}-\frac{5}{2}\right]\nonumber\\
\left(\delta f_{+,3}^{K^+\pi^0}\right)^\mathrm{IR}_{e^2p^2}&=&-\frac{\alpha}{4\sqrt{2}\pi}\frac{p\cdot(p-p')}{2M_K^2}\left[\ln\frac{M_K^2}{M_\gamma^2}-\frac{5}{2}\right]~,
\end{eqnarray}
and:
\begin{eqnarray}
\left(\delta f_{+,3}^{K^{0}\pi^{-}}\right)_{e^2p^2}^{\mathrm{fin}}&=&-8\pi Z\alpha\left[\frac{1}{2}\bar{h}_{K^{+}\pi^{0}}(t)+\bar{h}_{K^{0}\pi^{-}}(t)+\frac{3}{2}\bar{h}_{K^{+}\eta}(t)\right]\nonumber\\
\left(\delta f_{+,3}^{K^{+}\pi^{0}}\right)_{e^2p^2}^{\mathrm{fin}}&=&-\frac{8\pi Z\alpha}{\sqrt{2}}\left[\frac{1}{2}\bar{h}_{K^{+}\pi^{0}}(t)+\bar{h}_{K^{0}\pi^{-}}(t)+\frac{3}{2}\bar{h}_{K^{+}\eta}(t)\right]\nonumber\\
&&+\frac{Z\alpha}{2\sqrt{2}\pi}\frac{M_{K}^{2}}{M_{\eta}^{2}-M_{\pi}^{2}}\left[1+\ln\frac{M_{K}^{2}}{\mu^{2}}\right]-\frac{4\pi\alpha}{\sqrt{2}}\left[-2K_{3}^{r}+K_{4}^{r}+\frac{2}{3}K_{5}^{r}+\frac{2}{3}K_{6}^{r}\right]\nonumber\\
&&-\frac{8\pi\alpha}{\sqrt{2}}\frac{M_{\pi}^{2}}{M_{\eta}^{2}-M_{\pi}^{2}}\left[-2K_{3}^{r}+K_{4}^{r}+\frac{2}{3}K_{5}^{r}+\frac{2}{3}K_{6}^{r}-\frac{2}{3}K_{9}^{r}-\frac{2}{3}K_{10}^{r}\right]~.\label{eq:3ptfinite}
\end{eqnarray}
The definition of the loop function $\bar{h}_{PQ}(t)$ can be found in Appendix~\ref{sec:loopfun}.

A direct numerical application of the results above is not possible due to the incomplete cancellation of IR-divergences. This is because all other IR-divergence terms (i.e. terms attached to $\ln M_\gamma^2$) in both virtual and real corrections (as we will discuss later) are calculated exactly, except that in $(\delta f_{+,3}^{K\pi})^\mathrm{IR}$ which is expanded to $\mathcal{O}(e^2p^2)$. Fortunately, it is easy to resum this term to all orders in chiral expansion by simply noticing that it originates from the convection term contribution~\cite{Seng:2021wcf}. The resummed version reads:
\begin{eqnarray}
\left(\delta f_{+}^{K^{0}\pi^{-}}\right)^\mathrm{IR}&=&-\frac{\alpha}{4\pi}\frac{p'\cdot(p-p')}{2M_{\pi}^{2}}\left[f_+^{K^0\pi^-}(t)+f_-^{K^0\pi^-}(t)\right]\left[\ln\frac{M_{\pi}^{2}}{M_{\gamma}^{2}}-\frac{5}{2}\right]\nonumber\\
\left(\delta f_{+}^{K^+\pi^0}\right)^\mathrm{IR}&=&\frac{\alpha}{4\pi}\frac{p\cdot(p-p')}{2M_{K}^{2}}\left[f_+^{K^+\pi^0}(t)-f_-^{K^+\pi^0}(t)\right]\left[\ln\frac{M_{K}^{2}}{M_{\gamma}^{2}}-\frac{5}{2}\right]~.
\end{eqnarray}

To summarize, we obtain:
\begin{equation}
\delta f_{+,3}^{K\pi}=\left(\delta f_{+,3}^{K\pi}\right)^\mathrm{IR}+\left\{ \left(\delta f_{+,3}^{K\pi}\right)^\mathrm{fin}_{e^2p^2}+\mathcal{O}(e^2p^4)\right\}~,
\end{equation}
where the first term at the right hand side is exact, and only the second term is associated to a chiral expansion uncertainty. Furthermore, it was realized that $(\delta f_{+,3}^{K\pi})^\mathrm{fin}_{e^2p^2}$ consists of terms proportional to $Z$ which are assigned to $I_{K\ell}^{(0)}$, as well as counterterms for the electromagnetically-induced $\pi^0-\eta$ mixing which are assigned to $\delta_\mathrm{SU(2)}^{K^+\pi^0}$, in the standard ChPT convention (recall our discussion in Sec.\ref{sec:Kl3ChPT}). This means that only $(\delta f_{+,3}^{K\pi})^\mathrm{IR}$ contributes to $\delta_\mathrm{EM}^{Ke}$. 

\subsubsection{Bremsstrahlung contribution}

Finally, we briefly discuss the treatment of the bremsstrahlung contribution depicted in Fig.\ref{fig:brem}, of which amplitude is given by Eq.\eqref{eq:Mbrems}. In traditional ChPT treatment, the entire amplitude is evaluated to LO in the chiral expansion. But here we can do much better. First of all, we retain the full charged weak matrix element $F_\mu$. Next, we split $T^{\mu\nu}$ into two terms, which separately satisfies the EM Ward identity:
\begin{equation}
T^{\mu\nu}=T_{\mathrm{conv}}^{\mu\nu}+\left\{(T^{\mu\nu}-T^{\mu\nu}_{\mathrm{conv}})_{p^2}+\mathcal{O}(p^4)\right\}~.
\end{equation}
The first term at the right hand side is the full convection term (see Eq.\eqref{eq:Tmunuconv}) that contains the full information of the IR-divergence, 
while the second is the difference which is evaluated to $\mathcal{O}(p^2)$:
\begin{eqnarray}
(T^{\mu\nu}_{K^0\pi^-})_{p^2}&=&iV_{us}^*\left[\frac{(2p'+k)^\mu (p+p'+k)^\nu}{(p'+k)^2-M_\pi^2}-g^{\mu\nu}\right]\nonumber\\
(T^{\mu\nu}_{K^+\pi^0})_{p^2}&=&-\frac{iV_{us}^*}{\sqrt{2}}\left[\frac{(2p-k)^\mu (p+p'-k)^\nu}{(p-k)^2-M_K^2}-g^{\mu\nu}\right]
\end{eqnarray}
and
\begin{eqnarray}
(T^{\mu\nu}_{K^0\pi^-})_{p^2}^\mathrm{conv}&=&iV_{us}^*\left[\frac{(2p'+k)^\mu (p+p')^\nu}{(p'+k)^2-M_\pi^2}\right]\nonumber\\
(T^{\mu\nu}_{K^+\pi^0})_{p^2}^\mathrm{conv}&=&-\frac{iV_{us}^*}{\sqrt{2}}\left[\frac{(2p-k)^\mu (p+p')^\nu}{(p-k)^2-M_K^2}\right]~.
\end{eqnarray}
In this way, we can split the bremsstrahlung amplitude into two pieces:
\begin{equation}
\mathfrak{M}_\mathrm{brems}=\mathfrak{M}_A+\mathfrak{M}_B~,
\end{equation}
where $\mathfrak{M}_B$ consists of terms that depend on $(T^{\mu\nu}-T^{\mu\nu}_\mathrm{conv})_{p^2}$, and all other terms go to $\mathfrak{M}_A$. 
This means $\mathfrak{M}_A$ is exact, and only $\mathfrak{M}_B$ involves a chiral expansion. Hence, when we calculate the squared amplitude:
\begin{equation}
|\mathfrak{M}_\mathrm{brems}|^2=|\mathfrak{M}_A|^2+2\mathfrak{Re}\{\mathfrak{M}_A^*\mathfrak{M}_B\}+|\mathfrak{M}_B|^2~,
\end{equation}
the only IR-divergent and the numerically-largest contribution $|\mathfrak{M}_A|^2$ will not carry a chiral expansion uncertainty. This single change improves substantially the precision of the bremsstrahlung contribution. We summarize in Appendix~\ref{sec:bremsInt} all the relevant phase-space
integrals for the benefits of interested readers.

\subsubsection{Results and discussions}

With the methods outlined above, Refs.\cite{Seng:2021boy,Seng:2021wcf} reported an improved determination of the long-distance EMRC in $K_{e3}$:
\begin{eqnarray}
\delta_\mathrm{EM}^{K^+e}&=&0.21(2)_\mathrm{sg}(1)_{ r_K}(1)_\mathrm{lat}(4)_\mathrm{NF}(1)_{e^2p^4}\times 10^{-2}\nonumber\\
\delta_\mathrm{EM}^{K^0e}&=&1.16(2)_\mathrm{sg}(1)_\mathrm{lat}(1)_\mathrm{NF}(2)_{e^2p^4}\times 10^{-2}~,\label{eq:deltaEMnew}
\end{eqnarray}
The sources of uncertainties are as follows:
\begin{itemize}
	\item ``sg'' represents the very conservative estimation of the seagull diagram contribution in $(\delta \mathfrak{M}_2+\delta\mathfrak{M}_{\gamma W}^a)_\mathrm{int}+\delta \mathfrak{M}_{\gamma W}^{b,V}$.
	\item The uncertainties due to the form factor parameterization are mostly small. The only piece that brings a $\sim 10^{-4}$ uncertainty is the uncertainty of the $K^+$ charge radius $r_K$ that only enters $\delta_\mathrm{EM}^{K^+e}$.
	\item ``lat'' represents the lattice uncertainty in the evaluation of the forward $\gamma W$-box diagrams.
	\item ``NF'' is the estimation of the non-forward correction in $\delta\mathfrak{M}_{\gamma W}^{b,A}$.
	\item Finally, ``$e^2p^4$'' represents the (reduced) chiral uncertainty in the bremsstrahlung contribution. 
\end{itemize}

These new results are consistent with the ChPT determinations in Table~\ref{tab:deltaChPT}, but show a substantial increase in precision by nearly an order of magnitude. They indicate the great success of the Sirlin's representation of EWRC, not only in the nearly-degenerate decays but also in non-degenerate decay processes. However, its success in $K_{e3}$ relies critically on the fact that only $\delta f_+^{K\pi}$ is relevant in the discussion. In its future application to $K_{\mu 3}$ where $\delta f_-^{K\pi}$ also starts to be relevant, a more complicated analysis of the theory uncertainties is expected. A good news is that, even in $K_{\mu 3}$, the $\delta f_-^{K\pi}$ contribution to the squared amplitude is still suppressed by $r_\mu\approx 0.05$, so a possible way to proceed is to calculate the full $\delta f_-^{K\pi}$ within ChPT, and estimate its associated uncertainty by the usual power-counting argument. This is an interesting topic for future research.

\section{\label{sec:summary}Summary}

In this work we reviewed some recent progress in the calculation of the $\mathcal{O}(G_F\alpha)$ EWRCs to semileptonic beta decays of strongly-interacting systems, which are important avenues for precision tests of the SM. Combining the classical Sirlin's current algebra representation of EWRCs and non-perturbative techniques such as DR and lattice QCD, a number of breakthroughs were made in pion, neutron, nuclei and kaon by reaching an unprecedented level of $10^{-4}$ in precision:
\begin{itemize}
	\item The first lattice calculation of the forward axial $\gamma W$-box diagram $\pi_{e3}$ has removed the major theory uncertainty in its ERWCs, making it 3 times as precise as the previous state-of-the-art determination from ChPT. This makes $\pi_{e3}$ one of the theoretically cleanest avenues to measure $|V_{ud}|$. The major limitation is from the experimental error of its branching ratio, which will be improved in next-generation experiments of rare pion decays.
	\item The first DR analysis of the single-nucleon axial $\gamma W$-box diagram with neutrino-nucleus scattering data as inputs improved its precision by a factor 2 and revealed a large systematic effect that was not captured in previous studies. It led to a reduction of the $|V_{ud}|$ central value.
	\item A new theory framework for the EWRCs in $K_{\ell 3}$ was constructed to overcome the natural limitations from the traditional ChPT formalism. Its first application to $K_{e3}$ has improved the existing theory precision by almost an order of magnitude.
\end{itemize}

Some of the progress above revealed new anomalies in beta decays associated to the top-row CKM matrix elements, and some of them sharpened the existing ones. They turned beta decays into a promising avenue to search for BSM physics. There are, however, a number of unresolved problems that must be investigated in the future in order to make further progress along this direction:
\begin{itemize}
	\item The current DR treatment of the free neutron inner corrections suffers from low precision of the experimental inputs and a residual model dependence in the matching between the data and the nucleon $\gamma W$-box diagram, so one still cannot claim with full confidence that the outcome is free from detectable systematics. To resolve these ambiguities, one should perform direct lattice QCD calculations of the nucleon $\gamma W$-box diagram.
	\item A similar DR analysis of the nuclear $\gamma W$-box diagram revealed several potentially large nuclear structure corrections that were failed to be incorporated in previous nuclear model calculations. Ab-initio studies of these new effects are urgently needed. Also, the current results of the nuclear isospin breaking corrections should be scrutinized by model-independent approaches to make sure no large systematic uncertainties are hidden.
	\item The new theory analysis of the $K_{e3}$ EWRC should also be applied to $K_{\mu 3}$ to complete the story. One expects a more complicated error analysis in the latter due to the non-negligible contributions from $\delta f_-^{K\pi}$.
	\item So far the Sirlin's representation has only been applied to semileptonic decays, but in principle it is also valid for leptonic decays. Particularly interesting examples are $K_{\mu 2}$ and $\pi_{\mu 2}$ which measure the ratio $|V_{us}/V_{ud}|$. The EWRCs in these channels were previously studied with ChPT and lattice QCD, and it is interesting to see if the new method is able to further improve its precision. This also serves as another cross-check of the $K_{\mu 2}-K_{\ell 3}$ discrepancy in the $|V_{us}|$ measurement.  
\end{itemize}  
Future progress in both theories and experiments of these topics could further sharpen the existing beta decay anomalies  and eventually lead to one of the earliest confirmed signals of BSM physics, of which impact to the fundamental science is far-reaching.

\begin{acknowledgments}

The author thanks Jens Erler, Xu Feng, Daniel Galviz, Mikhail Gorchtein, Charles J. Horowitz, Lu-Chang Jin, Oleksandr Koshchii, Peng-Xiang Ma, William J. Marciano, Ulf-G. Mei{\ss}ner, Hiren H. Patel, Jorge Piekarewicz, Michael J. Ramsey-Musolf, Xavier Roca-Maza and Hubert Spiesberger for collaborations in related topics. The author is also grateful to the supermarkets in Bonn, Germany for keep opening during the COVID-19 pandemic. 
This work is supported in
part by the Deutsche Forschungsgemeinschaft (DFG, German Research Foundation) and the NSFC through the funds provided to the Sino-German Collaborative Research Center TRR110 ``Symmetries and the Emergence of Structure in QCD'' (DFG Project-ID 196253076 -
TRR 110, NSFC Grant No. 12070131001), and by the Alexander von Humboldt Foundation
through the Humboldt Research Fellowship.

\end{acknowledgments}

\begin{appendix}

\section{\label{sec:OPEderive} Derivation of the free-field OPE relation}

Here we use a simple example to illustrate how the leading-twist, free-field OPE in Eq.\eqref{eq:OPE} can be derived in a free-quark picture. Let us consider a $\beta^-$ decay involving only $u$ and $d$ quarks, so the electroweak currents read:
\begin{equation}
J_\mathrm{em}^\mu=e_u\bar{u}\gamma^\mu u+e_d\bar{d}\gamma^\mu d~,\quad J^\mu=V_{ud}\bar{u}\gamma^\mu(1-\gamma_5)d~.
\end{equation}
Now we study the following time-ordered current product at very large $q'$:
\begin{eqnarray}
&&\int d^4xe^{iq'\cdot x}T\{J_\mathrm{em}^\mu(x)J^\nu(0)\}\nonumber\\
&=&\int d^4xe^{iq'\cdot x}T\{(e_u\bar{u}(x)\gamma^\mu u(x)+e_d\bar{d}(x)\gamma^\mu d(x))V_{ud}\bar{u}(0)\gamma^\nu(1-\gamma_5)d(0)\}\nonumber\\
&=&V_{ud}\int d^4xe^{iq'\cdot x}\left\{e_u \bar{u}(0)\gamma^\mu\langle u(x)\bar{u}(0)\rangle\gamma^\nu(1-\gamma_5) d(0)\right.\nonumber\\
&&\left.+e_d \bar{u}(0)\gamma^\nu(1-\gamma_5)\langle d(0)\bar{d}(x)\rangle\gamma^\mu d(0)\right\}+...~,
\end{eqnarray}
where we have used the Wick's theorem in the last equality, and display only terms with one contraction. Also, we have set $x\rightarrow 0$ in the uncontracted field operators, because $q'$ is large so the integral at large $x$ vanishes due to the oscillating exponential factor. Next, we use the free propagator of a massless quark:
\begin{equation}
\int d^4x e^{iq'\cdot x}\langle \psi(x)\bar{\psi}(0)\rangle=\frac{i\slashed{q}'}{q^{\prime 2}}~
\end{equation}
to get:
\begin{eqnarray}
&&\int d^4xe^{iq'\cdot x}T\{J_\mathrm{em}^\mu(x)J^\nu(0)\}\nonumber\\
&=&\frac{i}{q^{\prime 2}}V_{ud}\bar{u}(0)\left[e_u\gamma^\mu\slashed{q}'\gamma^\nu-e_d\gamma^\nu\slashed{q}'\gamma^\mu\right](1-\gamma_5)d(0)+...\nonumber\\
&=&\frac{i}{q^{\prime 2}}V_{ud}\bar{u}(0)\left[(e_u-e_d)(q^{\prime\mu}\gamma^\nu-g^{\mu\nu}\slashed{q}'+q^{\prime \nu}\gamma^\mu)-i(e_u+e_d)\epsilon^{\mu\nu\alpha\beta}q'_\alpha\gamma_\beta\right](1-\gamma_5)d(0)\nonumber\\
&&+...\nonumber\\
&=&\frac{i}{q^{\prime 2}}\left\{\eta\left[g^{\mu\nu}q^{\prime\lambda}-g^{\nu\lambda}q^{\prime\mu}-g^{\mu\lambda}q^{\prime\nu}\right]-2i\bar{Q}\epsilon^{\mu\nu\alpha\lambda}q'_\alpha\right\}J_\lambda(0)+...~,
\end{eqnarray}
where in the third line we have used the identity in Eq.\eqref{eq:3Gamma}, and in the last line we have used $e_u-e_d=1$, $e_u+e_d=2\bar{Q}$ and $\eta=-1$ for $\beta^-$ decay. Finally, taking the matrix element at both sides of the equation above returns the OPE relation of $T_{(\gamma)}^{\mu\nu}(q';p',p)$ in Eq.\eqref{eq:OPE}. Other OPE relations can be derived accordingly.

\section{\label{sec:pQCDmatch}pQCD corrections in $\Box_{\gamma W}^>$}

In this Appendix we prove that the pQCD corrections to $\Box_{\gamma W}^>$ is identical to that to the polarized Bjorken sum rule.

First, from the definition of $\Box_{\gamma W}$ it is clear that only the component of $T^{\mu\nu}$ with an $\epsilon$-tensor is probed. The polarized Bjorken sum rule originates from the following pQCD-corrected OPE, where we only display the structure with an $\epsilon$-tensor (See, e.g. Ref.\cite{Larin:1991tj}, but take into account the definition of $\gamma_5$):
\begin{equation}
\int d^4x q^{iq\cdot x}T\{V_\mu^aV_\nu^b(0)\}=-\epsilon_{\mu\nu\rho\sigma}\frac{q^\rho}{q^2}C_\mathrm{Bj}(Q^2)d^{abc}A^{c\sigma}(0)+...~.\label{eq:BjOPE}
\end{equation}
In the above we have defined the vector and axial SU(3) currents:
\begin{equation}
V_\mu^a\equiv\bar{\psi}\gamma_\mu\frac{\lambda^a}{2}\psi~,\quad A_\mu^a\equiv\bar{\psi}\gamma_\mu\gamma_5\frac{\lambda^a}{2}\psi~,
\end{equation}
from which the electroweak currents can be reconstructed (for notational simplicity, let us concentrate on $u\leftrightarrow d$ transition, and neglect the factor $V_{ud}$):
\begin{equation}
J_\mathrm{em}^\mu=V_3^{\mu}+\frac{1}{\sqrt{3}}V_8^{\mu}~,\quad J_V^\mu=V_1^\mu-i\eta V_2^\mu~,\quad J_A^\mu=-A_1^\mu+i\eta A_2^\mu~,\label{eq:JemJW}
\end{equation}
and $C_\mathrm{Bj}(Q^2)$ is the pQCD correction factor. Substituting Eq.\eqref{eq:JemJW} into Eq.\eqref{eq:BjOPE} gives:
\begin{equation}
\int d^4x e^{iq\cdot x}T\{J^\mathrm{em}_\mu J^V_\nu(0)\}=\frac{1}{3}\epsilon_{\mu\nu\rho\sigma}\frac{q^\rho}{q^2}C_\mathrm{Bj}(Q^2)J_A^\sigma(0)+...~.\label{eq:JVOPE}
\end{equation}

To further proceed, we perform the following chiral rotation to the quark field:
\begin{equation}
\psi_R\rightarrow \psi_R~,\quad \psi_L\rightarrow L\psi_L~,
\end{equation}
where $L=\mathrm{diag}(1,-1,-1)$ is an SU(3)$_\mathrm{L}$ matrix. This rotation is equivalent to $d\rightarrow \gamma_5 d$, $s\rightarrow\gamma_5 s$, which transforms the currents as:
\begin{equation}
J^\mathrm{em}_\mu\rightarrow J^\mathrm{em}_\mu~,\quad J^V_\mu\rightarrow-J^A_\mu~,\quad J^A_\mu\rightarrow -J^V_\mu.
\end{equation}
Now suppose chiral symmetry is a good symmetry, then Eq.\eqref{eq:JVOPE} should hold even after the chiral rotation. That means, we should similarly obtain:
\begin{equation}
\int d^4x e^{iq\cdot x}T\{J^\mathrm{em}_\mu J^A_\nu(0)\}=\frac{1}{3}\epsilon_{\mu\nu\rho\sigma}\frac{q^\rho}{q^2}C_\mathrm{Bj}(Q^2)J_V^\sigma(0)+...~.\label{eq:JAOPE}
\end{equation}
So, adding Eqs.\eqref{eq:JVOPE} and \eqref{eq:JAOPE} gives:
\begin{equation}
\int d^4x e^{iq\cdot x}T\{J^\mathrm{em}_\mu J_\nu(0)\}=\frac{1}{3}\epsilon_{\mu\nu\rho\sigma}\frac{q^\rho}{q^2}C_\mathrm{Bj}(Q^2)J^\sigma(0)+...~,
\end{equation}
which reproduces the structure with an $\epsilon$-tensor of our previous OPE relation in Eq.\eqref{eq:OPE}, together with the pQCD correction factor $C_\mathrm{Bj}(Q^2)$ as a bonus. This completes our proof. 

\section{\label{sec:crossing}Crossing symmetry of the invariant amplitudes}

In this Appendix we outline a derivation of the crossing behavior of the invariant amplitudes $T_3$, $S_1$ and $S_2$ under $\nu\rightarrow-\nu$ using time-reversal invariance and isospin symmetry. Notice that a similar derivation was shown in Appendix A of Ref.\cite{Seng:2018qru}. Also, for definiteness we take $\phi_i=n$ and $\phi_f=p$, but the derivation obviously applies to general $I=1/2$ external states. 

We start with the following identity involving the time-reversal operator $\mathbb{T}$ acting on a matrix element of a generic current operator:
\begin{equation}
\langle\beta|J^\mu(\vec{x},t)|\alpha\rangle=\langle\tilde{\alpha}|\mathbb{T}J_\mu^\dagger(\vec{x},t)\mathbb{T}^{-1}|\tilde{\beta}\rangle=\langle\tilde{\alpha}|J^{\mu\dagger}(\vec{x},-t)|\tilde{\beta}\rangle~, 
\end{equation} 
where $|\tilde{\alpha}\rangle$ and $|\tilde{\beta}\rangle$ are the time-reversed states of $|\alpha\rangle$ and $|\beta\rangle$ respectively. The momentum and spin in the external states should also transformed as $p^\mu\rightarrow\tilde{p}^\mu\equiv p_\mu$ and $s^\mu\rightarrow\tilde{s}^\mu\equiv s_\mu$. Furthermore, we decompose the electromagnetic current into isosinglet and isotriplet components: $J_\mathrm{em}^\mu=J_\mathrm{em}^{(0)\mu}+J_\mathrm{em}^{(1)\mu}$ and subsequently $T_{\mu\nu}=T_{\mu\nu}^{(0)}+T_{\mu\nu}^{(1)}$. With all the above, we obtain the following identity for the forward generalized Compton tensor at $q\rightarrow-q$:
\begin{eqnarray}
T^{(I)\mu\nu}_{ss}(-q;p,p)&=&\int d^4xe^{-iq\cdot x}\langle p(p,s)|T[J_\mathrm{em}^{(I)\mu}(x)J^\nu(0)]|n(p,s)\rangle\nonumber\\
&=&\int d^4xe^{i\tilde{q}\cdot x}\langle n(\tilde{p},\tilde{s})|T[J^{(I)\mathrm{em}}_\mu(x)J^\dagger_\nu(0)]|p(\tilde{p},\tilde{s})\rangle\nonumber\\
&=&-\xi^{(I)}\int d^4xe^{i\tilde{q}\cdot x}\langle p(\tilde{p},\tilde{s})|T[J^{(I)\mathrm{em}}_\mu(x)J_\nu(0)]|n(\tilde{p},\tilde{s})\rangle~,
\end{eqnarray}
where $\xi^{I}=-1(+1)$ for $I=0(1)$, and the third line is obtained through a simple isospin rotation of the second line (or by applying the Wigner-Eckart theorem to the isospin space).

Now we may express both sides of the equation above in terms of invariant functions. That gives (we suppress the $Q^2$-dependence for simplicity):
\begin{eqnarray}
&&-i\epsilon^{\mu\nu\alpha\beta}\frac{q_\alpha p_\beta}{2p\cdot q}T_3^{(I)}(-\nu)+i\epsilon^{\mu\nu\alpha\beta}\frac{q_\alpha}{p\cdot q}\left[s_\beta S_1^{(I)}(-\nu)+\left(s_\beta-\frac{s\cdot q}{p\cdot q}p_\beta\right)S_2^{(I)}(-\nu)\right]\nonumber\\
&=&-\xi^{(I)}\left\{-i\epsilon_{\mu\nu\alpha\beta}\frac{\tilde{q}^\alpha \tilde{p}^\beta}{2p\cdot q}T_3^{(I)}(\nu)+i\epsilon_{\mu\nu\alpha\beta}\frac{\tilde{q}^\alpha}{p\cdot q}\left[\tilde{s}^\beta S_1^{(I)}(\nu)+\left(\tilde{s}^\beta-\frac{s\cdot q}{p\cdot q}\tilde{p}^\beta\right)S_2^{(I)}(\nu)\right]\right\}\nonumber\\
&=&\xi^{(I)}\left\{-i\epsilon^{\mu\nu\alpha\beta}\frac{q_\alpha p_\beta}{2p\cdot q}T_3^{(I)}(\nu)+i\epsilon^{\mu\nu\alpha\beta}\frac{q_\alpha}{p\cdot q}\left[s_\beta S_1^{(I)}(\nu)+\left(s_\beta-\frac{s\cdot q}{p\cdot q}p_\beta\right)S_2^{(I)}(\nu)\right]\right\}~,\nonumber\\
\end{eqnarray}
where we have used $\tilde{a}\cdot\tilde{b}=a\cdot b$ and $\epsilon^{\mu\nu\alpha\beta}=-\epsilon_{\mu\nu\alpha\beta}$. The equation above implies:
\begin{equation}
T_3^{(I)}(-\nu,Q^2)=\xi^{(I)}T_3(\nu,Q^2)~,\quad S_{1,2}^{(I)}(-\nu,Q^2)=\xi^{(I)}S_{1,2}^{(I)}(\nu,Q^2)~,
\end{equation}
which is our required crossing relation.

\section{\label{sec:loopfun}ChPT loop functions}

The loop functions $\bar{h}_{PQ}(t)$ that appear in Eq.\eqref{eq:3ptfinite} are defined as:
\begin{eqnarray}
\bar{h}_{K^+\pi^0}(t)&=&\frac{1}{64\pi^2(M_K^2-M_\pi^2)}\left[M_\pi^2-M_K^2+M_\pi^2\ln\frac{M_K^2}{M_\pi^2}\right]+\frac{M_K^2-M_\pi^2-t}{4t}\bar{J}_{K\pi}(t)\nonumber\\
\bar{h}_{K^0\pi^-}(t)&=&\frac{1}{64\pi^2(M_K^2-M_\pi^2)}\left[M_\pi^2-M_K^2+M_K^2\ln\frac{M_K^2}{M_\pi^2}\right]+\frac{M_\pi^2-M_K^2-t}{4t}\bar{J}_{K\pi}(t)\nonumber\\
\bar{h}_{K^+\eta}(t)&=&\frac{1}{64\pi^2(M_K^2-M_\eta^2)}\left[M_\eta^2-M_K^2+M_\eta^2\ln\frac{M_K^2}{M_\eta^2}\right]+\frac{M_K^2-M_\eta^2-t}{4t}\bar{J}_{K\eta}(t)~,\nonumber\\
\end{eqnarray}
where $\bar{J}_{PQ}$ is the standard mesonic loop function~\cite{Gasser:1984gg}:
\begin{eqnarray}
\bar{J}_{PQ}(t)&=&\frac{1}{32\pi^2}\left[2+\frac{\Delta_{PQ}}{t}\ln\frac{M_Q^2}{M_P^2}-\frac{\Sigma_{PQ}}{\Delta_{PQ}}\ln\frac{M_Q^2}{M_P^2}-\frac{\lambda^{1/2}(t,M_P^2,M_Q^2)}{t}\right.\nonumber\\
&&\left.\times\ln\left(\frac{[t+\lambda^{1/2}(t,M_P^2,M_Q^2)]^2-\Delta_{PQ}^2}{[t-\lambda^{1/2}(t,M_P^2,M_Q^2)]^2-\Delta_{PQ}^2}\right)\right]~,
\end{eqnarray}
with $\Sigma_{PQ}=M_P^2+M_Q^2$, $\Delta_{PQ}=M_P^2-M_Q^2$, and $\lambda(x,y,z)=x^2+y^2+z^2-2(xy+yz+xz)$ the K\"{a}ll\'{e}n function.

\section{\label{sec:bremsInt}IR-divergent and IR-finite integrals in the bremsstrahlung contribution}

In the bremsstrahlung process $K(p)\rightarrow \pi(p')e^+(p_e)\nu(p_\nu)\gamma(k)$, there is only one IR-divergent integral in the squared amplitude:
\begin{equation}
I_i(y,z)\equiv \int_0^{\alpha_+(y,z)}dx\int\frac{d^3k}{(2\pi)^32E_k}\frac{d^3p_\nu}{(2\pi)^32E_\nu}(2\pi)^4\delta^{(4)}(P-k-p_\nu)\left(\frac{p_e}{p_e\cdot k}-\frac{p_i}{p_i\cdot k}\right)^2~,
\end{equation}
where $i=K,\pi$. Regularizing the IR-divergence with a photon mass, one can write:
\begin{equation}
I_i(y,z)=I_i^\mathrm{IR}(y,z)+I_i^\mathrm{fin}(y,z)~,
\end{equation}
where the IR-divergent piece reads:
\begin{equation}
I_i^\mathrm{IR}(y,z)=\frac{1}{2\pi M_K^2}\left\{\left(1-\frac{1}{\beta_i(0)}\tanh^{-1}\beta_i(0)\right)\ln\left[\frac{M_K^2}{M_\gamma^2}\right]-\frac{1}{2}\ln\left[\frac{M_K^2}{m_e^2}\right]\right\}~,
\end{equation}
and the IR-finite piece reads:
\begin{eqnarray}
I_i^\mathrm{fin}(y,z)&=&\frac{1}{4\pi M_K^2}\left\{\left(1-\frac{2}{\beta_i(0)}\tanh^{-1}\beta_i(0)\right)\ln\left[\frac{M_K^2\alpha_+^2}{4P_0^2(0) }\right]+\ln\left[\frac{\alpha_+^2(y,z)}{(1-z+r_\pi-r_e)^2}\right]\right.\nonumber\\
&&-\frac{1}{\beta_i(0)}\mathrm{Li}_2\left[\frac{2\beta_i(0)}{1+\beta_i(0)}\right]+\frac{1}{\beta_i(0)}\mathrm{Li}_2\left[-\frac{2\beta_i(0)}{1-\beta_i(0)}\right]\nonumber\\
&&\left.+\frac{2}{\beta_i(0)}\mathrm{Li}_2\left[\frac{\beta_i(0)}{1+\beta_i(0)}\left(\frac{P_1(0)}{P_0(0)}+1\right)\right]-\frac{2}{\beta_i(0)}\mathrm{Li}_2\left[\frac{\beta_i(0)}{1-\beta_i(0)}\left(\frac{P_1(0)}{P_0(0)}-1\right)\right]\right\}\nonumber\\
&&-\frac{1}{2\pi M_K^2}\int_0^{\alpha_+(y,z)}dx\frac{1}{x}\left\{\frac{1}{\beta_i(x)}\ln\left[\frac{1+\beta_i(x)}{1-\beta_i(x)}\right]-\frac{1}{\beta_i(0)}\ln\left[\frac{1+\beta_i(0)}{1-\beta_i(0)}\right]\right\}~,\nonumber\\
\label{eq:Iifin}
\end{eqnarray}
with
\begin{equation}
\beta_i(x)\equiv\sqrt{1-\frac{M_i^2m_e^2}{(p_i\cdot p_e)^2}}~,\:\:P_0(x)\equiv\frac{p_i\cdot P}{M_i}~,\:\:P_1(x)\equiv\frac{1}{\beta_i(x)}\left(P_0(x)-\frac{p_e\cdot P}{p_i\cdot p_e}M_i\right)~.\label{eq:betaP}
\end{equation} 
Of course all the quantities in Eq.\eqref{eq:betaP} are functions of $\{y,z\}$ as well. Notice that $I_i^\mathrm{fin}$ includes a residual integral over $x$ that exists only for $i=\pi$ and can be easily performed numerically. 

There are also IR-finite integrals, which can be cast in the following form:
\begin{equation}
I_{m,n}(p_1,p_2)\equiv\frac{1}{2\pi}\int\frac{d^3k}{E_k}\frac{d^3p_\nu}{E_\nu}\frac{\delta^{(4)}(P-k-p_\nu)}{(p_1\cdot k)^m(p_2\cdot k)^n}~,
\end{equation}
where $m,n$ are integers. The analytic expressions of those appear in $K_{\ell 3}$ were first derived in Ref.\cite{Ginsberg:1969jh}:
\begin{eqnarray}
I_{0,0}(p_1,p_2)&=&1\nonumber\\
I_{1,0}(p_1,p_2)&=&\frac{1}{\beta_1}\ln\left(\frac{\alpha_1+\beta_1}{\alpha_1-\beta_1}\right)\nonumber\\
I_{2,0}(p_1,p_2)&=&\frac{4}{M_1^2M_K^2x}\nonumber\\
I_{1,1}(p_1,p_2)&=&\frac{2}{\gamma_{12}M_K^2x}\ln\left(\frac{p_1\cdot p_2+\gamma_{12}}{p_1\cdot p_2-\gamma_{12}}\right)\nonumber\\
I_{1,-1}(p_1,p_2)&=&\frac{(p_1p_2:P)}{\beta_1^2}+\frac{M_K^2x(p_2P:p_1)}{2\beta_1^3}\ln\left(\frac{\alpha_1+\beta_1}{\alpha_1-\beta_1}\right)\nonumber\\
I_{2,-1}(p_1,p_2)&=&\frac{2(p_2P:p_1)}{M_1^2\beta_1^2}+\frac{(p_1p_2:P)}{\beta_1^3}\ln\left(\frac{\alpha_1+\beta_1}{\alpha_1-\beta_1}\right)\nonumber\\
I_{1,-2}(p_1,p_2)&=&\frac{M_K^4x^2(p_2P:p_1)^2}{4\beta_1^5}\ln\left(\frac{\alpha_1+\beta_1}{\alpha_1-\beta_1}\right)+\frac{M_K^2x(p_1p_2:P)(p_2P:p_1)}{\beta_1^4}\nonumber\\
&&+\frac{\alpha_1(p_1p_2:P)^2}{2\beta_1^4}+\left[\frac{\beta_1^2\beta_2^2-(p_1p_2:P)^2}{4\beta_1^4}\right]\left[\alpha_1-\frac{M_1^2M_K^2x}{2\beta_1}\ln\left(\frac{\alpha_1+\beta_1}{\alpha_1-\beta_1}\right)\right]\nonumber\\
I_{2,-2}(p_1,p_2)&=&\frac{M_K^2x(p_2P:p_1)^2}{M_1^2\beta_1^4}+\frac{M_K^2x(p_2P:p_1)(p_1p_2:P)}{\beta_1^5}\ln\left(\frac{\alpha_1+\beta_1}{\alpha_1-\beta_1}\right)\nonumber\\
&&+\frac{(p_1p_2:P)^2}{\beta_1^4}-\left[\frac{\beta_1^2\beta_2^2-(p_1p_2:P)^2}{2\beta_1^4}\right]\left[2-\frac{\alpha_1}{\beta_1}\ln\left(\frac{\alpha_1+\beta_1}{\alpha_1-\beta_1}\right)\right]\nonumber\\
I_{-1,0}(p_1,p_2)&=&\frac{\alpha_1}{2}~,
\end{eqnarray}
where 
\begin{eqnarray}
p_i^2&=&M_i^2\nonumber\\
\alpha_i&=&p_i\cdot P\nonumber\\
\beta_i&=&(\alpha_i^2-M_i^2M_K^2x)^{1/2}\nonumber\\
\gamma_{ij}&=&[(p_i\cdot p_j)^2-M_i^2M_j^2]^{1/2}\nonumber\\
(ab:c)&=&(a\cdot c)(b\cdot c)-c^2(a\cdot b)~.
\end{eqnarray}

\end{appendix}

\bibliographystyle{JHEP-2}
\bibliography{RC_ref}

\providecommand{\href}[2]{#2}\begingroup\raggedright\begin{thebibliography}{100}

\bibitem{Brown:1978pb}
L.~M. Brown, {\it {The idea of the neutrino}},  {\em Phys. Today} {\bf 31N9}
  (1978) 23--28.

\bibitem{Fermi:1934hr}
E.~Fermi, {\it {An attempt of a theory of beta radiation. 1.}},  {\em Z. Phys.}
  {\bf 88} (1934) 161--177.

\bibitem{Wu:1957my}
C.~S. Wu, E.~Ambler, R.~W. Hayward, D.~D. Hoppes and R.~P. Hudson, {\it
  {Experimental Test of Parity Conservation in $\beta$ Decay}},  {\em Phys.
  Rev.} {\bf 105} (1957) 1413--1414.

\bibitem{Lee:1956qn}
T.~D. Lee and C.-N. Yang, {\it {Question of Parity Conservation in Weak
  Interactions}},  {\em Phys. Rev.} {\bf 104} (1956) 254--258.

\bibitem{Feynman:1958ty}
R.~P. Feynman and M.~Gell-Mann, {\it {Theory of Fermi interaction}},  {\em
  Phys. Rev.} {\bf 109} (1958) 193--198.

\bibitem{Sudarshan:1958vf}
E.~C.~G. Sudarshan and R.~e. Marshak, {\it {Chirality invariance and the
  universal Fermi interaction}},  {\em Phys. Rev.} {\bf 109} (1958) 1860--1860.

\bibitem{Cabibbo:1963yz}
N.~Cabibbo, {\it {Unitary Symmetry and Leptonic Decays}},  {\em Phys. Rev.
  Lett.} {\bf 10} (1963) 531--533.

\bibitem{Kobayashi:1973fv}
M.~Kobayashi and T.~Maskawa, {\it {CP Violation in the Renormalizable Theory of
  Weak Interaction}},  {\em Prog. Theor. Phys.} {\bf 49} (1973) 652--657.

\bibitem{Christenson:1964fg}
J.~H. Christenson, J.~W. Cronin, V.~L. Fitch and R.~Turlay, {\it {Evidence for
  the $2\pi$ Decay of the $K_2^0$ Meson}},  {\em Phys. Rev. Lett.} {\bf 13}
  (1964) 138--140.

\bibitem{Aghanim:2018eyx}
{\bf Planck} Collaboration, N.~Aghanim {\em et.~al.}, {\it {Planck 2018
  results. VI. Cosmological parameters}},  {\em Astron. Astrophys.} {\bf 641}
  (2020) A6 [\href{http://arXiv.org/abs/1807.06209}{{\tt 1807.06209}}].

\bibitem{Simon:2019nxf}
J.~D. Simon, {\it {The Faintest Dwarf Galaxies}},  {\em Ann. Rev. Astron.
  Astrophys.} {\bf 57} (2019), no.~1 375--415
  [\href{http://arXiv.org/abs/1901.05465}{{\tt 1901.05465}}].

\bibitem{Salucci:2018hqu}
P.~Salucci, {\it {The distribution of dark matter in galaxies}},  {\em Astron.
  Astrophys. Rev.} {\bf 27} (2019), no.~1 2
  [\href{http://arXiv.org/abs/1811.08843}{{\tt 1811.08843}}].

\bibitem{Allen:2011zs}
S.~W. Allen, A.~E. Evrard and A.~B. Mantz, {\it {Cosmological Parameters from
  Observations of Galaxy Clusters}},  {\em Ann. Rev. Astron. Astrophys.} {\bf
  49} (2011) 409--470 [\href{http://arXiv.org/abs/1103.4829}{{\tt 1103.4829}}].

\bibitem{Riess:1998cb}
{\bf Supernova Search Team} Collaboration, A.~G. Riess {\em et.~al.}, {\it
  {Observational evidence from supernovae for an accelerating universe and a
  cosmological constant}},  {\em Astron. J.} {\bf 116} (1998) 1009--1038
  [\href{http://arXiv.org/abs/astro-ph/9805201}{{\tt astro-ph/9805201}}].

\bibitem{Perlmutter:1998np}
{\bf Supernova Cosmology Project} Collaboration, S.~Perlmutter {\em et.~al.},
  {\it {Measurements of $\Omega$ and $\Lambda$ from 42 high redshift
  supernovae}},  {\em Astrophys. J.} {\bf 517} (1999) 565--586
  [\href{http://arXiv.org/abs/astro-ph/9812133}{{\tt astro-ph/9812133}}].

\bibitem{Sakharov:1967dj}
A.~D. Sakharov, {\it {Violation of CP Invariance, C asymmetry, and baryon
  asymmetry of the universe}},  {\em Sov. Phys. Usp.} {\bf 34} (1991), no.~5
  392--393.

\bibitem{Mossa:2020gjc}
V.~Mossa {\em et.~al.}, {\it {The baryon density of the Universe from an
  improved rate of deuterium burning}},  {\em Nature} {\bf 587} (2020),
  no.~7833 210--213.

\bibitem{Chatrchyan:2012ufa}
{\bf CMS} Collaboration, S.~Chatrchyan {\em et.~al.}, {\it {Observation of a
  New Boson at a Mass of 125 GeV with the CMS Experiment at the LHC}},  {\em
  Phys. Lett. B} {\bf 716} (2012) 30--61
  [\href{http://arXiv.org/abs/1207.7235}{{\tt 1207.7235}}].

\bibitem{Aad:2012tfa}
{\bf ATLAS} Collaboration, G.~Aad {\em et.~al.}, {\it {Observation of a new
  particle in the search for the Standard Model Higgs boson with the ATLAS
  detector at the LHC}},  {\em Phys. Lett. B} {\bf 716} (2012) 1--29
  [\href{http://arXiv.org/abs/1207.7214}{{\tt 1207.7214}}].

\bibitem{ATLASPublic}
Central scientific results page of the ATLAS Collaboration:
  \url{https://twiki.cern.ch/twiki/bin/view/AtlasPublic/}.

\bibitem{CMSPublic}
Central scientific results page of the CMS Collaboration:
  \url{http://cms-results.web.cern.ch/cms-results/public-results/publications/}.

\bibitem{Abada:2019lih}
{\bf FCC} Collaboration, A.~Abada {\em et.~al.}, {\it {FCC Physics
  Opportunities}: {Future Circular Collider Conceptual Design Report Volume
  1}},  {\em Eur. Phys. J. C} {\bf 79} (2019), no.~6 474.

\bibitem{Abada:2019zxq}
{\bf FCC} Collaboration, A.~Abada {\em et.~al.}, {\it {FCC-ee: The Lepton
  Collider}: {Future Circular Collider Conceptual Design Report Volume 2}},
  {\em Eur. Phys. J. ST} {\bf 228} (2019), no.~2 261--623.

\bibitem{Benedikt:2018csr}
{\bf FCC} Collaboration, A.~Abada {\em et.~al.}, {\it {FCC-hh: The Hadron
  Collider}: {Future Circular Collider Conceptual Design Report Volume 3}},
  {\em Eur. Phys. J. ST} {\bf 228} (2019), no.~4 755--1107.

\bibitem{CEPCStudyGroup:2018rmc}
{\bf CEPC Study Group} Collaboration, W.~Chou {\em et.~al.}, {\it {CEPC
  Conceptual Design Report: Volume 1 - Accelerator}},
  \href{http://arXiv.org/abs/1809.00285}{{\tt 1809.00285}}.

\bibitem{CEPCStudyGroup:2018ghi}
{\bf CEPC Study Group} Collaboration, M.~Dong {\em et.~al.}, {\it {CEPC
  Conceptual Design Report: Volume 2 - Physics \textbackslash{}\& Detector}},
  \href{http://arXiv.org/abs/1811.10545}{{\tt 1811.10545}}.

\bibitem{Zyla:2020zbs}
{\bf Particle Data Group} Collaboration, P.~Zyla {\em et.~al.}, {\it {Review of
  Particle Physics}},  {\em PTEP} {\bf 2020} (2020), no.~8 083C01.

\bibitem{Bryman:2019ssi}
D.~Bryman and R.~Shrock, {\it {Improved Constraints on Sterile Neutrinos in the
  MeV to GeV Mass Range}},  {\em Phys. Rev. D} {\bf 100} (2019), no.~5 053006
  [\href{http://arXiv.org/abs/1904.06787}{{\tt 1904.06787}}].

\bibitem{Bryman:2019bjg}
D.~Bryman and R.~Shrock, {\it {Constraints on Sterile Neutrinos in the MeV to
  GeV Mass Range}},  {\em Phys. Rev. D} {\bf 100} (2019) 073011
  [\href{http://arXiv.org/abs/1909.11198}{{\tt 1909.11198}}].

\bibitem{Kirk:2020wdk}
M.~Kirk, {\it {Cabibbo anomaly versus electroweak precision tests: An
  exploration of extensions of the standard model}},  {\em Phys. Rev. D} {\bf
  103} (2021), no.~3 035004 [\href{http://arXiv.org/abs/2008.03261}{{\tt
  2008.03261}}].

\bibitem{Grossman:2019bzp}
Y.~Grossman, E.~Passemar and S.~Schacht, {\it {On the Statistical Treatment of
  the Cabibbo Angle Anomaly}},  {\em JHEP} {\bf 07} (2020) 068
  [\href{http://arXiv.org/abs/1911.07821}{{\tt 1911.07821}}].

\bibitem{Belfatto:2019swo}
B.~Belfatto, R.~Beradze and Z.~Berezhiani, {\it {The CKM unitarity problem: A
  trace of new physics at the TeV scale?}},  {\em Eur. Phys. J. C} {\bf 80}
  (2020), no.~2 149 [\href{http://arXiv.org/abs/1906.02714}{{\tt 1906.02714}}].

\bibitem{Cheung:2020vqm}
K.~Cheung, W.-Y. Keung, C.-T. Lu and P.-Y. Tseng, {\it {Vector-like Quark
  Interpretation for the CKM Unitarity Violation, Excess in Higgs Signal
  Strength, and Bottom Quark Forward-Backward Asymmetry}},  {\em JHEP} {\bf 05}
  (2020) 117 [\href{http://arXiv.org/abs/2001.02853}{{\tt 2001.02853}}].

\bibitem{Jho:2020jsa}
Y.~Jho, S.~M. Lee, S.~C. Park, Y.~Park and P.-Y. Tseng, {\it {Light gauge boson
  interpretation for (g-2)$_\mu$ and the $K_L\rightarrow \pi^0$ + (invisible)
  anomaly at the J-PARC KOTO experiment}},  {\em JHEP} {\bf 04} (2020) 086
  [\href{http://arXiv.org/abs/2001.06572}{{\tt 2001.06572}}].

\bibitem{Yue:2020wkj}
C.~X. Yue and X.~J. Cheng, {\it {Constraints on the charged-current
  non-standard neutrino interactions induced by the gauge boson W'}},  {\em
  Nucl. Phys. B} {\bf 963} (2021) 115280
  [\href{http://arXiv.org/abs/2008.10027}{{\tt 2008.10027}}].

\bibitem{Endo:2020tkb}
M.~Endo and S.~Mishima, {\it {Muon g-2 and CKM unitarity in extra lepton
  models}},  {\em JHEP} {\bf 08} (2020), no.~08 004
  [\href{http://arXiv.org/abs/2005.03933}{{\tt 2005.03933}}].

\bibitem{Capdevila:2020rrl}
B.~Capdevila, A.~Crivellin, C.~A. Manzari and M.~Montull, {\it {Explaining
  $b\to s\ell^+\ell^-$ and the Cabibbo angle anomaly with a vector triplet}},
  {\em Phys. Rev. D} {\bf 103} (2021), no.~1 015032
  [\href{http://arXiv.org/abs/2005.13542}{{\tt 2005.13542}}].

\bibitem{Eberhardt:2020dat}
O.~Eberhardt, A.~P.~n. Mart\'\i{}nez and A.~Pich, {\it {Global fits in the
  Aligned Two-Higgs-Doublet model}},
  \href{http://arXiv.org/abs/2012.09200}{{\tt 2012.09200}}.

\bibitem{Crivellin:2020lzu}
A.~Crivellin and M.~Hoferichter, {\it {\ensuremath{\beta} Decays as Sensitive
  Probes of Lepton Flavor Universality}},  {\em Phys. Rev. Lett.} {\bf 125}
  (2020), no.~11 111801 [\href{http://arXiv.org/abs/2002.07184}{{\tt
  2002.07184}}].

\bibitem{Coutinho:2019aiy}
A.~M. Coutinho, A.~Crivellin and C.~A. Manzari, {\it {Global Fit to Modified
  Neutrino Couplings and the Cabibbo-Angle Anomaly}},  {\em Phys. Rev. Lett.}
  {\bf 125} (2020), no.~7 071802 [\href{http://arXiv.org/abs/1912.08823}{{\tt
  1912.08823}}].

\bibitem{Gonzalez-Alonso:2018omy}
M.~Gonzalez-Alonso, O.~Naviliat-Cuncic and N.~Severijns, {\it {New physics
  searches in nuclear and neutron $\beta$ decay}},  {\em Prog. Part. Nucl.
  Phys.} {\bf 104} (2019) 165--223 [\href{http://arXiv.org/abs/1803.08732}{{\tt
  1803.08732}}].

\bibitem{Falkowski:2019xoe}
A.~Falkowski, M.~Gonz\'alez-Alonso and Z.~Tabrizi, {\it {Reactor neutrino
  oscillations as constraints on Effective Field Theory}},  {\em JHEP} {\bf 05}
  (2019) 173 [\href{http://arXiv.org/abs/1901.04553}{{\tt 1901.04553}}].

\bibitem{Cirgiliano:2019nyn}
V.~Cirigliano, A.~Garcia, D.~Gazit, O.~Naviliat-Cuncic, G.~Savard and A.~Young,
  {\it {Precision Beta Decay as a Probe of New Physics}},
  \href{http://arXiv.org/abs/1907.02164}{{\tt 1907.02164}}.

\bibitem{Falkowski:2020pma}
A.~Falkowski, M.~Gonz\'alez-Alonso and O.~Naviliat-Cuncic, {\it {Comprehensive
  analysis of beta decays within and beyond the Standard Model}},  {\em JHEP}
  {\bf 04} (2021) 126 [\href{http://arXiv.org/abs/2010.13797}{{\tt
  2010.13797}}].

\bibitem{Becirevic:2020rzi}
D.~Be\v{c}irevi\'c, F.~Jaffredo, A.~Pe\~nuelas and O.~Sumensari, {\it {New
  Physics effects in leptonic and semileptonic decays}},
  \href{http://arXiv.org/abs/2012.09872}{{\tt 2012.09872}}.

\bibitem{Crivellin:2021njn}
A.~Crivellin, M.~Hoferichter and C.~A. Manzari, {\it {The Fermi constant from
  muon decay versus electroweak fits and CKM unitarity}},
  \href{http://arXiv.org/abs/2102.02825}{{\tt 2102.02825}}.

\bibitem{Tan:2019yqp}
W.~Tan, {\it {Laboratory tests of the ordinary-mirror particle oscillations and
  the extended CKM matrix}},  \href{http://arXiv.org/abs/1906.10262}{{\tt
  1906.10262}}.

\bibitem{Crivellin:2021bkd}
A.~Crivellin, M.~Hoferichter, M.~Kirk, C.~A. Manzari and L.~Schnell, {\it
  {First-Generation New Physics in Simplified Models: From Low-Energy Parity
  Violation to the LHC}},  \href{http://arXiv.org/abs/2107.13569}{{\tt
  2107.13569}}.

\bibitem{Crivellin:2020ebi}
A.~Crivellin, F.~Kirk, C.~A. Manzari and M.~Montull, {\it {Global Electroweak
  Fit and Vector-Like Leptons in Light of the Cabibbo Angle Anomaly}},  {\em
  JHEP} {\bf 12} (2020) 166 [\href{http://arXiv.org/abs/2008.01113}{{\tt
  2008.01113}}].

\bibitem{Crivellin:2020klg}
A.~Crivellin, F.~Kirk, C.~A. Manzari and L.~Panizzi, {\it {Searching for lepton
  flavor universality violation and collider signals from a singly charged
  scalar singlet}},  {\em Phys. Rev. D} {\bf 103} (2021), no.~7 073002
  [\href{http://arXiv.org/abs/2012.09845}{{\tt 2012.09845}}].

\bibitem{Dekens:2021bro}
W.~Dekens, L.~Andreoli, J.~de~Vries, E.~Mereghetti and F.~Oosterhof, {\it {A
  low-energy perspective on the minimal left-right symmetric model}},
  \href{http://arXiv.org/abs/2107.10852}{{\tt 2107.10852}}.

\bibitem{Fermigm2}
{\bf Muon g-2} Collaboration, B.~Abi {\em et.~al.}, {\it {Measurement of the
  Positive Muon Anomalous Magnetic Moment to 0.46 ppm}},  {\em Phys. Rev.
  Lett.} {\bf 126} (2021) 141801.

\bibitem{Aoyama:2020ynm}
T.~Aoyama {\em et.~al.}, {\it {The anomalous magnetic moment of the muon in the
  Standard Model}},  {\em Phys. Rept.} {\bf 887} (2020) 1--166
  [\href{http://arXiv.org/abs/2006.04822}{{\tt 2006.04822}}].

\bibitem{Miller:2007kk}
J.~P. Miller, E.~de~Rafael and B.~L. Roberts, {\it {Muon (g-2): Experiment and
  theory}},  {\em Rept. Prog. Phys.} {\bf 70} (2007) 795
  [\href{http://arXiv.org/abs/hep-ph/0703049}{{\tt hep-ph/0703049}}].

\bibitem{Miller:2012opa}
J.~P. Miller, E.~de~Rafael, B.~L. Roberts and D.~St\"ockinger, {\it {Muon
  (g-2): Experiment and Theory}},  {\em Ann. Rev. Nucl. Part. Sci.} {\bf 62}
  (2012) 237--264.

\bibitem{Jegerlehner:2009ry}
F.~Jegerlehner and A.~Nyffeler, {\it {The Muon g-2}},  {\em Phys. Rept.} {\bf
  477} (2009) 1--110 [\href{http://arXiv.org/abs/0902.3360}{{\tt 0902.3360}}].

\bibitem{Aaij:2019wad}
{\bf LHCb} Collaboration, R.~Aaij {\em et.~al.}, {\it {Search for
  lepton-universality violation in $B^+\to K^+\ell^+\ell^-$ decays}},  {\em
  Phys. Rev. Lett.} {\bf 122} (2019), no.~19 191801
  [\href{http://arXiv.org/abs/1903.09252}{{\tt 1903.09252}}].

\bibitem{Aaij:2014ora}
{\bf LHCb} Collaboration, R.~Aaij {\em et.~al.}, {\it {Test of lepton
  universality using $B^{+}\rightarrow K^{+}\ell^{+}\ell^{-}$ decays}},  {\em
  Phys. Rev. Lett.} {\bf 113} (2014) 151601
  [\href{http://arXiv.org/abs/1406.6482}{{\tt 1406.6482}}].

\bibitem{Aaij:2015yra}
{\bf LHCb} Collaboration, R.~Aaij {\em et.~al.}, {\it {Measurement of the ratio
  of branching fractions $\mathcal{B}(\bar{B}^0 \to
  D^{*+}\tau^{-}\bar{\nu}_{\tau})/\mathcal{B}(\bar{B}^0 \to
  D^{*+}\mu^{-}\bar{\nu}_{\mu})$}},  {\em Phys. Rev. Lett.} {\bf 115} (2015),
  no.~11 111803 [\href{http://arXiv.org/abs/1506.08614}{{\tt 1506.08614}}].
  [Erratum: Phys.Rev.Lett. 115, 159901 (2015)].

\bibitem{Aaij:2015oid}
{\bf LHCb} Collaboration, R.~Aaij {\em et.~al.}, {\it {Angular analysis of the
  $B^{0} \to K^{*0} \mu^{+} \mu^{-}$ decay using 3 fb$^{-1}$ of integrated
  luminosity}},  {\em JHEP} {\bf 02} (2016) 104
  [\href{http://arXiv.org/abs/1512.04442}{{\tt 1512.04442}}].

\bibitem{Behrends:1955mb}
R.~E. Behrends, R.~J. Finkelstein and A.~Sirlin, {\it {Radiative corrections to
  decay processes}},  {\em Phys. Rev.} {\bf 101} (1956) 866--873.

\bibitem{Kinoshita:1957zz}
T.~Kinoshita and A.~Sirlin, {\it {Muon Decay with Parity Nonconserving
  Interactions and Radiative Corrections in the Two-Component Theory}},  {\em
  Phys. Rev.} {\bf 107} (1957) 593--599.

\bibitem{Kinoshita:1958ru}
T.~Kinoshita and A.~Sirlin, {\it {Radiative corrections to Fermi
  interactions}},  {\em Phys. Rev.} {\bf 113} (1959) 1652--1660.

\bibitem{Sirlin:1967zza}
A.~Sirlin, {\it {General Properties of the Electromagnetic Corrections to the
  Beta Decay of a Physical Nucleon}},  {\em Phys. Rev.} {\bf 164} (1967)
  1767--1775.

\bibitem{Kallen:1967wfa}
G.~Kallen, {\it {Radiative Corrections to $\beta$-Decay and Nucleon Form
  Factors}},  {\em Nucl. Phys. B} {\bf 1} (1967), no.~5 225--263.

\bibitem{Wilkinson:1970cdv}
D.~Wilkinson and B.~Macefield, {\it {The numerical evaluation of radiative
  corrections of order \ensuremath{\alpha} to allowed nuclear
  \ensuremath{\beta}-decay}},  {\em Nucl. Phys. A} {\bf 158} (1970) 110--116.

\bibitem{Glashow:1961tr}
S.~L. Glashow, {\it {Partial Symmetries of Weak Interactions}},  {\em Nucl.
  Phys.} {\bf 22} (1961) 579--588.

\bibitem{Weinberg:1967tq}
S.~Weinberg, {\it {A Model of Leptons}},  {\em Phys. Rev. Lett.} {\bf 19}
  (1967) 1264--1266.

\bibitem{Salam:1968rm}
A.~Salam, {\it {Weak and Electromagnetic Interactions}},  {\em Conf. Proc. C}
  {\bf 680519} (1968) 367--377.

\bibitem{Sirlin:1974ni}
A.~Sirlin, {\it {Radiative corrections to g(v)/g(mu) in simple extensions of
  the su(2) x u(1) gauge model}},  {\em Nucl. Phys. B} {\bf 71} (1974) 29--51.

\bibitem{Sirlin:1977sv}
A.~Sirlin, {\it {Current Algebra Formulation of Radiative Corrections in Gauge
  Theories and the Universality of the Weak Interactions}},  {\em Rev. Mod.
  Phys.} {\bf 50} (1978) 573. [Erratum: Rev. Mod. Phys.50,905(1978)].

\bibitem{Gross:1973id}
D.~J. Gross and F.~Wilczek, {\it {Ultraviolet Behavior of Nonabelian Gauge
  Theories}},  {\em Phys. Rev. Lett.} {\bf 30} (1973) 1343--1346.

\bibitem{Politzer:1973fx}
H.~D. Politzer, {\it {Reliable Perturbative Results for Strong Interactions?}},
   {\em Phys. Rev. Lett.} {\bf 30} (1973) 1346--1349.

\bibitem{Sirlin:1981ie}
A.~Sirlin, {\it {Large m(W), m(Z) Behavior of the O(alpha) Corrections to
  Semileptonic Processes Mediated by W}},  {\em Nucl. Phys.} {\bf B196} (1982)
  83--92.

\bibitem{Ginsberg:1966zz}
E.~S. Ginsberg, {\it {Radiative Corrections to Kl-3 + /- Decays}},  {\em Phys.
  Rev.} {\bf 142} (1966) 1035--1040.

\bibitem{Ginsberg:1968pz}
E.~S. Ginsberg, {\it {Radiative corrections to k-e-3-neutral decays and the
  delta-i=1/2 rule. (erratum)}},  {\em Phys. Rev.} {\bf 171} (1968) 1675.
  [Erratum: Phys.Rev. 174, 2169 (1968)].

\bibitem{Ginsberg:1969jh}
E.~S. Ginsberg, {\it {Radiative corrections to the k-l-3 +- dalitz plot}},
  {\em Phys. Rev.} {\bf 162} (1967) 1570. [Erratum: Phys.Rev. 187, 2280
  (1969)].

\bibitem{Ginsberg:1970vy}
E.~Ginsberg, {\it {Radiative corrections to k-mu-3 decays}},  {\em Phys. Rev.
  D} {\bf 1} (1970) 229--239.

\bibitem{Becherrawy:1970ah}
T.~Becherrawy, {\it {Radiative Correction to K(l3) Decay}},  {\em Phys. Rev. D}
  {\bf 1} (1970) 1452--1468.

\bibitem{Bytev:2002nx}
V.~Bytev, E.~Kuraev, A.~Baratt and J.~Thompson, {\it {Radiative corrections to
  the K+-(e3) decay revised}},  {\em Eur. Phys. J. C} {\bf 27} (2003) 57--71
  [\href{http://arXiv.org/abs/hep-ph/0210049}{{\tt hep-ph/0210049}}]. [Erratum:
  Eur.Phys.J.C 34, 523--524 (2004)].

\bibitem{Andre:2004tk}
T.~C. Andre, {\it {Radiative corrections in K0(l3) decays}},  {\em Annals
  Phys.} {\bf 322} (2007) 2518--2544
  [\href{http://arXiv.org/abs/hep-ph/0406006}{{\tt hep-ph/0406006}}].

\bibitem{Garcia:1981it}
A.~Garcia and M.~Maya, {\it {MODEL INDEPENDENT RADIATIVE CORRECTIONS TO M+-(l3)
  DECAYS}},  {\em Phys. Rev. D} {\bf 23} (1981) 2603.

\bibitem{JuarezLeon:2010tj}
C.~Juarez-Leon, A.~Martinez, M.~Neri, J.~Torres and R.~Flores-Mendieta, {\it
  {Radiative corrections to the Dalitz plot of $K_{l3}^\pm$ decays}},  {\em
  Phys. Rev. D} {\bf 83} (2011) 054004
  [\href{http://arXiv.org/abs/1010.5547}{{\tt 1010.5547}}]. [Erratum:
  Phys.Rev.D 86, 059901 (2012)].

\bibitem{Torres:2012ge}
J.~Torres, A.~Martinez, M.~Neri, C.~Juarez-Leon and R.~Flores-Mendieta, {\it
  {Radiative corrections to the Dalitz plot of $K_{l3}^\pm$ decays:
  Contribution of the four-body region}},  {\em Phys. Rev. D} {\bf 86} (2012)
  077501 [\href{http://arXiv.org/abs/1209.5759}{{\tt 1209.5759}}].

\bibitem{Neri:2015eba}
M.~Neri, A.~Mart\'\i{}nez, C.~Ju\'arez-Le\'on, J.~Torres and
  R.~Flores-Mendieta, {\it {Radiative corrections to the Dalitz plot of
  $K_{l3}^0$ decays}},  {\em Phys. Rev. D} {\bf 92} (2015), no.~7 074022
  [\href{http://arXiv.org/abs/1510.00401}{{\tt 1510.00401}}].

\bibitem{Urech:1994hd}
R.~Urech, {\it {Virtual photons in chiral perturbation theory}},  {\em Nucl.
  Phys.} {\bf B433} (1995) 234--254
  [\href{http://arXiv.org/abs/hep-ph/9405341}{{\tt hep-ph/9405341}}].

\bibitem{Knecht:1999ag}
M.~Knecht, H.~Neufeld, H.~Rupertsberger and P.~Talavera, {\it {Chiral
  perturbation theory with virtual photons and leptons}},  {\em Eur. Phys. J.}
  {\bf C12} (2000) 469--478 [\href{http://arXiv.org/abs/hep-ph/9909284}{{\tt
  hep-ph/9909284}}].

\bibitem{Ananthanarayan:2004qk}
B.~Ananthanarayan and B.~Moussallam, {\it {Four-point correlator constraints on
  electromagnetic chiral parameters and resonance effective Lagrangians}},
  {\em JHEP} {\bf 06} (2004) 047
  [\href{http://arXiv.org/abs/hep-ph/0405206}{{\tt hep-ph/0405206}}].

\bibitem{DescotesGenon:2005pw}
S.~Descotes-Genon and B.~Moussallam, {\it {Radiative corrections in weak
  semi-leptonic processes at low energy: A Two-step matching determination}},
  {\em Eur. Phys. J.} {\bf C42} (2005) 403--417
  [\href{http://arXiv.org/abs/hep-ph/0505077}{{\tt hep-ph/0505077}}].

\bibitem{Cirigliano:2002ng}
V.~Cirigliano, M.~Knecht, H.~Neufeld and H.~Pichl, {\it {The Pionic beta decay
  in chiral perturbation theory}},  {\em Eur. Phys. J. C} {\bf 27} (2003)
  255--262 [\href{http://arXiv.org/abs/hep-ph/0209226}{{\tt hep-ph/0209226}}].

\bibitem{Cirigliano:2001mk}
V.~Cirigliano, M.~Knecht, H.~Neufeld, H.~Rupertsberger and P.~Talavera, {\it
  {Radiative corrections to K(l3) decays}},  {\em Eur. Phys. J.} {\bf C23}
  (2002) 121--133 [\href{http://arXiv.org/abs/hep-ph/0110153}{{\tt
  hep-ph/0110153}}].

\bibitem{Cirigliano:2004pv}
V.~Cirigliano, H.~Neufeld and H.~Pichl, {\it {K(e3) decays and CKM unitarity}},
   {\em Eur. Phys. J. C} {\bf 35} (2004) 53--65
  [\href{http://arXiv.org/abs/hep-ph/0401173}{{\tt hep-ph/0401173}}].

\bibitem{Cirigliano:2008wn}
V.~Cirigliano, M.~Giannotti and H.~Neufeld, {\it {Electromagnetic effects in
  K(l3) decays}},  {\em JHEP} {\bf 11} (2008) 006
  [\href{http://arXiv.org/abs/0807.4507}{{\tt 0807.4507}}].

\bibitem{Brown:1970dd}
L.~S. Brown, {\it {Perturbation theory and selfmass insertions}},  {\em Phys.
  Rev.} {\bf 187} (1969) 2260--2265.

\bibitem{Adler:1969ei}
S.~L. Adler and W.-K. Tung, {\it {Breakdown of asymptotic sum rules in
  perturbation theory}},  {\em Phys. Rev. Lett.} {\bf 22} (1969) 978--981.

\bibitem{Seng:2019lxf}
C.-Y. Seng, D.~Galviz and U.-G. Mei\ss{}ner, {\it {A New Theory Framework for
  the Electroweak Radiative Corrections in $K_{l3}$ Decays}},  {\em JHEP} {\bf
  02} (2020) 069 [\href{http://arXiv.org/abs/1910.13208}{{\tt 1910.13208}}].

\bibitem{Seng:2020jtz}
C.-Y. Seng, X.~Feng, M.~Gorchtein, L.-C. Jin and U.-G. Mei\ss{}ner, {\it {New
  method for calculating electromagnetic effects in semileptonic beta-decays of
  mesons}},  {\em JHEP} {\bf 10} (2020) 179
  [\href{http://arXiv.org/abs/2009.00459}{{\tt 2009.00459}}].

\bibitem{Kinoshita:1962ur}
T.~Kinoshita, {\it {Mass singularities of Feynman amplitudes}},  {\em J. Math.
  Phys.} {\bf 3} (1962) 650--677.

\bibitem{Lee:1964is}
T.~D. Lee and M.~Nauenberg, {\it {Degenerate Systems and Mass Singularities}},
  {\em Phys. Rev.} {\bf 133} (1964) B1549--B1562.

\bibitem{Erler:2002mv}
J.~Erler, {\it {Electroweak radiative corrections to semileptonic tau decays}},
   {\em Rev. Mex. Fis.} {\bf 50} (2004) 200--202
  [\href{http://arXiv.org/abs/hep-ph/0211345}{{\tt hep-ph/0211345}}].

\bibitem{Czarnecki:2004cw}
A.~Czarnecki, W.~J. Marciano and A.~Sirlin, {\it {Precision measurements and
  CKM unitarity}},  {\em Phys. Rev. D} {\bf 70} (2004) 093006
  [\href{http://arXiv.org/abs/hep-ph/0406324}{{\tt hep-ph/0406324}}].

\bibitem{Cirigliano:2011ny}
V.~Cirigliano, G.~Ecker, H.~Neufeld, A.~Pich and J.~Portoles, {\it {Kaon Decays
  in the Standard Model}},  {\em Rev. Mod. Phys.} {\bf 84} (2012) 399
  [\href{http://arXiv.org/abs/1107.6001}{{\tt 1107.6001}}].

\bibitem{Scherer:2012xha}
S.~Scherer and M.~R. Schindler, {\em {A Primer for Chiral Perturbation
  Theory}}, vol.~830.
\newblock 2012.

\bibitem{Bernard:1995dp}
V.~Bernard, N.~Kaiser and U.-G. Mei\ss{}ner, {\it {Chiral dynamics in nucleons
  and nuclei}},  {\em Int. J. Mod. Phys. E} {\bf 4} (1995) 193--346
  [\href{http://arXiv.org/abs/hep-ph/9501384}{{\tt hep-ph/9501384}}].

\bibitem{Bernard:2007zu}
V.~Bernard, {\it {Chiral Perturbation Theory and Baryon Properties}},  {\em
  Prog. Part. Nucl. Phys.} {\bf 60} (2008) 82--160
  [\href{http://arXiv.org/abs/0706.0312}{{\tt 0706.0312}}].

\bibitem{Gell-Mann:1961omu}
M.~Gell-Mann.
\newblock \textit{The Eightfold Way: A Theory of strong interaction symmetry},
  \textbf{1961}.

\bibitem{Neeman:1961jhl}
Y.~Ne'eman, {\it {Derivation of strong interactions from a gauge invariance}},
  {\em Nucl. Phys.} {\bf 26} (1961) 222--229.

\bibitem{Nambu:1960tm}
Y.~Nambu, {\it {Quasiparticles and Gauge Invariance in the Theory of
  Superconductivity}},  {\em Phys. Rev.} {\bf 117} (1960) 648--663.

\bibitem{Goldstone:1961eq}
J.~Goldstone, {\it {Field Theories with Superconductor Solutions}},  {\em Nuovo
  Cim.} {\bf 19} (1961) 154--164.

\bibitem{Gasser:1983yg}
J.~Gasser and H.~Leutwyler, {\it {Chiral Perturbation Theory to One Loop}},
  {\em Annals Phys.} {\bf 158} (1984) 142.

\bibitem{Gasser:1984gg}
J.~Gasser and H.~Leutwyler, {\it {Chiral Perturbation Theory: Expansions in the
  Mass of the Strange Quark}},  {\em Nucl. Phys.} {\bf B250} (1985) 465--516.

\bibitem{Fearing:1994ga}
H.~W. Fearing and S.~Scherer, {\it {Extension of the chiral perturbation theory
  meson Lagrangian to order p(6)}},  {\em Phys. Rev. D} {\bf 53} (1996)
  315--348 [\href{http://arXiv.org/abs/hep-ph/9408346}{{\tt hep-ph/9408346}}].

\bibitem{Bijnens:1999sh}
J.~Bijnens, G.~Colangelo and G.~Ecker, {\it {The Mesonic chiral Lagrangian of
  order p**6}},  {\em JHEP} {\bf 02} (1999) 020
  [\href{http://arXiv.org/abs/hep-ph/9902437}{{\tt hep-ph/9902437}}].

\bibitem{Jenkins:1990jv}
E.~E. Jenkins and A.~V. Manohar, {\it {Baryon chiral perturbation theory using
  a heavy fermion Lagrangian}},  {\em Phys. Lett. B} {\bf 255} (1991) 558--562.

\bibitem{Bernard:1992qa}
V.~Bernard, N.~Kaiser, J.~Kambor and U.~G. Mei\ss{}ner, {\it {Chiral structure
  of the nucleon}},  {\em Nucl. Phys. B} {\bf 388} (1992) 315--345.

\bibitem{Becher:1999he}
T.~Becher and H.~Leutwyler, {\it {Baryon chiral perturbation theory in
  manifestly Lorentz invariant form}},  {\em Eur. Phys. J. C} {\bf 9} (1999)
  643--671 [\href{http://arXiv.org/abs/hep-ph/9901384}{{\tt hep-ph/9901384}}].

\bibitem{Fuchs:2003qc}
T.~Fuchs, J.~Gegelia, G.~Japaridze and S.~Scherer, {\it {Renormalization of
  relativistic baryon chiral perturbation theory and power counting}},  {\em
  Phys. Rev. D} {\bf 68} (2003) 056005
  [\href{http://arXiv.org/abs/hep-ph/0302117}{{\tt hep-ph/0302117}}].

\bibitem{Gegelia:1999gf}
J.~Gegelia and G.~Japaridze, {\it {Matching heavy particle approach to
  relativistic theory}},  {\em Phys. Rev. D} {\bf 60} (1999) 114038
  [\href{http://arXiv.org/abs/hep-ph/9908377}{{\tt hep-ph/9908377}}].

\bibitem{Ando:2004rk}
S.~Ando, H.~W. Fearing, V.~P. Gudkov, K.~Kubodera, F.~Myhrer, S.~Nakamura and
  T.~Sato, {\it {Neutron beta decay in effective field theory}},  {\em Phys.
  Lett. B} {\bf 595} (2004) 250--259
  [\href{http://arXiv.org/abs/nucl-th/0402100}{{\tt nucl-th/0402100}}].

\bibitem{Bernard:2004cm}
V.~Bernard, S.~Gardner, U.~G. Mei\ss{}ner and C.~Zhang, {\it {Radiative neutron
  $\beta$-decay in effective field theory}},  {\em Phys. Lett. B} {\bf 593}
  (2004) 105--114 [\href{http://arXiv.org/abs/hep-ph/0403241}{{\tt
  hep-ph/0403241}}]. [Erratum: Phys.Lett.B 599, 348--348 (2004)].

\bibitem{Behrends:1960nf}
R.~E. Behrends and A.~Sirlin, {\it {Effect of mass splittings on the conserved
  vector current}},  {\em Phys. Rev. Lett.} {\bf 4} (1960) 186--187.

\bibitem{Ademollo:1964sr}
M.~Ademollo and R.~Gatto, {\it {Nonrenormalization Theorem for the Strangeness
  Violating Vector Currents}},  {\em Phys. Rev. Lett.} {\bf 13} (1964)
  264--265.

\bibitem{Pocanic:2003pf}
D.~Pocanic {\em et.~al.}, {\it {Precise measurement of the pi+
  ---\ensuremath{>} pi0 e+ nu branching ratio}},  {\em Phys. Rev. Lett.} {\bf
  93} (2004) 181803 [\href{http://arXiv.org/abs/hep-ex/0312030}{{\tt
  hep-ex/0312030}}].

\bibitem{Czarnecki:2019iwz}
A.~Czarnecki, W.~J. Marciano and A.~Sirlin, {\it {Pion beta decay and
  Cabibbo-Kobayashi-Maskawa unitarity}},  {\em Phys. Rev. D} {\bf 101} (2020),
  no.~9 091301 [\href{http://arXiv.org/abs/1911.04685}{{\tt 1911.04685}}].

\bibitem{Meister:1963zz}
N.~Meister and D.~Yennie, {\it {Radiative Corrections to High-Energy Scattering
  Processes}},  {\em Phys. Rev.} {\bf 130} (1963) 1210--1229.

\bibitem{Cirigliano:2003yr}
V.~Cirigliano, {\it {K(e3) and pi(e3) decays: Radiative corrections and CKM
  unitarity}},  in {\em {38th Rencontres de Moriond on Electroweak Interactions
  and Unified Theories}}, 5, 2003.
\newblock \href{http://arXiv.org/abs/hep-ph/0305154}{{\tt hep-ph/0305154}}.

\bibitem{Moussallam:1997xx}
B.~Moussallam, {\it {A Sum rule approach to the violation of Dashen's
  theorem}},  {\em Nucl. Phys. B} {\bf 504} (1997) 381--414
  [\href{http://arXiv.org/abs/hep-ph/9701400}{{\tt hep-ph/9701400}}].

\bibitem{Knecht:2004xr}
M.~Knecht, {\it {Chiral perturbation theory confronted with experiment}},  {\em
  Frascati Phys. Ser.} {\bf 36} (2004) 397--404
  [\href{http://arXiv.org/abs/hep-ph/0409089}{{\tt hep-ph/0409089}}].

\bibitem{Feng:2020zdc}
X.~Feng, M.~Gorchtein, L.-C. Jin, P.-X. Ma and C.-Y. Seng, {\it
  {First-principles calculation of electroweak box diagrams from lattice QCD}},
   {\em Phys. Rev. Lett.} {\bf 124} (2020), no.~19 192002
  [\href{http://arXiv.org/abs/2003.09798}{{\tt 2003.09798}}].

\bibitem{Gross:1969jf}
D.~J. Gross and C.~H. Llewellyn~Smith, {\it {High-energy neutrino - nucleon
  scattering, current algebra and partons}},  {\em Nucl. Phys.} {\bf B14}
  (1969) 337--347.

\bibitem{Marciano:2005ec}
W.~J. Marciano and A.~Sirlin, {\it {Improved calculation of electroweak
  radiative corrections and the value of V(ud)}},  {\em Phys. Rev. Lett.} {\bf
  96} (2006) 032002 [\href{http://arXiv.org/abs/hep-ph/0510099}{{\tt
  hep-ph/0510099}}].

\bibitem{Bjorken:1966jh}
J.~D. Bjorken, {\it {Applications of the Chiral U(6) x (6) Algebra of Current
  Densities}},  {\em Phys. Rev.} {\bf 148} (1966) 1467--1478.

\bibitem{Bjorken:1969mm}
J.~D. Bjorken, {\it {Inelastic Scattering of Polarized Leptons from Polarized
  Nucleons}},  {\em Phys. Rev. D} {\bf 1} (1970) 1376--1379.

\bibitem{Baikov:2010iw}
P.~A. Baikov, K.~G. Chetyrkin and J.~H. Kuhn, {\it {Adler Function, DIS sum
  rules and Crewther Relations}},  {\em Nucl. Phys. B Proc. Suppl.} {\bf
  205-206} (2010) 237--241 [\href{http://arXiv.org/abs/1007.0478}{{\tt
  1007.0478}}].

\bibitem{Baikov:2010je}
P.~Baikov, K.~Chetyrkin and J.~Kuhn, {\it {Adler Function, Bjorken Sum Rule,
  and the Crewther Relation to Order $\alpha^4_s$ in a General Gauge Theory}},
  {\em Phys. Rev. Lett.} {\bf 104} (2010) 132004
  [\href{http://arXiv.org/abs/1001.3606}{{\tt 1001.3606}}].

\bibitem{Chetyrkin:2000yt}
K.~G. Chetyrkin, J.~H. Kuhn and M.~Steinhauser, {\it {RunDec: A Mathematica
  package for running and decoupling of the strong coupling and quark masses}},
   {\em Comput. Phys. Commun.} {\bf 133} (2000) 43--65
  [\href{http://arXiv.org/abs/hep-ph/0004189}{{\tt hep-ph/0004189}}].

\bibitem{Seng:2021qdx}
C.-Y. Seng, {\it {$V_{ud}$ radiative corrections with lattice input}},  in {\em
  {55th Rencontres de Moriond on Electroweak Interactions and Unified
  Theories}}, 4, 2021.
\newblock \href{http://arXiv.org/abs/2104.02586}{{\tt 2104.02586}}.

\bibitem{ArevaloSnowmass}
A.~Aguilar-Arevalo {\em et.~al.}
\newblock \textit{Testing Lepton Flavor Universality and CKM Unitarity with
  Rare Pion Decay}.
  [\href{https://www.snowmass21.org/docs/files/summaries/RF/SNOWMASS21-RF2_RF3-048.pdf}{{\tt
  Link}}].

\bibitem{Weinberg:1958ut}
S.~Weinberg, {\it {Charge symmetry of weak interactions}},  {\em Phys. Rev.}
  {\bf 112} (1958) 1375--1379.

\bibitem{Jackson:1957zz}
J.~D. Jackson, S.~B. Treiman and H.~W. Wyld, {\it {Possible tests of time
  reversal invariance in Beta decay}},  {\em Phys. Rev.} {\bf 106} (1957)
  517--521.

\bibitem{Holstein:1974zf}
B.~R. Holstein, {\it {Recoil Effects in Allowed beta Decay: The Elementary
  Particle Approach}},  {\em Rev. Mod. Phys.} {\bf 46} (1974) 789. [Erratum:
  Rev.Mod.Phys. 48, 673--673 (1976)].

\bibitem{Wilkinson:1982hu}
D.~H. Wilkinson, {\it {ANALYSIS OF NEUTRON BETA DECAY}},  {\em Nucl. Phys. A}
  {\bf 377} (1982) 474--504.

\bibitem{Gudkov:2008pf}
V.~P. Gudkov, {\it {Asymmetry of recoil protons in neutron beta-decay}},  {\em
  Phys. Rev. C} {\bf 77} (2008) 045502
  [\href{http://arXiv.org/abs/0801.4896}{{\tt 0801.4896}}].

\bibitem{Ivanov:2012qe}
A.~N. Ivanov, M.~Pitschmann and N.~I. Troitskaya, {\it {Neutron $\beta^-$decay
  as a laboratory for testing the standard model}},  {\em Phys. Rev. D} {\bf
  88} (2013), no.~7 073002 [\href{http://arXiv.org/abs/1212.0332}{{\tt
  1212.0332}}].

\bibitem{Ivanov:2020ybx}
A.~N. Ivanov, R.~H\"ollwieser, N.~I. Troitskaya, M.~Wellenzohn and Y.~A.
  Berdnikov, {\it {Corrections of order $O(E^2_e/m^2_N)$, caused by weak
  magnetism and proton recoil, to the neutron lifetime and correlation
  coefficients of the neutron beta decay}},  {\em Results Phys.} {\bf 21}
  (2021) 103806 [\href{http://arXiv.org/abs/2010.14336}{{\tt 2010.14336}}].

\bibitem{Burkhardt:1970ti}
H.~Burkhardt and W.~N. Cottingham, {\it {Sum rules for forward virtual Compton
  scattering}},  {\em Annals Phys.} {\bf 56} (1970) 453--463.

\bibitem{Bowman:2014nsk}
J.~D. Bowman {\em et.~al.}, {\it {Determination of the Free Neutron Lifetime}},
   \href{http://arXiv.org/abs/1410.5311}{{\tt 1410.5311}}.

\bibitem{Bopp:1986rt}
P.~Bopp, D.~Dubbers, L.~Hornig, E.~Klemt, J.~Last, H.~Schutze, S.~J. Freedman
  and O.~Scharpf, {\it {The Beta Decay Asymmetry of the Neutron and
  $g_A/g_V$}},  {\em Phys. Rev. Lett.} {\bf 56} (1986) 919. [Erratum:
  Phys.Rev.Lett. 57, 1192 (1986)].

\bibitem{Erozolimsky:1997wi}
B.~Erozolimsky, I.~Kuznetsov, I.~Stepanenko and Y.~A. Mostovoi, {\it
  {Corrigendum: Corrected value of the beta-emission asymmetry in the decay of
  polarized neutrons measured in 1990}}, . [Erratum: Phys.Lett.B 412, 240--241
  (1997)].

\bibitem{Liaud:1997vu}
P.~Liaud, K.~Schreckenbach, R.~Kossakowski, H.~Nastoll, A.~Bussiere, J.~P.
  Guillaud and L.~Beck, {\it {The measurement of the beta asymmetry in the
  decay of polarized neutrons}},  {\em Nucl. Phys. A} {\bf 612} (1997) 53--81.

\bibitem{Mostovoi:2001ye}
Y.~A. Mostovoi {\em et.~al.}, {\it {Experimental value of G(A)/G(V) from a
  measurement of both P-odd correlations in free-neutron decay}},  {\em Phys.
  Atom. Nucl.} {\bf 64} (2001) 1955--1960.

\bibitem{Schumann:2007hz}
M.~Schumann, M.~Kreuz, M.~Deissenroth, F.~Gluck, J.~Krempel, B.~Markisch,
  D.~Mund, A.~Petoukhov, T.~Soldner and H.~Abele, {\it {Measurement of the
  Proton Asymmetry Parameter C in Neutron Beta Decay}},  {\em Phys. Rev. Lett.}
  {\bf 100} (2008) 151801 [\href{http://arXiv.org/abs/0712.2442}{{\tt
  0712.2442}}].

\bibitem{Mund:2012fq}
D.~Mund, B.~Maerkisch, M.~Deissenroth, J.~Krempel, M.~Schumann, H.~Abele,
  A.~Petoukhov and T.~Soldner, {\it {Determination of the Weak Axial Vector
  Coupling from a Measurement of the Beta-Asymmetry Parameter A in Neutron Beta
  Decay}},  {\em Phys. Rev. Lett.} {\bf 110} (2013) 172502
  [\href{http://arXiv.org/abs/1204.0013}{{\tt 1204.0013}}].

\bibitem{Darius:2017arh}
G.~Darius {\em et.~al.}, {\it {Measurement of the Electron-Antineutrino Angular
  Correlation in Neutron $\beta$ Decay}},  {\em Phys. Rev. Lett.} {\bf 119}
  (2017), no.~4 042502.

\bibitem{Brown:2017mhw}
{\bf UCNA} Collaboration, M.~A.~P. Brown {\em et.~al.}, {\it {New result for
  the neutron $\beta$-asymmetry parameter $A_0$ from UCNA}},  {\em Phys. Rev.
  C} {\bf 97} (2018), no.~3 035505 [\href{http://arXiv.org/abs/1712.00884}{{\tt
  1712.00884}}].

\bibitem{Markisch:2018ndu}
B.~M\"arkisch {\em et.~al.}, {\it {Measurement of the Weak Axial-Vector
  Coupling Constant in the Decay of Free Neutrons Using a Pulsed Cold Neutron
  Beam}},  {\em Phys. Rev. Lett.} {\bf 122} (2019), no.~24 242501
  [\href{http://arXiv.org/abs/1812.04666}{{\tt 1812.04666}}].

\bibitem{Czarnecki:2018okw}
A.~Czarnecki, W.~J. Marciano and A.~Sirlin, {\it {Neutron Lifetime and Axial
  Coupling Connection}},  {\em Phys. Rev. Lett.} {\bf 120} (2018), no.~20
  202002 [\href{http://arXiv.org/abs/1802.01804}{{\tt 1802.01804}}].

\bibitem{Pattie:2019brb}
R.~W. Pattie {\em et.~al.}, {\it {Status of the UCN$\tau$ experiment}},  {\em
  EPJ Web Conf.} {\bf 219} (2019) 03004.

\bibitem{Wietfeldt:2014gia}
F.~E. Wietfeldt, G.~Darius, M.~S. Dewey, N.~Fomin, G.~L. Greene, J.~Mulholland,
  W.~M. Snow and A.~T. Yue, {\it {A Path to a 0.1 s Neutron Lifetime
  Measurement Using the Beam Method}},  {\em Phys. Procedia} {\bf 51} (2014)
  54--58.

\bibitem{Fry:2018kvq}
J.~Fry {\em et.~al.}, {\it {The Nab Experiment: A Precision Measurement of
  Unpolarized Neutron Beta Decay}},  {\em EPJ Web Conf.} {\bf 219} (2019) 04002
  [\href{http://arXiv.org/abs/1811.10047}{{\tt 1811.10047}}].

\bibitem{Dubbers:2007st}
D.~Dubbers, H.~Abele, S.~Baessler, B.~Maerkisch, M.~Schumann, T.~Soldner and
  O.~Zimmer, {\it {A Clean, bright, and versatile source of neutron decay
  products}},  {\em Nucl. Instrum. Meth. A} {\bf 596} (2008) 238--247
  [\href{http://arXiv.org/abs/0709.4440}{{\tt 0709.4440}}].

\bibitem{Wang:2019pts}
{\bf PERC} Collaboration, X.~Wang {\em et.~al.}, {\it {Design of the magnet
  system of the neutron decay facility PERC}},  {\em EPJ Web Conf.} {\bf 219}
  (2019) 04007 [\href{http://arXiv.org/abs/1905.10249}{{\tt 1905.10249}}].

\bibitem{Marciano:1985pd}
W.~J. Marciano and A.~Sirlin, {\it {Radiative Corrections to beta Decay and the
  Possibility of a Fourth Generation}},  {\em Phys. Rev. Lett.} {\bf 56} (1986)
  22.

\bibitem{Hardy:2004id}
J.~C. Hardy and I.~S. Towner, {\it {Superallowed 0+ ---\ensuremath{>} 0+
  nuclear beta decays: A Critical survey with tests of CVC and the standard
  model}},  {\em Phys. Rev. C} {\bf 71} (2005) 055501
  [\href{http://arXiv.org/abs/nucl-th/0412056}{{\tt nucl-th/0412056}}].

\bibitem{Seng:2018yzq}
C.-Y. Seng, M.~Gorchtein, H.~H. Patel and M.~J. Ramsey-Musolf, {\it {Reduced
  Hadronic Uncertainty in the Determination of $V_{ud}$}},  {\em Phys. Rev.
  Lett.} {\bf 121} (2018), no.~24 241804
  [\href{http://arXiv.org/abs/1807.10197}{{\tt 1807.10197}}].

\bibitem{Nachtmann:1973mr}
O.~Nachtmann, {\it {Positivity constraints for anomalous dimensions}},  {\em
  Nucl. Phys. B} {\bf 63} (1973) 237--247.

\bibitem{Nachtmann:1974aj}
O.~Nachtmann, {\it {Is There Evidence for Large Anomalous Dimensions?}},  {\em
  Nucl. Phys. B} {\bf 78} (1974) 455--467.

\bibitem{Abramowicz:2015mha}
{\bf H1, ZEUS} Collaboration, H.~Abramowicz {\em et.~al.}, {\it {Combination of
  measurements of inclusive deep inelastic ${e^{\pm }p}$ scattering cross
  sections and QCD analysis of HERA data}},  {\em Eur. Phys. J. C} {\bf 75}
  (2015), no.~12 580 [\href{http://arXiv.org/abs/1506.06042}{{\tt
  1506.06042}}].

\bibitem{Argento:1983dj}
A.~Argento {\em et.~al.}, {\it {Measurement of the Interference Structure
  Function Xg(3) (X) in Muon - Nucleon Scattering}},  {\em Phys. Lett. B} {\bf
  140} (1984) 142--144.

\bibitem{Onengut:2005kv}
{\bf CHORUS} Collaboration, G.~Onengut {\em et.~al.}, {\it {Measurement of
  nucleon structure functions in neutrino scattering}},  {\em Phys. Lett. B}
  {\bf 632} (2006) 65--75.

\bibitem{Ye:2017gyb}
Z.~Ye, J.~Arrington, R.~J. Hill and G.~Lee, {\it {Proton and Neutron
  Electromagnetic Form Factors and Uncertainties}},  {\em Phys. Lett. B} {\bf
  777} (2018) 8--15 [\href{http://arXiv.org/abs/1707.09063}{{\tt 1707.09063}}].

\bibitem{Lorenz:2012tm}
I.~T. Lorenz, H.~W. Hammer and U.-G. Mei\ss{}ner, {\it {The size of the proton
  - closing in on the radius puzzle}},  {\em Eur. Phys. J. A} {\bf 48} (2012)
  151 [\href{http://arXiv.org/abs/1205.6628}{{\tt 1205.6628}}].

\bibitem{Lorenz:2014yda}
I.~T. Lorenz, U.-G. Mei\ss{}ner, H.~W. Hammer and Y.~B. Dong, {\it {Theoretical
  Constraints and Systematic Effects in the Determination of the Proton Form
  Factors}},  {\em Phys. Rev. D} {\bf 91} (2015), no.~1 014023
  [\href{http://arXiv.org/abs/1411.1704}{{\tt 1411.1704}}].

\bibitem{Lin:2021umk}
Y.-H. Lin, H.-W. Hammer and U.-G. Mei\ss{}ner, {\it {High-precision
  determination of the electric and magnetic radius of the proton}},  {\em
  Phys. Lett. B} {\bf 816} (2021) 136254
  [\href{http://arXiv.org/abs/2102.11642}{{\tt 2102.11642}}].

\bibitem{Lin:2021umz}
Y.-H. Lin, H.-W. Hammer and U.-G. Mei\ss{}ner, {\it {Dispersion-theoretical
  analysis of the electromagnetic form factors of the nucleon: Past, present
  and future}},  \href{http://arXiv.org/abs/2106.06357}{{\tt 2106.06357}}.

\bibitem{Bernard:2001rs}
V.~Bernard, L.~Elouadrhiri and U.-G. Mei\ss{}ner, {\it {Axial structure of the
  nucleon: Topical Review}},  {\em J. Phys. G} {\bf 28} (2002) R1--R35
  [\href{http://arXiv.org/abs/hep-ph/0107088}{{\tt hep-ph/0107088}}].

\bibitem{Bhattacharya:2011ah}
B.~Bhattacharya, R.~J. Hill and G.~Paz, {\it {Model independent determination
  of the axial mass parameter in quasielastic neutrino-nucleon scattering}},
  {\em Phys. Rev. D} {\bf 84} (2011) 073006
  [\href{http://arXiv.org/abs/1108.0423}{{\tt 1108.0423}}].

\bibitem{Seng:2018qru}
C.~Y. Seng, M.~Gorchtein and M.~J. Ramsey-Musolf, {\it {Dispersive evaluation
  of the inner radiative correction in neutron and nuclear $\beta$ decay}},
  {\em Phys. Rev.} {\bf D100} (2019), no.~1 013001
  [\href{http://arXiv.org/abs/1812.03352}{{\tt 1812.03352}}].

\bibitem{Lalakulich:2005cs}
O.~Lalakulich and E.~A. Paschos, {\it {Resonance production by neutrinos. I. J
  = 3/2 resonances}},  {\em Phys. Rev. D} {\bf 71} (2005) 074003
  [\href{http://arXiv.org/abs/hep-ph/0501109}{{\tt hep-ph/0501109}}].

\bibitem{Baikov:2012zn}
P.~A. Baikov, K.~G. Chetyrkin, J.~H. Kuhn and J.~Rittinger, {\it {Adler
  Function, Sum Rules and Crewther Relation of Order $\mathcal{O}(\alpha^4_s)$:
  the Singlet Case}},  {\em Phys. Lett. B} {\bf 714} (2012) 62--65
  [\href{http://arXiv.org/abs/1206.1288}{{\tt 1206.1288}}].

\bibitem{Kashevarov:2017vyl}
V.~L. Kashevarov, M.~Ostrick and L.~Tiator, {\it {Regge phenomenology in
  $\pi^0$ and $\eta$ photoproduction}},  {\em Phys. Rev. C} {\bf 96} (2017),
  no.~3 035207 [\href{http://arXiv.org/abs/1706.07376}{{\tt 1706.07376}}].

\bibitem{Lichard:1997ya}
P.~Lichard, {\it {Some implications of meson dominance in weak interactions}},
  {\em Phys. Rev. D} {\bf 55} (1997) 5385--5407
  [\href{http://arXiv.org/abs/hep-ph/9702345}{{\tt hep-ph/9702345}}].

\bibitem{deSwart:1963pdg}
J.~J. de~Swart, {\it {The Octet model and its Clebsch-Gordan coefficients}},
  {\em Rev. Mod. Phys.} {\bf 35} (1963) 916--939. [Erratum: Rev.Mod.Phys. 37,
  326--326 (1965)].

\bibitem{Kataev:1994rj}
A.~L. Kataev and A.~V. Sidorov, {\it {The Jacobi polynomials QCD analysis of
  the CCFR data for xF3 and the Q**2 dependence of the Gross-Llewellyn-Smith
  sum rule}},  {\em Phys. Lett.} {\bf B331} (1994) 179--186
  [\href{http://arXiv.org/abs/hep-ph/9402342}{{\tt hep-ph/9402342}}].

\bibitem{Kim:1998kia}
J.~H. Kim {\em et.~al.}, {\it {A Measurement of alpha(s)(Q**2) from the
  Gross-Llewellyn Smith sum rule}},  {\em Phys. Rev. Lett.} {\bf 81} (1998)
  3595--3598 [\href{http://arXiv.org/abs/hep-ex/9808015}{{\tt
  hep-ex/9808015}}].

\bibitem{Bolognese:1982zd}
{\bf Aachen-Bonn-CERN-Democritos-London-Oxford-Saclay} Collaboration,
  T.~Bolognese, P.~Fritze, J.~Morfin, D.~H. Perkins, K.~Powell and W.~G. Scott,
  {\it {Data on the Gross-llewellyn Smith Sum Rule as a Function of $q^2$}},
  {\em Phys. Rev. Lett.} {\bf 50} (1983) 224.

\bibitem{Allasia:1985hw}
D.~Allasia {\em et.~al.}, {\it {Q**2 Dependence of the Proton and Neutron
  Structure Functions from Neutrino and anti-neutrinos Scattering in
  Deuterium}},  {\em Z. Phys.} {\bf C28} (1985) 321.

\bibitem{Czarnecki:2019mwq}
A.~Czarnecki, W.~J. Marciano and A.~Sirlin, {\it {Radiative Corrections to
  Neutron and Nuclear Beta Decays Revisited}},
  \href{http://arXiv.org/abs/1907.06737}{{\tt 1907.06737}}.

\bibitem{Hayen:2020cxh}
L.~Hayen, {\it {Standard Model $\mathcal{O}(\alpha)$ renormalization of $g_A$
  and its impact on new physics searches}},
  \href{http://arXiv.org/abs/2010.07262}{{\tt 2010.07262}}.

\bibitem{Hayen:2021iga}
L.~Hayen, {\it {Radiative corrections to nucleon weak charges and Beyond
  Standard Model impact}},  \href{http://arXiv.org/abs/2102.03458}{{\tt
  2102.03458}}.

\bibitem{Seng:2020wjq}
C.-Y. Seng, X.~Feng, M.~Gorchtein and L.-C. Jin, {\it {Joint lattice
  QCD--dispersion theory analysis confirms the quark-mixing top-row unitarity
  deficit}},  {\em Phys. Rev. D} {\bf 101} (2020), no.~11 111301
  [\href{http://arXiv.org/abs/2003.11264}{{\tt 2003.11264}}].

\bibitem{Shiells:2020fqp}
K.~Shiells, P.~Blunden and W.~Melnitchouk, {\it {Electroweak axial structure
  functions and improved extraction of the $V_{ud}$ CKM matrix element}},
  \href{http://arXiv.org/abs/2012.01580}{{\tt 2012.01580}}.

\bibitem{Khan:2006de}
A.~A. Khan {\em et.~al.}, {\it {Axial coupling constant of the nucleon for two
  flavours of dynamical quarks in finite and infinite volume}},  {\em Phys.
  Rev. D} {\bf 74} (2006) 094508
  [\href{http://arXiv.org/abs/hep-lat/0603028}{{\tt hep-lat/0603028}}].

\bibitem{Lin:2008uz}
H.-W. Lin, T.~Blum, S.~Ohta, S.~Sasaki and T.~Yamazaki, {\it {Nucleon structure
  with two flavors of dynamical domain-wall fermions}},  {\em Phys. Rev. D}
  {\bf 78} (2008) 014505 [\href{http://arXiv.org/abs/0802.0863}{{\tt
  0802.0863}}].

\bibitem{Capitani:2012gj}
S.~Capitani, M.~Della~Morte, G.~von Hippel, B.~Jager, A.~Juttner,
  B.~Knippschild, H.~B. Meyer and H.~Wittig, {\it {The nucleon axial charge
  from lattice QCD with controlled errors}},  {\em Phys. Rev. D} {\bf 86}
  (2012) 074502 [\href{http://arXiv.org/abs/1205.0180}{{\tt 1205.0180}}].

\bibitem{Horsley:2013ayv}
R.~Horsley, Y.~Nakamura, A.~Nobile, P.~E.~L. Rakow, G.~Schierholz and J.~M.
  Zanotti, {\it {Nucleon axial charge and pion decay constant from two-flavor
  lattice QCD}},  {\em Phys. Lett. B} {\bf 732} (2014) 41--48
  [\href{http://arXiv.org/abs/1302.2233}{{\tt 1302.2233}}].

\bibitem{Bali:2014nma}
G.~S. Bali, S.~Collins, B.~Gl\"assle, M.~G\"ockeler, J.~Najjar, R.~H. R\"odl,
  A.~Sch\"afer, R.~W. Schiel, W.~S\"oldner and A.~Sternbeck, {\it {Nucleon
  isovector couplings from $N_f=2$ lattice QCD}},  {\em Phys. Rev. D} {\bf 91}
  (2015), no.~5 054501 [\href{http://arXiv.org/abs/1412.7336}{{\tt
  1412.7336}}].

\bibitem{Abdel-Rehim:2015owa}
A.~Abdel-Rehim {\em et.~al.}, {\it {Nucleon and pion structure with lattice QCD
  simulations at physical value of the pion mass}},  {\em Phys. Rev. D} {\bf
  92} (2015), no.~11 114513 [\href{http://arXiv.org/abs/1507.04936}{{\tt
  1507.04936}}]. [Erratum: Phys.Rev.D 93, 039904 (2016)].

\bibitem{Alexandrou:2017hac}
C.~Alexandrou, M.~Constantinou, K.~Hadjiyiannakou, K.~Jansen, C.~Kallidonis,
  G.~Koutsou and A.~Vaquero Aviles-Casco, {\it {Nucleon axial form factors
  using $N_f$ = 2 twisted mass fermions with a physical value of the pion
  mass}},  {\em Phys. Rev. D} {\bf 96} (2017), no.~5 054507
  [\href{http://arXiv.org/abs/1705.03399}{{\tt 1705.03399}}].

\bibitem{Capitani:2017qpc}
S.~Capitani, M.~Della~Morte, D.~Djukanovic, G.~M. von Hippel, J.~Hua,
  B.~J\"ager, P.~M. Junnarkar, H.~B. Meyer, T.~D. Rae and H.~Wittig, {\it
  {Isovector axial form factors of the nucleon in two-flavor lattice QCD}},
  {\em Int. J. Mod. Phys. A} {\bf 34} (2019), no.~02 1950009
  [\href{http://arXiv.org/abs/1705.06186}{{\tt 1705.06186}}].

\bibitem{Edwards:2005ym}
{\bf LHPC} Collaboration, R.~G. Edwards, G.~T. Fleming, P.~Hagler, J.~W.
  Negele, K.~Orginos, A.~V. Pochinsky, D.~B. Renner, D.~G. Richards and
  W.~Schroers, {\it {The Nucleon axial charge in full lattice QCD}},  {\em
  Phys. Rev. Lett.} {\bf 96} (2006) 052001
  [\href{http://arXiv.org/abs/hep-lat/0510062}{{\tt hep-lat/0510062}}].

\bibitem{Yamazaki:2008py}
{\bf RBC+UKQCD} Collaboration, T.~Yamazaki, Y.~Aoki, T.~Blum, H.~W. Lin, M.~F.
  Lin, S.~Ohta, S.~Sasaki, R.~J. Tweedie and J.~M. Zanotti, {\it {Nucleon axial
  charge in 2+1 flavor dynamical lattice QCD with domain wall fermions}},  {\em
  Phys. Rev. Lett.} {\bf 100} (2008) 171602
  [\href{http://arXiv.org/abs/0801.4016}{{\tt 0801.4016}}].

\bibitem{Yamazaki:2009zq}
T.~Yamazaki, Y.~Aoki, T.~Blum, H.-W. Lin, S.~Ohta, S.~Sasaki, R.~Tweedie and
  J.~Zanotti, {\it {Nucleon form factors with 2+1 flavor dynamical domain-wall
  fermions}},  {\em Phys. Rev. D} {\bf 79} (2009) 114505
  [\href{http://arXiv.org/abs/0904.2039}{{\tt 0904.2039}}].

\bibitem{Bratt:2010jn}
{\bf LHPC} Collaboration, J.~D. Bratt {\em et.~al.}, {\it {Nucleon structure
  from mixed action calculations using 2+1 flavors of asqtad sea and domain
  wall valence fermions}},  {\em Phys. Rev. D} {\bf 82} (2010) 094502
  [\href{http://arXiv.org/abs/1001.3620}{{\tt 1001.3620}}].

\bibitem{Green:2012ud}
J.~R. Green, M.~Engelhardt, S.~Krieg, J.~W. Negele, A.~V. Pochinsky and S.~N.
  Syritsyn, {\it {Nucleon Structure from Lattice QCD Using a Nearly Physical
  Pion Mass}},  {\em Phys. Lett. B} {\bf 734} (2014) 290--295
  [\href{http://arXiv.org/abs/1209.1687}{{\tt 1209.1687}}].

\bibitem{Yamanaka:2018uud}
{\bf JLQCD} Collaboration, N.~Yamanaka, S.~Hashimoto, T.~Kaneko and H.~Ohki,
  {\it {Nucleon charges with dynamical overlap fermions}},  {\em Phys. Rev. D}
  {\bf 98} (2018), no.~5 054516 [\href{http://arXiv.org/abs/1805.10507}{{\tt
  1805.10507}}].

\bibitem{Liang:2018pis}
J.~Liang, Y.-B. Yang, T.~Draper, M.~Gong and K.-F. Liu, {\it {Quark spins and
  Anomalous Ward Identity}},  {\em Phys. Rev. D} {\bf 98} (2018), no.~7 074505
  [\href{http://arXiv.org/abs/1806.08366}{{\tt 1806.08366}}].

\bibitem{Ishikawa:2018rew}
{\bf PACS} Collaboration, K.-I. Ishikawa, Y.~Kuramashi, S.~Sasaki,
  N.~Tsukamoto, A.~Ukawa and T.~Yamazaki, {\it {Nucleon form factors on a large
  volume lattice near the physical point in 2+1 flavor QCD}},  {\em Phys. Rev.
  D} {\bf 98} (2018), no.~7 074510 [\href{http://arXiv.org/abs/1807.03974}{{\tt
  1807.03974}}].

\bibitem{Ottnad:2018fri}
K.~Ottnad, T.~Harris, H.~Meyer, G.~von Hippel, J.~Wilhelm and H.~Wittig, {\it
  {Nucleon charges and quark momentum fraction with $N_f=2+1$ Wilson
  fermions}},  {\em PoS} {\bf LATTICE2018} (2018) 129
  [\href{http://arXiv.org/abs/1809.10638}{{\tt 1809.10638}}].

\bibitem{Bhattacharya:2016zcn}
T.~Bhattacharya, V.~Cirigliano, S.~Cohen, R.~Gupta, H.-W. Lin and B.~Yoon, {\it
  {Axial, Scalar and Tensor Charges of the Nucleon from 2+1+1-flavor Lattice
  QCD}},  {\em Phys. Rev. D} {\bf 94} (2016), no.~5 054508
  [\href{http://arXiv.org/abs/1606.07049}{{\tt 1606.07049}}].

\bibitem{Berkowitz:2017gql}
E.~Berkowitz {\em et.~al.}, {\it {An accurate calculation of the nucleon axial
  charge with lattice QCD}},  \href{http://arXiv.org/abs/1704.01114}{{\tt
  1704.01114}}.

\bibitem{Chang:2018uxx}
C.~C. Chang {\em et.~al.}, {\it {A per-cent-level determination of the nucleon
  axial coupling from quantum chromodynamics}},  {\em Nature} {\bf 558} (2018),
  no.~7708 91--94 [\href{http://arXiv.org/abs/1805.12130}{{\tt 1805.12130}}].

\bibitem{Gupta:2018qil}
R.~Gupta, Y.-C. Jang, B.~Yoon, H.-W. Lin, V.~Cirigliano and T.~Bhattacharya,
  {\it {Isovector Charges of the Nucleon from 2+1+1-flavor Lattice QCD}},  {\em
  Phys. Rev. D} {\bf 98} (2018) 034503
  [\href{http://arXiv.org/abs/1806.09006}{{\tt 1806.09006}}].

\bibitem{Walker-Loud:2019cif}
A.~Walker-Loud {\em et.~al.}, {\it {Lattice QCD Determination of $g_A$}},  {\em
  PoS} {\bf CD2018} (2020) 020 [\href{http://arXiv.org/abs/1912.08321}{{\tt
  1912.08321}}].

\bibitem{Gonzalez-Alonso:2016etj}
M.~Gonz\'alez-Alonso and J.~Martin~Camalich, {\it {Global
  Effective-Field-Theory analysis of New-Physics effects in (semi)leptonic kaon
  decays}},  {\em JHEP} {\bf 12} (2016) 052
  [\href{http://arXiv.org/abs/1605.07114}{{\tt 1605.07114}}].

\bibitem{Alioli:2017ces}
S.~Alioli, V.~Cirigliano, W.~Dekens, J.~de~Vries and E.~Mereghetti, {\it
  {Right-handed charged currents in the era of the Large Hadron Collider}},
  {\em JHEP} {\bf 05} (2017) 086 [\href{http://arXiv.org/abs/1703.04751}{{\tt
  1703.04751}}].

\bibitem{Anthony:1993uf}
{\bf E142} Collaboration, P.~L. Anthony {\em et.~al.}, {\it {Determination of
  the neutron spin structure function.}},  {\em Phys. Rev. Lett.} {\bf 71}
  (1993) 959--962.

\bibitem{Abe:1994cp}
{\bf E143} Collaboration, K.~Abe {\em et.~al.}, {\it {Precision measurement of
  the proton spin structure function g1(p)}},  {\em Phys. Rev. Lett.} {\bf 74}
  (1995) 346--350.

\bibitem{Abe:1997cx}
{\bf E154} Collaboration, K.~Abe {\em et.~al.}, {\it {Precision determination
  of the neutron spin structure function g1(n)}},  {\em Phys. Rev. Lett.} {\bf
  79} (1997) 26--30 [\href{http://arXiv.org/abs/hep-ex/9705012}{{\tt
  hep-ex/9705012}}].

\bibitem{Adams:1994zd}
{\bf Spin Muon (SMC)} Collaboration, D.~Adams {\em et.~al.}, {\it {Measurement
  of the spin dependent structure function $g_1(x)$ of the proton}},  {\em
  Phys. Lett. B} {\bf 329} (1994) 399--406
  [\href{http://arXiv.org/abs/hep-ph/9404270}{{\tt hep-ph/9404270}}]. [Erratum:
  Phys.Lett.B 339, 332--333 (1994)].

\bibitem{Alexakhin:2006oza}
{\bf COMPASS} Collaboration, V.~Y. Alexakhin {\em et.~al.}, {\it {The Deuteron
  Spin-dependent Structure Function g1(d) and its First Moment}},  {\em Phys.
  Lett. B} {\bf 647} (2007) 8--17
  [\href{http://arXiv.org/abs/hep-ex/0609038}{{\tt hep-ex/0609038}}].

\bibitem{Alekseev:2010hc}
{\bf COMPASS} Collaboration, M.~G. Alekseev {\em et.~al.}, {\it {The
  Spin-dependent Structure Function of the Proton $g_1^p$ and a Test of the
  Bjorken Sum Rule}},  {\em Phys. Lett. B} {\bf 690} (2010) 466--472
  [\href{http://arXiv.org/abs/1001.4654}{{\tt 1001.4654}}].

\bibitem{Aghasyan:2017vck}
{\bf COMPASS} Collaboration, M.~Aghasyan {\em et.~al.}, {\it {Longitudinal
  double-spin asymmetry $A_1^{\rm p}$ and spin-dependent structure function
  $g_1^{\rm p}$ of the proton at small values of $x$ and $Q^2$}},  {\em Phys.
  Lett. B} {\bf 781} (2018) 464--472
  [\href{http://arXiv.org/abs/1710.01014}{{\tt 1710.01014}}].

\bibitem{Ackerstaff:1997ws}
{\bf HERMES} Collaboration, K.~Ackerstaff {\em et.~al.}, {\it {Measurement of
  the neutron spin structure function g1(n) with a polarized He-3 internal
  target}},  {\em Phys. Lett. B} {\bf 404} (1997) 383--389
  [\href{http://arXiv.org/abs/hep-ex/9703005}{{\tt hep-ex/9703005}}].

\bibitem{Deur:2004ti}
A.~Deur {\em et.~al.}, {\it {Experimental determination of the evolution of the
  Bjorken integral at low Q**2}},  {\em Phys. Rev. Lett.} {\bf 93} (2004)
  212001 [\href{http://arXiv.org/abs/hep-ex/0407007}{{\tt hep-ex/0407007}}].

\bibitem{Wesselmann:2006mw}
{\bf RSS} Collaboration, F.~R. Wesselmann {\em et.~al.}, {\it {Proton spin
  structure in the resonance region}},  {\em Phys. Rev. Lett.} {\bf 98} (2007)
  132003 [\href{http://arXiv.org/abs/nucl-ex/0608003}{{\tt nucl-ex/0608003}}].

\bibitem{Deur:2008ej}
A.~Deur {\em et.~al.}, {\it {Experimental study of isovector spin sum rules}},
  {\em Phys. Rev. D} {\bf 78} (2008) 032001
  [\href{http://arXiv.org/abs/0802.3198}{{\tt 0802.3198}}].

\bibitem{Guler:2015hsw}
{\bf CLAS} Collaboration, N.~Guler {\em et.~al.}, {\it {Precise determination
  of the deuteron spin structure at low to moderate $Q^2$ with CLAS and
  extraction of the neutron contribution}},  {\em Phys. Rev. C} {\bf 92}
  (2015), no.~5 055201 [\href{http://arXiv.org/abs/1505.07877}{{\tt
  1505.07877}}].

\bibitem{Fersch:2017qrq}
{\bf CLAS} Collaboration, R.~Fersch {\em et.~al.}, {\it {Determination of the
  Proton Spin Structure Functions for $0.05 < Q^{2} < 5 GeV^{2}$ using CLAS}},
  {\em Phys. Rev. C} {\bf 96} (2017), no.~6 065208
  [\href{http://arXiv.org/abs/1706.10289}{{\tt 1706.10289}}].

\bibitem{Zheng:2021yrn}
{\bf CLAS} Collaboration, X.~Zheng {\em et.~al.}, {\it {Measurement of the
  proton spin structure at long distances}},  {\em Nature Phys.} {\bf 17}
  (2021), no.~6 736--741 [\href{http://arXiv.org/abs/2102.02658}{{\tt
  2102.02658}}].

\bibitem{Anthony:1999py}
{\bf E155} Collaboration, P.~L. Anthony {\em et.~al.}, {\it {Measurement of the
  proton and deuteron spin structure functions g(2) and asymmetry A(2)}},  {\em
  Phys. Lett. B} {\bf 458} (1999) 529--535
  [\href{http://arXiv.org/abs/hep-ex/9901006}{{\tt hep-ex/9901006}}].

\bibitem{Anthony:2002hy}
{\bf E155} Collaboration, P.~L. Anthony {\em et.~al.}, {\it {Precision
  measurement of the proton and deuteron spin structure functions g(2) and
  asymmetries A(2)}},  {\em Phys. Lett. B} {\bf 553} (2003) 18--24
  [\href{http://arXiv.org/abs/hep-ex/0204028}{{\tt hep-ex/0204028}}].

\bibitem{Amarian:2003jy}
{\bf Jefferson Lab E94-010} Collaboration, M.~Amarian {\em et.~al.}, {\it {Q**2
  evolution of the neutron spin structure moments using a He-3 target}},  {\em
  Phys. Rev. Lett.} {\bf 92} (2004) 022301
  [\href{http://arXiv.org/abs/hep-ex/0310003}{{\tt hep-ex/0310003}}].

\bibitem{Kramer:2005qe}
K.~Kramer {\em et.~al.}, {\it {The Q**2-dependence of the neutron spin
  structure function g**n(2) at low Q**2}},  {\em Phys. Rev. Lett.} {\bf 95}
  (2005) 142002 [\href{http://arXiv.org/abs/nucl-ex/0506005}{{\tt
  nucl-ex/0506005}}].

\bibitem{Gorchtein:2021qdf}
M.~Gorchtein and C.-Y. Seng, {\it {Dispersion relation analysis of the
  radiative corrections to $g_A$ in the neutron $\beta$-decay}},
  \href{http://arXiv.org/abs/2106.09185}{{\tt 2106.09185}}.

\bibitem{Deur:2014vea}
A.~Deur, Y.~Prok, V.~Burkert, D.~Crabb, F.~X. Girod, K.~A. Griffioen, N.~Guler,
  S.~E. Kuhn and N.~Kvaltine, {\it {High precision determination of the $Q^2$
  evolution of the Bjorken Sum}},  {\em Phys. Rev. D} {\bf 90} (2014), no.~1
  012009 [\href{http://arXiv.org/abs/1405.7854}{{\tt 1405.7854}}].

\bibitem{Kotlorz:2017wpu}
D.~Kotlorz, S.~V. Mikhailov, O.~V. Teryaev and A.~Kotlorz, {\it {Cut moments
  approach in the analysis of DIS data}},  {\em Phys. Rev. D} {\bf 96} (2017),
  no.~1 016015 [\href{http://arXiv.org/abs/1704.04253}{{\tt 1704.04253}}].

\bibitem{Ayala:2018ulm}
C.~Ayala, G.~Cveti\v{c}, A.~V. Kotikov and B.~G. Shaikhatdenov, {\it {Bjorken
  polarized sum rule and infrared-safe QCD couplings}},  {\em Eur. Phys. J. C}
  {\bf 78} (2018), no.~12 1002 [\href{http://arXiv.org/abs/1812.01030}{{\tt
  1812.01030}}].

\bibitem{Wandzura:1977qf}
S.~Wandzura and F.~Wilczek, {\it {Sum Rules for Spin Dependent
  Electroproduction: Test of Relativistic Constituent Quarks}},  {\em Phys.
  Lett. B} {\bf 72} (1977) 195--198.

\bibitem{Shuryak:1981pi}
E.~V. Shuryak and A.~I. Vainshtein, {\it {Theory of Power Corrections to Deep
  Inelastic Scattering in Quantum Chromodynamics. 2. Q**4 Effects: Polarized
  Target}},  {\em Nucl. Phys. B} {\bf 201} (1982) 141.

\bibitem{Jaffe:1989xx}
R.~L. Jaffe, {\it {$g_{2}$-The Nucleon's Other Spin-Dependent Structure
  Function}},  {\em Comments Nucl. Part. Phys.} {\bf 19} (1990), no.~5
  239--257.

\bibitem{Alarcon:2020icz}
J.~M. Alarc\'on, F.~Hagelstein, V.~Lensky and V.~Pascalutsa, {\it {Forward
  doubly-virtual Compton scattering off the nucleon in chiral perturbation
  theory: II. Spin polarizabilities and moments of polarized structure
  functions}},  {\em Phys. Rev. D} {\bf 102} (2020), no.~11 114026
  [\href{http://arXiv.org/abs/2006.08626}{{\tt 2006.08626}}].

\bibitem{Acciarri:2016crz}
{\bf DUNE} Collaboration, R.~Acciarri {\em et.~al.}, {\it {Long-Baseline
  Neutrino Facility (LBNF) and Deep Underground Neutrino Experiment (DUNE)}:
  {Conceptual Design Report, Volume 1: The LBNF and DUNE Projects}},
  \href{http://arXiv.org/abs/1601.05471}{{\tt 1601.05471}}.

\bibitem{Alvarez-Ruso:2017oui}
{\bf NuSTEC} Collaboration, L.~Alvarez-Ruso {\em et.~al.}, {\it {NuSTEC White
  Paper: Status and challenges of neutrino\textendash{}nucleus scattering}},
  {\em Prog. Part. Nucl. Phys.} {\bf 100} (2018) 1--68
  [\href{http://arXiv.org/abs/1706.03621}{{\tt 1706.03621}}].

\bibitem{BhattacharyaSnowmass}
T.~Bhattacharya {\em et.~al.}
\newblock \textit{Unitarity of CKM Matrix, $|V_{ud}|$, Radiative Corrections
  and Semi-leptonic Form Factors}.
  [\href{https://www.snowmass21.org/docs/files/summaries/EF/SNOWMASS21-EF4_EF5-RF2_RF3_Rajan_Gupta-249.pdf}{{\tt
  Link}}].

\bibitem{Feynman:1939zza}
R.~P. Feynman, {\it {Forces in Molecules}},  {\em Phys. Rev.} {\bf 56} (1939)
  340--343.

\bibitem{Hellmann}
H.~Hellmann.
\newblock {\em Einf\"uhrung in die Quantenchemie,} Deuticke, Leipzig und Wien
  (1937).

\bibitem{Seng:2019plg}
C.-Y. Seng and U.-G. Mei\ss{}ner, {\it {Toward a First-Principles Calculation
  of Electroweak Box Diagrams}},  {\em Phys. Rev. Lett.} {\bf 122} (2019),
  no.~21 211802 [\href{http://arXiv.org/abs/1903.07969}{{\tt 1903.07969}}].

\bibitem{Chambers:2017dov}
A.~Chambers, R.~Horsley, Y.~Nakamura, H.~Perlt, P.~Rakow, G.~Schierholz,
  A.~Schiller, K.~Somfleth, R.~Young and J.~Zanotti, {\it {Nucleon Structure
  Functions from Operator Product Expansion on the Lattice}},  {\em Phys. Rev.
  Lett.} {\bf 118} (2017), no.~24 242001
  [\href{http://arXiv.org/abs/1703.01153}{{\tt 1703.01153}}].

\bibitem{Chambers:2017tuf}
{\bf QCDSF, UKQCD, CSSM} Collaboration, A.~J. Chambers {\em et.~al.}, {\it
  {Electromagnetic form factors at large momenta from lattice QCD}},  {\em
  Phys. Rev. D} {\bf 96} (2017), no.~11 114509
  [\href{http://arXiv.org/abs/1702.01513}{{\tt 1702.01513}}].

\bibitem{Agadjanov:2016cjc}
A.~Agadjanov, U.-G. Mei\ss{}ner and A.~Rusetsky, {\it {Nucleon in a periodic
  magnetic field}},  {\em Phys. Rev. D} {\bf 95} (2017), no.~3 031502
  [\href{http://arXiv.org/abs/1610.05545}{{\tt 1610.05545}}].

\bibitem{Agadjanov:2018yxh}
A.~Agadjanov, U.-G. Mei\ss{}ner and A.~Rusetsky, {\it {Nucleon in a periodic
  magnetic field: Finite-volume aspects}},  {\em Phys. Rev. D} {\bf 99} (2019),
  no.~5 054501 [\href{http://arXiv.org/abs/1812.06013}{{\tt 1812.06013}}].

\bibitem{RuizdeElvira:2017aet}
J.~Ruiz~de Elvira, U.~G. Mei\ss{}ner, A.~Rusetsky and G.~Schierholz, {\it
  {Feynman\textendash{}Hellmann theorem for resonances and the quest for QCD
  exotica}},  {\em Eur. Phys. J. C} {\bf 77} (2017), no.~10 659
  [\href{http://arXiv.org/abs/1706.09015}{{\tt 1706.09015}}].

\bibitem{Borsanyi:2020mff}
S.~Borsanyi {\em et.~al.}, {\it {Leading hadronic contribution to the muon
  magnetic moment from lattice QCD}},  {\em Nature} {\bf 593} (2021), no.~7857
  51--55 [\href{http://arXiv.org/abs/2002.12347}{{\tt 2002.12347}}].

\bibitem{Towner:1973yrc}
I.~S. Towner and J.~C. Hardy, {\it {Superallowed 0 + \textrightarrow{} 0 +
  nuclear \ensuremath{\beta}-decays}},  {\em Nucl. Phys. A} {\bf 205} (1973)
  33--55.

\bibitem{Hardy:1975eq}
J.~C. Hardy and I.~S. Towner, {\it {Superallowed 0+ --\ensuremath{>} 0+ Nuclear
  beta Decays and Cabibbo Universality}},  {\em Nucl. Phys. A} {\bf 254} (1975)
  221--240.

\bibitem{Hardy:1990sz}
J.~C. Hardy, I.~S. Towner, V.~T. Koslowsky, E.~Hagberg and H.~Schmeing, {\it
  {Superallowed 0+ ---\ensuremath{>} 0+ nuclear beta decays: a Critical survey
  with tests of CVC and the standard model}},  {\em Nucl. Phys. A} {\bf 509}
  (1990) 429--460.

\bibitem{Hardy:2008gy}
J.~C. Hardy and I.~S. Towner, {\it {Superallowed 0+ ---\ensuremath{>} 0+
  nuclear beta decays: A New survey with precision tests of the conserved
  vector current hypothesis and the standard model}},  {\em Phys. Rev. C} {\bf
  79} (2009) 055502 [\href{http://arXiv.org/abs/0812.1202}{{\tt 0812.1202}}].

\bibitem{Towner:2010zz}
I.~S. Towner and J.~C. Hardy, {\it {The evaluation of V(ud) and its impact on
  the unitarity of the Cabibbo-Kobayashi-Maskawa quark-mixing matrix}},  {\em
  Rept. Prog. Phys.} {\bf 73} (2010) 046301.

\bibitem{Hardy:2014qxa}
J.~C. Hardy and I.~S. Towner, {\it {Superallowed $0^+\to 0^+$ nuclear
  \ensuremath{\beta} decays: 2014 critical survey, with precise results for
  $V_{ud}$ and CKM unitarity}},  {\em Phys. Rev. C} {\bf 91} (2015), no.~2
  025501 [\href{http://arXiv.org/abs/1411.5987}{{\tt 1411.5987}}].

\bibitem{Hardy:2020qwl}
J.~C. Hardy and I.~S. Towner, {\it {Superallowed $0^+ \to 0^+$ nuclear $\beta$
  decays: 2020 critical survey, with implications for V$_{ud}$ and CKM
  unitarity}},  {\em Phys. Rev. C} {\bf 102} (2020), no.~4 045501.

\bibitem{Sirlin:1987sy}
A.~Sirlin, {\it {Remarks Concerning the O(z alpha**2) Corrections to Fermi
  Decays, Conserved Vector Current Predictions and Universality}},  {\em Phys.
  Rev. D} {\bf 35} (1987) 3423.

\bibitem{Sirlin:1986cc}
A.~Sirlin and R.~Zucchini, {\it {Accurate Verification of the Conserved Vector
  Current and Standard Model Predictions}},  {\em Phys. Rev. Lett.} {\bf 57}
  (1986) 1994--1997.

\bibitem{Towner:2007np}
I.~S. Towner and J.~C. Hardy, {\it {An Improved calculation of the
  isospin-symmetry-breaking corrections to superallowed Fermi beta decay}},
  {\em Phys. Rev. C} {\bf 77} (2008) 025501
  [\href{http://arXiv.org/abs/0710.3181}{{\tt 0710.3181}}].

\bibitem{Towner:1994mw}
I.~S. Towner, {\it {Quenching of spin operators in the calculation of radiative
  corrections for nuclear beta decay}},  {\em Phys. Lett. B} {\bf 333} (1994)
  13--16 [\href{http://arXiv.org/abs/nucl-th/9405031}{{\tt nucl-th/9405031}}].

\bibitem{Towner:2002rg}
I.~S. Towner and J.~C. Hardy, {\it {Calculated corrections to superallowed
  Fermi beta decay: New evaluation of the nuclear structure dependent terms}},
  {\em Phys. Rev. C} {\bf 66} (2002) 035501
  [\href{http://arXiv.org/abs/nucl-th/0209014}{{\tt nucl-th/0209014}}].

\bibitem{Brown:1983zzc}
B.~A. Brown and B.~H. Wildenthal, {\it {Corrections to the free-nucleon values
  of the single-particle matrix elements of the M-1 and Gamow-Teller operators,
  from a comparison of shell-model predictions with sd-shell data}},  {\em
  Phys. Rev. C} {\bf 28} (1983) 2397--2413.

\bibitem{Brown:1987obh}
B.~A. Brown and B.~H. Wildenthal, {\it {Empirically optimum M1 operator for
  sd-shell nuclei}},  {\em Nucl. Phys. A} {\bf 474} (1987) 290--306.

\bibitem{Towner:1987zz}
I.~S. Towner, {\it {Quenching of spin matrix elements in nuclei}},  {\em Phys.
  Rept.} {\bf 155} (1987) 263--377.

\bibitem{Jaus:1989dh}
W.~Jaus and G.~Rasche, {\it {Nuclear Structure Dependence of O ($\alpha$)
  Corrections to Fermi Decays and the Value of the {Kobayashi-Maskawa} Matrix
  Element $V$ (U $D$)}},  {\em Phys. Rev. D} {\bf 41} (1990) 166--176.

\bibitem{Towner:1992xm}
I.~S. Towner, {\it {The Nuclear structure dependence of radiative corrections
  in superallowed Fermi beta decay}},  {\em Nucl. Phys. A} {\bf 540} (1992)
  478--500.

\bibitem{Barker:1991tw}
F.~C. Barker, B.~A. Brown, W.~Jaus and G.~Rasche, {\it {Determination of V (ud)
  from Fermi decays and the unitarity of the KM mixing matrix}},  {\em Nucl.
  Phys. A} {\bf 540} (1992) 501--519.

\bibitem{Weinberg:1990rz}
S.~Weinberg, {\it {Nuclear forces from chiral Lagrangians}},  {\em Phys. Lett.
  B} {\bf 251} (1990) 288--292.

\bibitem{Weinberg:1991um}
S.~Weinberg, {\it {Effective chiral Lagrangians for nucleon - pion interactions
  and nuclear forces}},  {\em Nucl. Phys. B} {\bf 363} (1991) 3--18.

\bibitem{Weinberg:1992yk}
S.~Weinberg, {\it {Three body interactions among nucleons and pions}},  {\em
  Phys. Lett. B} {\bf 295} (1992) 114--121
  [\href{http://arXiv.org/abs/hep-ph/9209257}{{\tt hep-ph/9209257}}].

\bibitem{vanKolck:1999mw}
U.~van Kolck, {\it {Effective field theory of nuclear forces}},  {\em Prog.
  Part. Nucl. Phys.} {\bf 43} (1999) 337--418
  [\href{http://arXiv.org/abs/nucl-th/9902015}{{\tt nucl-th/9902015}}].

\bibitem{Epelbaum:2005pn}
E.~Epelbaum, {\it {Few-nucleon forces and systems in chiral effective field
  theory}},  {\em Prog. Part. Nucl. Phys.} {\bf 57} (2006) 654--741
  [\href{http://arXiv.org/abs/nucl-th/0509032}{{\tt nucl-th/0509032}}].

\bibitem{Machleidt:2011zz}
R.~Machleidt and D.~R. Entem, {\it {Chiral effective field theory and nuclear
  forces}},  {\em Phys. Rept.} {\bf 503} (2011) 1--75
  [\href{http://arXiv.org/abs/1105.2919}{{\tt 1105.2919}}].

\bibitem{Bernard:2007sp}
V.~Bernard, E.~Epelbaum, H.~Krebs and U.-G. Mei\ss{}ner, {\it {Subleading
  contributions to the chiral three-nucleon force. I. Long-range terms}},  {\em
  Phys. Rev. C} {\bf 77} (2008) 064004
  [\href{http://arXiv.org/abs/0712.1967}{{\tt 0712.1967}}].

\bibitem{Bernard:2011zr}
V.~Bernard, E.~Epelbaum, H.~Krebs and U.~G. Mei\ss{}ner, {\it {Subleading
  contributions to the chiral three-nucleon force II: Short-range terms and
  relativistic corrections}},  {\em Phys. Rev. C} {\bf 84} (2011) 054001
  [\href{http://arXiv.org/abs/1108.3816}{{\tt 1108.3816}}].

\bibitem{Girlanda:2011fh}
L.~Girlanda, A.~Kievsky and M.~Viviani, {\it {Subleading contributions to the
  three-nucleon contact interaction}},  {\em Phys. Rev. C} {\bf 84} (2011),
  no.~1 014001 [\href{http://arXiv.org/abs/1102.4799}{{\tt 1102.4799}}].
  [Erratum: Phys.Rev.C 102, 019903 (2020)].

\bibitem{Krebs:2012yv}
H.~Krebs, A.~Gasparyan and E.~Epelbaum, {\it {Chiral three-nucleon force at
  N$^4$LO I: Longest-range contributions}},  {\em Phys. Rev. C} {\bf 85} (2012)
  054006 [\href{http://arXiv.org/abs/1203.0067}{{\tt 1203.0067}}].

\bibitem{Krebs:2013kha}
H.~Krebs, A.~Gasparyan and E.~Epelbaum, {\it {Chiral three-nucleon force at
  $N^4LO$ II: Intermediate-range contributions}},  {\em Phys. Rev. C} {\bf 87}
  (2013), no.~5 054007 [\href{http://arXiv.org/abs/1302.2872}{{\tt
  1302.2872}}].

\bibitem{Epelbaum:2014sza}
E.~Epelbaum, H.~Krebs and U.~G. Mei\ss{}ner, {\it {Precision nucleon-nucleon
  potential at fifth order in the chiral expansion}},  {\em Phys. Rev. Lett.}
  {\bf 115} (2015), no.~12 122301 [\href{http://arXiv.org/abs/1412.4623}{{\tt
  1412.4623}}].

\bibitem{Entem:2015xwa}
D.~R. Entem, N.~Kaiser, R.~Machleidt and Y.~Nosyk, {\it {Dominant contributions
  to the nucleon-nucleon interaction at sixth order of chiral perturbation
  theory}},  {\em Phys. Rev. C} {\bf 92} (2015), no.~6 064001
  [\href{http://arXiv.org/abs/1505.03562}{{\tt 1505.03562}}].

\bibitem{Reinert:2017usi}
P.~Reinert, H.~Krebs and E.~Epelbaum, {\it {Semilocal momentum-space
  regularized chiral two-nucleon potentials up to fifth order}},  {\em Eur.
  Phys. J. A} {\bf 54} (2018), no.~5 86
  [\href{http://arXiv.org/abs/1711.08821}{{\tt 1711.08821}}].

\bibitem{Entem:2017gor}
D.~R. Entem, R.~Machleidt and Y.~Nosyk, {\it {High-quality two-nucleon
  potentials up to fifth order of the chiral expansion}},  {\em Phys. Rev. C}
  {\bf 96} (2017), no.~2 024004 [\href{http://arXiv.org/abs/1703.05454}{{\tt
  1703.05454}}].

\bibitem{Hammer:2019poc}
H.~W. Hammer, S.~K\"onig and U.~van Kolck, {\it {Nuclear effective field
  theory: status and perspectives}},  {\em Rev. Mod. Phys.} {\bf 92} (2020),
  no.~2 025004 [\href{http://arXiv.org/abs/1906.12122}{{\tt 1906.12122}}].

\bibitem{Lee:2004si}
D.~Lee, B.~Borasoy and T.~Sch\"afer, {\it {Nuclear lattice simulations with
  chiral effective field theory}},  {\em Phys. Rev. C} {\bf 70} (2004) 014007
  [\href{http://arXiv.org/abs/nucl-th/0402072}{{\tt nucl-th/0402072}}].

\bibitem{Borasoy:2006qn}
B.~Borasoy, E.~Epelbaum, H.~Krebs, D.~Lee and U.-G. Mei\ss{}ner, {\it {Lattice
  Simulations for Light Nuclei: Chiral Effective Field Theory at Leading
  Order}},  {\em Eur. Phys. J. A} {\bf 31} (2007) 105--123
  [\href{http://arXiv.org/abs/nucl-th/0611087}{{\tt nucl-th/0611087}}].

\bibitem{Lee:2008fa}
D.~Lee, {\it {Lattice simulations for few- and many-body systems}},  {\em Prog.
  Part. Nucl. Phys.} {\bf 63} (2009) 117--154
  [\href{http://arXiv.org/abs/0804.3501}{{\tt 0804.3501}}].

\bibitem{Lahde:2019npb}
T.~A. L\"ahde and U.-G. Mei\ss{}ner, {\em {Nuclear Lattice Effective Field
  Theory}: {An introduction}}, vol.~957.
\newblock Springer, 2019.

\bibitem{Gorchtein:2018fxl}
M.~Gorchtein, {\it {$\gamma W$ Box Inside Out: Nuclear Polarizabilities Distort
  the Beta Decay Spectrum}},  {\em Phys. Rev. Lett.} {\bf 123} (2019), no.~4
  042503 [\href{http://arXiv.org/abs/1812.04229}{{\tt 1812.04229}}].

\bibitem{Ormand:1989hm}
W.~E. Ormand and B.~A. Brown, {\it {Corrections to the Fermi Matrix Element for
  Superallowed Beta Decay}},  {\em Phys. Rev. Lett.} {\bf 62} (1989) 866--869.

\bibitem{Ormand:1995df}
W.~E. Ormand and B.~A. Brown, {\it {Isospin-mixing corrections for fp-shell
  Fermi transitions}},  {\em Phys. Rev. C} {\bf 52} (1995) 2455--2460
  [\href{http://arXiv.org/abs/nucl-th/9504017}{{\tt nucl-th/9504017}}].

\bibitem{Satula:2011br}
W.~Satula, J.~Dobaczewski, W.~Nazarewicz and M.~Rafalski, {\it {Microscopic
  calculations of isospin-breaking corrections to superallowed beta-decay}},
  {\em Phys. Rev. Lett.} {\bf 106} (2011) 132502
  [\href{http://arXiv.org/abs/1101.0939}{{\tt 1101.0939}}].

\bibitem{Liang:2009pf}
H.~Liang, N.~Van~Giai and J.~Meng, {\it {Isospin corrections for superallowed
  Fermi beta decay in self-consistent relativistic random-phase approximation
  approaches}},  {\em Phys. Rev. C} {\bf 79} (2009) 064316
  [\href{http://arXiv.org/abs/0904.3673}{{\tt 0904.3673}}].

\bibitem{Auerbach:2008ut}
N.~Auerbach, {\it {Coulomb corrections to superallowed beta decay in nuclei}},
  {\em Phys. Rev. C} {\bf 79} (2009) 035502
  [\href{http://arXiv.org/abs/0811.4742}{{\tt 0811.4742}}].

\bibitem{Damgaard:1969yyx}
J.~Damgaard, {\it {Corrections to the $ft$-values of $0^+ → 0^+$ superallowed
  \ensuremath{\beta}-decays}},  {\em Nucl. Phys. A} {\bf 130} (1969) 233--240.

\bibitem{Koshchii:2020qkr}
O.~Koshchii, J.~Erler, M.~Gorchtein, C.~J. Horowitz, J.~Piekarewicz,
  X.~Roca-Maza, C.-Y. Seng and H.~Spiesberger, {\it {Weak charge and weak
  radius of $^{12}$C}},  {\em Phys. Rev. C} {\bf 102} (2020), no.~2 022501
  [\href{http://arXiv.org/abs/2005.00479}{{\tt 2005.00479}}].

\bibitem{Angeli:2013epw}
I.~Angeli and K.~P. Marinova, {\it {Table of experimental nuclear ground state
  charge radii: An update}},  {\em Atom. Data Nucl. Data Tabl.} {\bf 99}
  (2013), no.~1 69--95.

\bibitem{Souder:2016xcn}
P.~A. Souder, {\it {Parity Violation in Deep Inelastic Scattering with the
  SoLID Spectrometer at JLab}},  {\em Int. J. Mod. Phys. Conf. Ser.} {\bf 40}
  (2016) 1660077.

\bibitem{Becker:2018ggl}
D.~Becker {\em et.~al.}, {\it {The P2 experiment}},
  \href{http://arXiv.org/abs/1802.04759}{{\tt 1802.04759}}.

\bibitem{Sher:2003fb}
A.~Sher {\em et.~al.}, {\it {New, high statistics measurement of the K+
  ---\ensuremath{>} pi0 e+ nu (K+(e3)) branching ratio}},  {\em Phys. Rev.
  Lett.} {\bf 91} (2003) 261802
  [\href{http://arXiv.org/abs/hep-ex/0305042}{{\tt hep-ex/0305042}}].

\bibitem{Alexopoulos:2004sw}
{\bf KTeV} Collaboration, T.~Alexopoulos {\em et.~al.}, {\it {A Determination
  of the CKM parameter |V(us)|}},  {\em Phys. Rev. Lett.} {\bf 93} (2004)
  181802 [\href{http://arXiv.org/abs/hep-ex/0406001}{{\tt hep-ex/0406001}}].

\bibitem{Alexopoulos:2004sx}
{\bf KTeV} Collaboration, T.~Alexopoulos {\em et.~al.}, {\it {Measurements of
  K(L) branching fractions and the CP violation parameter |eta+-|}},  {\em
  Phys. Rev. D} {\bf 70} (2004) 092006
  [\href{http://arXiv.org/abs/hep-ex/0406002}{{\tt hep-ex/0406002}}].

\bibitem{Alexopoulos:2004sy}
{\bf KTeV} Collaboration, T.~Alexopoulos {\em et.~al.}, {\it {Measurements of
  semileptonic K(L) decay form-factors}},  {\em Phys. Rev. D} {\bf 70} (2004)
  092007 [\href{http://arXiv.org/abs/hep-ex/0406003}{{\tt hep-ex/0406003}}].

\bibitem{Abouzaid:2006ir}
{\bf KTeV} Collaboration, E.~Abouzaid {\em et.~al.}, {\it {Improved K(L)
  ---\ensuremath{>} pi+- e-+ nu form factor and phase space integral with
  reduced model uncertainty}},  {\em Phys. Rev. D} {\bf 74} (2006) 097101
  [\href{http://arXiv.org/abs/hep-ex/0608058}{{\tt hep-ex/0608058}}].

\bibitem{KTeV:2010sng}
{\bf KTeV} Collaboration, E.~Abouzaid {\em et.~al.}, {\it {Precise Measurements
  of Direct CP Violation, CPT Symmetry, and Other Parameters in the Neutral
  Kaon System}},  {\em Phys. Rev. D} {\bf 83} (2011) 092001
  [\href{http://arXiv.org/abs/1011.0127}{{\tt 1011.0127}}].

\bibitem{Ambrosino:2005vx}
{\bf KLOE} Collaboration, F.~Ambrosino {\em et.~al.}, {\it {Measurement of the
  K(L) meson lifetime with the KLOE detector}},  {\em Phys. Lett. B} {\bf 626}
  (2005) 15--23 [\href{http://arXiv.org/abs/hep-ex/0507088}{{\tt
  hep-ex/0507088}}].

\bibitem{Ambrosino:2006up}
{\bf KLOE} Collaboration, F.~Ambrosino {\em et.~al.}, {\it {Measurement of the
  branching ratio of the $K(L) \to \pi^+ \pi^-$ decay with the KLOE detector}},
   {\em Phys. Lett. B} {\bf 638} (2006) 140--145
  [\href{http://arXiv.org/abs/hep-ex/0603041}{{\tt hep-ex/0603041}}].

\bibitem{Ambrosino:2005ec}
{\bf KLOE} Collaboration, F.~Ambrosino {\em et.~al.}, {\it {Measurements of the
  absolute branching ratios for the dominant K(L) decays, the K(L) lifetime,
  and V(us) with the KLOE detector}},  {\em Phys. Lett. B} {\bf 632} (2006)
  43--50 [\href{http://arXiv.org/abs/hep-ex/0508027}{{\tt hep-ex/0508027}}].

\bibitem{Ambrosino:2006si}
{\bf KLOE} Collaboration, F.~Ambrosino {\em et.~al.}, {\it {Study of the
  branching ratio and charge asymmetry for the decay $K(s) \to \pi e \nu$ with
  the KLOE detector}},  {\em Phys. Lett. B} {\bf 636} (2006) 173--182
  [\href{http://arXiv.org/abs/hep-ex/0601026}{{\tt hep-ex/0601026}}].

\bibitem{Sciascia:2005zz}
{\bf KLOE} Collaboration, B.~Sciascia, {\it {KLOE extraction of $V_{us}$ from
  kaon decays and lifetimes}},  {\em PoS} {\bf HEP2005} (2006) 287
  [\href{http://arXiv.org/abs/hep-ex/0510028}{{\tt hep-ex/0510028}}].

\bibitem{Ambrosino:2006gn}
{\bf KLOE} Collaboration, F.~Ambrosino {\em et.~al.}, {\it {Measurement of the
  form-factor slopes for the decay $K(L) \to \pi^\pm e^\mp \nu$ with the KLOE
  detector}},  {\em Phys. Lett. B} {\bf 636} (2006) 166--172
  [\href{http://arXiv.org/abs/hep-ex/0601038}{{\tt hep-ex/0601038}}].

\bibitem{Ambrosino:2007ac}
{\bf KLOE} Collaboration, F.~Ambrosino {\em et.~al.}, {\it {Measurement of the
  charged kaon lifetime with the KLOE detector}},  {\em JHEP} {\bf 01} (2008)
  073 [\href{http://arXiv.org/abs/0712.1112}{{\tt 0712.1112}}].

\bibitem{KLOE:2007jte}
{\bf KLOE} Collaboration, F.~Ambrosino {\em et.~al.}, {\it {Measurement of the
  absolute branching ratios for semileptonic $K^\pm$ decays with the KLOE
  detector}},  {\em JHEP} {\bf 02} (2008) 098
  [\href{http://arXiv.org/abs/0712.3841}{{\tt 0712.3841}}].

\bibitem{KLOE:2010yit}
{\bf KLOE} Collaboration, F.~Ambrosino {\em et.~al.}, {\it {Precision
  Measurement of $K_S$ Meson Lifetime with the KLOE detector}},  {\em Eur.
  Phys. J. C} {\bf 71} (2011) 1604 [\href{http://arXiv.org/abs/1011.2668}{{\tt
  1011.2668}}].

\bibitem{KLOE-2:2019rev}
{\bf KLOE-2} Collaboration, D.~Babusci {\em et.~al.}, {\it {Measurement of the
  branching fraction for the decay $K_S \to \pi \mu \nu$ with the KLOE
  detector}},  {\em Phys. Lett. B} {\bf 804} (2020) 135378
  [\href{http://arXiv.org/abs/1912.05990}{{\tt 1912.05990}}].

\bibitem{Lai:2004bt}
{\bf NA48} Collaboration, A.~Lai {\em et.~al.}, {\it {Measurement of the
  branching ratio of the decay K(L) ---\ensuremath{>} pi+- e-+ nu and
  extraction of the CKM parameter |V(us)|}},  {\em Phys. Lett. B} {\bf 602}
  (2004) 41--51 [\href{http://arXiv.org/abs/hep-ex/0410059}{{\tt
  hep-ex/0410059}}].

\bibitem{Lai:2006cf}
{\bf NA48} Collaboration, A.~Lai {\em et.~al.}, {\it {Measurement of the ratio
  Gamma(KL ---\ensuremath{>} pi+ pi-) / Gamma(KL ---\ensuremath{>} pi e nu) and
  extraction of the CP violation parameter |eta(+-)|}},  {\em Phys. Lett. B}
  {\bf 645} (2007) 26--35 [\href{http://arXiv.org/abs/hep-ex/0611052}{{\tt
  hep-ex/0611052}}].

\bibitem{Batley:2006cj}
{\bf NA48/2} Collaboration, J.~R. Batley {\em et.~al.}, {\it {Measurements of
  Charged Kaon Semileptonic Decay Branching Fractions K+- ---\ensuremath{>} pi0
  mu+- nu and K+- ---\ensuremath{>} pi0 e+- nu and Their Ratio}},  {\em Eur.
  Phys. J. C} {\bf 50} (2007) 329--340
  [\href{http://arXiv.org/abs/hep-ex/0702015}{{\tt hep-ex/0702015}}]. [Erratum:
  Eur.Phys.J.C 52, 1021--1023 (2007)].

\bibitem{Lai:2004kb}
{\bf NA48} Collaboration, A.~Lai {\em et.~al.}, {\it {Measurement of K0(e3)
  form-factors}},  {\em Phys. Lett. B} {\bf 604} (2004) 1--10
  [\href{http://arXiv.org/abs/hep-ex/0410065}{{\tt hep-ex/0410065}}].

\bibitem{Romanovsky:2007qb}
V.~I. Romanovsky {\em et.~al.}, {\it {Measurement of K- ---\ensuremath{>} pi0
  e- anti-nu(gamma) branching ratio}},
  \href{http://arXiv.org/abs/0704.2052}{{\tt 0704.2052}}.

\bibitem{Yushchenko:2004zs}
O.~P. Yushchenko {\em et.~al.}, {\it {High statistic measurement of the K-
  ---\ensuremath{>} pi0 e- nu decay form-factors}},  {\em Phys. Lett. B} {\bf
  589} (2004) 111--117 [\href{http://arXiv.org/abs/hep-ex/0404030}{{\tt
  hep-ex/0404030}}].

\bibitem{Gamiz:2013xxa}
{\bf Fermilab Lattice, MILC} Collaboration, E.~G\'amiz {\em et.~al.}, {\it
  {Kaon Semileptonic Form Factors with $N_f$ = 2 + 1 + 1 HISQ Fermions and
  Physical Light Quark Masses}},  {\em PoS} {\bf LATTICE2013} (2014) 395
  [\href{http://arXiv.org/abs/1311.7264}{{\tt 1311.7264}}].

\bibitem{Bazavov:2013maa}
A.~Bazavov {\em et.~al.}, {\it {Determination of $|V_{us}|$ from a Lattice-QCD
  Calculation of the $K\to\pi\ell\nu$ Semileptonic Form Factor with Physical
  Quark Masses}},  {\em Phys. Rev. Lett.} {\bf 112} (2014), no.~11 112001
  [\href{http://arXiv.org/abs/1312.1228}{{\tt 1312.1228}}].

\bibitem{Bazavov:2018kjg}
{\bf Fermilab Lattice, MILC} Collaboration, A.~Bazavov {\em et.~al.}, {\it
  {$|V_{us}|$ from $K_{\ell 3}$ decay and four-flavor lattice QCD}},  {\em
  Phys. Rev.} {\bf D99} (2019), no.~11 114509
  [\href{http://arXiv.org/abs/1809.02827}{{\tt 1809.02827}}].

\bibitem{Carrasco:2016kpy}
N.~Carrasco, P.~Lami, V.~Lubicz, L.~Riggio, S.~Simula and C.~Tarantino, {\it
  {$K \to \pi$ semileptonic form factors with $N_f=2+1+1$ twisted mass
  fermions}},  {\em Phys. Rev. D} {\bf 93} (2016), no.~11 114512
  [\href{http://arXiv.org/abs/1602.04113}{{\tt 1602.04113}}].

\bibitem{FlavourLatticeAveragingGroup:2019iem}
{\bf Flavour Lattice Averaging Group} Collaboration, S.~Aoki {\em et.~al.},
  {\it {FLAG Review 2019: Flavour Lattice Averaging Group (FLAG)}},  {\em Eur.
  Phys. J. C} {\bf 80} (2020), no.~2 113
  [\href{http://arXiv.org/abs/1902.08191}{{\tt 1902.08191}}].

\bibitem{Kakazu:2019ltq}
{\bf PACS} Collaboration, J.~Kakazu, K.-i. Ishikawa, N.~Ishizuka, Y.~Kuramashi,
  Y.~Nakamura, Y.~Namekawa, Y.~Taniguchi, N.~Ukita, T.~Yamazaki and
  T.~Yoshi\'e, {\it {$K_{l3}$ form factors at the physical point on a $(10.9
  fm)^3$ volume}},  {\em Phys. Rev. D} {\bf 101} (2020), no.~9 094504
  [\href{http://arXiv.org/abs/1912.13127}{{\tt 1912.13127}}].

\bibitem{Antonelli:2010yf}
{\bf FlaviaNet Working Group on Kaon Decays} Collaboration, M.~Antonelli {\em
  et.~al.}, {\it {An Evaluation of $|V_{us}|$ and precise tests of the Standard
  Model from world data on leptonic and semileptonic kaon decays}},  {\em Eur.
  Phys. J. C} {\bf 69} (2010) 399--424
  [\href{http://arXiv.org/abs/1005.2323}{{\tt 1005.2323}}].

\bibitem{Abouzaid:2009ry}
{\bf KTeV} Collaboration, E.~Abouzaid {\em et.~al.}, {\it {Dispersive analysis
  of K (L mu3) and K (L e3) scalar and vector form factors using KTeV data}},
  {\em Phys. Rev. D} {\bf 81} (2010) 052001
  [\href{http://arXiv.org/abs/0912.1291}{{\tt 0912.1291}}].

\bibitem{Hill:2006bq}
R.~J. Hill, {\it {Constraints on the form factors for K ---\ensuremath{>} pi l
  nu and implications for |V(us)|}},  {\em Phys. Rev. D} {\bf 74} (2006) 096006
  [\href{http://arXiv.org/abs/hep-ph/0607108}{{\tt hep-ph/0607108}}].

\bibitem{Bernard:2006gy}
V.~Bernard, M.~Oertel, E.~Passemar and J.~Stern, {\it {K(mu3)**L decay: A
  Stringent test of right-handed quark currents}},  {\em Phys. Lett.} {\bf
  B638} (2006) 480--486 [\href{http://arXiv.org/abs/hep-ph/0603202}{{\tt
  hep-ph/0603202}}].

\bibitem{Bernard:2009zm}
V.~Bernard, M.~Oertel, E.~Passemar and J.~Stern, {\it {Dispersive
  representation and shape of the K(l3) form factors: Robustness}},  {\em Phys.
  Rev.} {\bf D80} (2009) 034034 [\href{http://arXiv.org/abs/0903.1654}{{\tt
  0903.1654}}].

\bibitem{NA482:2018rgv}
{\bf NA48/2} Collaboration, J.~R. Batley {\em et.~al.}, {\it {Measurement of
  the form factors of charged kaon semileptonic decays}},  {\em JHEP} {\bf 10}
  (2018) 150 [\href{http://arXiv.org/abs/1808.09041}{{\tt 1808.09041}}].

\bibitem{Seng:2021nar}
C.-Y. Seng, D.~Galviz, W.~J. Marciano and U.-G. Mei\ss{}ner, {\it {An update on
  $|V_{us}|$ and $|V_{us}/V_{ud}|$ from semileptonic kaon and pion decays}},
  \href{http://arXiv.org/abs/2107.14708}{{\tt 2107.14708}}.

\bibitem{Colangelo:2018jxw}
G.~Colangelo, S.~Lanz, H.~Leutwyler and E.~Passemar, {\it {Dispersive analysis
  of $\eta \rightarrow 3 \pi $}},  {\em Eur. Phys. J. C} {\bf 78} (2018),
  no.~11 947 [\href{http://arXiv.org/abs/1807.11937}{{\tt 1807.11937}}].

\bibitem{Bijnens:1996kk}
J.~Bijnens and J.~Prades, {\it {Electromagnetic corrections for pions and
  kaons: Masses and polarizabilities}},  {\em Nucl. Phys.} {\bf B490} (1997)
  239--271 [\href{http://arXiv.org/abs/hep-ph/9610360}{{\tt hep-ph/9610360}}].

\bibitem{Bijnens:2014lea}
J.~Bijnens and G.~Ecker, {\it {Mesonic low-energy constants}},  {\em Ann. Rev.
  Nucl. Part. Sci.} {\bf 64} (2014) 149--174
  [\href{http://arXiv.org/abs/1405.6488}{{\tt 1405.6488}}].

\bibitem{Giusti:2018guw}
D.~Giusti, V.~Lubicz, G.~Martinelli, C.~Sachrajda, F.~Sanfilippo, S.~Simula and
  N.~Tantalo, {\it {Radiative corrections to decay amplitudes in lattice QCD}},
   {\em PoS} {\bf LATTICE2018} (2019) 266
  [\href{http://arXiv.org/abs/1811.06364}{{\tt 1811.06364}}].

\bibitem{Sachrajda:2019uhh}
C.~Sachrajda, M.~Di~Carlo, G.~Martinelli, D.~Giusti, V.~Lubicz, F.~Sanfilippo,
  S.~Simula and N.~Tantalo, {\it {Radiative corrections to semileptonic decay
  rates}},  in {\em {37th International Symposium on Lattice Field Theory}},
  10, 2019.
\newblock \href{http://arXiv.org/abs/1910.07342}{{\tt 1910.07342}}.

\bibitem{Cirigliano:2019jig}
{\bf USQCD} Collaboration, V.~Cirigliano, Z.~Davoudi, T.~Bhattacharya,
  T.~Izubuchi, P.~E. Shanahan, S.~Syritsyn and M.~L. Wagman, {\it {The Role of
  Lattice QCD in Searches for Violations of Fundamental Symmetries and Signals
  for New Physics}},  {\em Eur. Phys. J. A} {\bf 55} (2019), no.~11 197
  [\href{http://arXiv.org/abs/1904.09704}{{\tt 1904.09704}}].

\bibitem{Carrasco:2015xwa}
N.~Carrasco, V.~Lubicz, G.~Martinelli, C.~Sachrajda, N.~Tantalo, C.~Tarantino
  and M.~Testa, {\it {QED Corrections to Hadronic Processes in Lattice QCD}},
  {\em Phys. Rev. D} {\bf 91} (2015), no.~7 074506
  [\href{http://arXiv.org/abs/1502.00257}{{\tt 1502.00257}}].

\bibitem{Lubicz:2016xro}
V.~Lubicz, G.~Martinelli, C.~Sachrajda, F.~Sanfilippo, S.~Simula and
  N.~Tantalo, {\it {Finite-Volume QED Corrections to Decay Amplitudes in
  Lattice QCD}},  {\em Phys. Rev. D} {\bf 95} (2017), no.~3 034504
  [\href{http://arXiv.org/abs/1611.08497}{{\tt 1611.08497}}].

\bibitem{Giusti:2017dwk}
D.~Giusti, V.~Lubicz, G.~Martinelli, C.~T. Sachrajda, F.~Sanfilippo, S.~Simula,
  N.~Tantalo and C.~Tarantino, {\it {First lattice calculation of the QED
  corrections to leptonic decay rates}},  {\em Phys. Rev. Lett.} {\bf 120}
  (2018), no.~7 072001 [\href{http://arXiv.org/abs/1711.06537}{{\tt
  1711.06537}}].

\bibitem{DiCarlo:2019thl}
M.~Di~Carlo, D.~Giusti, V.~Lubicz, G.~Martinelli, C.~T. Sachrajda,
  F.~Sanfilippo, S.~Simula and N.~Tantalo, {\it {Light-meson leptonic decay
  rates in lattice QCD+QED}},  {\em Phys. Rev.} {\bf D100} (2019), no.~3 034514
  [\href{http://arXiv.org/abs/1904.08731}{{\tt 1904.08731}}].

\bibitem{BoyleSnowmass}
P.~Boyle {\em et.~al.}
\newblock \textit{High-precision determination of $V_{us}$ and $V_{ud}$ from
  lattice QCD}.
  [\href{https://www.snowmass21.org/docs/files/summaries/RF/SNOWMASS21-RF2_RF0-TF5_TF0-CompF2_CompF0-054.pdf}{{\tt
  Link}}].

\bibitem{Seng:2021boy}
C.-Y. Seng, D.~Galviz, M.~Gorchtein and U.~G. Mei\ss{}ner, {\it {High-precision
  determination of the Ke3 radiative corrections}},  {\em Phys. Lett. B} {\bf
  820} (2021) 136522 [\href{http://arXiv.org/abs/2103.00975}{{\tt
  2103.00975}}].

\bibitem{Seng:2021wcf}
C.-Y. Seng, D.~Galviz, M.~Gorchtein and U.-G. Mei\ss{}ner, {\it {Improved
  $K_{e3}$ radiative corrections sharpen the $K_{\mu 2}$--$K_{l3}$
  discrepancy}},  \href{http://arXiv.org/abs/2103.04843}{{\tt 2103.04843}}.

\bibitem{Amendolia:1986wj}
{\bf NA7} Collaboration, S.~Amendolia {\em et.~al.}, {\it {A Measurement of the
  Space - Like Pion Electromagnetic Form-Factor}},  {\em Nucl. Phys. B} {\bf
  277} (1986) 168.

\bibitem{Amendolia:1986ui}
S.~Amendolia {\em et.~al.}, {\it {A Measurement of the Kaon Charge Radius}},
  {\em Phys. Lett. B} {\bf 178} (1986) 435--440.

\bibitem{Ananthanarayan:2017efc}
B.~Ananthanarayan, I.~Caprini and D.~Das, {\it {Electromagnetic charge radius
  of the pion at high precision}},  {\em Phys. Rev. Lett.} {\bf 119} (2017),
  no.~13 132002 [\href{http://arXiv.org/abs/1706.04020}{{\tt 1706.04020}}].

\bibitem{Colangelo:2018mtw}
G.~Colangelo, M.~Hoferichter and P.~Stoffer, {\it {Two-pion contribution to
  hadronic vacuum polarization}},  {\em JHEP} {\bf 02} (2019) 006
  [\href{http://arXiv.org/abs/1810.00007}{{\tt 1810.00007}}].

\bibitem{Lazzeroni:2018glh}
{\bf NA48/2} Collaboration, J.~R. Batley {\em et.~al.}, {\it {Measurement of
  the form factors of charged kaon semileptonic decays}},  {\em JHEP} {\bf 10}
  (2018) 150 [\href{http://arXiv.org/abs/1808.09041}{{\tt 1808.09041}}].

\bibitem{Passarino:1978jh}
G.~Passarino and M.~J.~G. Veltman, {\it {One Loop Corrections for e+ e-
  Annihilation Into mu+ mu- in the Weinberg Model}},  {\em Nucl. Phys. B} {\bf
  160} (1979) 151--207.

\bibitem{Ecker:1988te}
G.~Ecker, J.~Gasser, A.~Pich and E.~de~Rafael, {\it {The Role of Resonances in
  Chiral Perturbation Theory}},  {\em Nucl. Phys. B} {\bf 321} (1989) 311--342.

\bibitem{Ecker:1989yg}
G.~Ecker, J.~Gasser, H.~Leutwyler, A.~Pich and E.~de~Rafael, {\it {Chiral
  Lagrangians for Massive Spin 1 Fields}},  {\em Phys. Lett. B} {\bf 223}
  (1989) 425--432.

\bibitem{Cirigliano:2006hb}
V.~Cirigliano, G.~Ecker, M.~Eidemuller, R.~Kaiser, A.~Pich and J.~Portoles,
  {\it {Towards a consistent estimate of the chiral low-energy constants}},
  {\em Nucl. Phys. B} {\bf 753} (2006) 139--177
  [\href{http://arXiv.org/abs/hep-ph/0603205}{{\tt hep-ph/0603205}}].

\bibitem{Ma:2021azh}
P.-X. Ma, X.~Feng, M.~Gorchtein, L.-C. Jin and C.-Y. Seng, {\it {Lattice QCD
  calculation of the electroweak box diagrams for the kaon semileptonic
  decays}},  {\em Phys. Rev. D} {\bf 103} (2021) 114503
  [\href{http://arXiv.org/abs/2102.12048}{{\tt 2102.12048}}].

\bibitem{Larin:1991tj}
S.~A. Larin and J.~A.~M. Vermaseren, {\it {The alpha-s**3 corrections to the
  Bjorken sum rule for polarized electroproduction and to the Gross-Llewellyn
  Smith sum rule}},  {\em Phys. Lett. B} {\bf 259} (1991) 345--352.

\end{thebibliography}\endgroup

\end{document}